\documentclass[aos]{imsart}

\RequirePackage{amsthm,amsmath,amsfonts,amssymb}
\RequirePackage[numbers,sort&compress]{natbib}
\RequirePackage[colorlinks,citecolor=blue,urlcolor=blue]{hyperref}
\RequirePackage{graphicx}

\startlocaldefs
\theoremstyle{plain}

\newtheorem{theorem}{Theorem}
\newtheorem{corollary}[theorem]{Corollary}
\newtheorem{lemma}{Lemma}
\theoremstyle{definition}

\allowdisplaybreaks

\usepackage{multirow}
\usepackage{longtable}
\usepackage{booktabs}

\newcommand{\E}{\operatorname{E}}

\newcommand{\var}{\operatorname{Var}}

\newcommand{\tr}{\operatorname{tr}}
\newcommand{\diag}{\operatorname{diag}}
\newcommand{\bA}{\mathbf{A}}
\newcommand{\bB}{\mathbf{B}}
\newcommand{\bC}{\mathbf{C}}

\newcommand{\bK}{\mathbf{K}}

\newcommand{\bc}{\mathbf{c}} 

\newcommand{\bm}{\mathbf{m}}

\newcommand{\bt}{\mathbf{t}}

\newcommand{\bx}{\mathbf{x}}
\newcommand{\by}{\mathbf{y}} 

\newcommand{\btheta}{\boldsymbol{\theta}}
\newcommand{\bbeta}{\boldsymbol{\beta}}
\newcommand{\siga}{\sigma_{\alpha}}
\newcommand{\sige}{\sigma_e}
\newcommand{\bsigma}{\boldsymbol{\sigma}}
\newcommand{\bbetab}{\boldsymbol{\beta}^{(b)}}
\newcommand{\bbetabt}{\boldsymbol{\beta}^{(b)\top}}
\newcommand{\bbetaw}{\boldsymbol{\beta}^{(w)}}
\newcommand{\bbetawt}{\boldsymbol{\beta}^{(w)\top}}
\newcommand{\bxwij}{\mathbf{x}_{ij}^{(w)}}
\newcommand{\bxbi}{\mathbf{x}_{i}^{(b)}}
\newcommand{\bxwijt}{\mathbf{x}_{ij}^{(w)\top}}
\newcommand{\bxbit}{\mathbf{x}_{i}^{(b)\top}}
\newcommand{\balpha}{\boldsymbol{\alpha}}

\newcommand{\hbtheta}{\hat{\boldsymbol{\theta}}}
\newcommand{\hbbeta}{\hat{\boldsymbol{\beta}}}

\newcommand{\hbsigma}{\hat{\boldsymbol{\sigma}}}

\newcommand{\dbsigma}{\dot{\boldsymbol{\sigma}}}

\newcommand{\dbtheta}{\dot{\boldsymbol{\theta}}}

\newcommand{\dsiga}{\dot{\sigma}_{\alpha}}
\newcommand{\dsige}{\dot{\sigma}_e}

\newcommand{\bPsi}{\boldsymbol{\Psi}}
\newcommand{\bpsi}{\boldsymbol{\psi}}

\newcommand{\sumig}{\sum_{i=1}^g}
\newcommand{\sumjmi}{\sum_{j=1}^{m_i}}
\newcommand{\bxi}{\boldsymbol{\xi}}

\makeatletter
\def\journal@name{}
\makeatother

\endlocaldefs

\begin{document}

\begin{frontmatter}
\title{Asymptotic Theory for Restricted Maximum Likelihood Estimators in Generalized Linear Mixed Models}
\runtitle{Asymptotics for ML and REML estimators in GLMMs}

\begin{aug}
\author[A]{\fnms{Zhanzhongyu}~\snm{Gao}\ead[label=e1]{Zhanzhongyu.Gao@unsw.edu.au} }
\author[A]{\fnms{Huadong}~\snm{Mo}\ead[label=e2]{Huadong.Mo@unsw.edu.au}\orcid{0000-0002-7782-2884}}
\author[B]{\fnms{Ziyang}~\snm{Lyu}\ead[label=e3]{Ziyang.Lyu@unsw.edu.au}\orcid{0000-0003-3307-4148}}
\address[A]{School of Systems and Computing,
University of New South Wales\printead[presep={,\ }]{e1,e2}}

\address[B]{UNSW Data Science Hub, and School of Mathematics and Statistics,
University of New South Wales\printead[presep={,\ }]{e3}}
 
\end{aug}

\begin{abstract}
Developing an asymptotic theory for restricted maximum likelihood (REML) estimators in generalized linear mixed models (GLMMs) has remained an open problem for decades. In this paper, we establish the asymptotic distributions of the REML and maximum likelihood (ML) estimators for GLMMs when both the number of clusters and the cluster sizes tend to infinity. Under very mild conditions, requiring only finite-moment assumptions on the random effects rather than normality and imposing no restriction on the relative rates at which the number of clusters and the cluster sizes diverge, both estimators are proved to be asymptotically normal with an explicit block-diagonal covariance structure, and the bias-correction property of REML is established. Simulation studies support the asymptotic results, and a genetic data analysis illustrates the methodology.
\end{abstract}



\end{frontmatter}
 \section{Introduction}
\label{sec:Introduction}


Generalized linear mixed models (GLMMs) extend linear mixed models (LMMs) to accommodate non-Gaussian responses arising from clustered, longitudinal, and multilevel data structures \citep{jiang2007linear, mcculloch2008generalized, stroup2024generalized}. They have been widely used in a variety of applications, including the analysis of periodontal disease \citep{neuhaus2011estimation}, neonatal jaundice and repeated birth-weight data \citep{neuhaus2006separating}, and ecological and evolutionary studies involving non-Gaussian and hierarchically structured data \citep{bolker2009generalized}.

Unlike in LMMs, likelihood-based inference for GLMMs is analytically intractable, as the marginal likelihood is generally unavailable in closed form \citep{ning2026asymptotic}.
Early developments therefore largely relied on approximating the marginal likelihood, including numerical integration methods \citep{hinde1982compound,crouch1990evaluation}, Bayesian approaches based on sampling schemes \citep{raghunathan1994monte, besag1991bayesian, zeger1991generalized}, and quasi-likelihood methods such as penalized and marginal quasi-likelihood \citep{laird1978empirical, goldstein1991nonlinear, breslow1993approximate}. Rigorous results for maximum likelihood (ML) estimators defined from the original marginal likelihood were developed subsequently. \cite{jiang2001mixed} established asymptotic theory for ML estimators in GLMMs as the number of clusters tends to infinity with finite random cluster sizes, while \cite{nie2007convergence} studied likelihood-based asymptotics for GLMMs with independent clusters and hierarchical random effects. \cite{jiang2013subset} later introduced the subset argument to establish consistency of ML estimators in GLMMs with crossed random effects and extended the approach to more general GLMM settings. More recently, \cite{jiang2022usable} developed asymptotic theory for GLMMs with hierarchical structure when both the number of clusters and cluster sizes diverge, subject to rate restrictions on their relative growth. 
\cite{maestrini2024second} further refined this theory by deriving second-order approximations to the variances and covariances of the ML estimator. 

Despite recent advances in asymptotics for ML estimation, practical likelihood inference for GLMMs still relies heavily on numerical integration or likelihood approximations. These procedures can, however, produce appreciable finite-sample bias in variance-component estimation. For example, penalized quasi-likelihood is known to underestimate the random-effect variance, particularly for binary responses with small cluster size \citep{breslow1993approximate, breslow1995bias}. This has motivated the development of alternative procedures aimed at reducing the bias of variance-component estimators in GLMMs. A natural choice is restricted maximum likelihood (REML), which is a standard method for variance-component estimation in LMMs that accounts for the loss of degrees of freedom due to fixed-effect estimation and typically reduces finite-sample bias. However, extending the REML principle to GLMMs is considerably more difficult because the marginal likelihood already involves generally intractable integration over the random effects, and eliminating the fixed effects requires an additional layer of marginalization. The additional complexity also makes the theoretical analysis of REML in GLMMs substantially challenging. \cite[Section 3.7.2]{jiang2017asymptotic} points out that establishing a rigorous asymptotic theory for REML in GLMMs has remained an important open problem. To the best of our knowledge, no rigorous asymptotic theory has yet been established for REML estimators defined directly from the original, rather than approximated, marginal likelihood.

Despite the difficulty of constructing a REML criterion for GLMMs, considerable effort has been devoted to developing REML-type estimators through approximation and adjustment. \cite{maestrini2024restricted} systematically reviewed these developments and grouped the existing approaches into four broad classes: approximate linearization methods, which construct a working LMM through local linearization and then apply REML-type estimation \citep{schall1991estimation,wolfinger1993generalized,breslow1993approximate}; integrated likelihood methods, which construct REML-type criteria by integrating out the fixed effects, with the resulting integrals evaluated or approximated to obtain practical finite-sample estimators \citep{stiratelli1984random,mcculloch1994maximum,meza2009estimation}; modified profile likelihood methods, which adjust the variance-component profile likelihood for the estimation of nuisance fixed effects \citep{bellio2011restricted}; and direct bias-correction methods, which modify the variance-component profile score or the corresponding estimating equations \citep{liao2002type}. Although these approaches differ substantially in motivation and construction, numerical comparisons in \cite{maestrini2024restricted} suggest that they often produce similar, if not identical, REML estimates. These developments provide a variety of practical strategies for constructing REML-type estimators in GLMMs. However, their asymptotic properties still remain largely unknown. The recent precise asymptotics for ML estimation developed by \cite{jiang2022usable} and refined by \cite{maestrini2024second} assume that the average cluster sample size grows more slowly than the number of clusters, specifically requiring their ratio to converge to zero. This condition can be restrictive in applications with large cluster sizes; for example, the breast single-cell study of \cite{reed2024single} contains 55 donors, with several thousand to tens of thousands of cells observed within each donor. Motivated by these two limitations, the purpose of this paper is to develop asymptotic theory for ML and REML estimation in GLMMs with hierarchical random effects, allowing both the number of clusters and the minimum cluster size to tend to infinity without imposing any rate restrictions.

Let $[y_{ij}, \bx_{ij}^\top]$ denote the $j$th observed vector, $j=1,\ldots,m_i$, from the $i$th cluster, $i=1,\ldots,g$,
where $y_{ij}$ is a scalar response variable and $\mathbf{x}_{ij}$ is a $p$-vector of explanatory variables or covariates. Let $n=\sum_{i=1}^g m_i$ denote the total sample size and
$m_L=\min_{1\le i\le g} m_i$ the minimum cluster size. Assume that the
$g$ clusters are mutually independent. Conditional on a cluster-specific random effect $\alpha_i$, the responses
$y_{ij}$ are independent and follow an exponential-family
distribution with density 
\begin{equation}
\label{eq:exp density}
f(y_{ij}\mid \alpha_i; \bbeta,\sige^2)
=
\exp\left\{
\frac{y_{ij}\zeta_{ij}-b(\zeta_{ij})}{\sige^2}
+c(y_{ij},\sige^2)
\right\},
\end{equation}
where $\zeta_{ij}$ is the natural parameter and $\sige^2$ is a dispersion parameter, and
$b(\cdot)$ and $c(\cdot,\cdot)$ are known functions determined by the
underlying distribution. We write $b^{(k)}(\cdot)$ for the $k$th derivative of $b(\cdot)$. It follows that $\mu_{ij}=\E(y_{ij}\mid\alpha_i) =b^{(1)}(\zeta_{ij})$,
and
$\var(y_{ij}\mid\alpha_i)=\sige^2 b^{(2)}(\zeta_{ij}).$ The conditional mean is linked to a linear predictor through
\begin{equation}
\label{init link function}
g(\mu_{ij})=g\{b^{(1)}(\zeta_{ij})\}=\eta_{ij}=\bx_{ij}^{\top}\bbeta+\alpha_i,
\end{equation}
where $g(\cdot)$ is a known link function,
$\eta_{ij}$ denotes the linear predictor, and
$\bbeta$ is a $p$-vector of unknown fixed-effect coefficients. We consider the canonical link throughout this paper, so that $\eta_{ij}=\zeta_{ij}$. The random effects $\{\alpha_i\}$ are assumed to be independent across clusters with mean zero and variance $\siga^2$; we do not assume normality for the random effects.
 
\cite{lyu2022asymptotics,lyu2022increasing} established asymptotic theory for ML, REML, and estimated best linear unbiased predictors (EBLUPs) in LMMs, allowing both the number of clusters $g$ and the minimum cluster size $m_L$ to diverge without imposing restrictions on their relative rates of growth. Motivated by their results, we distinguish between within-cluster covariates, which vary across units within a cluster, and between-cluster covariates, which remain constant within a cluster. Let $\bxwij$ denote a $p_w$-vector of within-cluster covariates and $\bxbi$ denote a $p_b$-vector of between-cluster covariates. Then \eqref{init link function} can be rewritten as
\begin{align}
g(\mu_{ij})=g\{b^{(1)}(\zeta_{ij})\}=\eta_{ij}=\bxbit\bbetab+\bxwijt\bbetaw+\alpha_i,
\label{new link function}
\end{align}
where  $\bbetab$ is a $p_b$-vector of between-cluster regression coefficients and $\bbetaw$ is a $p_w$-vector of within-cluster regression coefficients, with $p=p_b+p_w$.
Either kind of covariate can be absent in the data; the form \eqref{new link function} is general enough to include the three cases with either or both in the model.  

Let
$\btheta=(\bbetabt,\siga^2,\bbetawt,\sige^2)^\top$.
The joint likelihood of
$\by=(y_{11},\ldots,y_{gm_g})^\top$
and
$\balpha=(\alpha_1,\ldots,\alpha_g)^\top$
under the GLMM formulated above is
\begin{equation}
\label{eq:joint likelihood}
\begin{aligned}
L(\by,\balpha\mid\btheta)
=
\prod_{i=1}^g
\left[
\prod_{j=1}^{m_i}
f(y_{ij}\mid\eta_{ij},\sige^2)
\right]
f(\alpha_i\mid\siga^2),
\end{aligned}
\end{equation}
where $f(\alpha_i\mid\siga^2)$ denotes the density function of the random effect $\alpha_i$. The marginal log-likelihood is obtained by integrating out the random effect
\begin{equation}
\label{eq:marginal likelihood}
\begin{aligned}
\ell(\btheta)
&=
\log\left\{\int
L(\by,\balpha\mid\btheta)
\,d\balpha \right\}
\\&
= -\frac{g}{2} \log 2\pi- \frac{g}{2} \log\siga^2 + \sum_{i=1}^g \log \int \exp\{h_i(\by_i,\alpha_i)\} d\alpha_i,
\end{aligned}
\end{equation}
where
\begin{align*}
h_i(\by_i,\alpha_i\mid\btheta)
=
\sum_{j=1}^{m_i}
\left[
\frac{y_{ij}\eta_{ij}-b(\eta_{ij})}{\sige^2}
+
c(y_{ij},\sige^2)
\right]
-
\frac{\alpha_i^2}{2\siga^2}.
\end{align*}
The expression in \eqref{eq:marginal likelihood} is derived by adopting a Gaussian density for $f(\alpha_i\mid\siga^2)$. Under normality of the random effects, it is the exact marginal likelihood; otherwise, for random effects satisfying the required finite-moment conditions, \eqref{eq:marginal likelihood} is treated as a working objective function.
For simplicity, we continue to refer to this objective function as the likelihood and to its maximizer $\hat{\btheta} = \arg\max_{\btheta} \ell(\btheta)$ as the ML estimator.

To describe REML, we partition the parameter vector $\btheta$ into the regression parameter
$\bbeta=(\bbetabt,\bbetawt)^\top$
and the variance component
$\bsigma=(\siga^2,\sige^2)^\top$,
so that $\ell(\btheta)=\ell(\bbeta,\bsigma)$. For each fixed $\bsigma$, define the profile map $\bbeta(\bsigma) = \arg\max_{\bbeta} \ell(\bbeta, \bsigma)$ and the corresponding profile criterion $\ell_P(\bsigma) = \ell(\bbeta(\bsigma), \bsigma).$ {The restricted likelihood for $\bsigma$ is defined by integrating the likelihood over the fixed effect
\begin{equation}
\label{eq:reml criterion}
\ell_R(\bsigma)= \int\exp\{\ell(\bbeta, \bsigma)\}d\bbeta.
\end{equation}
}
The REML estimator of $\bsigma$ is defined by $\hbsigma_R=\arg\max_{\bsigma}\ell_R(\bsigma)$,
and then the corresponding estimator of the regression parameter is $\hbbeta_R=\bbeta(\hbsigma_R)$.
The REML estimator of $\btheta$ is denoted by $\hbtheta_R=(\hbbeta_R^\top,\hbsigma_R^\top)^\top$.

The marginalization in \eqref{eq:marginal likelihood} originally involves a $g$-dimensional integral over $\balpha$. Independence across clusters reduces it to $g$ one-dimensional integrals, but these integrals are generally unavailable in closed form. Nevertheless, the marginal likelihood and the corresponding ML and REML estimators remain exactly defined through the original integrals. {Our asymptotic analysis starts from differentiating \eqref{eq:marginal likelihood} and \eqref{eq:reml criterion} to obtain the corresponding implicit equations and derivatives,
and uses Laplace expansion \citep{tierney1986accurate, shun1995laplace, miyata2004fully} to obtain asymptotic representations of these quantities; details of derivations are provided in Supplementary, Section~S.3.1 and S.4.1, respectively.} The leading terms in the asymptotic representations then determine the limiting behavior of the ML and REML estimators.
Importantly, the use of Laplace expansion is purely analytical and does not require the estimators themselves to be computed by Laplace expansion. In practice, the underlying integrals may be evaluated using more accurate numerical methods, such as adaptive Gaussian quadrature \citep{stringer2022asymptotics}, as in our numerical experiments.

The main contributions of this paper are as follows. First, we establish rigorous asymptotic theory for the REML and ML estimators in GLMMs through asymptotic representations of the corresponding estimating equations and their derivatives. In particular, we show that the REML estimator of the variance components is asymptotically equivalent to the corresponding variance-component block of the ML estimator, which yields its asymptotic normality and provides a basis for likelihood-based inference. We further establish the bias-correction property of REML, whereby its adjustment offsets the dominant profiling bias and restores unbiasedness of the variance-component score. Second, the theory is developed under mild distributional and design conditions. We require finite moments of the random effects rather than normality, allow the explanatory variables to be either fixed or random, and permit both the number of clusters $g$ and the minimum cluster size $m_L$ to diverge without restrictions on their relative rates. Third, the asymptotic theory reveals two distinct rates of information accumulation. The information for the between-cluster regression coefficients and the random-effect variance accumulates at rate $g$, whereas that for the within-cluster regression coefficients and the dispersion parameter accumulates at rate $n$. Consequently, the two parameter blocks require different normalizations, leading to a simple block-diagonal limiting covariance structure.

The remainder of the paper is organized as follows. Section~\ref{sec:asymptotic results} presents the main asymptotic results for the ML and REML estimators. Section~\ref{sec:interpretation} explains how these results can be used to construct asymptotically valid confidence intervals and develops finite-sample corrections for covariance estimation. Section~\ref{sec:numerical study} reports numerical experiments and a case study illustrating the finite-sample performance and practical applicability of the proposed inference procedures. Section~\ref{sec:discussion} concludes the paper and discusses directions for future research.

The Appendix contains the lemmas and proofs for the main asymptotic results. The Supplementary includes the detailed derivations of the ML and REML estimating equations and their derivatives, the corresponding asymptotic expansions, and the technical calculations underlying the proofs of the theorems and lemmas.

\section{Asymptotic Results}
\label{sec:asymptotic results}
To simplify notation, a hat denotes an estimator and a dot denotes the corresponding true parameter. 
Let $\mathcal{N}$ denote a sufficiently small neighborhood of $\dbtheta$ contained in the interior of the parameter space
$\boldsymbol{\Theta}
=\{ \btheta:\bbetab \in\mathbb{R}^{p_b}, \siga^2>0, \bbetaw\in\mathbb{R}^{p_w}, \sige^2>0\}.$ For the local analysis, define the profile neighborhood $\mathcal{N}_{\bsigma}=\left\{\bsigma:(\bbeta(\bsigma)^\top, \bsigma^\top)^\top \in \mathcal{N}\right\}$. Thus, $\mathcal{N}_{\bsigma}$ contains those variance-component values whose corresponding profiled parameter points lie in $\mathcal{N}$. Consequently, all uniform expansions established over $\btheta \in \mathcal{N}$ remain valid when evaluated along the profile path uniformly over $\bsigma \in \mathcal{N}_{\bsigma}.$ 

To control the estimating functions and derive the asymptotic properties of the maximum likelihood estimator $\hbtheta$ and the REML estimator $\hbtheta_R$, we introduce the following notation.
Let $\hat{\bc}_1=1/g\sumig\bxbi$, $\hat{\bC}_2=1/g\sumig\bxbi\bxbit$, 
\begin{align*}
&
\hat{\bC}_3(\btheta)=\frac{1}{n}\sumig\left[\sumjmi\frac{b^{(2)}(\eta_{ij})}{\sige^2}\bxwij\bxwijt-\left\{
\sumjmi\frac{b^{(2)}(\eta_{ij})}{\sige^2}+\siga^{-2}
\right\}\bm_i(\btheta)\bm_i^\top(\btheta)
\right],
\\&
\hat{c}_4(\btheta)= -\frac{1}{n}\sumig\sumjmi\left[
\frac{
2\{y_{ij}\eta_{ij}-b(\eta_{ij})\}
}
{\sige^6}
+
\frac{\partial^2}
{\partial(\sige^2)^2}
c(y_{ij},\sige^2)
\right],
\\&\text{and}\qquad
\bm_i(\btheta)=\frac{\sumjmi\bxwij b^{(2)}(\eta_{ij})/\sige^2}{\sumjmi b^{(2)}(\eta_{ij})/\sige^2+\siga^{-2}}.
\end{align*}
These quantities are introduced to summarize the information contributed by the between- and within-cluster covariates. Their limits enter the asymptotic covariance matrices of the ML and REML estimators and help characterize the different information scales associated with the corresponding parameter blocks. One distinction between the GLMM and the LMM setting concerns the limiting design quantities. Under the present setting, we define $\hat{\bc}_1, \hat{\bC}_2, \hat{\bC}_3(\btheta)$, and $\hat{c}_4(\btheta)$, where the latter two quantities depend on $\btheta$. However, in the LMM considered by \cite{lyu2022increasing}, the corresponding quantities depend only on the covariate structure and are free of the model parameters. In fact, the LMM is a special case of the present GLMM framework where the conditional distribution of $\by_i$ is Gaussian and the identity link is adopted. In that case, $b^{(2)}(\eta_{ij}) = 1$ so the factor $1/\sige^2$ can be taken outside the summation in the first part of $\hat{\bC}_3(\btheta)$, leaving the within-cluster design of the form $\sumig\sumjmi \bxwij \bxwijt$; The same simplification applies to $\bm_i(\btheta)$ and $\hat{c}_4(\btheta)$, where all model-dependent factors can be taken outside the relevant summations, leaving only design quantities that are free of the model parameters. For general exponential-family responses, however, $b^{(2)}(\eta_{ij})$  varies with the linear predictor and covariates, so these limiting quantities inherit dependence on $\btheta$.

To control the estimating equations and derive the asymptotic properties of the REML estimator $\hat{\btheta}_R$ and ML estimator $\hat{\btheta}$, we impose the following conditions.

\medskip
	\noindent
	\textbf{Condition A}

\begin{enumerate}
\item
The model in \eqref{eq:exp density} holds with true parameter $\dbtheta$ that is, an interior point of the parameter space $\Theta$; the number of clusters $g \rightarrow \infty$ and the minimum cluster size $m_L\to\infty$. 

\item 
The score functions satisfy $    \E[\bPsi(\dbtheta)]=\boldsymbol{0}_{[p+2:1]}$,  where $\bPsi(\btheta)=\nabla_{\btheta}\ell(\btheta)$, whose $k$th component is
$\partial\ell(\btheta)/\partial\theta_k$, $k=1,\ldots,p+2$, and $\boldsymbol{0}_{[p+2:1]}$ represents $(p+2)$-dimensional zero vector.

\item  There exists a neighborhood $\mathcal{N}$ of $\dot{\btheta}$ such that, uniformly for $\btheta \in\mathcal{N}$, the corresponding linear predictors remain in a compact set on which $b(\cdot)$ is at least three times continuously differentiable with bounded derivatives up to the third order.

\item
The random effects $\{\alpha_i\}$ are independent and identically distributed, and there is a $\delta>0$, such that  $\E|\alpha_1|^{4+\delta}<\infty$. Condition on $\alpha_i$, the responses within cluster $i$ are independent and, for any $i$, $j$, $\E |y_{ij}-b^{(1)}(\eta_{ij})|^{4+\delta}<\infty$.


\item
Suppose that the limits
$\bc_1=\lim_{g\to\infty}\hat{\bc}_1$,
$\bC_2=\lim_{g\to\infty}\hat{\bC}_2$,
$\bC_3(\btheta)=\lim_{g,m_L\to\infty}\hat{\bC}_3(\btheta)$,
and
$c_4(\btheta)=\lim_{g,m_L\to\infty}\hat c_4(\btheta)$
exist, where the latter two limits hold uniformly for
$\btheta\in\mathcal N$.
Assume that $\bC_2$ is finite and positive definite,
$\bC_3(\btheta)$ is finite and positive definite for all
$\btheta\in\mathcal N$, and
$\inf_{\btheta\in\mathcal N}c_4(\btheta)>0$.
Furthermore, the quantities
$g^{-1}\sum_{i=1}^g\|\bm_i(\btheta)\|^2$
and
$n^{-1}\sum_{i=1}^g\sum_{j=1}^{m_i}
\|\bxwij-\bm_i(\btheta)\|^{2+\delta}$
are uniformly bounded over $\mathcal N$, and
\\
$\lim_{g\to\infty}
g^{-1}\sum_{i=1}^g\|\bxbi\|^{2+\delta}<\infty$.
\end{enumerate}

As noted in Section~\ref{sec:Introduction}, these are mild conditions. Condition A2 follows the unbiased-score condition used in \cite[Theorem 1, Condition (iii)]{jiang2001mixed}, requiring the score evaluated at the true parameter to have mean zero.
Condition A3 is a local regularity condition on the cumulant function $b(\cdot)$. It prevents the linear predictors from approaching the boundary of the natural parameter space in a neighborhood of $\dot{\btheta}$,  ensuring that the score and information matrices remain uniformly well-behaved. This condition is mild for regular exponential family models, since $b^{(2)}(\eta)$ represents the conditional variance and $b(\cdot)$ is typically smooth in the interior of the natural parameter space. For the expansions used later, it suffices to assume that $b(\cdot)$ is three times continuously differentiable, with bounded higher-order derivatives required only when they are explicitly invoked. Conditions A4 and A5 provide the moment and design conditions required for the limiting covariance matrix and the central limit theorem for the estimating functions. The finite $4+\delta$ moment condition ensures sufficient control of the higher-order moments needed for the Lyapunov condition.
They also ensure that the negative of the appropriately normalized second derivative of the estimating function converges to a matrix $\bB$ given in \eqref{matrix B}.

Our main results are the following asymptotic theorems for REML and ML estimators, with detailed proofs in the Appendix.

\begin{theorem}
\label{theorem1}
Suppose Condition A holds and let $\bK_\sigma=\operatorname{diag}(g,n)$. There exists a solution $\hbsigma_R$ to $\nabla_{\bsigma} \ell_R(\bsigma)  = \boldsymbol{0}$, satisfying $\|\bK_{\bsigma}^{1/2} \left(\hbsigma_R - \dbsigma\right)\|=O_p(1)$, that has the asymptotic representation
$$
\bK_\sigma^{1/2}
(\hbsigma_R-\dbsigma)
=\bB_{\bsigma}^{-1}\bK_\sigma^{-1/2}\bxi_{\bsigma}
+
o_p(1),
$$
where  $\bB_{\bsigma}=\diag\{1/(2\dsiga^4),c_4(\dbtheta)\}$ and $\bxi_{\bsigma}=(\xi_{\siga^2},\xi_{\sige^2})^\top$  has components
\begin{align*}
\xi_{\siga^2}
=
\frac{1}{2\dot\siga^4}
\sum_{i=1}^g
\left(
\alpha_i^2
-
\dot\siga^2
\right),
\qquad
\xi_{\sige^2}
=
\sum_{i=1}^g
\sum_{j=1}^{m_i}
\left[
\frac{
y_{ij}\dot\eta_{ij}
-
b(\dot\eta_{ij})
}
{\dot\sigma_e^4}
+
\frac{\partial}{\partial\sige^2}
c(y_{ij},\dot\sige^2)
\right].
\end{align*}
Moreover, we have
$$
\mathbf K_\sigma^{1/2}
(\hat{\boldsymbol{\sigma}}_R-\dot{\boldsymbol{\sigma}})
\overset{D}{\longrightarrow}
N(\mathbf 0,\mathbf C_R), 
$$
where $\bC_R= \bB_{\bsigma}^{-1}\bA_{\sigma}\bB_{\bsigma}^{-1}$ with $\bA_{\sigma}=\lim_{g,m_L\to\infty} \var(\bK_{\sigma}^{-1/2}\bxi_{\bsigma})$.
\end{theorem}

\begin{theorem}
\label{theorem2}
Suppose Condition~A holds and let  $\bK=\diag (g\boldsymbol{1}_{p_b},g,n\boldsymbol{1}_{p_w},n)$ with $\boldsymbol{1}_p$ the $p$ vector of ones. There exists a solution
$\hbtheta$ to
$\nabla_{\btheta}\ell(\btheta)=\boldsymbol{0}$
satisfying $\|\bK^{1/2}(\hbtheta-\dbtheta)\|=O_p(1)$, that has the  asymptotic linear representation
$$
\bK^{1/2}(\hbtheta-\dbtheta)=\bB^{-1}\bK^{-1/2}\bxi+o_p(1),
$$
where $\bB$ is given in \eqref{matrix B}, and
$\boldsymbol{\xi}
=
\left(
\boldsymbol{\xi}_{\bbeta^{(b)}}^\top,
\xi_{\siga^2},
\boldsymbol{\xi}_{\bbeta^{(w)}}^\top,
\xi_{\sige^2}
\right)^\top$
has components
\begin{align*}
&\bxi_{\bbetab}=\frac{1}{\dsiga^2}\sumig\bxbi{\alpha}_i,
&&
\bxi_{\bbetaw}=\frac{1}{\dsige^2}\sumig\sumjmi\{\bxwij-\bm_i(\dbtheta)\}\{y_{ij} - b^{(1)}(\eta_{ij})\}
\\
&
\xi_{\siga^2}=\frac{1}{2\dsiga^4}\sumig(\alpha_i^2-\dsiga^2),
&&
\xi_{\sige^2}=\sumig\sumjmi
\left\{
\frac{
y_{ij}\dot\eta_{ij}
-
b(\dot\eta_{ij})
}
{\dsige^4}
+
\frac{\partial}{\partial\sige^2}
c(y_{ij},\dsige^2)
\right\}.
\end{align*}
It follows that
$$
\bK^{1/2}(\hbtheta-\dbtheta)\overset{D}{\longrightarrow}
N(\boldsymbol{0},\bC),
$$
where
$$
\bC
=
\begin{bmatrix}
\dsiga^2\mathbf{C}_2^{-1}
&
\mathbb{E}(\alpha_i^3)\mathbf{C}_2^{-1}\mathbf{c}_1
&
\boldsymbol{0}_{[p_b:p_w]}
&
\boldsymbol{0}_{[p_b:1]}
\\[0.8em]
\mathbb{E}(\alpha_i^3)\mathbf{c}_1^\top\mathbf{C}_2^{-1}
&
\mathbb{E}(\alpha_i^4)-\dsiga^4
&
\boldsymbol{0}_{[1:p_w]}
&
0
\\[0.8em]
\boldsymbol{0}_{[p_w:p_b]}
&
\boldsymbol{0}_{[p_w:1]}
&
\bC_3^{-1}(\dbtheta)
&
\boldsymbol{0}_{[p_w:1]}
\\[0.8em]
\boldsymbol{0}_{[1:p_b]}
&
0
&
\boldsymbol{0}_{[1:p_w]}
&
1/c_4(\dot{\btheta})
\end{bmatrix}.
$$
Here $\boldsymbol{0}_{[q:r]}$ denotes the $q \times r$ matrix of zeros.
\end{theorem}

\begin{corollary}
\label{corollary3}
Suppose Condition A holds. 
Then $\bC_R$ and $\bC$ can be consistently estimated by
\begin{align*}
\hat{\bC}_R=\begin{bmatrix}
\hat{\mu}_{4\alpha}&0\\
0&1/\hat{c}_4(\hbtheta)
\end{bmatrix}
\text{ and }
\hat{\bC} = \begin{bmatrix}
\hat\siga^2 \hat{\mathbf{C}}_2^{-1}
&
\hat{\mu}_{3\alpha}\hat{\mathbf{C}}_2^{-1}\hat{\mathbf{c}}_1
&
\boldsymbol{0}_{[p_b:p_w]}
&
\boldsymbol{0}_{[p_b:1]}
\\ 
\hat{\mu}_{3\alpha}\hat{\mathbf{c}}_1^\top\hat{\mathbf{C}}_2^{-1}
&
\hat{\mu}_{4\alpha}-\hat{\sigma}_\alpha^4
&
\boldsymbol{0}_{[1:p_w]}
&
0
\\ 
\boldsymbol{0}_{[p_w:p_b]}
&
\boldsymbol{0}_{[p_w:1]}
&
\hat{\bC}_3^{-1}(\hat{\btheta})
&
\boldsymbol{0}_{[p_w:1]}
\\ 
\boldsymbol{0}_{[1:p_b]}
&
0
&
\boldsymbol{0}_{[1:p_w]}
&
1/\hat{c}_4(\hat{\btheta})
\end{bmatrix},
\end{align*}
where
$\hat{\mu}_{k\alpha}
=
g^{-1}\sumig(\hat{\alpha}_i^*)^k, k=3,4,$
and $\hat{\alpha}_i^* = \arg\max_{\alpha_i}h_i(\by_i, \alpha_i \mid \hat{\btheta})$.
\end{corollary}

We now make some remarks concerning Theorem~\ref{theorem1} and~\ref{theorem2}, and Corollary~\ref{corollary3}.

\begin{enumerate}
    \item Theorem~\ref{theorem1} establishes the asymptotic property of $\hbsigma_R$. For the full REML estimator, it remains to characterize the profiled fixed-effect component. Under the corresponding normalization $\bK_{\bbeta} = \diag(g\boldsymbol{1}_{p_b}, n\boldsymbol{1}_{p_w})$, $$\bK_{
    \bbeta}^{1/2}\left(\hbbeta_R -\hbbeta_P\right) =o_p(1),$$ while the profile estimator $\{\hbbeta_P,\hbsigma_P\}$
    coincides with the ML estimator $\hbtheta$. These results are established in Supplementary, Sections. S.6.10-6.11. Combining the above asymptotic equivalence with Theorem~\ref{theorem2} gives 
    $$\bK^{1/2}
    \left(
    \hbtheta_R
    -
    \dbtheta
    \right)
    \overset{D}{\longrightarrow}
    N(\boldsymbol{0},\mathbf{C}).$$

    \item REML is asymptotically unbiased through its bias-correction property. The original marginal score is unbiased at the true parameter, whereas profiling the fixed effects and replacing $\bbeta$ with $\hbbeta$ introduces bias into the variance-component score. For $\siga^2$, the leading bias introduced by profiling is $-p_b/(2\siga^2)$, while the leading term of the corresponding REML adjustment contributes $p_b/(2\siga^2)$, so the two terms offset each other. Similarly, for $\sige^2$, the leading profile-score bias is $-p_w/(2\sige^2)$, and the REML adjustment contributes $p_w/(2\sige^2)$. Hence, for both $\siga^2$ and $\sige^2$, REML restores asymptotic unbiasedness. This parallels the classical bias-correction role of REML in the LMM, but here the correction separates according to the two information scales: the between-cluster fixed effects contribute to the correction for $\siga^2$, whereas the within-cluster fixed effects contribute to the correction for $\sige^2$. Detailed derivations are provided in Supplementary, Section S.7.

    \item The normalizing matrix $\bK$ reflects the different sources of information for the model parameters. The between-cluster regression coefficients $\bbetab$ and the random-effect variance $\siga^2$ converge at the $g^{1/2}$ rate, since their information is determined primarily by the number of independent clusters; the within-cluster regression coefficients $\bbetaw$ and the dispersion parameter $\sige^2$ converge at the $n^{1/2}$ rate, as their information accumulates over the individual observations. These rate distinctions alone do not imply that both $g$ and $m_L$ need to diverge. Indeed, when $g \rightarrow \infty$ with a finite $m_L$, the model parameters can be estimated consistently. However, the amount of information available within each cluster remains bounded, so the cluster-specific random effects $\alpha_i$ can not be recovered consistently. This is analogous to the result of \cite{jiang1998asymptotic} for the EBLUP of random effects in LMMs with bounded cluster sizes. Therefore, the regime $g, m_L \rightarrow \infty$ is adopted to accommodate both consistent estimation of the population parameters and consistent localization of the cluster-specific random effects.

    \item Asymptotic orthogonality between different model parameters is revealed by the form of $\bC$. The between-cluster block $(\bbeta^{(b)}, \siga^2)$ and the within-cluster block $(\bbeta^{(w)}, \sige^2)$ are asymptotically orthogonal, since the corresponding off-diagonal blocks of $\mathbf{C}$ are zero. Within the within-cluster block, $\bbeta^{(w)}$ and $\sige^2$ are asymptotically orthogonal. By contrast, $\bbeta^{(b)}$ and $\siga^2$ need not be asymptotically orthogonal. Their limiting covariance is governed by $\mathbb{E}(\dot\alpha_i^3)\mathbf{c}_1^\top\mathbf{C}_2^{-1}$, and hence this covariance vanishes when the random-effect distribution is symmetric about zero.
    
    \item The asymptotic covariance matrix in Theorem~\ref{theorem2} depends on the third and fourth moments of the random effects. If the distribution of $\alpha_i$ is specified, these moments can be substituted directly, leading to a simpler covariance expression. In particular, under $\alpha_i \sim N(0, \siga^2)$, we have $$\mathbb{E}(\alpha_i^3) = 0, \qquad \mathbb{E}(\alpha_i^4) = 3\siga^4.$$ Substituting these identities into $\bC$, the resulting asymptotic covariance matrix coincides with the corresponding covariance matrix of \cite{jiang2022usable} in the random-intercept setting. The plug-in covariance estimator in Corollary~\ref{corollary3} simplifies accordingly, with $$\hat{\mu}_{3\alpha} =0, \qquad \hat{\mu}_{4\alpha} - \hat{\sigma}_\alpha^4 = 2\hat{\sigma}_\alpha^4.$$
    
\end{enumerate}

\section{Interpreting and using the results}
\label{sec:interpretation}
Theorems~\ref{theorem1} and \ref{theorem2} establish the asymptotic equivalence, asymptotic linear representations, and limiting distributions of the REML and MLE estimators under the GLMM framework. These results characterize the large-sample behavior of the estimators and provide a basis for likelihood-based inference. In this section, we discuss their interpretation and practical implementation.

The asymptotic linear representations in Theorem~\ref{theorem1} and \ref{theorem2} show the influence functions of the REML and ML estimators. Since the REML estimator is asymptotically equivalent to the MLE, the two estimators have the same influence functions. These are obtained from the summand of $\mathbf{B}^{-1}\boldsymbol{\xi}.$ Explicitly, at a point $[\alpha_i, y_{ij} - b^{(1)}(\eta_{ij}), \bxbit,\bxwijt]^\top$ and suppressing the dot notation for quantities evaluated at the true parameter value, the influence function is the $(p_b+p_w+2)$-vector $\boldsymbol{\lambda} = [\boldsymbol{\lambda}_{\bbeta^{b}}^\top, \lambda_{\siga^2}, \boldsymbol{\lambda}_{\bbeta^{w}}^\top, \lambda_{\sige^2}]^\top$, where
\begin{align*}
    &\boldsymbol{\lambda}_{\bbeta^{b}} = \mathbf{C}_2^{-1} \bxbi\alpha_i, 
    &&\boldsymbol{\lambda}_{\bbeta^{w}} = \mathbf{C}_3^{-1}(\btheta) \frac{\left\{\bxwij - \boldsymbol{m}_i(\btheta)\right\}(y_{ij} - b^{(1)}(\eta_{ij}))}{\sige^2}, 
    \\& \lambda_{\siga^2} = \alpha_i^2 - \siga^2, 
    &&\lambda_{\sige^2} = c^{-1}_4(\btheta)\left[ 
    \frac{y_{ij}\eta_{ij}
    -b(\eta_{ij})}
    {\sigma_e^4}
    +\frac{\partial}{\partial\sige^2}
    c(y_{ij},\sige^2)\right].
\end{align*}
These expressions are generally unbounded in the covariates, random effects, conditional residuals, and responses. Hence, neither the ML estimator nor the REML estimator has a bounded influence function, and the estimators are not robust in the bounded-influence sense.

By the central limit theorem, we can construct asymptotic confidence intervals for the parameters in the model. Let $z_{1-\gamma/2} = \mathit{\Phi}^{-1}(1-\gamma/2)$, where $\mathit{\Phi}^{-1}$ is the $N(0,1)$ quantile function. For $k=1,\ldots, p_b$, an asymptotic $100(1-\gamma)\%$ confidence interval for $\dot{\bbeta}_k^{(b)}$ is 
\begin{align*}
    \left[\hat{\bbeta}_k^{(b)} - z_{1-\gamma/2} \hat{\sigma}_\alpha 
\left\{{(\hat{\mathbf C}_2^{-1})_{kk}}/{g}\right\}^{1/2},\quad \hat{\bbeta}_k^{(b)} + z_{1-\gamma/2} \hat{\sigma}_\alpha \left\{{(\hat{\mathbf C}_2^{-1})_{kk}}/{g}\right\}^{1/2}\right],
\end{align*}
where $(\hat{\mathbf C}_2^{-1})_{kk}$ is the $k$th diagonal element of $\hat{\mathbf{C}}_2^{-1}$. Similarly, an asymptotic $100(1-\gamma)\%$ confidence interval for $\dot{\bbeta}_l^{(w)}, l=1,\ldots, p_w$ is
\begin{align*}
\left[\hat{\bbeta}_l^{(w)} - z_{1-\gamma/2}\left\{{(\hat{\mathbf C}_3^{-1}(\hat{\btheta}))_{ll}}/{n}\right\}^{1/2},\quad \hat{\bbeta}_l^{(w)} + z_{1-\gamma/2}\left\{{(\hat{\mathbf C}_3^{-1}(\hat{\btheta}))_{ll}}/{n}\right\}^{1/2}\right],    
\end{align*}
where $(\hat{\mathbf C}_3^{-1}(\hat{\btheta}))_{ll}$ is the $l$th diagonal element of $\hat{\mathbf{C}}_3^{-1}(\hat{\btheta})$. 
Setting the confidence interval on the log scale and then back transforming, an asymptotic $100(1-\gamma)\%$ confidence interval for $\dot{\sigma}^2_\alpha$ is 
\begin{align*}
\left[\hat{\sigma}^2_\alpha\exp\left\{-z_{1-\gamma/2}{(\hat{\mu}_{4\sigma} - \hat{\sigma}^4_\alpha)^{1/2}}/{(g^{1/2}\hat{\sigma}^2_\alpha)}\right\}, \quad \hat{\sigma}^2_\alpha\exp\left\{z_{1-\gamma/2}{(\hat{\mu}_{4\sigma} - \hat{\sigma}^4_\alpha)^{1/2}}/{(g^{1/2}\hat{\sigma}^2_\alpha)}\right\} \right].
\end{align*}
Similarly, an asymptotic $100(1-\gamma)\%$ confidence interval for $\dot{\sigma}^2_e$ is 
\begin{align*}
\left[\hat{\sigma}^2_e\exp\left\{-z_{1-\gamma/2}/({n^{1/2}\hat{\sigma}^2_e \hat{c}^{1/2}_4(\hat{\btheta})})\right\}, \quad \hat{\sigma}^2_e\exp\left\{z_{1-\gamma/2}/({n^{1/2}\hat{\sigma}^2_e\hat{c}^{1/2}_4(\hat{\btheta})})\right\} \right].
\end{align*}

As noted in Section~\ref{sec:Introduction}, the Laplace expansion is applied here only as an analytical device for deriving tractable asymptotic representations of the implicit ML and REML score equations and their derivatives. It is not required for computing the estimators in practice. More accurate numerical integration methods, particularly adaptive Gaussian quadrature \citep{stringer2022asymptotics}, may instead be used to obtain $\hat{\btheta}$ and $\hat{\btheta}_R$. Adaptive Gaussian quadrature places multiple integration nodes around each cluster-specific conditional mode and scales them according to the local curvature, whereas the Laplace expansion retains only the leading local contribution around that mode. \cite{stringer2022asymptotics} show that the numerical approximation error decreases as the number of quadrature nodes increases, and that the resulting estimator has the same asymptotic property as the estimator based on the original marginal likelihood when this error is of smaller order than the sampling error.

In addition to computing the point estimators, practical implementation of the asymptotic theory also requires estimation of the covariance matrix in Corollary~\ref{corollary3}. This covariance matrix involves the third and fourth moments of the random effects, which must be estimated when their distribution is left unspecified. A direct empirical approach replaces these moments by averages of powers of the fitted conditional mode $\hat{\alpha}_i^*$. However, $\hat{\alpha}_i^*$ represents only the mode of the conditional distribution of $\alpha_i \mid \by_i$, and does not account for the local conditional uncertainty around the mode. This may lead to underestimation of higher-order moments such as $\mathbb{E}(\alpha_i^4)$ when $m_L$ is small. A simple correction can be obtained from a second-order local approximation to the conditional density of $\alpha_i \mid \by_i$. Specifically, since $\hat{\alpha}_i^{*}$ is the conditional mode, a Taylor expansion
of $h_i(\mathbf{y}_i,\alpha_i \mid \hat{\btheta})$ around
$\hat{\alpha}_i^{*}$ gives
$$
h_i(\by_i,\alpha_i \mid \hat{\btheta})
=
h_i(\by_i,\hat{\alpha}_i^{*} \mid \hat{\btheta})
+
\frac{1}{2}
\left.
\frac{\partial^2}{\partial \alpha_i^2}
h_i(\mathbf{y}_i,\alpha_i \mid \hat{\btheta})
\right|_{\alpha_i=\hat{\alpha}_i^{*}}
(\alpha_i-\hat{\alpha}_i^{*})^2
+
o\{(\alpha_i-\hat{\alpha}_i^{*})^2\}.
$$
Since $h_i$ is the conditional log-density, the second derivative at the
mode is positive. Therefore, locally around $\hat{\alpha}_i^{*}$,
$$
\exp\left\{h_i(\mathbf{y}_i,\alpha_i \mid \hat{\btheta})\right\}
\propto
\exp\left[
-\frac{1}{2}
\left\{\left.
\frac{\partial^2}{\partial \alpha_i^2}
h_i(\mathbf{y}_i,\alpha_i \mid \hat{\btheta})
\right|_{\alpha_i=\hat{\alpha}_i^{*}}\right\}
(\alpha_i-\hat{\alpha}_i^{*})^2
\right].
$$
This gives the local normal approximation
$$
\alpha_i \mid \mathbf{y}_i;\hat{\btheta}
\approx
N(\hat{\alpha}_i^{*},v_i),
$$
where
$$
v_i =
\left\{
\left.-
\frac{\partial^2}{\partial \alpha_i^2}
h_i(\mathbf{y}_i,\alpha_i \mid \hat{\btheta})
\right|_{\alpha_i=\hat{\alpha}_i^{*}}
\right\}^{-1}.
$$ 
By the law of total expectation, the plug-in estimator for empirical moments in Corollary~\ref{corollary3} can be replaced by the following corrected moment estimator
\begin{align}
\hat{\mu}_{3\alpha}^c = g^{-1}\sum_{i=1}^g\{(\hat{\alpha}_i^*)^3 + 3\hat{\alpha}_i^*v_i\},\quad \text{and}\quad
\hat{\mu}_{4\alpha}^c = g^{-1}\sum_{i=1}^g\{(\hat{\alpha}_i^*)^4 + 6(\hat{\alpha}_i^*)^2v_i + 3v_i^2\}.\label{correction moment}
\end{align}

These corrections account for the leading conditional uncertainty in predicting the random effects and can be substituted into the covariance estimator in Corollary~\ref{corollary3}. They mainly serve as finite-sample refinements for small $m_L$, where first-order approximations may underestimate the prediction uncertainty. When $m_L$ is sufficiently large so that Condition~A holds, these higher-order corrections become asymptotically negligible, and no such refinements are required. Their finite-sample performance is examined in the simulation studies of Section~\ref{sec:numerical study}.

\section{Numerical Study}
\label{sec:numerical study}

\subsection{Simulation}
\label{subsec:simulation}
We generated data from Poisson, binomial and Gamma GLMMs with
$g \in \{10, 20, 50, 100\}$ and
$m_L\in\{10,20,50,100\}$. For each $(g,m_L)$ setting, the cluster sizes were generated as $m_i-m_L \sim {\rm Poisson}(m_L+m_L^{1/2}),\quad i=1,\ldots,g,$
and were then held fixed throughout the Monte Carlo replications for that setting. The linear predictor was
$$
   \eta_{ij}
    =
    \beta_0^{(b)} +\beta_1^{(b)} \bar x_i
    +\beta_2^{(w)} (x_{ij} - \bar x_i)+\alpha_i,
    \qquad
    i=1,\ldots,g,\quad j=1,\ldots,m_i.
$$
The random effects were generated from either $F_{\alpha} = N(0,\dsiga^2)$ or $F_{\alpha} = 0.3N(0.5,0.5) + 0.7N(\mu, \{\dsiga^2 - 0.225 - 0.7\mu^2\})$ with $\mu = -0.3\times0.5/0.7$, so that $\mathbb{E}(\alpha_i) = 0$ and $\var(\alpha_i) = \dsiga^2$. The covariates were generated with a cluster structure by setting
$x_{ij}=4+\sqrt{2}u_i+2v_{ij}$, where $u_i$ and $v_{ij}$ are independent standard normal random variables. We center the covariates with their cluster mean $\bar{x}_i$ to obtain the within-cluster covariate $x_{ij} - \bar x_i$ and include $\bar x_i$ in the model as a between-cluster covariate. 
For the Poisson model, the conditional mean was linked to the linear predictor by the log link. For the Binomial model, we generated binomial counts with number of trials $k=5$ and the success probability was linked to the linear predictor through the logit link. For the Gamma model, the conditional mean was linked to the linear predictor by the log link with dispersion parameter $\sige^2$. Although the log link is not canonical for the Gamma family, we include the Gamma case as an additional numerical investigation, while the theoretical results are established for canonical links.
We considered $\dsiga^2\in\{1,4\}$ for the Poisson and binomial models, and $\dsiga^2 = \{0.5, 1\}$ and $\dsige^2 = \{0.5,1\}$ for the Gamma model. The true fixed-effect slopes $\dot{\bbeta} = [\beta_0^{(b)}, \beta_1^{(b)}, \beta_0^{(w)}]^\top$ for Poisson, Binomial and Gamma models are $[-1.5,0.3,0.5]^\top$, $[1,0.2,0.2]^\top$, and $[0.5, 0.03, 0.05]^\top$, respectively.

For each simulation setting, we generated 1000 replications. Poisson and binomial GLMMs were fitted using the \texttt{R} package \texttt{lme4} \textcolor{red}{\citep{Douglas2015lme4}}, and the Gamma GLMM was fitted using the \texttt{R} package \texttt{GLMMadaptive} \textcolor{red}{\citep{Rizopoulos2025adaptive}}. For each fitted model, we computed standard errors for all fixed-effect, variance-component and dispersion parameters, and used them to construct nominal 95\% Wald confidence intervals discussed in Section~\ref{sec:interpretation} (Monte Carlo standard error $< 0.007$). 
For $\siga^2$, we compared three versions of standard errors: SE, obtained by substituting the estimated random-effect modes into the empirical sample moment formula in Corollary~\ref{corollary3}; SE$_C$, based on the corrected empirical sample moments introduced in \eqref{correction moment}; and SE$_N$, computed under the normality assumption for the random effects.
\begin{table}[!h]
\centering
\caption{Simulated coverage probabilities and average confidence interval lengths for the Poisson GLMM When $\alpha_i$ has a Gaussian distribution with $\siga^2 = 1$.}
\label{tab:coverage_poisson}
\small
\setlength{\tabcolsep}{4pt}
\renewcommand{\arraystretch}{1.05}

\begin{tabular}{llcccccccc}
\toprule
\multirow{2}{*}{$g$}
& \multirow{2}{*}{Estimator}
& \multicolumn{2}{c}{$m_L=10$}
& \multicolumn{2}{c}{$m_L=20$}
& \multicolumn{2}{c}{$m_L=50$}
& \multicolumn{2}{c}{$m_L=100$} \\
\cmidrule(lr){3-4}
\cmidrule(lr){5-6}
\cmidrule(lr){7-8}
\cmidrule(lr){9-10}
&
& Cvge & Len
& Cvge & Len
& Cvge & Len
& Cvge & Len \\
\midrule

\multirow{6}{*}{$10$}
& SE$(\hat\beta^{(b)}_0)$ & 0.87 & 3.57 & 0.88 & 3.66 & 0.89 & 3.64 & 0.89 & 3.68 \\
& SE$(\hat\beta^{(b)}_1)$ & 0.87 & 0.85 & 0.88 & 0.87 & 0.87 & 0.87 & 0.89 & 0.87 \\
& SE$(\hat\beta^{(w)}_0)$ & 0.96 & 0.07 & 0.95 & 0.05 & 0.94 & 0.03 & 0.95 & 0.02 \\
& SE$(\hat\siga^2)$ & 0.65 & 1.09 & 0.69 & 1.19 & 0.72 & 1.25 & 0.71 & 1.24 \\
& SE$_{C}(\hat\siga^2)$ & 0.71 & 1.29 & 0.73 & 1.30 & 0.74 & 1.31 & 0.72 & 1.27 \\
& SE$_{N}(\hat\siga^2)$ & 0.81 & 1.60 & 0.84 & 1.58 & 0.83 & 1.59 & 0.84 & 1.59 \\
\midrule

\multirow{6}{*}{$20$}
& SE$(\hat\beta^{(b)}_0)$ & 0.89 & 2.58 & 0.92 & 2.57 & 0.90 & 2.56 & 0.91 & 2.65 \\
& SE$(\hat\beta^{(b)}_1)$ & 0.91 & 0.61 & 0.91 & 0.61 & 0.91 & 0.60 & 0.91 & 0.63 \\
& SE$(\hat\beta^{(w)}_0)$ & 0.95 & 0.05 & 0.95 & 0.04 & 0.95 & 0.02 & 0.95 & 0.02 \\
& SE$(\hat\siga^2)$ & 0.77 & 0.93 & 0.80 & 0.98 & 0.84 & 1.01 & 0.85 & 1.07 \\
& SE$_{C}(\hat\siga^2)$ & 0.82 & 1.06 & 0.83 & 1.06 & 0.85 & 1.05 & 0.85 & 1.09 \\
& SE$_{N}(\hat\siga^2)$ & 0.89 & 1.23 & 0.90 & 1.20 & 0.90 & 1.17 & 0.91 & 1.21 \\
\midrule

\multirow{6}{*}{$50$}
& SE$(\hat\beta^{(b)}_0)$ & 0.93 & 1.60 & 0.94 & 1.63 & 0.95 & 1.65 & 0.93 & 1.65 \\
& SE$(\hat\beta^{(b)}_1)$ & 0.93 & 0.38 & 0.94 & 0.38 & 0.95 & 0.39 & 0.93 & 0.39 \\
& SE$(\hat\beta^{(w)}_0)$ & 0.96 & 0.03 & 0.95 & 0.02 & 0.94 & 0.01 & 0.95 & 0.01 \\
& SE$(\hat\siga^2)$ & 0.83 & 0.64 & 0.87 & 0.67 & 0.88 & 0.72 & 0.91 & 0.72 \\
& SE$_{C}(\hat\siga^2)$ & 0.88 & 0.71 & 0.90 & 0.71 & 0.89 & 0.74 & 0.91 & 0.73 \\
& SE$_{N}(\hat\siga^2)$ & 0.91 & 0.77 & 0.93 & 0.77 & 0.91 & 0.77 & 0.93 & 0.77 \\
\midrule

\multirow{6}{*}{$100$}
& SE$(\hat\beta^{(b)}_0)$ & 0.94 & 1.13 & 0.94 & 1.15 & 0.93 & 1.17 & 0.94 & 1.17 \\
& SE$(\hat\beta^{(b)}_1)$ & 0.95 & 0.27 & 0.94 & 0.27 & 0.93 & 0.28 & 0.94 & 0.28 \\
& SE$(\hat\beta^{(w)}_0)$ & 0.94 & 0.02 & 0.95 & 0.01 & 0.96 & 0.01 & 0.95 & 0.01 \\
& SE$(\hat\siga^2)$ & 0.87 & 0.47 & 0.90 & 0.50 & 0.91 & 0.52 & 0.92 & 0.53 \\
& SE$_{C}(\hat\siga^2)$ & 0.90 & 0.51 & 0.92 & 0.53 & 0.92 & 0.53 & 0.92 & 0.53 \\
& SE$_{N}(\hat\siga^2)$ & 0.93 & 0.55 & 0.93 & 0.55 & 0.94 & 0.55 & 0.94 & 0.55 \\
\bottomrule
\end{tabular}
\end{table}

\begin{table}[!h]
\centering
\caption{Simulated coverage probabilities and average confidence interval lengths for the Gamma GLMM When $\alpha_i$ has a Gaussian distribution with $\siga^2 = 0.5$ and $\sige^2 = 0.5$.}
\label{tab:coverage_gamma}
\small
\setlength{\tabcolsep}{4pt}
\renewcommand{\arraystretch}{1.05}

\begin{tabular}{llcccccccc}
\toprule
\multirow{2}{*}{$g$}
& \multirow{2}{*}{Estimator}
& \multicolumn{2}{c}{$m_L=10$}
& \multicolumn{2}{c}{$m_L=20$}
& \multicolumn{2}{c}{$m_L=50$}
& \multicolumn{2}{c}{$m_L=100$} \\
\cmidrule(lr){3-4}
\cmidrule(lr){5-6}
\cmidrule(lr){7-8}
\cmidrule(lr){9-10}
&
& Cvge & Len
& Cvge & Len
& Cvge & Len
& Cvge & Len \\
\midrule

\multirow{7}{*}{$10$}
& SE$(\hat\beta^{(b)}_0)$ & 0.87 & 2.50 & 0.87 & 2.53 & 0.89 & 2.59 & 0.90 & 2.64 \\
& SE$(\hat\beta^{(b)}_1)$ & 0.88 & 0.59 & 0.87 & 0.60 & 0.90 & 0.62 & 0.90 & 0.63 \\
& SE$(\hat\beta^{(w)}_0)$ & 0.94 & 0.09 & 0.95 & 0.07 & 0.95 & 0.04 & 0.95 & 0.03 \\
& SE$(\hat\sige^2)$ & 0.96 & 0.18 & 0.94 & 0.12 & 0.95 & 0.08 & 0.96 & 0.06 \\
& SE$(\hat\siga^2)$ & 0.67 & 0.57 & 0.71 & 0.61 & 0.73 & 0.64 & 0.73 & 0.62 \\
& SE$_{C}(\hat\siga^2)$ & 0.73 & 0.65 & 0.74 & 0.66 & 0.75 & 0.66 & 0.74 & 0.62 \\
& SE$_{N}(\hat\siga^2)$ & 0.82 & 0.79 & 0.83 & 0.79 & 0.86 & 0.81 & 0.88 & 0.82 \\

\midrule

\multirow{7}{*}{$20$}
& SE$(\hat\beta^{(b)}_0)$ & 0.91 & 1.76 & 0.92 & 1.79 & 0.93 & 1.82 & 0.92 & 1.86 \\
& SE$(\hat\beta^{(b)}_1)$ & 0.91 & 0.42 & 0.92 & 0.42 & 0.92 & 0.43 & 0.92 & 0.44 \\
& SE$(\hat\beta^{(w)}_0)$ & 0.95 & 0.07 & 0.95 & 0.05 & 0.95 & 0.03 & 0.96 & 0.02 \\
& SE$(\hat\sige^2)$ & 0.96 & 0.12 & 0.97 & 0.09 & 0.95 & 0.06 & 0.94 & 0.04 \\
& SE$(\hat\siga^2)$ & 0.79 & 0.48 & 0.81 & 0.50 & 0.84 & 0.52 & 0.86 & 0.53 \\
& SE$_{C}(\hat\siga^2)$ & 0.84 & 0.53 & 0.84 & 0.53 & 0.85 & 0.53 & 0.86 & 0.53 \\
& SE$_{N}(\hat\siga^2)$ & 0.89 & 0.58 & 0.89 & 0.59 & 0.90 & 0.59 & 0.92 & 0.60 \\
\midrule

\multirow{7}{*}{$50$}
& SE$(\hat\beta^{(b)}_0)$ & 0.93 & 1.12 & 0.92 & 1.15 & 0.94 & 1.16 & 0.93 & 1.16 \\
& SE$(\hat\beta^{(b)}_1)$ & 0.93 & 0.26 & 0.93 & 0.27 & 0.95 & 0.27 & 0.94 & 0.28 \\
& SE$(\hat\beta^{(w)}_0)$ & 0.95 & 0.04 & 0.96 & 0.03 & 0.95 & 0.02 & 0.96 & 0.01 \\
& SE$(\hat\sige^2)$ & 0.95 & 0.08 & 0.95 & 0.06 & 0.95 & 0.04 & 0.95 & 0.03 \\
& SE$(\hat\siga^2)$ & 0.86 & 0.34 & 0.89 & 0.36 & 0.91 & 0.37 & 0.90 & 0.37 \\
& SE$_{C}(\hat\siga^2)$ & 0.89 & 0.37 & 0.91 & 0.37 & 0.92 & 0.38 & 0.90 & 0.37 \\
& SE$_{N}(\hat\siga^2)$ & 0.90 & 0.38 & 0.93 & 0.39 & 0.93 & 0.39 & 0.93 & 0.38 \\
\midrule

\multirow{7}{*}{$100$}
& SE$(\hat\beta^{(b)}_0)$ & 0.93 & 0.80 & 0.95 & 0.81 & 0.94 & 0.83 & 0.95 & 0.83 \\
& SE$(\hat\beta^{(b)}_1)$ & 0.94 & 0.19 & 0.94 & 0.19 & 0.95 & 0.19 & 0.95 & 0.20 \\
& SE$(\hat\beta^{(w)}_0)$ & 0.95 & 0.03 & 0.95 & 0.02 & 0.96 & 0.01 & 0.95 & 0.01 \\
& SE$(\hat\sige^2)$ & 0.97 & 0.06 & 0.96 & 0.04 & 0.95 & 0.03 & 0.96 & 0.02 \\
& SE$(\hat\siga^2)$ & 0.89 & 0.25 & 0.93 & 0.26 & 0.93 & 0.27 & 0.92 & 0.27 \\
& SE$_{C}(\hat\siga^2)$ & 0.91 & 0.27 & 0.94 & 0.27 & 0.93 & 0.27 & 0.92 & 0.27 \\
& SE$_{N}(\hat\siga^2)$ & 0.92 & 0.28 & 0.95 & 0.27 & 0.93 & 0.28 & 0.93 & 0.28 \\
\bottomrule
\end{tabular}
\end{table}

Tables~\ref{tab:coverage_poisson} and~\ref{tab:coverage_gamma} report the empirical coverage probabilities (Cvge) and average confidence interval lengths (Len) for two representative settings: the Poisson GLMM with $\siga^2=1$, and the Gamma GLMM with $\siga^2=0.5$ and $\sige^2=0.5$.  The tables show patterns consistent with the asymptotic theory. Parameters estimated at the faster $n$-rate behave well even when both $g$ and $m_L$ are small: the confidence intervals for $\beta_2^{(w)}$ attain coverage close to the nominal level across almost all settings, and the same holds for $\sige^2$ in the Gamma model. Moving down each column, with $m_L$ fixed and $g$ increasing, the coverage probabilities generally move towards the nominal level, especially for the between-cluster parameters and $\siga^2$. In contrast, moving across each row, with $g$ fixed and $m_L$ increasing, the coverage for $\beta_0^{(b)}$, $\beta_1^{(b)}$, and $\siga^2$ changes only mildly, reflecting that these parameters are mainly governed by between-cluster information. Along the diagonal, where both $g$ and $m_L$ increase, the coverage probabilities are the most stable and closest to the nominal level. The three standard error estimators for $\siga^2$ also highlight the finite-sample effect of estimating the random effects. The empirical sample-moment standard error uses estimated random-effect modes, which are shrinkage estimates in finite samples because the normal random-effect density pulls them towards zero. This tends to underestimate the variability of the latent random effects, producing narrow confidence intervals and undercoverage. The corrected empirical-moment standard error proposed in \eqref{correction moment} incorporates the posterior variance of the random effects and noticeably improves coverage relative to the uncorrected empirical-moment method. Nevertheless, in small and moderate samples, the corrected standard errors can still be smaller than those obtained under the normality-based calculation. The differences among the three methods diminish as $g$ and $m_L$ increase, since larger $m_L$ improves the accuracy of the random-effect modes and larger $g$ improves estimation of $\siga^2$.
The binomial results and the remaining variance-component and dispersion settings are provided in Supplementary  Section~S.8, with qualitatively similar conclusions

\subsection{Example Data}
We illustrate the use of our asymptotic results on the human breast
single-cell dataset from \cite{reed2024single}. The dataset contains breast
tissue samples from 55 female donors who underwent reduction mammoplasty
or prophylactic mastectomy, with individual cells classified as epithelial,
immune or stromal. We investigate whether donor characteristics are
associated with the relative abundance of immune cells in breast tissue.
The donor-level explanatory variables include age, parity and BRCA status (Pathogenic BRCA1 and BRCA2 variants confer increased breast cancer risk and are therefore clinically relevant characteristics of these donors).
We discarded six donors with unknown parity or BRCA status, leaving
\(g=49\) donors. We standardised age, treated parity, defined as the
number of previous births, as a numerical variable taking values from
0 to 4, and introduced indicator variables for BRCA1 and BRCA2 carriers,
with wild-type and assumed-wild-type donors forming the non-carrier
reference category. To reduce computational cost, we randomly sampled at
most 2,000 cells from each donor, retaining all available cells for donors
with fewer than 2,000 cells, leaving us with data on \(n=96{,}481\)
cells with \(m_L=1005\).

For cell $j$ from donor $i$, let $y_{ij}$ denote the binary
response, where $y_{ij}=1$ if the cell is classified as an immune cell
and $y_{ij}=0$ otherwise. We assumed
$y_{ij}\mid\alpha_i\sim\operatorname{Bernoulli}(p_{ij})$, with
$p_{ij}$ defined through the logit link by
$$
\operatorname{logit}(p_{ij}) = \beta_0 +\beta_1\operatorname{age}_i + \beta_2\operatorname{parity}_i + \beta_3 I(\operatorname{BRCA1}_i) + \beta_4 I(\operatorname{BRCA2}_i) +\alpha_i,
$$
where the independent random variables
$\alpha_i\sim N(0,\siga^2)$ represent residual variation between donors.

We fitted the model by REML using the R package \texttt{glmmTMB} \textcolor{red}{\citep{brooks2017glmmtmb}}. The
REML estimates of the parameters are shown in
Table~\ref{tab:breast_data}. Across both the fixed-effect parameters and
the donor variance component, the three sets of standard errors are
generally of similar magnitude, although somewhat larger differences are
seen for the intercept and BRCA2 parameters. All these results are
broadly as expected, given that $m_L$ is large, while $g$ is only
moderately large.

\begin{table}[t]
\centering
\caption{REML parameter estimates and standard errors for the breast
single-cell model.}
\label{tab:breast_data}
\setlength{\tabcolsep}{12pt}
\begin{tabular}{lrrrr}
\hline
& Estimate & $\mathrm{SE}$ & $\mathrm{SE}_{C}$ & $\mathrm{SE}_{N}$ \\
\hline
Intercept
    & $-4.321$ & $0.345$ & $0.352$ & $0.402$ \\
Age
    & $-0.239$ & $0.255$ & $0.260$ & $0.233$ \\
Parity
    & $ 0.197$ & $0.194$ & $0.198$ & $0.187$ \\
BRCA1
    & $ 0.507$ & $0.425$ & $0.432$ & $0.446$ \\
BRCA2
    & $ 0.349$ & $0.442$ & $0.450$ & $0.516$ \\
\hline
Donor ($\siga^2$)
    & $1.869$ & $0.327$ & $0.338$ & $0.378$ \\
\hline
\end{tabular}
\end{table}

\section{Discussion}
\label{sec:discussion}
In this paper, we establish rigorous asymptotics for ML and REML estimation in GLMMs defined from the original, rather than approximated, marginal likelihood.
Although the corresponding score functions do not admit closed-form expressions, Laplace expansion provides explicit asymptotic representations of the scores and their derivatives, from which the asymptotic normality of both the ML and REML estimators is established. Here, the Laplace expansion serves only as an analytical device and is not required for numerical computation of the estimators, so the results remain applicable when the marginal integrals are evaluated by more accurate numerical methods. Within this framework, the REML estimator of the variance components is asymptotically equivalent to the corresponding ML estimator, and its adjustment removes the dominant profiling bias in the variance-component score. The established theory requires only finite $4+\delta$ moment conditions on the random effects rather than normality, allows the covariates to be fixed or random, and imposes no restriction on the relative divergence rates of $g$ and $m_L$. It further reveals two distinct convergence scales: the between-cluster regression coefficients and the random-effect variance are normalized at rate $g^{1/2}$, whereas the within-cluster regression coefficients and the dispersion parameter are normalized at rate $n^{1/2}$, leading to a block-diagonal limiting covariance structure.

The present analysis is confined to random-intercept GLMMs with independent clusters and a fixed-dimensional parameter vector. Extending the theory to hierarchical random-slope models would introduce multivariate random effects, additional covariance parameters, and higher-dimensional cluster-specific integrals. The corresponding likelihood equations would generally remain implicit, and the main challenge would be to obtain suitable asymptotic representations of the score functions and their derivatives from which the limiting behavior of the resulting estimators can be characterized. The framework developed here therefore provides a natural starting point for studying ML and REML estimation in more general hierarchical GLMMs.

Further extensions arise when the assumptions on the parameter dimension or variance components are relaxed. One such setting allows the fixed-effect dimension to increase with the sample size. Under the fixed-dimensional framework considered here, the ML and REML estimators are asymptotically equivalent, whereas possible asymptotic advantages of REML may emerge when $p$ increases with the sample size. In mixed ANOVA models, \cite{jiang1996reml} showed that REML can remain consistent and asymptotically normal under weaker growth conditions than ML when the dimension of fixed effects $p$ diverges. Whether a similar separation occurs in GLMMs remains open. Another interesting setting arises when variance components approach the boundary of the parameter space. The present theory assumes nondegenerate variance components bounded away from zero. When variance components are allowed to approach zero, the information structure may become singular, and the standard asymptotic approximations established here need not remain uniform \citep{ekvall2022confidence}. These extensions would broaden the present theory to more general GLMM settings and therefore constitute important directions for future research.



\begin{appendix}
\section{Proofs}\label{appn}
The proofs of Theorems~\ref{theorem1} and \ref{theorem2} is given in Sections~\ref{subsec:proof of theorem 1} and \ref{subsec:proof of theorem 2}, respectively. Since the proof of Theorem~\ref{theorem1} relies on the local asymptotic expansion established for the ML estimating equation, we first prove Theorem~\ref{theorem2}, which gives the asymptotic linear representation and limiting distribution of the ML estimator. We then prove Theorem~\ref{theorem1} by showing that the REML estimating equation and its derivative are asymptotically equivalent to their ML counterparts.

\subsection{Proof of Theorem 2}
\label{subsec:proof of theorem 2}
Let $\bpsi(\btheta)=\{\bpsi_{\bbetab}^\top(\btheta),\psi_{\siga^2}(\btheta),\bpsi_{\bbetaw}^\top(\btheta),\psi_{\sige^2}(\btheta)\}^\top$ denote the ML score function, and let $\bxi$ denote its leading stochastic term evaluated at the true parameter. Their expressions and derivations are given in Supplementary, Sections S.3.4 and S.3.5.

Since $\bpsi(\btheta) = \bpsi(\boldsymbol{\dot{\theta}}) + \bpsi(\btheta) - \bpsi(\dot{\btheta})$, we write
\begin{equation*}
    \bK^{-1/2}\bpsi(\btheta) = \bK^{-1/2}\bxi - \bB\bK^{1/2}(\btheta -\dbtheta) + T_1 + T_2(\btheta) + T_3(\btheta),
\end{equation*}
where
\begin{align*}
    &\bB = \lim_{g, m_L \rightarrow \infty}\left\{-\bK^{-1/2}\mathbb{E}\nabla\bpsi(\dbtheta)\bK^{-1/2} \right\}, 
   \quad
   T_1 = \bK^{-1/2}\{\bpsi(\dbtheta)-\bxi\}, \\
    &T_2(\btheta) = \bK^{-1/2}\mathbb{E}\{\bpsi(\btheta) - \bpsi(\dbtheta)\} + \bB\bK^{-1/2}(\btheta -\dbtheta),
\\&
\text{and}\quad
T_3(\btheta) = \bK^{-1/2}\left[\bpsi(\btheta) - \bpsi(\dbtheta) - \mathbb{E}\left\{\bpsi(\btheta) - \bpsi(\dbtheta)\right\}\right].
\end{align*}
For $0<M<\infty$, define the local shrinking neighborhood $\mathcal{N} = \{\btheta:\|\bK^{1/2}(\btheta - \dot{\btheta})\| \leq M\}$. We show below that 
$$
\|T_1\| = o_p(1), \quad \sup_{\btheta\in\mathcal{N}}\|T_2(\btheta)\| = o(1), \quad
\sup_{\btheta\in\mathcal{N}}\|T_3(\btheta)\| = o(1).
$$
It then follows that, uniformly on $\mathcal{N}$,
\begin{equation}
\label{eq:asymptotic linear representation}
\bK^{-1/2}\bpsi(\btheta) = \bK^{-1/2}\boldsymbol{\xi} - \mathbf{B}\bK^{1/2}(\btheta - \dot{\btheta}) + o_p(1).
\end{equation}
Since $\mathbf{B}$ is positive definite, the right-hand side is negative for $|\bK^{1/2}(\btheta - \dot{\btheta})|=M$ sufficiently large. Then, according
to Result 6.3.4 of \cite{ortega2000iterative}, a solution to the estimating equations exists in probability and satisfies $|\bK^{-1/2}(\hat{\btheta} - \dot{\btheta})| = O_p(1)$, so $\hat{\btheta} \in \mathcal{N}$. This allows us to substitute $\hat{\btheta}$ for $\btheta$ in \eqref{eq:asymptotic linear representation} and rearrange the terms to obtain the asymptotic representation for $\hat{\btheta}$; the central limit theorem follows from the asymptotic representation and the central limit theorem for $\boldsymbol{\xi}$ that we establish in Lemma~\ref{lemma 1}.

It remains to show that the remainder terms $T_1$, $T_2(\btheta)$ and $T_3(\btheta)$
are of sufficiently
small order that they can be ignored. By Lemma~\ref{lemma 2}, $\|T_1\| = o_p(1)$. 
To handle $T_2(\btheta)$ and $T_3(\btheta)$, we adapt the one-step expansion argument originally used by \cite{bickel1975one} and later extended to maximum likelihood and REML estimation in LMMs by \cite{richardson1994asymptotic}. We further follow the technical procedure developed in \cite{lyu2022increasing}, where analogous plug-in terms are controlled in the score expansion for LMMs. For $T_2(\btheta)$, we have 
\begin{align*}
    \sup_{\btheta\in\mathcal{N}}\|T_2(\btheta)\|  &\leq  \sup_{\btheta\in\mathcal{N}} \left\| \bK^{-1/2}\left[\mathbb{E}\left\{\bpsi(\btheta) - \bpsi(\dot{\btheta})\right\} - \mathbb{E}\nabla \bpsi(\btheta)(\btheta - \dot{\btheta})\right] \right\| \\
    &+ \sup_{\btheta\in\mathcal{N}} \left\| \bK^{-1/2}\mathbb{E}\nabla \bpsi(\btheta)(\btheta - \dot{\btheta})+ \mathbf{B}\bK^{1/2}(\btheta - \dot{\btheta}) \right\| \\
    &\leq M \sup_{\btheta\in\mathcal{N}}\left\Vert \bK^{-1/2}\left\{\mathbb{E}\nabla\bpsi(\boldsymbol{\Omega}) - \mathbb{E}\nabla \bpsi(\dot{\btheta})\right\} \bK^{-1/2}\right\Vert \\ &
    + M \left\Vert -\bK^{-1/2}\mathbb{E}\nabla\bpsi(\dot{\btheta})\bK^{-1/2} - \mathbf{B} \right\Vert  \\
    &\leq M \sup_{\btheta\in\mathcal{N}}\left\Vert \bK^{-1/2}\left\{\mathbb{E}\nabla\bpsi(\boldsymbol{\Omega}) - \mathbb{E}\nabla \bpsi(\dot{\btheta})\right\} \bK^{-1/2}\right\Vert + M \left\Vert\mathbf{B}_n - \mathbf{B} \right\Vert,
\end{align*}
where each of $\boldsymbol{\Omega}$ lies between $\btheta$ and $\dot{\btheta}$ and $\mathbf{B}_n = -\bK^{-1/2}\mathbb{E}\nabla\bpsi(\dot{\btheta})\bK^{-1/2}$. We show that $\Vert\mathbf{B}_n - \mathbf{B}\Vert = o(1)$ in Lemma~\ref{lemma 3} and in Lemma~\ref{lemma 4},
$$\sup_{\btheta \in \mathcal{N}}\left\Vert\bK^{-1/2}\mathbb{E}\left\{ \nabla \bpsi(\boldsymbol{\Omega}) - \nabla \bpsi(\dot{\btheta})\right\}\bK^{-1/2}\right\Vert = o(1).$$

To handle $T_3(\btheta)$, we partition $\mathcal{N} = \{\btheta:|\bK^{1/2}(\btheta - \dot{\btheta})| \leq M\}$ into the set of $N = O(g^{1/4})$ smaller cubes $\mathcal{C} = \{\mathcal{C}(\bt)\}$, where $\mathcal{C}(\bt) = \{\btheta:\|\bK^{1/2}(\btheta - \bt)\| \leq Mg^{-1/4}\}$. Then, we have
\begin{equation}
\begin{aligned}
\label{eq:H2}
\sup_{\btheta\in\mathcal N}
\|T_3(\btheta)\|
&\leq
\max_{1\leq k\leq N}\left\|\bK^{-1/2}\left[\bpsi(\bt_k)-\bpsi(\dot{\btheta})-\mathbb{E}\left\{\bpsi(\bt_k)-\bpsi(\dot{\btheta})\right\}\right]\right\| \\
&\quad+
\max_{1\leq k\leq N}\sup_{\btheta\in \mathcal{C}(\bt_k)}\left\|\bK^{-1/2}\left[\bpsi(\btheta)-\bpsi(\bt_k)-\mathbb{E}\left\{\bpsi(\btheta)-\bpsi(\bt_k)\right\}\right]\right\|,\end{aligned}
\end{equation}
where $\bt_k = (\bt_{k\bbeta^{(b)}}, \bt_{k\siga^2}, \bt_{k\bbeta^{(w)}}, \bt_{k\sige^2})^\top$ is the set of indices  for the cubes in $\mathcal{C}$.

We first show that $|T_3(\btheta)| = o(1)$ holds over the set of  $\bt_k$ and then that the difference between taking the supremum over a fine grid of points and over $\mathcal{N}$ is small.

For the first part on the right-hand side of \eqref{eq:H2}, applying Chebyshev's inequality, for any $\varepsilon > 0$, we have

\begin{align*}
&\Pr\left(
\max_{1\leq k\leq N}\left\|\bK^{-1/2}\left[\bpsi(\bt_k)-\bpsi(\dot{\btheta})-\mathbb{E}\left\{\bpsi(\bt_k)-\bpsi(\dot{\btheta})\right\}\right]\right\|>\varepsilon\right) \\
&\leq
\varepsilon^{-2}\sum_{k=1}^{N}\mathbb{E}\left\|\bK^{-1/2}\left[\bpsi(\bt_k)-\bpsi(\dot{\btheta})-\mathbb{E}\left\{\bpsi(\bt_k)-\bpsi(\dot{\btheta})\right\}\right]\right\|^2 \\
&= \varepsilon^{-2}g^{-1}
\sum_{k=1}^{N}\operatorname{tr}\left[\operatorname{Var}\left\{\bpsi_{\bbeta^{(b)}}(\bt_k)-\bpsi_{\bbeta^{(b)}}(\dot{\btheta})\right\}\right] \\
&\quad+
\varepsilon^{-2}n^{-1}\sum_{k=1}^{N}\operatorname{tr}\left[\operatorname{Var}\left\{\bpsi_{\bbeta^{(w)}}(\bt_k)-\bpsi_{\bbeta^{(w)}}(\dot{\btheta})\right\}\right] \\
&\quad+
\varepsilon^{-2}g^{-1}\sum_{k=1}^{N}\operatorname{Var}\left\{\psi_{\siga^2}(\bt_k)-\psi_{\siga^2}(\dot{\btheta})\right\} \\
&\quad+
\varepsilon^{-2}n^{-1}\sum_{k=1}^{N}\operatorname{Var}\left\{\psi_{\sige^2}(\bt_k)-\psi_{\sige^2}(\dot{\btheta})\right\}.
\end{align*}

We show in Lemma~\ref{lemma 5} that all variance terms are uniformly bounded by a finite constant L.
Therefore,
\begin{align*}
&\Pr\left(\max_{1\leq k\leq N}\left\|\bK^{-1/2}\left[\bpsi(\bt_k)-\bpsi(\dot{\btheta})-\mathbb{E}\left\{\bpsi(\bt_k)-\bpsi(\dot{\btheta})\right\}\right]\right\|>\varepsilon\right) \\
&\leq
\varepsilon^{-2}LN
\left\{g^{-1}(p_b+1)+n^{-1}(p_w+1)\right\} \\
&=o(1)
\end{align*} 
by the fact that $N=O(g^{1/4})$. 

For the second part on the right-hand side of \eqref{eq:H2}, Using Taylor expansion, we get
\begin{align*}
&\max_{1\leq k\leq N}
\sup_{\btheta\in C(\bt_k)}
\left\|\bK^{-1/2}\left[\bpsi(\btheta)-\bpsi(\bt_k)-\mathbb{E}\left\{\bpsi(\btheta)-\bpsi(\bt_k)\right\}\right]\right\| \\
&\leq
\max_{1\leq k\leq N}\sup_{\btheta\in C(\bt_k)}\left\|\bK^{-1/2}\left\{\nabla\bpsi(\boldsymbol{\Omega}_k)-\mathbb{E}\nabla\bpsi(\boldsymbol{\Omega}_k)\right\}\bK^{-1/2}\right\|\left\|\bK^{1/2}(\btheta-\bt_k)\right\| \\
&\leq
M g^{-1/4}\sup_{\tilde{\btheta}_k\in\mathcal N}\left\|\bK^{-1/2}\left\{\nabla\bpsi(\boldsymbol{\Omega}_k)-\mathbb{E}\nabla\bpsi(\boldsymbol{\Omega}_k)\right\}\bK^{-1/2}\right\|,
\end{align*}
where the rows of $\boldsymbol{\omega}_k$ are between $\bt_k$ and $\btheta$. By Lemma~\ref{lemma 6}, the right-hand side is $o_p(1)$. Combining the two bounds gives
$$
\sup_{\btheta\in\mathcal N}
\|T_3(\btheta)\| = o_p(1).
$$

\subsection{Proof of Theorem 1}
\label{subsec:proof of theorem 1}
Let $\bpsi_P(\bsigma)$ and $\bpsi_R(\bsigma)$ denote the score functions associated with $\ell_P(\bsigma)$ and $\ell_R(\sigma)$, respectively. By definition,
$$\bpsi_P(\hat{\bsigma}_P) = 0, \qquad \bpsi_R(\hat{\bsigma}_R) = 0.$$

The proof proceeds in three steps. We first establish the asymptotic equivalence of $\hat{\bsigma}_R$ and $\hat{\bsigma}_P$. We then transfer this equivalence to the corresponding fixed-effect estimators through the profiling map $\bbeta(\bsigma)$. Finally, we compare $\hat{\btheta}_P$ with the ML estimator $\hat{\btheta}$.

Applying the row-wise mean-value expansion to $\bpsi_R(\hat{\bsigma}_R)$ around $\hat{\bsigma}_P$, we obtain
$$0 = \bpsi_R(\hat{\bsigma}_R) = \bpsi_R(\hat{\bsigma}_P) + \nabla_{\bsigma}\bpsi(\tilde{\bsigma})(\hat{\bsigma}_R - \hat{\bsigma}_P),$$
where each row of $\tilde{\bsigma}$ lies between $\hat{\bsigma}_R$ and $\hat{\bsigma}_P$.
Since $\bpsi_P(\hat{\bsigma}_P)$ is 0, we have 
$$
0 = \bpsi_R(\hat{\bsigma}_P) -  \bpsi_P(\hat{\bsigma}_P) + \nabla_{\bsigma}\bpsi(\tilde{\bsigma})(\hat{\bsigma}_R - \hat{\bsigma}_P).$$
After inserting the normalization matrices $\bK_{\bsigma}$, the above expansion becomes 
\begin{equation}
\label{eq:taylor expansion for REML}
-\left\{\bK_{\bsigma}^{-1/2} \nabla_{\bsigma}\bpsi_R(\tilde{\bsigma}) \bK_{\bsigma}^{-1/2}\right\} \bK_{\bsigma}^{1/2}(\hat{\bsigma}_R - \hat{\bsigma}_P) = \bK_{\bsigma}^{-1/2}\{\bpsi_R(\hat{\bsigma}_P) - \bpsi_P(\hat{\bsigma}_P) \}.
\end{equation}
By Lemma~\ref{lemma 7}, the norm of the right-hand side of \eqref{eq:taylor expansion for REML} is $o_p(1)$. For the normalized derivative matrix on the left-hand side, write
\begin{equation}
\label{eq:decomposition of REML derivative matrix}
\begin{aligned}
-\bK_{\bsigma}^{-1/2} \nabla_{\bsigma}\bpsi_R(\tilde{\bsigma}) \bK_{\bsigma}^{-1/2} &= -\bK_{\bsigma}^{-1/2} \nabla_{\bsigma}\bpsi_P(\tilde{\bsigma}) \bK_{\bsigma}^{-1/2} \\
&\quad - \bK_{\bsigma}^{-1/2} \nabla_{\bsigma}\{\bpsi_R(\tilde{\bsigma}) - \bpsi_P(\tilde{\bsigma})\}\mathbf K_{\bsigma}^{-1/2}.
\end{aligned}
\end{equation}
By Lemma~\ref{lemma 8}, the first term on the right-hand side of \eqref{eq:decomposition of REML derivative matrix} converges to a finite positive definite matrix, whereas Lemma~\ref{lemma 9} shows that the norm the second term is $o_p(1)$. Hence, the normalized derivative matrix is asymptotically nonsingular. It follows from \eqref{eq:taylor expansion for REML} that $$\bK_{\bsigma}^{1/2}(\hat{\bsigma}_R - \hat{\bsigma}_P) = o_p(1),$$ which implies the $\hat{\bsigma}_R$ is asymptotically equivalent to $\hat{\bsigma}_P$. 

We next compare the corresponding fixed-effect estimators. By construction,
$$\hat{\bbeta}_R = \hat{\bbeta}(\hat{\bsigma}_R), \qquad \hat{\bbeta}_P = \hat{\bbeta}(\hat{\bsigma}_P),$$ We establish in Lemma~\ref{lemma 10} that $$\bK_{\bbeta}^{1/2}(\hat{\bbeta}_R - \hat{\bbeta}_P) = o_p(1),$$ where $\bK_{\bbeta} = \operatorname{diag}(g\boldsymbol{1}_{p_b}, n\boldsymbol{1}_{p_w})$. Combining this with asymptotic equivalence between $\hat{\bsigma}_R$ and $\hat{\bsigma}_P$, we have $$\bK^{1/2}(\hat{\btheta}_R - \hat{\btheta}_P) = o_p(1).$$

It remains to compare $\hat{\btheta}_P$ with $\hat{\btheta}$. By Lemma~\ref{lemma 11}, maximizing the $\ell_P(\bsigma)$ over $\bsigma$ and recovering $\bbeta$ through the profiling map $\bbeta(\bsigma)$ is equivalent to jointly maximizing $\ell(\bbeta, \sigma)$ over $(\bbeta, \bsigma)$. Hence $\hat{\btheta}_P$ is the joint maximizer of $\ell(\bbeta, \bsigma)$, which implies $\hat{\btheta}_P = \hat{\btheta}$
Consequently,
$$
\bK^{1/2}\left(\hat{\btheta}_R - \hat{\btheta}\right) = \bK^{1/2}\left(\hat{\btheta}_R - \hat{\btheta}_P\right) = o_p(1).
$$
Thus, the REML estimator is asymptotically equivalent to the ML estimator. It follows from Theorem~\ref{theorem2} that
$$\bK^{1/2}(\hat{\btheta}_R - \dbtheta) = \bB^{-1}\bK^{-1/2}\bxi + o_p(1).$$
Restricting this relation to the variance-component block gives
$$\bK^{1/2}_{\bsigma}(\hat{\bsigma}_R - \dbsigma) = \bB^{-1}\bK^{-1/2}\bxi_{\sigma} + o_p(1).$$

\subsection{Lemmas for the Estimating Function}
We collect below the lemmas used in the proofs of Theorems~\ref{theorem1} and \ref{theorem2}. Lemma~\ref{lemma 1}-\ref{lemma 6} show that $\bpsi(\dbtheta)$ approximates $\bxi$, a central limit theorem holds for $\bxi$, $T_1 = o_p(1)$, and the variances of $\bpsi(\btheta) - \bpsi(\dbtheta)$ are uniformly bounded. Lemma~\ref{lemma 7}-\ref{lemma 11} establish the asymptotic equivalence between the REML and profile estimating equations and transfer this equivalence to the corresponding estimators. Detailed proofs are provided in the Supplementary, Sections S.6.1-S.6.11.

\begin{lemma}
\label{lemma 1}
Suppose Condition A holds. Then, as $g,m_L\to\infty$, $\bK^{-1/2}\boldsymbol{\xi}
\overset{D}{\longrightarrow}
N(\boldsymbol{0},\mathbf{A})$, where
$$
\mathbf{A}
=
\begin{pmatrix}
\bC_2/\dot\siga^2
&
\mathbb{E}(\dot\alpha_i^3)\bc_1 / 2\dot\siga^6

& \boldsymbol{0}_{[p_b:p_w]} & \boldsymbol{0}_{[p_b:1]}
\\
\mathbb{E}(\dot\alpha_i^3)\bc_1^\top / 2\dot{\siga}^6
& \{\mathbb{E}(\dot\alpha_i^4)-\dot\siga^4\}/
4\dot\siga^8
& \boldsymbol{0}_{[1:p_w]} & 0
\\
\boldsymbol{0}_{[p_w:p_b]}
& \boldsymbol{0}_{[p_w:1]}
& \bC_3(\dbtheta) & \boldsymbol{0}_{[p_w:1]}
\\
\boldsymbol{0}_{[1:p_b]} & 0 & \boldsymbol{0}_{[1:p_w]} & c_4(\dbtheta)
\end{pmatrix}.
$$
\end{lemma}

\begin{lemma}
\label{lemma 2}
Suppose Condition A holds. Then $\|T_1\| = o_p(1).$  
\end{lemma}

\begin{lemma}
\label{lemma 3}
Suppose Condition A holds. Then, as $g,m_L\to\infty, \left\|\mathbf B_n- \mathbf B\right\|=o(1),$ where
$\mathbf{B}_n = -\bK^{-1/2} \mathbb{E}\{\nabla_{\btheta} \bpsi(\dot{\btheta})\}\bK^{-1/2} $
and 
\begin{equation}\label{matrix B}
\mathbf B
=
\begin{pmatrix}
\bC_2 / \dot\siga^2 & \boldsymbol 0_{[p_b:1]} & \boldsymbol 0_{[p_b:p_w]} & \boldsymbol 0_{[p_b:1]}
\\
\boldsymbol 0_{[1:p_b]} & 1/2\dot\siga^{4} & \boldsymbol 0_{[1:p_w]} & 0
\\
\boldsymbol 0_{[p_w:p_b]} & \boldsymbol 0_{[p_w:1]} & \bC_3(\dbtheta) & \boldsymbol 0_{[p_w:1]}
\\
\boldsymbol 0_{[1:p_b]} & 0 & \boldsymbol 0_{[1:p_w]} & c_4(\dbtheta)
\end{pmatrix}.
\end{equation}
\end{lemma}

\begin{lemma}
\label{lemma 4}
Suppose condition A holds. Then, as $g,m_L \rightarrow \infty$,
$$\sup_{\btheta \in \mathcal{N}}\left\Vert\bK^{-1/2}\mathbb{E}\left\{\nabla_{\btheta} \bpsi(\boldsymbol{\Omega}) - \nabla_{\btheta} \bpsi(\dot{\btheta})\right\}\bK^{-1/2}\right\Vert = o(1).$$
\end{lemma}

\begin{lemma}
\label{lemma 5}
Suppose Condition A holds. Then there exists a finite constant $L$ such that
\begin{align*}
    &\sup_{\btheta\in\mathcal N}\tr\var\left\{\bpsi_{\bbeta^{(b)}}(\btheta)-\bpsi_{\bbeta^{(b)}}(\dot{\btheta})\right\}\leq L,
    &&\sup_{\btheta\in\mathcal N}\var\left\{\psi_{\siga^2}(\btheta)-\psi_{\siga^2}(\dot{\btheta})\right\}\leq L, 
    \\
    &\sup_{\btheta\in\mathcal N}\tr\var\left\{\bpsi_{\bbeta^{(w)}}(\btheta)-\bpsi_{\bbeta^{(w)}}(\dot{\btheta})\right\}\leq L, 
    &&\sup_{\btheta\in\mathcal{N}}\var\left\{\psi_{\sige^2}(\btheta)-\psi_{\sige^2}(\dot{\btheta})\right\}\leq L. 
\end{align*}
\end{lemma}

\begin{lemma}
\label{lemma 6}
Suppose Condition A holds. Then, as $g,m_L\to\infty$,
$$
\sup_{\tilde{\btheta}_k\in\mathcal N}
g^{-1/4}\left\|\bK^{-1/2}\left\{\nabla_{\btheta}\bpsi(\boldsymbol{\Omega}_k)-\mathbb{E}\nabla_{\btheta}\bpsi(\boldsymbol{\Omega}_k)\right\}\bK^{-1/2}\right\|=o_p(1).
$$
\end{lemma}

\begin{lemma}
\label{lemma 7}
Suppose Condition A holds. Then, as $g, m_L \rightarrow \infty$ 

$$\sup_{\bsigma \in \mathcal{N}_{\bsigma}}\left\|
\bK_{\bsigma}^{-1/2}
\left\{
\bpsi_R(\bsigma)
-
\bpsi_P(\bsigma)
\right\}\right\|
=
o_p(1).
$$
\end{lemma}

\begin{lemma}
\label{lemma 8}
Suppose Condition A holds. Define $\bB_{\bsigma n}(\bsigma)=-\bK_\sigma^{-1/2}\nabla_{\bsigma}\bpsi_P(\bsigma)\bK_\sigma^{-1/2}.$
Then, as $g,m_L\to\infty$,
$$
\sup_{\bsigma \in \mathcal{N}_{\bsigma}}\left\|\bB_{\bsigma n}(\bsigma)
-
\bB_{\bsigma}(\bsigma)\right\| = o_p(1),
$$
where $\bB_{\bsigma}(\bsigma)
= \diag\{1/(2\siga^4), c_4(\bbeta(\bsigma), \bsigma)\}$ is finite and positive definite.
\end{lemma}

\begin{lemma}
\label{lemma 9} Suppose Condition A holds. Then, as $g, m_L \rightarrow \infty$,
$$
\sup_{\bsigma \in \mathcal{N}_{\bsigma}} \left\|
\bK_{\bsigma}^{-1/2}
\nabla_{\bsigma}
\left\{
\bpsi_R(\bsigma) - \bpsi_P(\bsigma)
\right\}
\bK_{\bsigma}^{-1/2} \right\|
=
o_p(1).$$
\end{lemma}

\begin{lemma}
\label{lemma 10}
Suppose Condition A holds and the profile map
$\bbeta(\bsigma)$ is continuously differentiable on $\mathcal{N}_{\bsigma}$. 
Then,
$$
\bK_{\bbeta}^{1/2}
\left(\hat{\bbeta}_R-\hat{\bbeta}_P\right)
=o_p(1),
$$
provided that
$$
\bK_{\bsigma}^{1/2}
\left(\hat{\sigma}_R-\hat{\sigma}_P\right)
=o_p(1).
$$
\end{lemma}

\begin{lemma}
\label{lemma 11} For every fixed $\bsigma$, suppose that $\bbeta(\bsigma) = \arg\max_{\bbeta}\ell(\bbeta, \bsigma)$. Then  
$$\max_{\bsigma} \ell_P(\bsigma) = \max_{\bbeta, \bsigma}\ell(\bbeta, \bsigma).$$ If  $\hat{\bsigma} \in \arg\max_{\bsigma}\ell_P(\bsigma)$, then $\btheta_P = \left\{\hat{\bbeta}_P^\top, \hat{\bsigma}_P^\top\right\}^\top$ is a joint maximizer of $\ell(\bbeta, \bsigma)$, and hence is an ML estimator.
\end{lemma}

\end{appendix}






\begin{acks}[Acknowledgments]
The authors would like to thank the anonymous referees, an Associate
Editor and the Editor for their constructive comments that improved the
quality of this paper. Huadong Mo and Ziyang Lyu are joint corresponding authors.
\end{acks}

\begin{funding}
Ziyang Lyu is supported by the Australian Research Council Discovery Projects DE260101297.
\end{funding}

\begin{supplement}
\stitle{Additional information: Results, calculations, and proofs.}
\sdescription{Comprehensive calculations, detailed proofs, and formal statements support the findings presented in the main manuscript.}
\end{supplement}
\begin{supplement}
\stitle{Simulation resources:}
\sdescription{This part includes the R code used for simulations and real data application.} 
\end{supplement}
 
\bibliographystyle{imsart-number}
\bibliography{reference}

\end{document}


\title{
{\LARGE  Supplementary for:}\\[0.5em]
{\LARGE
Asymptotic Theory for Restricted Maximum Likelihood Estimators in Generalized Linear Mixed Models
}
}
\author{Zhanzhongyu Gao, Huadong Mo, Ziyang Lyu}
\date{}
\maketitle

\section{General Mean-Value Argument}
We repeatedly use the following consequence of the mean-value theorem
throughout the Supplementary. Let $f:\mathcal{I}\to\mathbb{R}$
be continuously differentiable on an interval $\mathcal{I}$. For any
$z,z_0\in\mathcal{I}$, there exists some $\tilde{z}$ between
$z$ and $z_0$ such that
$$
f(z)-f(z_0)
=
f'(\tilde{z})(z-z_0).
$$
Hence, if $f'$ is uniformly bounded on $\mathcal{I}$, then
$$
|f(z)-f(z_0)|
\leq
C|z-z_0|
$$
for some finite constant $C$, uniformly over $z,z_0\in\mathcal{I}$. More generally, suppose that $f$ is $q$-times continuously differentiable on $\mathcal{I}$. Then for any $k=0,\ldots,q-1$, there exists some $\tilde{z}_k$ between $z$ and $z_0$ such that
$$
f^{(k)}(z)-f^{(k)}(z_0)
=
f^{(k+1)}(\tilde{z}_k)(z-z_0).
$$ 
Hence, if $f^{(k+1)}$ is
uniformly bounded on $\mathcal{I}$, then
$$
|f^{(k)}(z)-f^{(k)}(z_0)|
=
O(|z-z_0|),
$$
uniformly over $z,z_0\in\mathcal{I}$. In what follows, we refer to applications of these relations simply as the mean-value argument.

\section{Notation and Stochastic Orders of Key Summation Quantities}
The quantities introduced in this section appear repeatedly in the score functions and their higher-order derivatives. For ease of reference, we collect their definitions and establish their stochastic orders. These preliminary results will be used throughout the Supplementary to identify the leading terms and control the remainder terms in the score and score-derivative expansions.

For a scalar-valued function $f$ of a $d$-dimensional argument  $\ba = (a_1, \ldots, a_d)^\top$, define $$\nabla_{\ba}f = \left(\frac{\partial f}{\partial a_1}, \ldots, \frac{\partial f}{\partial a_d}\right)^\top.$$
For a vector-valued function $\mathbf{f} = (f_1, \ldots, f_g)^\top$, define $\nabla_{\ba} \mathbf{f} = (\nabla_{\ba}f_1,\ldots,\nabla_{\ba}f_g)^\top$. Repeated differentiation with respect to the same argument is abbreviated as $$\nabla_{\ba}^k f = \underbrace{\nabla_{\ba} \ldots \nabla_{\ba}}_{k \quad \text{times}}f.$$ 
Derivatives with respect to different arguments are not combined. For example, we write $\nabla_{\ba} \nabla_{\bb} f$, where the rightmost differential operator acts first. Thus,
$$[\nabla_{\ba} \nabla_{\bb} f]_{rs} = \frac{\partial^2 f}{\partial \nabla_{a_r} \partial \nabla_{b_r}}.$$
For notational convenience, define the conditional residual at the general parameter $\btheta$ by $e_{ij} = y_{ij} - b^{(1)}(\eta_{ij})$, it satisfies that $$\mathbb{E}_{\btheta}(e_{ij}\mid \bx_{ij}, \alpha_i) = 0, \qquad \mathbb{E}_{\btheta}(e^2_{ij}\mid \bx_{ij}, \alpha_i) = \sige^2b^{(2)}(\eta_{ij}).$$

For each $1\le i\le g$, define 
$$
S_i=\sumjmi\frac{e_{ij}}{\sige^2},\qquad\boldsymbol{S}_i^{(b)}=\bxbi S_i, \qquad \boldsymbol{S}_i^{(w)}=\sumjmi\bxwij\frac{e_{ij}}{\sige^2}.
$$
For each integer $k\geq2$ such that $b^{(k)}(\cdot)$ exits, define 
\begin{align*}
&\tau_{i,k}=\sumjmi\frac{b^{(k)}(\eta_{ij})}{\sige^2}, \qquad \btau_{i,k}^{(b)} = \bxbi\tau_{i,k}, \qquad \btau_{i,k}^{(w)} = \sumjmi\bxwij\frac{b^{(k)}(\eta_{ij})}{\sige^2},\\
&\btau_{i,k}^{(bb)} = \bxbi \bxbit \tau_{i,k}, \quad \btau_{i,k}^{(bw)} = \bxbi \btau_{i,k}^{(w)}, \quad \btau_{i,k}^{(ww)}=\sumjmi\bxwij \bxwijt\frac{b^{(k)}(\eta_{ij})}{\sige^2}.
\end{align*}
Also define
$$
A_i=\tau_{i,2}+\siga^{-2}.
$$
The following conditional moment identities will be used repeatedly in this Supplementary:
\begin{align*}
\mathbb{E}_{\btheta}(S_i\mid \bxbi,\bxwij,\alpha_i)=0, \quad
\mathbb{E}_{\btheta}(\boldsymbol{S}_i^{(b)}\mid \bxbi,\bxwij,\alpha_i)=\boldsymbol{0}, \quad
\mathbb{E}_{\btheta}(\boldsymbol{S}_i^{(w)}\mid \bxbi,\bxwij,\alpha_i)=\boldsymbol{0}.
\end{align*}
Moreover,
\begin{align*}
&\mathbb{E}_{\btheta}(S_i^2\mid \bxbi,\bxwij,\alpha_i)=\tau_{i,2}, &&\mathbb{E}_{\btheta}\{S_i\boldsymbol{S}_i^{(b)\top}\mid \bxbi,\bxwij,\alpha_i\}=\btau_{i,2}^{(b)\top}, \\
&\mathbb{E}_{\btheta}\{S_i\boldsymbol{S}_i^{(w)\top}\mid \bxbi,\bxwij,\alpha_i\}
=\btau_{i,2}^{(w)\top}, &&\mathbb{E}_{\btheta}\{\boldsymbol{S}_i^{(b)}\boldsymbol{S}_i^{(b)\top}\mid \bxbi,\bxwij,\alpha_i\}=\btau_{i,2}^{(bb)}, \\
&\mathbb{E}_{\btheta}\{\boldsymbol{S}_i^{(w)}\boldsymbol{S}_i^{(w)\top}\mid \bxbi,\bxwij,\alpha_i\}
=\btau_{i,2}^{(ww)}, &&\mathbb{E}_{\btheta}\{\boldsymbol{S}_i^{(w)}\boldsymbol{S}_i^{(w)\top}\mid \bxbi,\bxwij,\alpha_i\}=\btau_{i,2}^{(ww)}.
\end{align*}

Throughout the Supplementary, for any column vector $\bv$, write
$$\bv^{\otimes 2}=\bv \bv^\top.$$
Under Condition A, the local linear predictors $\eta_{ij}$ remain in a compact set uniformly over $\btheta \in \mathcal{N}.$ For a nondegenerate exponential family, $b^{(2)}(\cdot)$ is continuous and strictly positive on this compact set as $\sige^2b^{(2)}(\eta_{ij})$ is the conditional variance of the response $y_{ij}$. Therefore, uniformly over $\btheta \in \mathcal{N}$, $b^{(2)}(\cdot)$ is bounded and away from zero. We also assumed that $b(\cdot)$ has bounded higher-order derivatives $(k \geq 3)$ when they are explicitly invoked. 

For each fixed $\btheta$,
$$
S_i=O_p(m_i^{1/2}),
\qquad
\boldsymbol{S}_i^{(b)}=O_p(m_i^{1/2})\boldsymbol{1}_{[p_b:1]},
\qquad
\boldsymbol{S}_i^{(w)}=O_p(m_i^{1/2})\boldsymbol{1}_{[p_w:1]},
$$
For each $k\geq2$, uniformly over $\btheta \in \mathcal{N}$
\begin{align*}
 &   \tau_{i,k}=O_p(m_i),
&&
\btau_{i,k}^{(b)}=O_p(m_i)\boldsymbol{1}_{[p_b:1]},
&&
\btau_{i,2}^{(w)}=O_p(m_i)\boldsymbol{1}_{[p_w:1]},
\\
&
\btau_{i,2}^{(bb)}
=
O_p(m_i)\boldsymbol{1}_{p_b}^{\otimes 2},  
&&
\btau_{i,2}^{(bw)}
=
O_p(m_i)\boldsymbol{1}_{[p_b:p_w]}, 
&&
\btau_{i,2}^{(ww)}
=
O_p(m_i)\boldsymbol{1}_{p_w}^{\otimes 2},
\end{align*}
and
$$
A_i=O_p(m_i),
\qquad
A_i^{-1}=O_p(m_i^{-1}).
$$

\newpage
\section{Derivation of the ML Score $\bpsi(\btheta)$ and $\nabla_{\btheta} \bpsi(\btheta)$}
\setcounter{equation}{0} 
\subsection{Cluster-wise Laplace Expansion of the Marginal Log-likelihood $\ell(\btheta)$} 
Recall the eaxct marginal log-likelihood defined in the main text $$\ell(\btheta) = -\frac{g}{2} \log \siga^2 + \sumig \log \int \exp\{h_i(\by_i,\alpha_i \mid \btheta)\} d\alpha_i,$$
where
\begin{align}
h_i(\by_i,\alpha_i\mid\btheta)
=
\sum_{j=1}^{m_i}
\left[
\frac{y_{ij}\eta_{ij}-b(\eta_{ij})}{\sige^2}
+
c(y_{ij},\sige^2)
\right]
-
\frac{\alpha_i^2}{2\siga^2}.\label{equ h_i}
\end{align}
The cluster-specific integrals are generally unavailable in closed form, except in the linear mixed model case \citep{jiang2001mixed}. We therefore apply a Laplace expansion to each integral. For $i = 1, \ldots, g$, let $\alpha_i^*$ denote the unique maximizer of $h_i(\by_i, \alpha_i \mid \btheta)$ satisfying $$\nabla_{\alpha_i}h_i(\by_i, \alpha_i^* \mid \btheta) = 0,$$
and 
$$
-\nabla_{\alpha_i^2} h_i(\by_i, \alpha_i \mid \btheta)\Big|_{\alpha_i = \alpha_i^*}  = \tau_{i,2}^* + \siga^{-2} = A_i^*.
$$
Throughout the Supplementary, a superscript $*$ indicates that the corresponding local quantity is evaluated at $\alpha_i = \alpha_i^*$.

Following the standard Laplace expansion, define the transformation $\alpha_i \mapsto u_i(\alpha_i)$ such that
$$h_i(\by_i, \alpha_i \mid \btheta) - h_i(\by_i, \alpha_i^* \mid \btheta) = -\frac12A_i^*u_i^2,$$
and let $J_i(u_i)$ denote the Jacobian of this transformation. Then,
$$\int \exp\{h_i(\by_i, \alpha_i \mid \btheta)\}d\alpha_i = \exp\{h_i(\by_i, \alpha_i^* \mid \btheta)\int\exp\left\{-\frac12A_i^*u_i^2\right\}J_i(u_i)du_i.$$
Expanding $J_i(u_i)$ around the origin gives the standard higher-order Laplace expansion \citep{shun1995laplace}
$$\int \exp\{h_i(\by_i, \alpha_i \mid \btheta)\}d\alpha_i  = \exp\{h_i(\by_i, \alpha_i^* \mid \btheta)\}\left(\frac{2\pi}{A_i^*}\right)^{1/2}(1+\epsilon_i),$$ where $\epsilon_i$ collects the higher-order terms generated by the Taylor expansion of $J_i(u_i)$. If the higher order derivative of $b(\cdot)$ exits, we have 
$$\epsilon_i = \epsilon_{i1} + \epsilon_{i2} + \ldots,$$
where the leading correction term is
$$
\epsilon_{i1} = -\frac{\tau_{i,4}^*}{8(A_i^*)^2}+\frac{5(\tau_{i,3}^*)^2}{24(A_i^*)^3}.
$$ 
Under Condition A and the mean-value argument, $$A_i^* = O(m_i), \quad \tau_{i,3}^* = O_p(m_i), \quad \tau_{i,4}^* = O_p(m_i).$$ Hence $\epsilon_{i1} = O_p(m_i^{-1})$. 
The remaining terms $\epsilon_{ir}$, $r\ge2$, are of smaller order. Further details on the higher-order expansion can be found in  \cite{tierney1986accurate, shun1995laplace}. Hence, as $m_L\to \infty$, we have 
$$\epsilon_i = O_p(m_i^{-1}), \qquad \log(1+\epsilon_i) = O_p(m_i^{-1}).$$
Substituting the cluster-wise expansions into the marginal log-likelihood yields 
\begin{equation}
\label{eq:LA decomposition}
\ell(\btheta) = \ell_L(\btheta) + R(\btheta),
\end{equation}
where the leading criterion is 
$$
\ell_L(\btheta) = -\frac{g}{2}\log\siga^2 +\frac{g}{2}\log(2\pi)+ \sum_{i=1}^g \left\{ h_i(\by_i, \alpha^* \mid \btheta) - \frac{1}{2}\log A_i^* \right\},
$$ 
and the Laplace remainder is defined by
$$
R(\btheta) = \sumig \log(1+\epsilon_i).
$$
Thus, $\ell_L(\btheta)$ contains the leading contribution from each cluster-specific integral, whereas $R(\btheta)$ collects all higher-order Laplace corrections.

\subsection{Expansion of the Conditional Mode $\alpha_i^*$}
\label{subsec:expansion of alpha_i^*}
The derivation of $\bpsi(\btheta)$ and $\nabla_{\btheta} \bpsi(\btheta)$ from $\ell(\btheta)$ requires of an expansion of $\alpha_i^*$. Let $\Delta_i = \alpha_i^* - \alpha_i$. By definition, $\alpha_i^*$ is the unique solution of
$$\nabla_{\alpha_i} h_i(\by_i, \alpha_i^* \mid \btheta) = 0.$$ By the mean-value argument, there exists some $\tilde{\alpha}_i$ between $\alpha_i$ and $\alpha_i^*$ such that
$$0 = \nabla_{\alpha_i}h_i(\by_i, \alpha_i \mid \btheta) + \nabla_{\alpha_i}^2 h_i(\by_i, \tilde{\alpha}_i \mid \btheta)\Delta_i.$$
Differentiating $h_i(\by_i,\alpha_i \mid \btheta)$ in \eqref{equ h_i} with respect to $\alpha_i$, we obtain
$$
\nabla_{\alpha_i} h_i(\by_i, \alpha_i \mid \btheta) = S_i - \sigma^{-2}_\alpha \alpha_i = O_p(m_i^{1/2}),
$$ 
with $S_i=\sumjmi{e_{ij}}/{\sige^2}$,
and 
$$-\frac{\partial}{\partial \alpha_i^2}h_i(\by_i, \tilde{\alpha}_i \mid \btheta) = \tilde{A}_i =O_p(m_i),$$
where $\tilde{A}_i$ is the value of $A_i$ estimated at $\alpha_i = \tilde{\alpha}_i$.
Thus, $$\Delta_i = (S_i - \siga^2 \alpha_i) \tilde{A}_i^{-1} = O_p(m_i^{-1/2}).$$

To obtain a more precise approximation, we consider the following higher-order Taylor expansion:
\begin{align*}
   \nabla_{\alpha_i} h_i(\by_i, \alpha_i^* \mid \btheta) &= \nabla_{\alpha_i} h_i(\by_i, \alpha_i+\alpha_i^*-\alpha_i \mid \btheta) \\
    &= \nabla_{\alpha_i} h_i(\by_i, \alpha_i \mid \btheta) + \nabla_{\alpha_i}^2 h_i(\by_i, \alpha_i \mid \btheta)\Delta_i \\
    &\qquad +\frac{1}{2}\nabla_{\alpha_i}^3h_i(\by_i, \alpha_i \mid \btheta)\Delta_i^2 + O_p\left\{\nabla_{\alpha_i}^4h_i(\by_i, \alpha_i \mid \btheta)\Delta_i^3\right\}.
\end{align*}
Following the standard calculus, we have
$$\nabla_{\alpha_i}^3 h_i(\by_i, \alpha_i \mid \btheta) = -\tau_{i,3} = O_p(m_i),$$ and $$\nabla_{\alpha_i}^4h_i(\by_i, \alpha_i \mid \btheta)\Delta_i^3 = -\tau_{i,4}\Delta_i^3 = O_p(m_i) O_P(m_i^{-3/2}) = O_p(m_i^{-1/2}).$$
We then have
\begin{equation}
\label{eq:E1}
S_i - \sigma^{-2}_\alpha \alpha_i = A_i\Delta_i + \frac12\tau_{i,3}\Delta_i^2 + O_p(m_i^{-1/2}).
\end{equation}
To align with the notations used in the power-series inverse expansion raised in equation (9.43) and (9.44) of \cite{pace1997principles}, set
$$y \equiv (S_i - \sigma^{-2}_\alpha \alpha_i)A_i^{-1},\qquad  x \equiv \Delta_i,$$
and rearrange \eqref{eq:E1} as
\begin{equation*}
    \underbrace{(S_i - \sigma^{-2}_\alpha \alpha_i)A_i^{-1}}_y = \underbrace{\Delta_i}_x + \underbrace{\frac{1}{2}\tau_{i,3}A_i^{-1}\Delta_i^2 }_{ax^2} + O_p(m_i^{-3/2}).
\end{equation*}
The inverse solution derived by \cite{pace1997principles} under second-order Taylor expansion satisfies the following relationship,
$$y = x + ax^2 + O_p(m_i^{-3/2}),$$
$$\Downarrow$$
$$x = y - ay^2 + O_p(m_i^{-3/2}).$$
This leads to an approximated expression of $\alpha_i^*$ as
\begin{equation}
\label{eq:approximate of alpha_i^*}
\begin{aligned}
   \alpha^*_i &= \alpha_i + \frac{S_i - \sigma^{-2}_\alpha \alpha_i}{A_i} - \frac{1}{2} \frac{\tau_{i,3}}{A_i} \left[\frac{S_i - \sigma^{-2}_\alpha \alpha_i}{A_i}\right]^2 + O_p(m_i^{-3/2}) \\
   &= \alpha_i + \frac{S_i}{A_i} - \frac{\sigma^{-2}_\alpha \alpha_i}{A_i} - \frac{1}{2}\frac{\tau_{i,3} S^2_i}{A_i^3} + O_p(m_i^{-3/2}),
\end{aligned}
\end{equation}
with terms of order smaller than $O_p(m_i^{-3/2})$ been absorbed into the remainder.

We next derive the derivatives of $\alpha_i^*$ with respect to components in $\btheta$.
For any component $\vartheta$ in $\btheta$, implicit differentiation of $\nabla_{\alpha_i}h_i(\by_i, \alpha_i^* \mid \btheta) = 0$ gives
$$\nabla_{\vartheta}\alpha_i^* = \frac{1}{A_i^*} \nabla_{\vartheta} \nabla_{\alpha_i} h_i(\by_i, \alpha_i \mid \btheta) \Big|_{\alpha_i = \alpha_i^*},$$ where the derivative of the right-hand side treats $\alpha_i$ as fixed. The detailed expressions of these derivatives are
\begin{align*}
    &\nabla{\bbetab} \alpha_i^* = -\frac{\btau_{i,2}^{(b)*}}{A_i^*},
    &&\nabla{\bbetaw} \alpha_i^* = -\frac{\btau_{i,2}^{(w)*}}{A_i^*}, 
    &&\nabla{\siga^2} \alpha_i^* = \frac{\alpha_i^*}{\siga^4 A_i^*}, 
    &&\nabla{\sige^2} \alpha_i^* = -\frac{S_i^*}{\sige^2 A_i^*}. 
\end{align*}
Under Condition A, uniformly over $\btheta \in \mathcal{N}$,
\begin{equation}
\label{eq:d alpha_i^*/d theta}
\begin{aligned}
    &\nabla{\bbetab} \alpha_i^* = O_p(1)\boldsymbol{1}_{p_b}, \qquad \nabla{\siga^2} \alpha_i^* = O_p(m_i^{-1}), \\
    &\nabla{\bbetaw} \alpha_i^* = O_p(1)\boldsymbol{1}_{p_w}, \qquad \nabla{\sige^2} \alpha_i^* = O_p(m_i^{-1}).
\end{aligned}
\end{equation}
It is worth noting that, given $\alpha_i^*$, $\nabla_{\alpha_i} h_i(\by_i, \alpha_i^* \mid \btheta) = 0$ leads to $S_i^* = \siga^{-2}\alpha_i^* = O_p(1)$, which implies $\nabla{\sige^2} \alpha_i^* = O_p(m_i^{-1})$ rather than $O_p(m_i^{-1/2})$.
These derivatives also determine the derivatives of all starred local quantities. For $k \geq 2$, 
\begin{align*}
    \nabla_{\bbetab}\btau_{1,k}^* &= \btau_{i,k+1}^{(b)*} + \tau_{i,k+1}^* \nabla_{\bbetab}\alpha_i^* 
    \\
    &= \btau_{i,k+1}^{(b)*} - \frac{\btau_{i,k+1}^{(b)*}\tau_{1,2}^*}{A_i^*} 
    \\&
    = \btau_{i,k+1}^{(b)*}\left(1 - \frac{\tau_{1,2}^*}{A_i^*}\right) =O_p(1),
\end{align*}
and 
$$\nabla_{\bbetaw}\btau_{1,k}^* = \btau_{i,k+1}^{(w)*} + \tau_{i,k+1}^* \nabla_{\bbetaw}\alpha_i^* = \btau_{i,k+1}^{(w)*} - \frac{\tau_{i,k+1}^*\btau_{1,2}^{(w)*}}{A_i^*} = O_p(m_i).$$

\subsection{First and Second Order Derivatives of Laplace Remainder $R(\btheta)$}\label{subsec:contribution of R(theta)}
By definition
$$\bpsi(\btheta) = \nabla_{\btheta} \ell(\btheta) = \nabla_{\btheta}\ell_L(\btheta) + \nabla_{\btheta}R(\btheta),$$ and 
$$\nabla_{\btheta} \bpsi(\btheta) = \nabla^2_{\btheta} \ell(\btheta) = \nabla^2_{\btheta}\ell_L(\btheta) + \nabla^2_{\btheta}R(\btheta),$$
with $R(\btheta) = \sumig \log(1+\epsilon_i).$
In this section, we examine the contribution of $\nabla_{\btheta}R(\btheta)$ and $\nabla^2_{\btheta}R(\btheta)$.
Since the calculations are analogous across parameter blocks, we present the details for the between-cluster regression parameter $\bbetab$.

The contribution of the $i$th cluster to the score function of $\bbetab$ is
$$
\nabla_{\bbetab} \log(1+\epsilon_i) = \frac{\nabla_{\bbetab}\epsilon_i}{1+\epsilon_i}.$$ For the dominant term $\epsilon_{i1}$, direct differentiation gives
\begin{equation}
\label{eq:first derivative of epsilon_i}
\begin{aligned}
\nabla_{\bbeta^{(b)}}\epsilon_{i1} &= -\frac18 \nabla_{\bbeta^{(b)}}\tau_{i,4}^*(A_i^*)^{-2} + \frac14\tau_{i,4}^*(A_i^*)^{-3}\nabla_{\bbeta^{(b)}}A_i^*\\
&+\frac{5}{12}\tau_{i,3}^*\nabla_{\bbeta^{(b)}}\tau_{i,3}^*(A_i^*)^{-3} - \frac{5}{8}(\tau_{i,3}^*)^2(A_i^*)^{-4}\nabla_{\bbeta^{(b)}}A_i^*
\\
&= \sigma^{-2}_\alpha \left\{\frac{\btau_{i,5}^{(b)*}}{8(A_i^*)^3} - \frac{\tau_{i,4}^*\btau_{i,3}^{(b)*}}{4(A_i^*)^4} + \frac{5\tau_{i,3}^*\btau_{i,4}^{(b)*}}{12(A_i^*)^4} - \frac{5(\tau_{i,3}^*)^2\btau_{i,3}^{(b)*}}{8(A_i^*)^5}\right\} \\
&= O_p(m_i^{-2})\boldsymbol{1}_{[p_b:1]}.
\end{aligned}
\end{equation}
The derivatives of the subsequent corrections $\epsilon_{i2}, \epsilon_{i3}, \ldots$ are of small order. Thus,
$\nabla_{\bbetab} \epsilon_i = O_p(m_i^{-2})\boldsymbol{1}_{[p_b:1]}$. Since $1+\epsilon_i = O_p(1)$, it follows that $$\nabla_{\bbetab} \log(1 + \epsilon_i) = O_p(m_i^{-2})\boldsymbol{1}_{[p_b:1]},$$ 
then
$$\nabla_{\bbetab} R(\btheta) = \sumig O_p(m_i^{-2})\boldsymbol{1}_{[p_b:1]}.$$
The same argument leads to \begin{align*}
    &\nabla_{\bbetaw} R(\btheta) = \sumig O_p(m_i^{-1})\boldsymbol{1}_{[p_w:1]}, 
    && \nabla_{\siga^2} R(\btheta) = \sumig O_p(m_i^{-2}), 
    && \nabla_{\sige^2} R(\btheta) = \sumig O_p(m_i^{-1}).
\end{align*}

The contribution of the $i$th cluster to the score derivatives of $\bbetab$ is $$\nabla_{\bbetab}^2 \log(1+\epsilon_i) = \frac{\nabla^2_{\bbetab} \epsilon_i}{1+\epsilon_i} - \frac{\nabla_{\bbetab}\epsilon_i \left(\nabla_{\bbetab}\epsilon_i\right)^\top}{(1+\epsilon_i)^2}.$$
Differentiating \eqref{eq:first derivative of epsilon_i} with respect to $\bbetab$ gives
\begin{equation}
\label{eq:second derivative of epsilon_i}
\begin{aligned}
\nabla_{\bbeta^{(b)}}^2\epsilon_{i1}
=
\siga^{-4}
\Bigg\{
&
-\frac{\btau_{i,6}^{(bb)*}}{8(A_i^*)^4}
+
\frac{3\btau_{i,5}^{(b)*}\btau_{i,3}^{(b)*\top}}{8(A_i^*)^5}
+
\frac{\tau_{i,4}^*\btau_{i,4}^{(bb)*}}{4(A_i^*)^5}
+
\frac{\btau_{i,3}^{(b)*}\btau_{i,5}^{(b)*\top}}{4(A_i^*)^5}
\\
&-
\frac{\tau_{i,4}^*\btau_{i,3}^{(b)*}\btau_{i,3}^{(b)*\top}}{(A_i^*)^6}
+
\frac{5\tau_{i,3}^*\btau_{i,5}^{(bb)*}}{12(A_i^*)^5}
+
\frac{5\btau_{i,4}^{(b)*}\btau_{i,4}^{(b)*\top}}{12(A_i^*)^5}
\\
&-
\frac{5\tau_{i,3}^*\btau_{i,4}^{(b)*}\btau_{i,3}^{(b)*\top}}{3(A_i^*)^6}
-
\frac{5(\tau_{i,3}^*)^2\btau_{i,4}^{(bb)*}}{8(A_i^*)^6}
-
\frac{5\tau_{i,3}^*\btau_{i,3}^{(b)*}\btau_{i,4}^{(b)*\top}}{4(A_i^*)^6}
\\
&+
\frac{25(\tau_{i,3}^*)^2\btau_{i,3}^{(b)*}\btau_{i,3}^{(b)*\top}}{8(A_i^*)^7}
\Bigg\} \\
= O_p(m_i&^{-3})\boldsymbol{1}_{p_b}^{\otimes 2}.
\end{aligned}
\end{equation}
Consequently, $\nabla_{\bbetab}^2\epsilon_i = O_p(m_i^{-3})\boldsymbol{1}_{p_b}^{\otimes 2}.$ With the fact that $\nabla_{\bbetab}\epsilon_i \left(\nabla_{\bbetab}\epsilon_i\right)^\top = O_p(m_i^{-4})\boldsymbol{1}_{p_b}^{\otimes 2}$, we obtain $$\nabla_{\bbetab}^2 \log(1 + \epsilon_i) = O_p(m_i^{-3})\boldsymbol{1}_{p_b}^{\otimes 2},$$ 
and $$\nabla_{\bbetab}^2 R(\btheta) = \sumig O_p(m_i^{-3})\boldsymbol{1}_{p_b}^{\otimes 2}.$$ The same argument leads to
$$
\nabla_{\btheta}^{2} R(\btheta)
=
\sum_{i=1}^{g}
\begin{pmatrix}
O_p(m_i^{-3})\boldsymbol{1}_{p_b}^{\otimes 2}
&
O_p(m_i^{-3})\boldsymbol{1}_{[p_b:1]}
&
O_p(m_i^{-2})\boldsymbol{1}_{[p_b:p_w]}
&
O_p(m_i^{-2})\boldsymbol{1}_{[p_b:1]}
\\[1ex]
O_p(m_i^{-3})\boldsymbol{1}_{[1:p_b]}
&
O_p(m_i^{-3})
&
O_p(m_i^{-2})\boldsymbol{1}_{[1:p_w]}
&
O_p(m_i^{-2})
\\[1ex]
O_p(m_i^{-2})\boldsymbol{1}_{[p_w:p_b]}
&
O_p(m_i^{-2})\boldsymbol{1}_{[p_w:1]}
&
O_p(m_i^{-1})\boldsymbol{1}_{p_w}^{\otimes 2}
&
O_p(m_i^{-1})\boldsymbol{1}_{[p_w:1]}
\\[1ex]
O_p(m_i^{-2})\boldsymbol{1}_{[1:p_b]}
&
O_p(m_i^{-2})
&
O_p(m_i^{-1})\boldsymbol{1}_{[1:p_w]}
&
O_p(m_i^{-1})
\end{pmatrix}.
$$

\subsection{Exact and Asymptotic Expressions of $\bpsi(\btheta)$}
\label{subsec:expression of ML score}
By definition, 
$$\bpsi(\btheta) = \sumig \bpsi_i(\btheta),$$
where $$\bpsi_i(\btheta) = \left\{\bpsi_{i, \bbetab}^\top(\btheta), \psi_{i, \siga^2}(\btheta), \bpsi_{i, \bbetaw}^\top(\btheta), \psi_{i,\sige^2}(\btheta) \right\}^\top.$$
We first record the exact cluster-wise score expressions.
Differentiating \eqref{eq:LA decomposition} with respect to the components of $\btheta$ gives the following exact cluster-wise score contributions:
\begin{align}
&\bpsi_{i, \bbetab}(\btheta) = \boldsymbol{S}_i^{(b)*} - \frac12\frac{\sigma^{-2}_\alpha \btau^{(b)*}_{i,3}}{(A_i^*)^2} + \nabla_{\bbeta^{(b)}}\log(1+\epsilon_i), \label{eq:raw score of betab} \\
&\psi_{i, \siga^2}(\btheta) = -\frac{1}{2\siga^2} + \frac{(\alpha_i^*)^2}{2\sigma^4_\alpha} + \frac{1}{2\sigma^4_\alpha A_i^*} - \frac{\tau_{i,3}^*\alpha_i^*}{2\sigma^4_\alpha(A_i^*)^2} + \nabla_{\siga^2}\log(1+\epsilon_i), \label{eq:raw score of siga} \\
&\bpsi_{i, \bbetaw}(\btheta) = \boldsymbol{S}_i^{(w)*}-\frac{1}{2}\left\{\frac{\btau_{i,3}^{(w)*}}{A_i^*} - \frac{\tau^*_{i,3} \btau_i^{(w)*}}{(A_i^*)^2} \right\} + \nabla_{\bbeta^{(w)}}\log(1+\epsilon_i), \label{eq:raw score of betaw} \\
& \psi_{i,\sige^2}(\btheta) = \sum_{j=1}^{m_i} \left[-\frac{y_{ij}\eta_{ij}^* - b(\eta_{ij}^*)}{\sigma^4_e} + \nabla_{\sige^2} c(y_{ij}, \sige^2)\right] + \frac{1}{2}\frac{\tau_{i,2}^*}{\sige^2 A_i^*} \notag \\ &\qquad \qquad + \frac{1}{2}\frac{\tau_{i,3}^* S_i^*}{\sige^2(A_i^*)^2} + \nabla_{\sige^2}\log(1+\epsilon_i). \label{eq:raw score of sige}
\end{align}
To obtain tractable asymptotic representations, we substitute \eqref{eq:approximate of alpha_i^*} into these exact score expressions. Since these expressions are originally evaluated at $\alpha_i^*$, the resulting terms involve local quantities such as $S_i^*$, $A_i^*$, and $\tau_{i,k}^*$. For the asymptotic representation, we further replace these quantities by their counterparts evaluated at the general latent variable $\alpha_i$. For brevity, we illustrate this replacement only for $\bpsi_{i, \bbetab}$; the other score components are handled in the same way. The derivatives of the Laplace correction are controlled by the bounds established in Section~\ref{subsec:contribution of R(theta)}. We present the details for each parameter block below.

For $\bbetab$, the equation $\partial h(\by_i, \alpha_i^* \mid \btheta) / \partial \alpha_i = 0$ leads to $S_i^* = \sigma^{-2}_\alpha \alpha^*_i$. Thus, \eqref{eq:raw score of betab} can be rewrite as
$$
\bpsi_{i, \bbeta^{(b)}} = \bxbi\sigma^{-2}_\alpha \alpha^*_i - \frac12\frac{\sigma^{-2}_\alpha \btau^{(b)*}_{i,3}}{(A_i^*)^2} + \nabla_{\bbeta^{(b)}}\log(1+\epsilon_i).
$$
Substitute \eqref{eq:approximate of alpha_i^*} into the above equation, the first term  
is  \begin{align*}
\sigma^{-2}_\alpha\left\{\bxbi\alpha_i + \frac{\boldsymbol{S}_i^{(b)}}{A_i} - \frac{ \bxbi\sigma^{-2}_\alpha \alpha_i} {A_i} - \frac{1}{2}\frac{\btau^{(b)}_{i,3} S_i^2}{A_i^3} \right\} + O_p(m_i^{-3/2})\boldsymbol{1}_{[p_b:1]}.
\end{align*}
For the second term, since $\Delta_i = O_p(m_i^{-1/2})$, the mean-value argument gives $$\btau_{i,3}^* - \btau_{i,3} = \tilde{\tau}_{i,4}\Delta_i= O_p(m_i^{1/2})\boldsymbol{1}_{[p_b:1]},$$ where $\tilde{\tau}_{i,4}$ is the value of $\tau_{i,4}$ evaluated at $\alpha_i = \tilde{\alpha_i}$ for some $\tilde{\alpha}_i$ between $\alpha_i$ and $\alpha_i^*$. Similarly, $A_i^* - A_i = O_p(m_i^{1/2})$. Then we have 
$$\frac{\btau_{i,3}^{(b)*}}{(A_i)^*} - \frac{\btau_{i,3}^{(b)}}{A_i} =\frac{\btau_{i,3}^{(b)*} - \btau_{i,3}^{(b)}}{(A_i^*)^2} + \btau_{i,3}^{(b)}\frac{(A_i - A_i^*)(A_i + A_i^*)}{(A_i^*)^2A_i^2} = O_p(m_i^{-3/2})\boldsymbol{1}_{[p_b:1]}.$$
Therefore, the contribution of the second term is $$\frac12\frac{\sigma^{-2}_\alpha \btau^{(b)}_{i,3}}{A_i^2} + O_p(m_i^{-3/2})\boldsymbol{1}_{[p_b:1]}.$$
Combining the contribution established above with the contribution of Laplace remainder in Section~\ref{subsec:contribution of R(theta)}, we have
\begin{equation}\label{eq:asymptotic score of betab}
    \begin{split}    
\bpsi_{i, \bbeta^{(b)}}(\btheta) 
=& \sigma^{-2}_\alpha\left\{\bxbi\alpha_i + \frac{\boldsymbol{S}_i^{(b)}}{A_i} - \frac{ \bxbi\sigma^{-2}_\alpha \alpha_i} {A_i} - \frac{1}{2}\frac{\btau_{i,3}^{(b)} S_i^2}{A_i^3} - \frac12\frac{ \btau_{i,3}^{(b)}}{A_i^2} \right\} 
\\&
+ O_p(m_i^{-3/2})\boldsymbol{1}_{[p_b:1]}.
    \end{split}
\end{equation}

For $\siga^2$, substituting \eqref{eq:approximate of alpha_i^*} into the first two terms of \eqref{eq:raw score of siga}, and applying the mean-value argument to replace the starred local quantities, we have
$$\frac{1}{2\sigma^4_\alpha} \left\{(\alpha_i^2-\siga^2) + \frac{2 S_i\alpha_i}{A_i} - \frac{\sigma^{-2}_\alpha \alpha_i^2}{A_i} - \frac{\tau_{i,3}\alpha_i S_i^2}{A_i^3} - \frac{S_i^2}{A_i^2}\right\} + O_p(m_i^{-3/2}).$$
By the mean-value argument, 
$$\frac{1}{2\sigma^4_\alpha A_i^*} = \frac{1}{2\sigma^4_\alpha A_i} + O_p(m_i^{-3/2}), \qquad -\frac{\tau_{i,3}^*\alpha_i^*}{2\sigma^4_\alpha A_i^*} = -\frac{\tau_{i,3}\alpha_i}{2\sigma^4_\alpha A_i} + O_p(m_i^{-3/2}).$$ Combining the contribution established above with the contribution of Laplace remainder in Section~\ref{subsec:contribution of R(theta)}, we have
\begin{equation}
\label{eq:asymptotic score of siga}
\begin{aligned}
\psi_{i,\siga^2}(\btheta) = \frac{1}{2\siga^2}&\left\{(\alpha_i^2-\siga^2) + \frac{2 S_i\alpha_i}{A_i} - \frac{\sigma^{-2}_\alpha \alpha_i^2}{A_i} - \frac{\tau_{i,3}\alpha_i S_i^2}{A_i^3} - \frac{S_i^2}{A_i^2} \right. \\ 
&\left.+ \frac{1}{A_i} - \frac{\alpha_i\tau_{i,3}}{A_i^2}\right\} + O_p(m_i^{-3/2}).
\end{aligned}    
\end{equation}

For $\bbetaw$, we consider the following second-order Taylor expansion,
$$e_{ij}^* = e_{ij} - b^{(2)}(\eta_{ij})\Delta_i - \frac12b^{(3)}\Delta_i^2 + O_p(m_i^{-3/2}).$$ Therefore, \begin{align*}
\boldsymbol{S}_i^{(w)*} &= \sum_{j=1}^{m_i}\bxwij\frac{e_{ij}^*}{\sige^2} \\
&= \sum_{j=1}^{m_i}\bxwij\left\{\frac{e_{ij}}{\sige^2} - \frac{b^{(2)}(\eta_{ij})}{\sige^2}\Delta_i - \frac{b^{(3)}(\eta_{ij})}{2\sige^2}\Delta_i^2 + O_p(m_i^{-3/2})\boldsymbol{1}_{[p_w:1]} \right\} \\
&= \boldsymbol{S}_i^{(w)} - \btau_{i,2}^{(w)}\Delta_i - \frac{1}{2}\btau_{i,3}^{(w)}\Delta_i^2 + O_p(m_i^{-1/2})\boldsymbol{1}_{[p_w:1]}.
\end{align*}
Substituting \eqref{eq:approximate of alpha_i^*} into $\Delta_i$ in the above equation and applying the mean-value argument gives
$$\boldsymbol{S}_i^{(w)} -\frac{\boldsymbol{S}_i \btau_{i,2}^{(w)}}{A_i} + \frac{\siga^{-2} \alpha_i \btau_{i,2}^{(w)}}{A_i} + \frac{\tau_{i,3} S_i^2\btau_{i,2}^{(w)}}{2A_i^3} - \frac{S_i^2\btau_{i,3}^{(w)}}{2A_i^2} + O_p(m_i^{-1/2})\boldsymbol{1}_{[p_w:1]},$$ and contribution of the second term is $$-\frac{1}{2}\left\{\frac{\btau_{i,3}^{(w)*}}{A_i^*} - \frac{\tau_{i,3}^*\btau_{i,2}^{(w)*}}{(A_i^*)^2}\right\} = -\frac{1}{2}\left\{\frac{\btau_{i,3}^{(w)}}{A_i} - \frac{\tau_{i,3}\btau_{i,2}^{(w)}}{A_i^2}\right\} + O_p(m_i^{-1/2})\boldsymbol{1}_{[p_w:1]}.$$  Combining the contribution established above with the contribution of Laplace remainder in Section~\ref{subsec:contribution of R(theta)}, we have
\begin{equation}
\label{eq:asymptotic score of betaw}
\begin{aligned}
    \bpsi_{i, \bbeta^{(w)}}(\btheta) = &\boldsymbol{S}_i^{(w)} -\frac{\boldsymbol{S}_i \btau_{i,2}^{(w)}}{A_i} + \frac{\siga^{-2} \alpha_i \btau_{i,2}^{(w)}}{A_i} + \frac{\tau_{i,3} S_i^2\btau_{i,2}^{(w)}}{2A_i^3} - \frac{S_i^2\btau_{i,3}^{(w)}}{2A_i^2} \\
    & - \frac{\btau_{i,3}^{(w)}}{2A_i} + \frac{\tau_{i,3}\btau_{i,2}^{(w)}}{2A_i^2} + O_p(m_i^{-1/2})\boldsymbol{1}_{[p_w:1]}. 
\end{aligned}
\end{equation}

For $\sige^2$, applying the mean-value argument, the contribution of the first term is 
$$\sumjmi \left\{\frac{y_{ij}\eta_{ij} - b(\eta_{ij})}{\sigma^4_e} + \nabla_{\sige^2}c(y_{ij}, \sige^2)\right\} + O_p(m_i^{-1/2}).$$ Similarly, 
$$\frac{\tau_{i,2}^*}{2\sige^2A_i^*} = \frac{\tau_{i,2}}{2\sige^2A_i} + O_p(m_i^{-1/2}),$$ and $$\frac{\tau_{i,3}^*S_i^*}{2\sige^2(A_i^*)^2} + O_p(m_i^{-1/2}).$$ Combining the contribution established above with the contribution of Laplace remainder in Section~\ref{subsec:contribution of R(theta)}, we have
\begin{equation}
\label{eq:asymptotic score of sige}
\begin{split}
 \psi_{i, \sige^2}(\btheta) = &\sum_{j=1}^{m_i}\left[-\frac{y_{ij}\eta_{ij} - b(\eta_{ij})}{\sigma^4_e} + \nabla_{\sige^2} c(y_{ij}, \sige^2)\right] + \frac{\tau_{i,2}}{2\sige^2A_i} + \frac{\tau_{i,3}S_i}{2\sige^2A_i^2} \\&+ O_p(m_i^{-1/2}).        
\end{split}\end{equation}

\subsection{Expression of $\bxi$}
\label{subsec:expression of xi}
Evaluating the score expansions \eqref{eq:asymptotic score of betab}-\eqref{eq:asymptotic score of sige}at $\dbtheta$, we define the leading stochastic term
$$\bxi = \left(\bxi_{\bbetab}^\top, \xi_{\siga^2}, \bxi_{\bbetaw}^\top, \xi_{\sige^2} \right)^\top,$$
where 
\begin{equation}
\label{eq:expression of xi}
\begin{aligned}
&\bxi_{\bbetab}=\frac{1}{\dsiga^2}\sumig\bxbi{\alpha}_i, 
&&\bxi_{\bbetaw}=\sumig\left\{\dot{\boldsymbol{S}}_i^{(w)} - \frac{\dot{S}_i \dot{\btau}_{i,2}^{(w)}}{\dot{A}_i}\right\},
\\&\xi_{\siga^2}=\frac{1}{2\dsiga^4}\sumig(\alpha_i^2-\dsiga^2),
&&\xi_{\sige^2}=\sumig\sumjmi
\left\{\frac{y_{ij}\dot\eta_{ij}-b(\dot\eta_{ij})}{\dsige^4}+\frac{\partial}{\partial\sige^2}c(y_{ij},\dsige^2)
\right\},
\end{aligned}
\end{equation}
and dotted quantities representing the evaluation at $\btheta =\dbtheta$.

To connect the expression of $\bxi_{\bbetaw}$ with the notation used in the main text, note that
\begin{align*}
&\bm_i(\btheta) = \frac{\btau_{i,2}^{(w)}}{A_i}, 
&&S_i = \frac{1}{\sige^2}\sumjmi \left\{y_{ij} -b^{(1)}(\eta_{ij})\right\},
&&\boldsymbol{S}_i^{(w)} = \frac{1}{\sige^2}\sumjmi \bxwij \left\{y_{ij} -b^{(1)}(\eta_{ij})\right\}.
\end{align*} It follows that
\begin{align*}
    \boldsymbol{S}_i^{(w)} - \frac{S_i\btau_{i,2}^{(w)}}{A_i} &= \boldsymbol{S}_i^{(w)} - \bm_i(\btheta) S_i  
    = \frac{1}{\sige^2}\sumjmi \left\{\bxwij - \bm_i(\btheta)\right\}\left\{y_{ij} -b^{(1)}(\eta_{ij})\right\}.
\end{align*}
Hence, $$\bxi_{\bbetaw} = \frac{1}{\dsige^2}\sumig\sumjmi\{\bxwij-\bm_i(\dbtheta)\}\{y_{ij} - b^{(1)}(\eta_{ij})\},$$ which is the representation used in the main text. The approximation of $\bpsi(\dbtheta)$ by $\bxi$ is established in Section~\ref{subsec:proof of lemma 2}.

\subsection{Exact and Asymptotic Expressions of $\nabla_{\btheta} \bpsi(\btheta)$}
\label{subsec:expression of ML score derivatuve}
We now consider the derivatives of score functions $\nabla_{\btheta} \bpsi(\btheta)$. By definition, 
$$
\nabla_{\btheta}^{2}\ell(\btheta) = \sumig \nabla_{\btheta}\bpsi_i(\btheta),
$$
where
$$
\nabla_{\btheta}\bpsi_i(\btheta)
=
\begin{pmatrix}
\nabla_{\bbetab}\psi_{i,\bbetab}^\top
&
\nabla_{\bbetab}\psi_{i,\siga^2}
&
\nabla_{\bbetab}\psi_{i,\bbetaw}^\top
&
\nabla_{\bbetab}\psi_{i,\sige^2}
\\
\nabla_{\siga^2}\psi_{i,\bbetab}^\top
&
\nabla_{\siga^2}\psi_{i,\siga^2}
&
\nabla_{\siga^2}\psi_{i,\bbetaw}^top
&
\nabla_{\siga^2}\psi_{i,\sige^2}
\\
\nabla_{\bbetaw}\psi_{i,\bbetab}^\top
&
\nabla_{\bbetaw}\psi_{i,\siga^2}
&
\nabla_{\bbetaw}\psi_{i,\bbetaw}^\top
&
\nabla_{\bbetaw}\psi_{i,\sige^2}
\\
\nabla_{\sige^2}\psi_{i,\bbetab}^\top
&
\nabla_{\sige^2}\psi_{i,\siga^2}
&
\nabla_{\sige^2}\psi_{i,\bbetaw}^\top
&
\nabla_{\sige^2}\psi_{i,\sige^2}
\end{pmatrix}.
$$
For notational simplicity, the dependence of the individual score-derivative blocks of $\btheta$ is suppressed.

We first record the exact cluster-wise derivative expressions. Differentiating the cluster-wise score expressions in \eqref{eq:raw score of betab}-\eqref{eq:raw score of sige} gives the blocks of $\nabla_{\btheta} \bpsi_{i}(\btheta)$. Using the symmetry of the score derivative, it suffices to report the diagonal and upper off-diagonal blocks. The exact expression of score derivatives are

\begin{align}
&\nabla_{\bbetab}\psi_{i,\bbetab}^\top = -\siga^{-2}\left\{\frac{\btau_{i,2}^{(bb)*}}{A_i^*} - \frac{\btau_{i,4}^{(bb)*}}{2(A_i^*)^3} - \frac{\btau_{i,3}^{(b)*}\btau_{i,3}^{(b)*\top}}{(A_i^*)^4} \right\} + \nabla^2_{\bbeta^{(b)}}\log(1+\epsilon_i), \notag\\    
&\nabla_{\bbetab}\psi_{i,\siga^2}=\bxbi\left\{-\siga^{-4}\alpha_i^*+\frac{\siga^{-6}\alpha_i^*}{A_i^*}\right\}+\frac{1}{2}\siga^{-4}\frac{\btau_{i,3}^{(b)*}}{(A_i^*)^2} \notag\\
&\qquad \qquad-\frac{1}{2}\siga^{-6}\frac{\alpha_i^*\btau_{i,4}^{(b)*}}{(A_i^*)^3}+\siga^{-6}\frac{\btau_{i,3}^{(b)*}(\alpha_i^*\tau_{i,3}^*-A_i^*)}{(A_i^*)^4}+\nabla_{\bbeta^{(b)}\siga^2}\log(1+\epsilon_i), \notag\\
&\nabla_{\bbetab}\psi_{i,\bbetaw}^\top=-\siga^{-2}\frac{\bxbi\btau_{i,2}^{(w)*\top}}{A_i^*}-\frac{1}{2}\siga^{-2}\left\{\frac{\bxbi\left(A_i^*\btau_{i,4}^{(w)*}\tau_{i,4}^*\btau_{i,2}^{(w)*}\right)^\top}{(A_i^*)^3}\right. \notag\\
&\qquad\qquad\left.-\frac{2\btau_{i,3}^{(b)*}\left(A_i^*\btau_{i,3}^{(w)*}\tau_{i,3}^*\btau_{i,2}^{(w)*}\right)^\top}{(A_i^*)^4}\right\}+\nabla_{\bbeta^{(b)\bbeta^{(w)}}}\log(1+\epsilon_i), \notag \\
&\nabla_{\bbetab}\psi_{i,\sige^2}=-\frac{\siga^{-4}\alpha_i^*\bxbi}{\sige^2 A_i^*}-\frac{1}{2}\siga^{-2}\left\{-\frac{A_i^*\btau_{i,3}^{(b)*}+\siga^{-2}\alpha_i^*\btau_{i,4}^{(b)*}}{\sige^2(A_i^*)^3}\right. \notag\\
&\qquad\qquad\left.+\frac{2\btau_{i,3}^{(b)*}\left(A_i^*\tau_{i,2}^*+\siga^{-2}\alpha_i^*\tau_{i,3}^*\right)}{\sige^2(A_i^*)^4}\right\}+\nabla_{\bbeta^{(b)}\sige^2}\log(1+\epsilon_i), \notag\\
&\nabla_{\siga^2}\psi_{i,\siga^2}=\frac{1}{2}\siga^{-4}-\siga^{-6}(\alpha_i^*)^2+\frac{\siga^{-8}(\alpha_i^*)^2}{A_i^*}-\frac{\siga^{-6}}{A_i^*}-\frac{1}{2}\siga^{-8}\frac{\alpha_i^*\tau_{i,3}^*-A_i^*}{(A_i^*)^3} \notag\\
&\qquad \qquad +\siga^{-6}\frac{\alpha_i^*\tau_{i,3}^*}{(A_i^*)^2}-\frac{1}{2}\siga^{-8}\frac{(\alpha_i^*)^2\tau_{i,4}^*+\alpha_i^*\tau_{i,3}^*}{(A_i^*)^3} \notag\\
&\qquad \qquad +\siga^{-8}\frac{\alpha_i^*\tau_{i,3}^*(\alpha_i^*\tau_{i,3}^*-A_i^*)}{(A_i^*)^4}+\nabla_{\siga^2}^2\log(1+\epsilon_i), \notag\\
&\nabla_{\siga^2}\psi_{i,\bbetaw}^\top
=-\siga^{-4}\frac{\alpha_i^*\btau_{i,2}^{(w)*\top}}{A_i^*}-\frac{1}{2}\siga^{-4}\frac{\left(A_i^*\btau_{i,3}^{(w)*}-\tau_{i,3}^*\btau_{i,2}^{(w)*}\right)^\top}{(A_i^*)^3} \notag\\
&\qquad \qquad
-\frac{1}{2}\siga^{-4}\left\{\frac{\alpha_i^*\left(A_i^*\btau_{i,4}^{(w)*}-\tau_{i,4}^*\btau_{i,2}^{(w)*}\right)^\top-\tau_{i,3}^*\btau_{i,2}^{(w)*\top}}{(A_i^*)^3}\right. \notag\\
&\qquad\qquad\left.-\frac{2\alpha_i^*\tau_{i,3}^*\left(A_i^*\btau_{i,3}^{(w)*}-\tau_{i,3}^*\btau_{i,2}^{(w)*}\right)^\top}{(A_i^*)^4}\right\}+\nabla_{\siga^2\bbeta^{(w)}}\log(1+\epsilon_i), \notag\\
&\nabla_{\siga^2}\psi_{i,\sige^2}
=-\siga^{-6}\frac{(\alpha_i^*)^2}{\sige^2 A_i^*}+\frac{1}{2}\siga^{-4}\frac{A_i^*\tau_{i,2}^*+\siga^{-2}\alpha_i^*\tau_{i,3}^*}{\sige^2(A_i^*)^3} \notag\\
&\qquad \qquad +\frac{1}{2}\siga^{-4}\left\{\frac{A_i^*\alpha_i^*\tau_{i,3}^*+\siga^{-2}(\alpha_i^*)^2\tau_{i,4}^*+\siga^{-2}\alpha_i^*\tau_{i,3}^*}{\sige^2(A_i^*)^3}\right. \notag\\
&\qquad \qquad \left.-\frac{2\alpha_i^*\tau_{i,3}^*\left(A_i^*\tau_{i,2}^*+\siga^{-2}\alpha_i^*\tau_{i,3}^*\right)}{\sige^2(A_i^*)^4}\right\}+\nabla_{\siga^2\sige^2}\log(1+\epsilon_i). \notag\\
&\nabla_{\bbetaw}\bpsi_{i,\bbetaw}^\top
=-\btau_{i,2}^{(ww)*}+\frac{\btau_{i,2}^{(w)*}\btau_{i,2}^{(w)*\top}}{A_i^*} \notag\\
&\qquad \qquad -\frac{1}{2}\left\{\frac{(A_i^*)^2\btau_{i,4}^{(ww)*}-A_i^*\btau_{i,4}^{(w)*}\btau_{i,2}^{(w)*\top}-A_i^*\btau_{i,3}^{(w)*}\btau_{i,3}^{(w)*\top}+\tau_{i,3}^*\btau_{i,3}^{(w)*}\btau_{i,2}^{(w)*\top}}{(A_i^*)^3}\right.\notag\\
&\qquad \qquad \left.-\frac{\tau_{i,3}^*(A_i^*)^2\btau_{i,3}^{(ww)*}-\tau_{i,3}^*A_i^*\btau_{i,3}^{(w)*}\btau_{i,2}^{(w)*\top}+(A_i^*)^2\btau_{i,2}^{(w)*}\btau_{i,4}^{(w)*\top}}{(A_i^*)^4}\right.\notag\\
&\qquad \qquad \left.-\frac{A_i^*\tau_{i,4}^*\btau_{i,2}^{(w)*}\btau_{i,2}^{(w)*\top}+2A_i^*\tau_{i,3}^*\btau_{i,2}^{(w)*}\btau_{i,3}^{(w)*\top}-2(\tau_{i,3}^*)^2\btau_{i,2}^{(w)*}\btau_{i,2}^{(w)*\top}}{(A_i^*)^4}\right\} \notag\\
&\qquad \qquad +\nabla_{\bbeta^{(w)}}^2\log(1+\epsilon_i), \notag\\
&\nabla_{\bbetaw}\bpsi_{i,\sige^2}=-\frac{\boldsymbol{S}_i^{(w)*}}{\sige^2}+\frac{\siga^{-2}\alpha_i^*\btau_{i,2}^{(w)*}}{\sige^2 A_i^*} \notag\\
&\qquad \qquad -\frac{1}{2}\left\{-\frac{A_i^*\btau_{i,3}^{(w)*}+\siga^{-2}\alpha_i^*\btau_{i,4}^{(w)*}}{\sige^2(A_i^*)^2}+\frac{\btau_{i,3}^{(w)*}\left(A_i^*\tau_{i,2}^*+\siga^{-2}\alpha_i^*\tau_{i,3}^*\right)}{\sige^2(A_i^*)^3}\right. \notag\\
&\qquad \qquad \left.+\frac{\btau_{i,2}^{(w)*}\left(A_i^*\tau_{i,3}^*+\siga^{-2}\alpha_i^*\tau_{i,4}^*\right)+\tau_{i,3}^*\left(A_i^*\btau_{i,2}^{(w)*}+\siga^{-2}\alpha_i^*\btau_{i,3}^{(w)*}\right)}{\sige^2(A_i^*)^3}\right. \notag\\
&\qquad \qquad \left.-\frac{2\tau_{i,3}^*\btau_{i,2}^{(w)*}\left(A_i^*\tau_{i,2}^*+\siga^{-2}\alpha_i^*\tau_{i,3}^*\right)}{\sige^2(A_i^*)^4}\right\}+\nabla_{\bbeta^{(w)}\sige^2}\log(1+\epsilon_i), \notag\\
&\nabla_{\sige^2}\psi_{i,\sige^2}=\sum_{j=1}^{m_i}\left[2\frac{y_{ij}\eta_{ij}^*-b(\eta_{ij}^*)}{\sigma_e^6}+\nabla_{\sige^2}^2c(y_{ij},\sige^2)\right]-\frac{\siga^{-4}(\alpha_i^*)^2}{\sigma_e^4 A_i^*}\notag\\
&\qquad \qquad +\frac{1}{2}\frac{-2\tau_{i,2}^*(A_i^*)^2+(\tau_{i,2}^*)^2A_i^*-\siga^{-2}\alpha_i^*\tau_{i,3}^*A_i^*+\siga^{-2}\alpha_i^*\tau_{i,2}^*\tau_{i,3}^*}{\sigma_e^4(A_i^*)^3} \notag\\
&\qquad \qquad +\frac{1}{2}\siga^{-2}\frac{-\siga^{-2}\alpha_i^*\tau_{i,3}^*A_i^*-2\alpha_i^*\tau_{i,3}^*(A_i^*)^2-\siga^{-2}(\alpha_i^*)^2\tau_{i,4}^*A_i^*}{\sigma_e^4(A_i^*)^4} \notag\\
&\qquad \qquad +\frac{1}{2}\siga^{-2}\frac{2\alpha_i^*\tau_{i,3}^*\tau_{i,2}^*A_i^*+2\siga^{-2}(\alpha_i^*)^2(\tau_{i,3}^*)^2}{\sigma_e^4(A_i^*)^4}+\nabla_{\sige^2}^2\log(1+\epsilon_i). \label{eq:exact expression of ML score derivative}
\end{align}

We next derive the asymptotic expressions for the derivatives of the cluster-wise score functions. The derivation is conducted by substituting the asymptotic expressions of $\alpha_i^*$ in \eqref{eq:approximate of alpha_i^*} into the above equations and applying the mean-value argument to replace the starred local summing quantities by their counterparts evaluated at $\alpha_i$. We display only the leading terms, which are sufficient for the proof of Theorem 2. The contributions of the Laplace remainder are of smaller order by the results of Section~\ref{subsec:contribution of R(theta)} and are therefore absorbed into the stated remainder terms. The derivation details are analogous to the derivation of the asymptotic expression of $\bpsi_i(\btheta)$ in Section~\ref{subsec:expression of ML score}, and therefore are omitted. The final asymptotic expressions are
\begin{align}
    &\nabla_{\bbetab}\bpsi_{i, \bbetab}^\top = -\frac{\siga^{-2} \btau_{i,2}^{(bb)}}{A_i} + O_p(m_i^{-3/2})\boldsymbol{1}_{p_b}^{\otimes 2}, \notag\\    
    &\nabla_{\bbetab}\bpsi_{i,\siga^2}=-\frac{\siga^{-4} \alpha_i \btau_{i,2}^{(b)}}{A_i} + O_p(m_i^{-1/2})\boldsymbol{1}_{[p_b:1]}, \notag\\
    &\nabla_{\bbetab}\bpsi_{i,\bbetawt}^\top=\frac{\siga^{-2}\btau_{i,2}^{(bw)}}{A_i} + O_p(m_i^{-1/2})\boldsymbol{1}_{[p_b:p_w]}, \notag\\
    &\nabla_{\bbetab}\bpsi_{i,\sige^2} = -\frac{\siga^{-4}\sige^{-2}\alpha_i \bxbi}{A_i} + O_p(m_i^{-1})\boldsymbol{1}_{[p_b:1]}, \notag\\
    &\nabla_{\siga^2}\psi_{i,\siga^2}= \frac{\siga^{-4}}{2} - \siga^{-6} \alpha_i^2 + O_p(m_i^{-1/2}), \notag\\
    &\nabla_{\siga^2}\bpsi_{i,\bbetaw}^\top=-\frac{\siga^{-4} \alpha_i \btau_{i,2}^{(w)\top}}{A_i} + O_p(m_i^{-1/2})\boldsymbol{1}_{[1:p_w]}, \notag\\
    &\nabla_{\siga^2}\psi_{i,\sige^2}=-\siga^{-4} \sige^{-2}\left\{\frac{\siga^{-2} \alpha_i^2}{A_i} - \frac{\tau_{i,2}}{2A_i^2} - \frac{\alpha_i\tau_{i,3}(3\siga^{-2} - \tau_{i,2})}{2A_i^3}\right\} + O_p(m_i^{-3/2}), \notag\\
    &\nabla_{\bbetaw}\bpsi_{i,\bbetaw}^\top = -\btau_{i,2}^{(ww)} + \frac{\btau_{i,2}^{(w)} \btau_{i,2}^{(w)\top}}{A_i} + O_p(m_i^{1/2})\boldsymbol{1}_{p_w}^{\otimes 2}, \notag\\
    &\nabla_{\bbetaw}\bpsi_{i,\sige^2} = -\sige^{-2} \boldsymbol{S}_i^{(w)} + O_p(1)\boldsymbol{1}_{[p_w:1]}, \notag\\
    &\nabla_{\sige^2}\psi_{i, \sige^2} = \sige^{-4} \sum_{j=1}^{m_i}\left[\frac{2\{y_{ij}\eta_{ij} - b(\eta_{ij})\}}{\sige^2}+\sige^4 \nabla_{\sige^2}^2 c(y_{ij}, \sige^2) \right] + O_p(1). \label{eq:asymptotic expression of ML score derivative}
\end{align}

\section{Construction and Preliminary Properties of the REML Criterion $\ell_R(\btheta)$}
\setcounter{equation}{0} 
\subsection{Construction of $\ell_R(\btheta)$}
Conceptually, REML removes the fixed effects by integrating the exact marginal log-likelihood $\ell(\bbeta, \bsigma)$ with respect to $\bbeta$, leading to the following form
$$\ell_R(\bsigma) = \log \int\exp\{\ell(\bbeta, \bsigma)\} d\bbeta.$$ The resulting criterion depends only on $\bsigma$. For each fixed $\bsigma$, define the profile map
$$\bbeta(\bsigma) = \arg\max_{\bbeta} \ell(\bbeta, \bsigma).$$
We then define the corresponding profile criterion by $$\ell_P(\bsigma) = \ell\left\{\bbeta(\bsigma), \bsigma\right\},$$
and the negative fixed-effect curvature evaluated along the profile path is $$\bD(\bsigma) = -\nabla_{\bbeta}^2 \ell(\bbeta, \bsigma)\big|_{\bbeta = \bbeta(\bsigma)}.$$
Since the dimension of $\bbeta$ is fixed, the integral over $\bbeta$ can be treated by an ordinary finite-dimensional Laplace expansion. Let
$\bv = \bbeta - \bbeta(\bsigma)$, expanding $\ell(\bbeta, \bsigma)$ around $\bbeta(\bsigma)$ gives
\begin{align*}
    \ell(\bbeta, \bsigma) &=  \ell(\bbeta(\bsigma), \bsigma) - \frac12 \mathbf{v}^\top \bD(\bsigma)\mathbf{v} + J_{\bbeta}(\mathbf{v}, \bsigma)\\
    &= \ell_P(\bsigma) - \frac12 \mathbf{v}^\top \bD(\bsigma)\mathbf{v} + J_{\bbeta}(\mathbf{v}, \bsigma),
\end{align*}
where $J_{\bbeta}(\mathbf{v}, \bsigma)$ collects the third- and higher-order terms in the Taylor expansion with respect to the fixed effects. Hence,
\begin{align*}
    \ell_R(\bsigma) &= \log \int \exp\left\{\ell_P(\bsigma) - \frac{1}{2}\mathbf{v}^\top \bD(\bsigma) \mathbf v + J_{\bbeta}(\mathbf{v}, \bsigma)\right\} d\bv \\
    &= \ell_P(\bsigma) - \frac{1}{2}\log|\bD(\bsigma)| + \log(1+\epsilon_{\bbeta}),
\end{align*}
where $\log(1+\epsilon_{\bbeta})$ is the fixed-effect Laplace remainder generated by the higher-order term $J_\beta(\mathbf v,\bsigma)$. Define 
$$U(\bsigma) = \log{(1+\epsilon_{\bbeta})}.$$
Then the exact REML criterion can be written as
$$\ell_R(\bsigma) = \ell_P(\bsigma) - \frac{1}{2}\log|\bD(\bsigma)| + U(\bsigma).$$
The Laplace remainder $U(\bsigma)$ plays a role analogous to the Laplace remainder $R(\btheta)$. The distinction is that $R(\btheta)$ arises from the collection of cluster-specific integrations over the random effects, whose total dimension increases with $\balpha$, whose total dimension increases with $g$, whereas $U(\bsigma)$ arises from the fixed-dimensional integration over the regression parameter $\bbeta$. In Section~\ref{subsec:contribution of U(theta)}, we derive the expression of the dominant term of $U(\bsigma)$ and establish the asymptotic orders of $\nabla_{\bsigma}U(\bsigma)$ and $\nabla_{\bsigma}^2U(\bsigma)$. These results will be used in the subsequent REML analysis to control the contribution of $U(\bsigma)$ under the relevant normalizations. For conciseness, the main text suppresses $U(\bsigma)$ when presenting the leading form of the REML criterion, which takes the form $\ell_P(\bsigma) - 1/2 \log|\bD(\bsigma)|$.

We denote the REML estimator by $\hat{\btheta}_R = \left(\hat{\bbeta}_R^\top, \hat{\bsigma}_R^\top \right)^\top$, where
$$\hat{\bsigma}_R = \arg\max_{\bsigma} \ell_R(\bsigma), \quad \hat{\bbeta}_R =\bbeta(\hat{\bsigma}_R).$$
 Analogously, denote the profile estimator by $\hat{\btheta}_P = \left(\hat{\bbeta}_P^\top, \hat{\bsigma}_P^\top \right)^\top$, where
$$\hat{\bsigma}_P = \arg\max_{\bsigma} \ell_P(\bsigma), \quad \hat{\bbeta}_P =\bbeta(\hat{\bsigma}_P).$$ 

For the local analysis below, define the profile neighborhood
$$\mathcal{N}_{\bsigma} = \left\{\bsigma:(\bbeta(\bsigma)^\top, \bsigma^\top)^\top \in \mathcal{N}\right\}.$$
Thus, $\mathcal{N}_{\bsigma}$ is the inverse image of $\mathcal{N}$ under the profile map and contains values of the variance component whose corresponding profiled parameter points lie in $\mathcal{N}$. In particular,
$$\hat{\bsigma}_P \in \mathcal{N}_{\bsigma} \Longleftrightarrow \hat{\btheta}_P \in \mathcal{N}, \qquad \hat{\bsigma}_R \in \mathcal{\bsigma} \Longleftrightarrow \hat{\btheta}_R \in \mathcal{N}.$$
Since $\bsigma \in \mathcal{N}_{\bsigma} \Rightarrow (\bbeta(\bsigma)^\top, \bsigma^\top) \in \mathcal{N}$, all uniform bounds previously established over $\btheta \in \mathcal{N}$ remain valid when evaluated along the profile path uniformly over $\bsigma \in \mathcal{N}_{\bsigma}$.
Unless otherwise stated, all uniform bounds in the subsequent REML analysis are taken over $\bsigma \in \mathcal{N}_{\bsigma}$.

\subsection{Property of $\bD(\sigma)$ and Its Inverse}
\label{subsec:property of D(sigma)}
Recall \eqref{eq:LA decomposition}, we have
$$\ell(\bbeta, \bsigma) = \ell_L(\bbeta, \bsigma) + R(\bbeta, \bsigma).$$ Thus, $$\bD(\sigma) = \bD_L(\bsigma) - \nabla_{\bbeta}^2 R(\bbeta(\bsigma),\bsigma)\big|_{\bbeta = \bbeta(\bsigma)},$$ where $$\bD_L(\bsigma) = -\nabla_{\bbeta}^2 \ell_L(\bbeta(\bsigma), \bsigma)\big|_{\bbeta = \bbeta(\bsigma)}.$$
Write
$$\bD_L(\bsigma) = \begin{pmatrix}
    \bD_{L, bb}(\bsigma) & \bD_{L, bw}(\bsigma) \\
    \bD_{L, wb}(\bsigma) & \bD_{L, ww}(\bsigma)
\end{pmatrix},$$
and let $\bK_{\bbeta} = \diag(g\boldsymbol{1}_{p_b}, n\boldsymbol{1}_{p_w})$. The asymptotic expansions of ML score-derivative in \eqref{eq:asymptotic expression of ML score derivative}, which hold uniformly over $\btheta \in \mathcal{N}$, remain valid when evaluated along the profile path uniformly for $\bsigma \in \mathcal{N}_{\bsigma}$. Consequently, 
$$g^{-1}\bD_{L, bb}(\bsigma) = g^{-1}\sumig \left\{ \frac{\siga^{-2}\btau_{i,2}^{(bb)}}{A_i} + O_p(m_i^{-3/2})\boldsymbol{1}_{p_b}^{\otimes 2}\right\} \overset{p}{\longrightarrow}  \siga^{-2}\bC_2,$$ where the convergence follows from Condition A. Similarly,
$$n^{-1}\bD_{L,ww}(\bsigma) = n^{-1}\sumig \left\{\btau_{1,2}^{(ww)} - \frac{\btau_{1,2}^{(w)}\btau_{1,2}^{(w)\top}}{A_i} + O_p(m_i^{-1/2})\boldsymbol{1}_{p_w}^{\otimes 2} \right\} \overset{p}{\longrightarrow} \bC_3\{\bbeta(\bsigma), \bsigma\}.$$
For the off-diagonal block, \begin{align*}
    (gn)^{-1/2} \bD_{L,bw}(\bsigma) &= (gn)^{-1/2}\sumig\left\{\frac{\sigma^{-2}_\alpha\btau_{i,2}^{(bw)}}{A_i} + O_p(m_i^{-1/2})\boldsymbol{1}_{[p_b:p_w]} \right\}  \\
    &= O_p(\sqrt{g/n})\boldsymbol{1}_{[p_w:p_b]} = o_p(1) \boldsymbol{1}_{[p_w:p_b]},
\end{align*} and by symmetry,
$$\bD_{L, wb}(\bsigma) = \bD_{L, bw}(\bsigma)^\top.$$
It follows that 
\begin{equation}
\label{eq:norm of D_L(sigma)}
\bK_{\bbeta}^{-1/2}\bD_{L}(\bsigma)\bK_{\bbeta}^{-1/2} \xrightarrow{p} \begin{pmatrix}
       \siga^{-2}\bC_2 &\boldsymbol{0}_{[p_b:p_w]}\\ 
        \boldsymbol{0}_{[p_w:p_b]} &
        \bC_3\{\bbeta(\bsigma), \bsigma\}
    \end{pmatrix}.
\end{equation}
The block-wise bounds we established in Section~\ref{subsec:contribution of R(theta)} gives
$$ \nabla_{\bbeta}^2 R(\hat{\bbeta}_P, \bsigma)\big|_{\bbeta = \bbeta(\bsigma)} = \sumig \begin{pmatrix}
    O_p(m_i^{-3})\boldsymbol{1}_{p_b}^{\otimes 2} & O_p(m_i^{-2})\boldsymbol{1}_{[p_b:p_w]} \\
    O_p(m_i^{-2})\boldsymbol{1}_{[p_w:p_b]} &
    O_p(m_i^{-1})\boldsymbol{1}_{p_w}^{\otimes 2}
\end{pmatrix}.$$
After applying the block-wise normalization, 
\begin{align*}
\bK_{\bbeta}^{-1/2} &\nabla_{\bbeta}^2 R(\hat{\bbeta}_P, \bsigma)\big|_{\bbeta = \bbeta(\bsigma)}\bK_{\bbeta}^{-1/2} \\
&= \begin{pmatrix}
g^{-1}\sumig O_p(m_i^{-3})\boldsymbol{1}_{p_b}^{\otimes 2} & (gn)^{-1/2}\sumig O_p(m_i^{-2})\boldsymbol{1}_{[p_b:p_w]} \\
(gn)^{-1/2}\sumig O_p(m_i^{-2})\boldsymbol{1}_{[p_w:p_b]} & n^{-1}\sumig O_p(m_i^{-1})\boldsymbol{1}_{p_w}^{\otimes 2}  
\end{pmatrix} \\
&= o_p(1)\boldsymbol{1}_p^{\otimes 2}.
\end{align*}
Thus,
\begin{equation}
\label{eq:norm of R(sigma)}
\sup_{\bsigma \in \mathcal{N}_{\bsigma}} \left\|\bK_{\bbeta}^{-1/2}\nabla_{\bbeta}^2 R(\hat{\bbeta}_P, \bsigma)\big|_{\bbeta = \bbeta(\bsigma)} \bK_{\bbeta}^{-1/2}\right\| = o_p(1).
\end{equation}
Combine \eqref{eq:norm of D_L(sigma)} and \eqref{eq:norm of R(sigma)}, we have 
$$\sup_{\bsigma \in \mathcal{N}_{\bsigma}}\left\|\bK_{\bbeta}^{-1/2}\bD(\bsigma)\bK_{\bbeta}^{-1/2}\right\| = O_p(1),$$
and
\begin{equation}
\label{eq:norm of D(sigma)}
\bK_{\bbeta}^{-1/2}\bD(\bsigma)\bK_{\bbeta}^{-1/2} \overset{p}{\longrightarrow} \begin{pmatrix}
       \siga^{-2}\bC_2 &\boldsymbol{0}_{[p_b:p_w]}\\ 
        \boldsymbol{0}_{[p_w:p_b]} &
        \bC_3\{\bbeta(\bsigma), \bsigma\}
    \end{pmatrix}.
\end{equation}
By Condition A, the limiting block-diagonal matrix in \eqref{eq:norm of D(sigma)} is finite and uniformly positive definite over $\mathcal{N}_{\bsigma}$. It follows that $\bD(\bsigma)$ in non-singular with probability tending to one. We next define $$\bH(\bsigma) = \bD(\bsigma)^{-1}.$$
Write $$\bD(\bsigma) = \begin{pmatrix}
    \bD_{bb}(\bsigma) & \bD_{bw}(\bsigma) \\
    \bD_{wb}(\bsigma) & \bD_{ww}(\bsigma)
\end{pmatrix}.$$
The preceding results imply, uniformly for $\bsigma \in \mathcal{N}_{\bsigma}$
\begin{align*}
   \bD_{bb}(\bsigma) & = \bD_{L, bb}(\bsigma) + o_p(1)\boldsymbol{1}_{p_b}^{\otimes 2} = O_p(g)\boldsymbol{1}_{p_b}^{\otimes 2}, \\
   \bD_{bw}(\bsigma) & = \bD_{L, bw}(\bsigma) + o_p(1)\boldsymbol{1}_{[p_b:p_w]} = O_p(g)\boldsymbol{1}_{[p_b:p_w]},\\
   \bD_{wb}(\bsigma) & = \bD_{L, wb}(\bsigma) + o_p(1)\boldsymbol{1}_{[p_w:p_b]} = O_p(g)\boldsymbol{1}_{[p_w:p_b]},\\
   \bD_{ww}(\bsigma) & = \bD_{L, ww}(\bsigma) + o_p(1)\boldsymbol{1}_{p_w}^{\otimes 2} = O_p(n)\boldsymbol{1}_{p_w}^{\otimes 2}.
\end{align*}
Define the Schur complement of $\bD_{bb}(\bsigma)$ by 
$$\bS_{ww}(\bsigma) = \bD_{ww}(\bsigma) - \bD_{wb}(\bsigma)\bD_{bb}(\bsigma)^{-1}\bD_{wb}(\bsigma) = O_p(n)\boldsymbol{1}_{p_w}^{\otimes 2}.$$
Partition
$$\bH_{\bsigma} = \begin{pmatrix}
    \bH_{bb}(\bsigma) & \bH_{bw}(\bsigma) \\
    \bH_{wb}(\bsigma) & \bH_{ww}(\bsigma)
\end{pmatrix}.$$
The block inverse formula gives 
\begin{align*}
    \bH_{bb}(\bsigma) &= \bD_{bb}(\bsigma)^{-1} + \bD_{bb}(\bsigma)^{-1}\bD_{bw}(\bsigma)\bD_{ww}(\bsigma)^{-1}\bD_{wb}(\bsigma)\bD_{bb}(\bsigma)^{-1} \\
    &= O_p(g^{-1})\boldsymbol{1}_{p_b}^{\otimes 2} +  O_p(n^{-1})\boldsymbol{1}_{p_b}^{\otimes} =  O_p(g^{-1})\boldsymbol{1}_{p_b}^{\otimes 2}, \\
    \bH_{bw}(\bsigma) &= -\bD_{bb}(\bsigma)^{-1} \bD_{bw}(\bsigma) \bD_{ww}(\bsigma)^{-1} = O_p(n^{-1})\boldsymbol{1}_{[p_b:p_w]}, \\
    \bH_{wb}(\bsigma) &= \bH_{bw}(\bsigma)^\top, \\
    \bH_{ww}(\bsigma) &= \bS_{ww}(\bsigma)^{-1} = O_p(n^{-1})\boldsymbol{1}_{p_w}^{\otimes 2}.
\end{align*}
As $g, m_L\to \infty$, the preceding bounds of each block imply
$$\sup_{\bsigma \in \mathcal{N}_{\bsigma}}\left\|\bK_{\bbeta}^{1/2}\bH(\bsigma)\bK_{\bbeta}^{1/2}\right\| = O_p(1).$$

\subsection{Orders of the First Derivatives of $\bD(\bsigma)$ and Its Inverse}
\label{subsec:first derivative of D}
In this section, we establish the corresponding bounds for the first derivatives of $\bD(\bsigma)$ and $\bH(\bsigma)$. We illustrate using the $\siga^2$ block; the results of $\sige^2$ block can be obtained by following the same argument.

Recall the exact expression for ML score derivatives in Section~\ref{subsec:expression of ML score derivatuve}, we have 
\begin{align*}
\bD_{bb}(\bsigma) &= \sumig \nabla_{\bbetab}\bpsi_{i, \bbetab}^\top \Bigg|_{\bbeta = \bbeta(\bsigma)} 
\\
&= \sumig\left[-\siga^{-2}\left\{\frac{\btau_{i,2}^{(bb)*}}{A_i^*} - \frac{\btau_{i,4}^{(bb)*}}{2(A_i^*)^3} - \frac{\btau_{i,3}^{(b)*}\btau_{i,3}^{(b)*\top}}{(A_i^*)^4} \right\} + \nabla^2_{\bbeta^{(b)}}\log(1+\epsilon_i)\right]\Bigg|_{\bbeta = \bbeta(\bsigma)}  \\
&= \bD_{L, bb}(\bsigma) + \nabla^2_{\bbeta^{(b)}}R(\btheta)\Big|_{\bbeta = \bbeta(\bsigma)},
\end{align*}
where the leading part is 
$$\bD_{L, bb}(\bsigma) = -\sumig \siga^{-2}\left\{\frac{\btau_{i,2}^{(bb)*}}{A_i^*} - \frac{\btau_{i,4}^{(bb)*}}{2(A_i^*)^3} - \frac{\btau_{i,3}^{(b)*}\btau_{i,3}^{(b)*\top}}{(A_i^*)^4} \right\}\Bigg|_{\bbeta = \bbeta(\bsigma)}.$$
By standard calculus, 
\begin{equation}
\label{eq:derivative of D_{L, bb}}
\begin{aligned}
&\nabla_{\siga^2} \bD_{L, bb}(\bsigma)\\& =  \nabla_{\siga^2}\left[-\sumig \siga^{-2}\left\{\frac{\btau_{i,2}^{(bb)*}}{A_i^*} - \frac{\btau_{i,4}^{(bb)*}}{2(A_i^*)^3} - \frac{\btau_{i,3}^{(b)*}\btau_{i,3}^{(b)*\top}}{(A_i^*)^4} \right\}\right]\Bigg|_{\bbeta = \bbeta(\bsigma)} \\
&+ \nabla_{\bbetab}\left[-\sumig \siga^{-2}\left\{\frac{\btau_{i,2}^{(bb)*}}{A_i^*} - \frac{\btau_{i,4}^{(bb)*}}{2(A_i^*)^3} - \frac{\btau_{i,3}^{(b)*}\btau_{i,3}^{(b)*\top}}{(A_i^*)^4} \right\}\right]\Bigg|_{\bbeta = \bbeta(\bsigma)} \nabla_{\siga^2} \bbetab(\bsigma) \\
&+ \nabla_{\bbetaw}\left[-\sumig \siga^{-2}\left\{\frac{\btau_{i,2}^{(bb)*}}{A_i^*} - \frac{\btau_{i,4}^{(bb)*}}{2(A_i^*)^3} - \frac{\btau_{i,3}^{(b)*}\btau_{i,3}^{(b)*\top}}{(A_i^*)^4} \right\}\right]\Bigg|_{\bbeta = \bbeta(\bsigma)} \nabla_{\siga^2} \bbetaw(\bsigma).
\end{aligned}
\end{equation}
For the first term, by standard calculus, we have
\begin{equation}
\label{eq:derivative of the first term in nabla_siga^2 D_L,dd}
\begin{aligned}
    \nabla_{\siga^2}&\left[-\sumig \siga^{-2}\left\{\frac{\btau_{i,2}^{(bb)*}}{A_i^*} - \frac{\btau_{i,4}^{(bb)*}}{2(A_i^*)^3} - \frac{\btau_{i,3}^{(b)*}\btau_{i,3}^{(b)*\top}}{(A_i^*)^4} \right\}\right]\Bigg|_{\bbeta = \hat{\bbeta}_P} \\
    &= \sumig \left[\siga^{-4}\left\{\frac{\btau^{(bb)*}_{i,2}}{A_i^*} - \frac{\btau^{(bb)*}_{i,4}}{2(A_i^*)3} - \frac{\btau_{i,3}^{(b)*}\btau_{i,3}^{(b)*\top}}{(A_i^*)^4}\right\}\right. \\
    &- \siga^{-2} \left\{\frac{1}{A_i^*}\nabla_{\siga^2} \btau_{i,2}^{(bb*)} - \frac{\btau_{i,2}^{(bb)*}}{(A_i^*)^2}\nabla_{\siga^2}A_i^*\right. \\
    &\qquad - \frac{1}{2(A_i^*)^3} \nabla_{\siga^2} \btau_{i,4}^{(bb)*} - \frac{3\btau_{i,4}^{(bb)*}}{2(A_i^*)^4}\nabla_{\siga^2}A_i^*\\
    &\qquad -\frac{\nabla_{\siga^2}\btau_{i,3}^{(b)*} \btau_{i,3}^{(b)*\top} + \btau_{i,3}^{(b)*} \left(\nabla_{\siga^2}\btau_{i,3}^{(b)*}\right)^\top}{(A_i^*)^4} \\
    &\qquad \left.\left. + \frac{4\btau_{i,3}^{(b)*}\btau_{i,3}^{(b)*\top}}{(A_i^*)^5}\nabla_{\siga^2} A_i^*\right\}\right]\Bigg|_{\bbeta = \bbeta(\bsigma)}.
\end{aligned}
\end{equation}
By \eqref{eq:d alpha_i^*/d theta}, uniformly for $\bsigma \in \mathcal{N}_{\bsigma}$, $\nabla_{\siga^2}\alpha_i^* = O_p(m_i^{-1})$, and $$\nabla_{\siga^2}A_i^* = -\siga^{-4}+ \tau_{i,3}^* \nabla_{\siga^2} \alpha_i^* = O_p(1).$$ The same argument yields
$$\nabla_{\siga^2} \btau_{i,2}^{(bb)*} = O_p(1), \qquad \nabla_{\siga^2} \btau_{i,3}^{(b)*} = O_p(1), \qquad \nabla_{\siga^2}\btau_{i,4}^{(bb)*} =O_p(1).$$
Substituting these bounds into \eqref{eq:derivative of the first term in nabla_siga^2 D_L,dd}, we have
$$ \nabla_{\siga^2}\left[-\sumig \siga^{-2}\left\{\frac{\btau_{i,2}^{(bb)*}}{A_i^*} - \frac{\btau_{i,4}^{(bb)*}}{2(A_i^*)^3} - \frac{\btau_{i,3}^{(b)*}\btau_{i,3}^{(b)*\top}}{(A_i^*)^4} \right\}\right]\Bigg|_{\bbeta = \bbeta(\bsigma)} = O_p(g)\boldsymbol{1}_{p_b}^{\otimes 2}.$$
The same argument yields
\begin{align*}
\nabla_{\bbetab}\left[-\sumig \siga^{-2}\left\{\frac{\btau_{i,2}^{(bb)*}}{A_i^*} - \frac{\btau_{i,4}^{(bb)*}}{2(A_i^*)^3} - \frac{\btau_{i,3}^{(b)*}\btau_{i,3}^{(b)*\top}}{(A_i^*)^4} \right\}\right]\Bigg|_{\bbeta = \bbeta(\bsigma)} &= O_p(gm_L^{-1})\boldsymbol{1}_{p_b \times p_b \times p_b}, \\
\nabla_{\bbetaw}\left[-\sumig \siga^{-2}\left\{\frac{\btau_{i,2}^{(bb)*}}{A_i^*} - \frac{\btau_{i,4}^{(bb)*}}{2(A_i^*)^3} - \frac{\btau_{i,3}^{(b)*}\btau_{i,3}^{(b)*\top}}{(A_i^*)^4} \right\}\right]\Bigg|_{\bbeta = \bbeta(\bsigma)} &= O_p(g)\boldsymbol{1}_{p_b \times p_b \times p_w}.
\end{align*}
We next establish the order of $\nabla_{\siga^2} \bbetab(\bsigma)$ and $\nabla_{\siga^2} \bbetaw(\bsigma)$.
By definition, the profile map satisfies 
$$\nabla_{\bbeta} \ell(\bbeta(\bsigma), \bsigma) = \boldsymbol{0}.$$ Differentiating the above equation with respect to $\siga^2$ gives 
\begin{equation}
\label{eq:derivative of profile score}
\nabla_{\bbeta}^2\ell(\bbeta, \bsigma)\big|_{\bbeta = \bbeta(\bsigma)} \left\{\nabla_{\siga^2}\bbeta(\bsigma)\right\}^\top + \nabla_{\siga^2}\nabla_{\bbeta} \ell(\bbeta, \bsigma)\big|_{\bbeta = \bbeta(\bsigma)} = \boldsymbol{0}.
\end{equation}
Replacing $-\nabla_{\bbeta}^2\ell(\bbeta, \bsigma)\big|_{\bbeta = \bbeta(\bsigma)}=\bD(\bsigma)$ and $\bH(\bsigma) = \bD(\bsigma)^{-1}$, we obtain
$$\left\{\nabla_{\siga^2}\bbeta(\bsigma)\right\}^\top = \bH(\sigma)\nabla_{\siga^2}\nabla_{\bbeta} \ell(\bbeta, \bsigma)\big|_{\bbeta = \beta(\bsigma)}.$$
By partition, 
\begin{align*}
&\left\{\nabla_{\siga^2}\bbetab(\bsigma)\right\}^\top = \bH_{bb}(\bsigma)\nabla_{\siga^2}\nabla_{\bbetab}\ell(\bbeta, \bsigma)\big|_{\bbeta = \bbeta(\bsigma)} + \bH_{bw}(\bsigma)\nabla_{\siga^2}\nabla_{\bbetaw}\ell(\bbeta, \bsigma)\big|_{\bbeta = \bbeta(\bsigma)}, \\
&\left\{\nabla_{\siga^2}\bbetaw(\bsigma)\right\}^\top = \bH_{wb}(\bsigma)\nabla_{\siga^2}\nabla_{\bbetab}\ell(\bbeta, \bsigma)\big|_{\bbeta = \bbeta(\bsigma)} + \bH_{ww}(\bsigma)\nabla_{\siga^2}\nabla_{\bbetaw}\ell(\bbeta, \bsigma)\big|_{\bbeta = \bbeta(\bsigma)}.
\end{align*}
By \eqref{eq:asymptotic expression of ML score derivative}, 
\begin{equation}
\label{eq:order of nabla_siga^2 D}
\begin{aligned}
\nabla_{\siga^2}\nabla_{\bbetab}\ell(\bbeta, \bsigma)\big|_{\bbeta = \bbeta(\bsigma)} &= \sumig \nabla_{\siga^2}\bpsi_{i,\bbetab}^\top \big|_{\bbeta = \bbeta(\bsigma)} = O_p(g)\boldsymbol{1}_{[p_b:1]},\\
\nabla_{\siga^2}\nabla_{\bbetaw}\ell(\bbeta, \bsigma)\big|_{\bbeta = \bbeta(\bsigma)} &= \sumig \nabla_{\siga^2}\bpsi_{i,\bbetaw}^\top \big|_{\bbeta = \bbeta(\bsigma)} = O_p(g)\boldsymbol{1}_{[p_w:1]}.
\end{aligned}
\end{equation}
Together with the block-wise bounds of $\bH(\bsigma)$, we obtain 
\begin{equation}
\label{eq:order of nabla beta}
\left\{\nabla_{\siga^2}\bbetab(\bsigma)\right\}^\top = O_p(1)\boldsymbol{1}_{[p_b:1]}, \qquad \left\{\nabla_{\siga^2}\bbetaw(\bsigma)\right\}^\top = O_p(1)\boldsymbol{1}_{[p_w:1]}.
\end{equation} 

Substituting these orders back into \eqref{eq:derivative of D_{L, bb}}, we have
$$\nabla_{\siga^2} \bD_{L, bb}(\bsigma) = O_p(g)\boldsymbol{1}_{p_b}^{\otimes 2}.$$
By the same argument, the results for the remaining blocks are 
\begin{align*}
  &  \nabla_{\siga^2} \bD_{L, bw}(\bsigma) = O_p(g)\boldsymbol{1}_{[p_b:p_w]}, \quad
    \nabla_{\siga^2} \bD_{L, wb}(\bsigma) = O_p(g)\boldsymbol{1}_{[p_w:p_b]}, \\&
    \nabla_{\siga^2} \bD_{L, ww}(\bsigma) = O_p(n)\boldsymbol{1}_{p_w}^{\otimes 2}. 
\end{align*}

For the remaining term $\nabla^2_{\bbeta^{(b)}}R(
\btheta)$ evaluated at $\bbeta(\bsigma)$, the first derivative with respect to $\siga^2$ are of smaller order than the corresponding leading-part terms. They are obtained by differentiating the dominant correction term in \eqref{eq:second derivative of epsilon_i}  with respect to $\siga^2$ and applying the same chain rule and arguments as above, but involve lengthy differentiation of the higher-order correction terms. We therefore omit the details.
Combining the leading and remainder contributions yields
\begin{equation}
\label{eq:order of nabla_siga^2 D}
\nabla_{\siga^2}\bD(\sigma) = \begin{pmatrix}
    O_p(g)\boldsymbol{1}_{p_b}^{\otimes 2} & O_p(g)\boldsymbol{1}_{[p_b:p_w]} \\
    O_p(g)\boldsymbol{1}_{[p_w:p_b]} & 
    O_p(n)\boldsymbol{1}_{p_w}^{\otimes 2}
\end{pmatrix}.
\end{equation}
Similarly, for $\sige^2$, 
\begin{equation}
\label{eq:order of nabla_sige^2 D}
\nabla_{\sige^2}\bD(\sigma) = \begin{pmatrix}
    O_p(g)\boldsymbol{1}_{p_b}^{\otimes 2} & O_p(g)\boldsymbol{1}_{[p_b:p_w]} \\
    O_p(g)\boldsymbol{1}_{[p_w:p_b]} & 
    O_p(n)\boldsymbol{1}_{p_w}^{\otimes 2}
\end{pmatrix}.
\end{equation}

For $\bH(\bsigma)$, the matrix inverse derivative identity gives
$$\nabla_{\siga^2}\bH(\bsigma) = -\bH(\bsigma)\{\nabla_{\siga^2}\bD(\bsigma)\}\bH(\bsigma),$$
where the expression of each block is
\begin{align*}
\nabla_{\siga^2}\bH_{bb}(\bsigma) = &-\bH_{bb}(\bsigma)\{\nabla_{\siga^2}\bD_{bb}(\bsigma)\} \bH_{bb}(\bsigma) 
-\bH_{bb}(\bsigma)\{\nabla_{\siga^2}\bD_{bw}(\bsigma)\} \bH_{wb}(\bsigma) \\
&-\bH_{bw}(\bsigma)\{\nabla_{\siga^2}\bD_{wb}(\bsigma)\} \bH_{bb}(\bsigma) 
-\bH_{bw}(\bsigma)\{\nabla_{\siga^2}\bD_{ww}(\bsigma)\} \bH_{wb}(\bsigma) \\
=& O_p(g^{-1})\boldsymbol{1}_{p_b}^{\otimes 2}, \\
\nabla_{\siga^2}\bH_{bw}(\bsigma) = &-\bH_{bb}(\bsigma)\{\nabla_{\siga^2}\bD_{bb}(\bsigma)\} \bH_{bw}(\bsigma) 
-\bH_{bb}(\bsigma)\{\nabla_{\siga^2}\bD_{bw}(\bsigma)\} \bH_{ww}(\bsigma) \\
&-\bH_{bw}(\bsigma)\{\nabla_{\siga^2}\bD_{wb}(\bsigma)\} \bH_{bw}(\bsigma) 
-\bH_{bw}(\bsigma)\{\nabla_{\siga^2}\bD_{ww}(\bsigma)\} \bH_{ww}(\bsigma) \\
=& O_p(n^{-1})\boldsymbol{1}_{[p_b:p_w]}, \\
\nabla_{\siga^2}\bH_{wb}(\bsigma) =& \{\nabla_{\siga^2}\bH_{wb}(\bsigma)\}^\top =  O_p(n^{-1})\boldsymbol{1}_{[p_w:p_b]}, \\
\nabla_{\siga^2} \bH_{ww}(\bsigma) =&-\bH_{wb}(\bsigma)\{\nabla_{\siga^2}\bD_{wb}(\bsigma)\} \bH_{bw}(\bsigma) 
-\bH_{wb}(\bsigma)\{\nabla_{\siga^2}\bD_{bw}(\bsigma)\} \bH_{ww}(\bsigma) \\
&-\bH_{ww}(\bsigma)\{\nabla_{\siga^2}\bD_{wb}(\bsigma)\} \bH_{bw}(\bsigma) 
-\bH_{ww}(\bsigma)\{\nabla_{\siga^2}\bD_{ww}(\bsigma)\} \bH_{ww}(\bsigma) \\
=& O_p(n^{-1})\boldsymbol{1}_{p_w}^{\otimes 2},
\end{align*} 
uniformly for $\bsigma \in \mathcal{N}_{\bsigma}$. Combining these results, we obtain 
\begin{equation}
\label{eq:order of nabla_siga^2 H}
\nabla_{\siga^2} \bH(\bsigma) = \begin{pmatrix}
    O_p(g^{-1})\boldsymbol{1}_{p_b}^{\otimes 2} & O_p(n^{-1})\boldsymbol{1}_{[p_b:p_w]} \\
    O_p(n^{-1})\boldsymbol{1}_{[p_w:p_b]} &
    O_P(n^{-1})\boldsymbol{1}_{p_w}^{\otimes 2}
\end{pmatrix}.
\end{equation}
Similarly, for $\sige^2$, 
\begin{equation}
\label{eq:order of nabla_sige^2 H}
\nabla_{\sige^2} \bH(\bsigma) = \begin{pmatrix}
    O_p(g^{-1})\boldsymbol{1}_{p_b}^{\otimes 2} & O_p(n^{-1})\boldsymbol{1}_{[p_b:p_w]} \\
    O_p(n^{-1})\boldsymbol{1}_{[p_w:p_b]} &
    O_P(n^{-1})\boldsymbol{1}_{p_w}^{\otimes 2}
\end{pmatrix}.
\end{equation}

\subsection{Orders of the Second Derivatives of $\bD(\bsigma)$ and Its Inverse}
\label{subsec:second derivatives of D}
In this section, we establish the corresponding bounds for the second derivatives of $\bD(\bsigma)$ and $\bH(\bsigma)$. We illustrate using the $\siga^2$ block; the results of $\sige^2$ block can be obtained by following the same argument. By definition,
$$\nabla_{\siga^2}^2 \bD_{bb}(\bsigma) = \nabla_{\siga^2}^2 \bD_{L,bb}(\bsigma) + \nabla_{\siga^2}^2 \nabla_{\bbetab}^2R(\btheta)\big|_{\bbeta = \beta(\bsigma)}.$$
For the leading part, differentiating \eqref{eq:derivative of D_{L, bb}} with respect to $\siga^2$ gives
\begin{equation}
\label{eq:second derivative of D_{L, bb}}
\begin{aligned}
&\nabla_{\siga^2}^{2}\bD_{L,bb}(\boldsymbol\sigma) \\
=&\left.\nabla_{\siga^2}^{2}\left[-\sum_{i=1}^{g}\siga^{-2}
\left\{\frac{\btau_{i,2}^{(bb)*}}{A_i^*}
-\frac{\btau_{i,4}^{(bb)*}}{2(A_i^*)^3}
-\frac{\btau_{i,3}^{(b)*}
\btau_{i,3}^{(b)*\top}}{(A_i^*)^4}\right\}\right]\right|_{\bbeta=\beta(\bsigma)}\\
&+2\left.\nabla_{\siga^2}\nabla_{\bbetab}
\left[-\sum_{i=1}^{g}\siga^{-2}
\left\{\frac{\btau_{i,2}^{(bb)*}}{A_i^*}
-\frac{\btau_{i,4}^{(bb)*}}{2(A_i^*)^3}
-\frac{\btau_{i,3}^{(b)*}\btau_{i,3}^{(b)*\top}}{(A_i^*)^4}\right\}\right]\right|_{\bbeta=\beta(\bsigma)} \nabla_{\siga^2} \bbetab(\bsigma)\\
&+2\left.\nabla_{\siga^2}\nabla_{\bbetaw}
\left[-\sum_{i=1}^{g}\siga^{-2}\left\{
\frac{\btau_{i,2}^{(bb)*}}{A_i^*}
-\frac{\btau_{i,4}^{(bb)*}}{2(A_i^*)^3}
-\frac{\btau_{i,3}^{(b)*}\btau_{i,3}^{(b)*\top}}{(A_i^*)^4}\right\}\right]\right|_{\bbeta=\beta(\bsigma)}\nabla_{\siga^2} \bbetaw(\bsigma)\\
&+\left.\nabla_{\bbetab}^{2}
\left[-\sum_{i=1}^{g}\siga^{-2}
\left\{\frac{\btau_{i,2}^{(bb)*}}{A_i^*}
-\frac{\btau_{i,4}^{(bb)*}}{2(A_i^*)^3}
-\frac{\btau_{i,3}^{(b)*}\btau_{i,3}^{(b)*\top}}{(A_i^*)^4}\right\}\right]\right|_{\bbeta=\beta(\bsigma)}
\left[\nabla_{\siga^2}\bbetab(\bsigma),\nabla_{\siga^2}\bbetab(\bsigma)\right]\\
&+2\left.
\nabla_{\bbetab}\nabla_{\bbetaw}
\left[-\sum_{i=1}^{g}\siga^{-2}
\left\{\frac{\btau_{i,2}^{(bb)*}}{A_i^*}
-\frac{\btau_{i,4}^{(bb)*}}{2(A_i^*)^3}
-\frac{\btau_{i,3}^{(b)*}\btau_{i,3}^{(b)*\top}}{(A_i^*)^4}\right\}\right]\right|_{\bbeta=\beta(\bsigma)}\left[\nabla_{\siga^2}\bbetab(\bsigma),\nabla_{\siga^2}\bbetaw(\bsigma)\right]\\
&+\left.\nabla_{\bbetaw}^{2}
\left[-\sum_{i=1}^{g}\siga^{-2}
\left\{\frac{\btau_{i,2}^{(bb)*}}{A_i^*}
-\frac{\btau_{i,4}^{(bb)*}}{2(A_i^*)^3}
-\frac{\btau_{i,3}^{(b)*}\btau_{i,3}^{(b)*\top}}{(A_i^*)^4}\right\}\right]\right|_{\bbeta=\beta(\bsigma)}\left[\nabla_{\siga^2}\bbetaw(\bsigma),\nabla_{\siga^2}\bbetaw(\bsigma)\right]\\
&+\left.\nabla_{\bbetab}\left[-\sum_{i=1}^{g}\siga^{-2}
\left\{\frac{\btau_{i,2}^{(bb)*}}{A_i^*}
-\frac{\btau_{i,4}^{(bb)*}}{2(A_i^*)^3}
-\frac{\btau_{i,3}^{(b)*}\btau_{i,3}^{(b)*\top}}{(A_i^*)^4}\right\}\right]\right|_{\bbeta=\beta(\bsigma)}\nabla_{\siga^2}^{2}\bbetab(\bsigma)\\
&+\left.\nabla_{\bbetaw}\left[
-\sum_{i=1}^{g}\siga^{-2}
\left\{\frac{\btau_{i,2}^{(bb)*}}{A_i^*}
-\frac{\btau_{i,4}^{(bb)*}}{2(A_i^*)^3}
-\frac{\btau_{i,3}^{(b)*}\btau_{i,3}^{(b)*\top}}{(A_i^*)^4}\right\}\right]\right|_{\bbeta=\beta(\bsigma)}
\nabla_{\siga^2}^{2}\bbetaw(\bsigma).
\end{aligned}
\end{equation}
Here, square brackets for $$\left[\nabla_{\siga^2}\bbetab(\bsigma),\nabla_{\siga^2}\bbetab(\bsigma)\right], \quad \left[\nabla_{\siga^2}\bbetab(\bsigma),\nabla_{\siga^2}\bbetaw(\bsigma)\right], \quad \left[\nabla_{\siga^2}\bbetaw(\bsigma),\nabla_{\siga^2}\bbetaw(\bsigma)\right]$$ denote contraction of the derivative tensor with the indicated profile-derivative vectors over its fixed-effect indices.

For the first term, by standard calculus, we have
\begin{equation}
\label{eq:derivative of the first term in nabla^2_siga^2 D_L,dd}
\begin{aligned}
\nabla_{\siga^2}^{2}&\left[-\sum_{i=1}^{g}\siga^{-2}
\left\{\frac{\btau_{i,2}^{(bb)*}}{A_i^*}
-\frac{\btau_{i,4}^{(bb)*}}{2(A_i^*)^3}
-\frac{\btau_{i,3}^{(b)*}
\btau_{i,3}^{(b)*\top}}{(A_i^*)^4}\right\}\right]\Bigg|_{\bbeta=\bbeta(\bsigma)} \\
=&-2\siga^{-6}\left[\frac{\btau_{i,2}^{(bb)*}}{A_i^*}
-\frac{\btau_{i,4}^{(bb)*}}{2(A_i^*)^3}
+\frac{\btau_{i,3}^{(b)*}\btau_{i,3}^{(b)*\top}}{(A_i^*)^4}\right]\\
&+\siga^{-4}\left[\frac{2}{A_i^*}\nabla_{\siga^2} \btau_{i,2}^{(bb)*}-\frac{2\btau_{i,2}^{(bb)*}}{(A_i^*)^2}\nabla_{\siga^2} A_i^*
-\frac{1}{(A_i^*)^3}\nabla_{\siga^2} \btau_{i,4}^{(bb)*}+\frac{3\btau_{i,4}^{(bb)*}}{(A_i^*)^4}\nabla_{\siga^2} A_i^*\right]
\\
&-\siga^{-2}\left[\frac{1}{A_i^*}\nabla_{\siga^2}^2\btau_{i,2}^{(bb)*}
-\frac{2}{(A_i^*)^2}\nabla_{\siga^2} \btau_{i,2}^{(bb)*}\nabla_{\siga^2} A_i^*
-\frac{\btau_{i,2}^{(bb)*}}{(A_i^*)^2}
\nabla_{\siga^2}^2 A_i^*
+\frac{2\btau_{i,2}^{(bb)*}}{(A_i^*)^3}
\left(\nabla_{\siga^2} A_i^*\right)^2\right.\\
&\qquad-\frac{1}{2(A_i^*)^3}
\nabla_{\siga^2} \btau_{i,4}^{(bb)*}+
\frac{3}{(A_i^*)^4}\nabla_{\siga^2} \btau_{i,4}^{(bb)*}\nabla_{\siga^2} A_i^* \\
&\qquad+
\frac{3\btau_{i,4}^{(bb)*}}{2(A_i^*)^4}
\nabla_{\siga^2}^2 A_i^*
-\frac{6\btau_{i,4}^{(bb)*}}{(A_i^*)^5}
\left(\nabla_{\siga^2}A_i^*\right)^2\\
&\qquad-\frac{\nabla_{\siga^2}^2 \btau_{i,3}^{(b)*}
\btau_{i,3}^{(b)*\top}
+2\nabla_{\siga^2}\btau_{i,3}^{(b)*}
\left(\nabla_{\siga^2}\btau_{i,3}^{(b)*}
\right)^\top
+\btau_{i,3}^{(b)*}\left(\nabla_{\siga^2}^2 \btau_{i,3}^{(b)*}
\right)^\top}{(A_i^*)^4}\\
&\qquad+\frac{8\nabla_{\siga^2} \btau_{i,3}^{(b)*}
\btau_{i,3}^{(b)*\top}+8\btau_{i,3}^{(b)*}
\left(\nabla_{\siga^2} \btau_{i,3}^{(b)*}
\right)^\top}{(A_i^*)^5}
\nabla_{\siga^2}A_i^*\\
&\qquad
\left.+\frac{4\btau_{i,3}^{(b)*}\btau_{i,3}^{(b)*\top}
}{(A_i^*)^5}\nabla_{\siga^2}^2 A_i^*-\frac{20\btau_{i,3}^{(b)*}\btau_{i,3}^{(b)*\top}
}{(A_i^*)^6}\left(\nabla_{\siga^2} A_i^*
\right)^2\right].
\end{aligned}
\end{equation}
For the derivatives involved in \eqref{eq:derivative of the first term in nabla^2_siga^2 D_L,dd}, uniformly for $\bsigma \in \mathcal{N}_{\bsigma}$, we have 
\begin{align*}
    \nabla_{\siga^2}^2 \alpha_i^* &= -\frac{2\siga^{-6}\alpha_i*}{A_i^*} + \frac{\siga^{-4}\nabla_{\siga^2}\alpha^*}{A_i^*} - \frac{\siga^{-4}\alpha_i^*\nabla_{\siga^2}A_i^*}{(A_i^*)^2} = O_p(m_i^{-1}), \\
    \nabla_{\siga^2}^2 A_i^* &= 2\siga^{-6} + (\nabla_{\siga^2}\tau_{i,3}^*)\nabla_{\siga^2}\alpha_i^* + \tau_{i,3}^*\nabla_{\siga^2}^2\alpha_i^* = O_p(1).
\end{align*}
Substituting these bounds back into \eqref{eq:derivative of the first term in nabla^2_siga^2 D_L,dd}, we have 
$$\nabla_{\siga^2}^{2}\left[-\sum_{i=1}^{g}\siga^{-2}
\left\{\frac{\btau_{i,2}^{(bb)*}}{A_i^*}
-\frac{\btau_{i,4}^{(bb)*}}{2(A_i^*)^3}
-\frac{\btau_{i,3}^{(b)*}
\btau_{i,3}^{(b)*\top}}{(A_i^*)^4}\right\}\right]\Bigg|_{\bbeta=\bbeta(\bsigma)} = O_p(g)\boldsymbol{1}_{p_b}^{\otimes 2}.$$ By the same argument, we have
\begin{align*}
&\left.\nabla_{\siga^2}\nabla_{\bbetab}
\left[-\sum_{i=1}^{g}\siga^{-2}
\left\{\frac{\btau_{i,2}^{(bb)*}}{A_i^*}
-\frac{\btau_{i,4}^{(bb)*}}{2(A_i^*)^3}
-\frac{\btau_{i,3}^{(b)*}\btau_{i,3}^{(b)*\top}}{(A_i^*)^4}\right\}\right]\right|_{\bbeta=\bbeta(\bsigma)} = O_p(gm_L^{-1})\boldsymbol{1}_{p_b \times p_b \times p_b}, \\
&\left.\nabla_{\siga^2}\nabla_{\bbetaw}
\left[-\sum_{i=1}^{g}\siga^{-2}\left\{
\frac{\btau_{i,2}^{(bb)*}}{A_i^*}
-\frac{\btau_{i,4}^{(bb)*}}{2(A_i^*)^3}
-\frac{\btau_{i,3}^{(b)*}\btau_{i,3}^{(b)*\top}}{(A_i^*)^4}\right\}\right]\right|_{\bbeta=\bbeta(\bsigma)} = O_p(g)\boldsymbol{1}_{p_b \times p_b \times p_w}, \\
&\left.\nabla_{\bbetab}^{2}
\left[-\sum_{i=1}^{g}\siga^{-2}
\left\{\frac{\btau_{i,2}^{(bb)*}}{A_i^*}
-\frac{\btau_{i,4}^{(bb)*}}{2(A_i^*)^3}
-\frac{\btau_{i,3}^{(b)*}\btau_{i,3}^{(b)*\top}}{(A_i^*)^4}\right\}\right]\right|_{\bbeta=\bbeta(\bsigma)} = O_p(gm_L^{-1})\boldsymbol{1}_{p_b \times p_b \times p_b \times p_b}, \\
&\left.
\nabla_{\bbetab}\nabla_{\bbetaw}
\left[-\sum_{i=1}^{g}\siga^{-2}
\left\{\frac{\btau_{i,2}^{(bb)*}}{A_i^*}
-\frac{\btau_{i,4}^{(bb)*}}{2(A_i^*)^3}
-\frac{\btau_{i,3}^{(b)*}\btau_{i,3}^{(b)*\top}}{(A_i^*)^4}\right\}\right]\right|_{\bbeta=\bbeta(\bsigma)} = O_p(g)\boldsymbol{1}_{p_b \times p_b \times p_b \times p_w}, \\
&\left.\nabla_{\bbetaw}^{2}
\left[-\sum_{i=1}^{g}\siga^{-2}
\left\{\frac{\btau_{i,2}^{(bb)*}}{A_i^*}
-\frac{\btau_{i,4}^{(bb)*}}{2(A_i^*)^3}
-\frac{\btau_{i,3}^{(b)*}\btau_{i,3}^{(b)*\top}}{(A_i^*)^4}\right\}\right]\right|_{\bbeta=\bbeta(\bsigma)} = O_p(g)\boldsymbol{1}_{p_b \times p_b \times p_w \times p_w}, \\
&\left.\nabla_{\bbetab}\left[-\sum_{i=1}^{g}\siga^{-2}
\left\{\frac{\btau_{i,2}^{(bb)*}}{A_i^*}
-\frac{\btau_{i,4}^{(bb)*}}{2(A_i^*)^3}
-\frac{\btau_{i,3}^{(b)*}\btau_{i,3}^{(b)*\top}}{(A_i^*)^4}\right\}\right]\right|_{\bbeta=\bbeta(\bsigma)} = O_p(gm_L^{-1})\boldsymbol{1}_{p_b \times p_b \times p_b}, \\
&\left.\nabla_{\bbetaw}\left[
-\sum_{i=1}^{g}\siga^{-2}
\left\{\frac{\btau_{i,2}^{(bb)*}}{A_i^*}
-\frac{\btau_{i,4}^{(bb)*}}{2(A_i^*)^3}
-\frac{\btau_{i,3}^{(b)*}\btau_{i,3}^{(b)*\top}}{(A_i^*)^4}\right\}\right]\right|_{\bbeta=\bbeta(\bsigma)} = O_p(g)\boldsymbol{1}_{p_b \times p_b \times p_w}.
\end{align*} 
Here, the third- and fourth-order derivatives are interpreted as tensors, with dimensions determined by the corresponding parameter blocks. For example, $O_p(gm_L^{-1})\boldsymbol{1}_{p_b \times p_b \times p_b}$ denotes a third-order tensor of dimension $p_b \times p_b \times p_b$ whose entries are all $O_p(gm_L^{-1})$.

We next establish the order of $\nabla_{\siga^2}^2 \bbetab(\bsigma)$ and $\nabla_{\siga^2}^2 \bbetaw(\bsigma)$. Differentiating \eqref{eq:derivative of profile score} with respect $\siga^2$ gives 
\begin{align*}
    -\nabla_{\siga^2}\bD(\bsigma)\left\{\nabla_{\siga^2}\bbeta(\bsigma)\right\}^\top -\bD(\bsigma)\left\{\nabla^2_{\siga^2}\bbeta(\bsigma)\right\}^\top + \nabla_{\siga^2}\left\{\nabla_{\siga^2}\bpsi_{\bbeta}^\top(
    \bbeta, \bsigma)\big|_{\bbeta = \bbeta(\bsigma)}\right\} = \boldsymbol{0}.
\end{align*}
Rearranging gives 
\begin{align*}
\left\{\nabla_{\siga^2}^2\bbeta(\bsigma)\right\}^\top &= \bH(\bsigma)\left[\nabla_{\siga^2}\left\{\nabla_{\siga^2}\bpsi_{\bbeta}^\top(
    \bbeta, \bsigma)\big|_{\bbeta = \bbeta(\bsigma)}\right\} - \nabla_{\siga^2} \bD(\bsigma) \left\{\nabla_{\siga^2}\bbeta(\bsigma)\right\}^\top\right] \\
&= \begin{pmatrix}
    \bH_{bb}(\bsigma) & \bH_{bw}(\bsigma) \\
    \bH_{bw}(\bsigma) & \bH_{ww}(\bsigma) 
\end{pmatrix} \left[\begin{pmatrix}
    \nabla_{\siga^2}\left\{\nabla_{\siga^2}\bpsi_{\bbetab}^\top(
    \bbeta, \bsigma)\big|_{\bbeta = \bbeta(\bsigma)}\right\} \\
   \nabla_{\siga^2} \left\{\nabla_{\siga^2}\bpsi_{\bbetaw}^\top(
    \bbeta, \bsigma)\big|_{\bbeta = \bbeta(\bsigma)}\right\}
\end{pmatrix}\right. \\
&\left.- \begin{pmatrix}
   \nabla_{\siga^2} \bD_{bb}(\bsigma) &
   \nabla_{\siga^2} \bD_{bw}(\bsigma) \\
   \nabla_{\siga^2} \bD_{wb}(\bsigma) &
   \nabla_{\siga^2} \bD_{ww}(\bsigma) 
\end{pmatrix} \begin{pmatrix}
    \left\{\nabla_{\siga^2} \bbetab(\bsigma)\right\}^\top \\
     \left\{\nabla_{\sige^2} \bbetab(\bsigma)\right\}^\top
\end{pmatrix} \right]  \\
&= \begin{pmatrix}
    \bH_{bb}(\bsigma) & \bH_{bw}(\bsigma) \\
    \bH_{bw}(\bsigma) & \bH_{ww}(\bsigma) 
\end{pmatrix} \begin{pmatrix}
    \bL_b(\bsigma) \\ \bL_w(\bsigma)
\end{pmatrix},
\end{align*}
where 
\begin{align*}
\bL_b(\bsigma) &= \nabla_{\siga^2}\left\{\nabla_{\siga^2}\bpsi_{\bbetab}^\top(
    \bbeta, \bsigma)\big|_{\bbeta = \bbeta(\bsigma)}\right\}
- 
\nabla_{\siga^2}\bD_{bb}(\bsigma) \left\{\nabla_{\siga^2}\bbetab(\bsigma)\right\}^\top \\
&-
\nabla_{\siga^2}\bD_{bw}(\bsigma) \left\{\nabla_{\siga^2}\bbetaw(\bsigma)\right\}^\top, 
\end{align*}
and 
\begin{align*}
\bL_w(\bsigma) &= \nabla_{\siga^2}\left\{\nabla_{\siga^2}\bpsi_{\bbetaw}^\top(
    \bbeta, \bsigma)\big|_{\bbeta = \bbeta(\bsigma)}\right\}
- 
\nabla_{\siga^2}\bD_{wb}(\bsigma) \left\{\nabla_{\siga^2}\bbetab(\bsigma)\right\}^\top \\
&-
\nabla_{\siga^2}\bD_{ww}(\bsigma) \left\{\nabla_{\siga^2}\bbetaw(\bsigma)\right\}^\top.
\end{align*}
The orders of terms $\nabla_{\siga^2}\bD_{bb}(\bsigma), \nabla_{\siga^2}\bD_{bw}(\bsigma), \nabla_{\siga^2}\bD_{wb}(\bsigma), \nabla_{\siga^2}\bD_{ww}(\bsigma)$ and $\left\{\nabla_{\siga^2}\bbetab(\bsigma)\right\}^\top, \left\{\nabla_{\siga^2}\bbetaw(\bsigma)\right\}^\top$ have been established in \eqref{eq:order of nabla beta} and \eqref{eq:order of nabla_siga^2 D}. It remains to establish the order of the first parts in $\bL_b(\bsigma)$ and $\bL_w(\bsigma)$; here we only illustrate for $\bL_b(\bsigma)$.
By the chain rule, we have
\begin{equation}
\label{eq:derivative of first term in L_b}
\begin{aligned}
\nabla_{\siga^2}\left\{\nabla_{\siga^2}\bpsi_{\bbetab}^\top(
    \bbeta, \bsigma)\big|_{\bbeta = \bbeta(\bsigma)}\right\} = \sumig &\left[ \nabla_{\siga^2}^2\bpsi_{i,\bbetab}^\top(\bbeta, \bsigma)\bigg|_{\bbeta = \bbeta(\bsigma)} \right. \\
&+ \nabla_{\bbetab}\nabla_{\siga^2}\bpsi_{i,\bbetab}^\top(\bbeta, \bsigma)\big|_{\bbeta = \bbeta(\bsigma)}\left\{\nabla_{\siga^2}\bbetab(\bsigma)\right\}^\top \\
&\left.+ \nabla_{\bbetaw}\nabla_{\siga^2}\bpsi_{i,\bbetab}^\top(\bbeta, \bsigma)\big|_{\bbeta = \bbeta(\bsigma)}\left\{\nabla_{\siga^2}\bbetaw(\bsigma)\right\}^\top \right].
\end{aligned}
\end{equation}
We evaluate these derivatives from the exact expression of $\nabla_{\siga^2}\bpsi_{i, \bbetab}^\top$ in \eqref{eq:exact expression of ML score derivative}. The leading contribution of $\nabla_{\siga^2}\bpsi_{i, \bbeta}^\top$ contains terms of the form $$-\frac{\siga^{-4}\alpha_i^*\bxbi}{A_i^*}\Big|_{\bbeta = \bbeta(\bsigma)}.$$
Differentiating the displayed term with respect to $\siga^2$ gives 
\begin{align*}
    \nabla_{\siga^2}\left(-\frac{\siga^{-4}\alpha_i^*\bxbi}{A_i^*}\right)\Bigg|_{\bbeta = \bbeta(\bsigma)} = \left(\frac{2\siga^{-6}\alpha_i^*\bxbi}{A_i^*} - \frac{\siga^{-4}\nabla_{\siga^2}\alpha_i^*\bxbi}{A_i*} + \frac{\siga^{-4} \alpha_i^*\bxbi \nabla_{\siga^2}A_i^*}{(A_i^*)^2}\right)\Bigg|_{\bbeta = \bbeta(\bsigma)}.
\end{align*}
We have shown that $\nabla_{\siga^2} \alpha_i^* = O_p(m_i^{-1})$ in \eqref{eq:d alpha_i^*/d theta}, and $$\nabla_{\siga^2}A_i^* = \tau_{i,3}^* \nabla_{\siga^2}\alpha_i^* - \siga^{-4} = O_p(1).$$
The remaining terms in $\nabla_{\siga^2}\bpsi_{i, \bbetab}^\top$ contain higher negative powers of $A_i^*$ and are therefore of smaller order than the leading part. Thus, uniformly for $\bsigma \in \mathcal{N}_{\bsigma}$, we have $$\nabla_{\siga^2}^2\bpsi_{i,\bbetab}(\bbeta, \bsigma)\bigg|_{\bbeta = \bbeta(\bsigma)} = O_p(m_i^{-1})\boldsymbol{1}_{[p_b:1]}.$$
The derivatives with respect to $\bbetab$ and $\bbetaw$ are handled similarly, and we have 
\begin{align*}
\nabla_{\bbetab}\nabla_{\siga^2}\bpsi_{i,\bbetab}^\top(\bbeta, \bsigma)\bigg|_{\bbeta = \bbeta(\bsigma)} &= O_p(m_i^{-1})\boldsymbol{1}_{p_b}^{\otimes 2}, \\
\nabla_{\bbetaw}\nabla_{\siga^2}\bpsi_{i,\bbetab}^\top(\bbeta, \bsigma)\bigg|_{\bbeta = \bbeta(\bsigma)} &= O_p(m_i^{-1})\boldsymbol{1}_{[p_b:p_w]}.
\end{align*}
Substituting these bounds into \eqref{eq:derivative of first term in L_b}, we have $$\nabla_{\siga^2}\left\{\nabla_{\siga^2}\bpsi_{\bbetab}^\top(\bbeta, \bsigma)\big|_{\bbeta = \bbeta(\bsigma)}\right\} = O_p(g)\boldsymbol{1}_{[p_b:1]}.$$ Thus, $\bL_b(\bsigma) = O_p(g)\boldsymbol{1}_{p_b}$. Similarly, $\bL_w(\bsigma) = O_p(n)\boldsymbol{1}_{[p_w:1]}$. Thus, 
\begin{align*}
    \nabla_{\siga^2}^2 \bbetab(\bsigma) &= \bH_{bb}(\bsigma)\bL_b(\bsigma) + \bH_{bw}(\bsigma)\bL_w(\bsigma) \\
    &= O_p(g^{-1})\boldsymbol{1}_{p_b}^{\otimes 2}O_p(g)\boldsymbol{1}_{[p_b:1]} + O_p(n^{-1})\boldsymbol{1}_{[p_b:p_w]}O_p(n)\boldsymbol{1}_{[p_w:1]} \\
    &= O_p(1)\boldsymbol{1}_{[p_b:1]}, \\
    \nabla_{\siga^2}^2 \bbetaw(\bsigma) &= \bH_{wb}(\bsigma)\bL_b(\bsigma) + \bH_{ww}(\bsigma)\bL_w(\bsigma) \\
    &= O_p(n^{-1})\boldsymbol{1}_{[p_w:p_b]}O_p(g)\boldsymbol{1}_{[p_b:1]} + O_p(n^{-1})\boldsymbol{1}_{p_w}^{\otimes 2}O_p(n)\boldsymbol{1}_{[p_w:1]}\\
    &= O_p(1)\boldsymbol{1}_{[p_w:1]}.
\end{align*}

Substituting these orders back into \eqref{eq:second derivative of D_{L, bb}}, we have
$$\nabla_{\siga^2}^2 \bD_{L, bb}(\bsigma) = O_p(g)\boldsymbol{1}_{p_b}^{\otimes 2}.$$ By the same argument, we have 
\begin{align*}
&\nabla_{\siga^2}^2 \bD_{L, bw}(\bsigma) = O_p(g)\boldsymbol{1}_{[p_b:p_w]}, \quad
\nabla_{\siga^2}^2 \bD_{L, wb}(\bsigma) = O_p(g)\boldsymbol{1}_{[p_w:p_b]}, 
\\&\nabla_{\siga^2}^2 \bD_{L, ww}(\bsigma) = O_p(n)\boldsymbol{1}_{p_w}^{\otimes 2}.
\end{align*}

For the remaining term $\nabla^2_{\bbeta^{(b)}}R(
\btheta)$ evaluated at $\bbeta(\bsigma)$, the second derivative with respect to $\siga^2$ are of smaller order than the corresponding leading-part terms. They are obtained by differentiating the dominant correction term in \eqref{eq:second derivative of epsilon_i} twice with respect to $\siga^2$ and applying the same chain rule and argument as above, but involve lengthy differentiation of the higher-order correction terms. We therefore omit the details. Combining the leading and remainder contributions, uniformly for $\bsigma \in \mathcal{N}$, we have
\begin{equation}
\label{eq:order of nabla^2_siga^2 D}
\nabla_{\siga^2}^2\bD(\sigma) = \begin{pmatrix}
    O_p(g)\boldsymbol{1}_{p_b}^{\otimes 2} & O_p(g)\boldsymbol{1}_{[p_b:p_w]} \\
    O_p(g)\boldsymbol{1}_{[p_w:p_b]} & 
    O_p(n)\boldsymbol{1}_{p_w}^{\otimes 2}
\end{pmatrix}.
\end{equation}
Similarly, for $\sige^2$,
\begin{equation}
\label{eq:order of nabla^2_sige^2 D}
\nabla_{\sige^2}^2\bD(\sigma) = \begin{pmatrix}
    O_p(g)\boldsymbol{1}_{p_b}^{\otimes 2} & O_p(g)\boldsymbol{1}_{[p_b:p_w]} \\
    O_p(g)\boldsymbol{1}_{[p_w:p_b]} & 
    O_p(n)\boldsymbol{1}_{p_w}^{\otimes 2}
\end{pmatrix}.
\end{equation}

For $\bH(\bsigma)$, differentiating the inverse identity once more gives \begin{align*}
    \nabla_{\siga^2}^2 \bH(\bsigma) =& 2\bH(\bsigma)\{\nabla_{\siga^2}\bD(\bsigma)\}\bH(\bsigma)\{\nabla_{\siga^2}\bD(\bsigma)\}\bH(\bsigma) \\
    &-\bH(\bsigma)\{\nabla_{\siga^2}^2\bD(\bsigma)\}\bH(\bsigma).
\end{align*}
The derivation of orders for each block is similar to the derivation for $\nabla_{\siga^2} \bH(\bsigma)$ in Section~\ref{subsec:first derivative of D}, so we omit the details and directly present the result. Uniformly for $\bsigma \in \mathcal{N}$, we have
\begin{equation}
\label{eq:order of nabla^2_siga^2 H}
\nabla_{\siga^2}^2\bH(\sigma) = \begin{pmatrix}
    O_p(g^{-1})\boldsymbol{1}_{p_b}^{\otimes 2} & O_p(n^{-1})\boldsymbol{1}_{[p_b:p_w]} \\
    O_p(n^{-1})\boldsymbol{1}_{[p_w:p_b]} & 
    O_p(n^{-1})\boldsymbol{1}_{p_w}^{\otimes 2}
\end{pmatrix}.
\end{equation}
Similarly, for $\sige^2$,
\begin{equation}
\label{eq:order of nabla^2_sige^2 H}
\nabla_{\sige^2}^2\bH(\sigma) = \begin{pmatrix}
    O_p(g^{-1})\boldsymbol{1}_{p_b}^{\otimes 2} & O_p(n^{-1})\boldsymbol{1}_{[p_b:p_w]} \\
    O_p(n^{-1})\boldsymbol{1}_{[p_w:p_b]} & 
    O_p(n^{-1})\boldsymbol{1}_{p_w}^{\otimes 2}
\end{pmatrix}.
\end{equation}

\subsection{Orders of the First and Second Derivatives of Laplace Remainder $U(\btheta)$}\label{subsec:contribution of U(theta)}
To express the dominant term in $U(\bsigma)$, we denote $$\bbeta = (\beta_1, \ldots, \beta_p)^\top,$$ where the first $p_b$ components corresponding to $\bbeta^{(b)}$ and the remaining $p_w$ components correspond to $\bbeta^{(w)}$. For $a,b=1,\ldots,p$, denote the entry of $ \bH(\bsigma)$ at the position $(a,b)$ by $$H_{ab}(\bsigma) = \{\bH(\bsigma)\}_{ab}.$$
Define the third- and fourth-order derivatives of the exact marginal log-likelihood with respect to the components of $\bbeta$ by
\begin{align*}
    d_{abc}(\bsigma) &= \frac{\partial^3 \ell(\bbeta, \bsigma)}{\partial \beta_a \partial \beta_b \partial \beta_c}\bigg|_{\bbeta=\bbeta(\bsigma)}, \\
    d_{abcd}(\bsigma) &= \frac{\partial^4 \ell(\bbeta, \bsigma)}{\partial \beta_a \partial \beta_b \partial \beta_c \partial \beta_d}\bigg|_{\bbeta=\bbeta(\bsigma)}.
\end{align*}

Refer to the result in equation (2) of \cite{shun1995laplace}, the fixed-effect Laplace correction $\epsilon_{\bbeta}$ admits the expansion
$$\epsilon_{\bbeta} = \epsilon_{\bbeta1} + \epsilon_{\bbeta2} + \cdots,$$
where the dominant term is 
\begin{equation}
\label{eq:expression of dominant term in U}
\begin{aligned}
    \epsilon_{\beta1} &= \frac{1}{24} \sum_{a,b,c,d=1}^p d_{abcd}\left(H_{ab}H_{cd}+H_{ac}H_{bd}+H_{ad}H_{bc}\right) \\
    &+ \frac{1}{72}\sum_{a,b,c,d,e,f=1}^p d_{abc} d_{def}\mathcal{W}_{abcdef}.
\end{aligned}
\end{equation}
Here, $\mathcal{W}_{abcdef}$ denotes the sum over the 15 combinations of the six indices $a,b,c,d,e,f$. Explicitly
\begin{align*}
    \mathcal{W}_{abcdef} &= H_{ab}H_{cd}H_{ef} + H_{ab}H_{ce}H_{df} + H_{ab}H_{cf}H_{de} \\
    &+ H_{ac}H_{bd}H_{ef} + H_{ac}H_{be}H_{df} + H_{ac}H_{bf}H_{de} \\
    &+ H_{ad}H_{bc}H_{ef} + H_{ad}H_{be}H_{cf} + H_{ad}H_{bf}H_{ce} \\
    &+ H_{ae}H_{bc}H_{df} + H_{ae}H_{bd}H_{cf} + H_{ae}H_{bf}H_{cd} \\
    &+ H_{af}H_{bc}H_{de} + H_{af}H_{bd}H_{ce} + H_{af}H_{be}H_{cd}. 
\end{align*}
For brevity, the argument $\bsigma$ in $d_{abc}(\bsigma)$ and $H_{ab}(\bsigma)$ has been suppressed in the display above. As the expressions of the remaining terms $\epsilon _{2, \beta}, \ldots$ are massive, and all are of smaller order than $\epsilon _{1, \beta}$, we omit their details in the Supplementary.

Since the dimension $p$ is fixed, the sums in \eqref{eq:expression of dominant term in U} contain only finitely many terms. Recall that $H_{ab}$ denotes the $(a,b)$ entry of $\bH(\bsigma)$. Therefore, the block-wise bounds established in \eqref{eq:order of nabla_siga^2 H}-\eqref{eq:order of nabla_sige^2 H} and \eqref{eq:order of nabla^2_siga^2 H}-\eqref{eq:order of nabla^2_sige^2 H} apply entry-wise. Similarly, $d_{abc}$ and $d_{abcd}$ are component entries of the third- and fourth-order fixed-effect derivative tensors, respectively. Their first and second derivatives with respect to $\bsigma$ are therefore component entries of the mixed derivative tensors controlled in \eqref{eq:order of nabla_siga^2 D}-\eqref{eq:order of nabla_sige^2 D} and \eqref{eq:order of nabla^2_siga^2 D}-\eqref{eq:order of nabla^2_sige^2 D}.

Differentiating the first contraction in \eqref{eq:expression of dominant term in U} gives 
\begin{equation}
\label{eq:second contraction in U}
\begin{aligned}
\nabla_{\siga^2}&\left\{d_{abcd}\left(H_{ab}H_{cd}+H_{ac}H_{bd}+H_{ad}H_{bc}\right)\right\} \\
&= \left(\nabla_{\siga^2}d_{abcd}\right)\left(H_{ab}H_{cd}+H_{ac}H_{bd}+H_{ad}H_{bc}\right) \\
&\quad+ d_{abcd}\left\{(\nabla_{\siga^2}H_{ab})H_{cd} + H_{ab}(\nabla_{\siga^2}H_{cd})\right. \\
&\qquad \qquad + (\nabla_{\siga^2}H_{ac})H_{bd} + H_{ac}(\nabla_{\siga^2}H_{bd}) \\
&\qquad \qquad \left. + (\nabla_{\siga^2}H_{ad})H_{bc} + H_{ad}(\nabla_{\siga^2}H_{bc}) \right\} \\
&=O_p(1).
\end{aligned}
\end{equation}
Similarly, a generic term in the second contraction in \eqref{eq:expression of dominant term in U} satisfies 
\begin{equation}
\label{eq:first contraction in U}
\begin{aligned}
\nabla_{\siga^2}(d_{abc}d_{def}H_{ab}H_{cd}H_{ef}) &= (\nabla_{\siga^2}^2 d_{abc})d_{def}H_{ab}H_{cd}H_{ef} \\
&\quad + d_{abc}(\nabla_{\siga^2}d_{def})H_{ab}H_{cd}H_{ef} \\
&\quad + d_{abc}d_{def}(\nabla_{\siga^2}H_{ab})H_{cd}H_{ef} \\
&\quad + d_{abc}d_{def}H_{ab}(\nabla_{\siga^2}H_{cd})H_{ef} \\
&\quad + d_{abc}d_{def}H_{ab}H_{cd}(\nabla_{\siga^2}H_{ef}) \\
&= O_p(1).
\end{aligned}
\end{equation}
Since $p$ is fixed, summation over the component indices does not alter the stochastic order. Hence, $$\nabla_{\siga^2}\epsilon_{\beta1} = O_p(1).$$
The higher-order correction terms $\epsilon_{\beta2}, \epsilon_{\beta3}, \ldots ,$ together with their first and second derivatives, are of smaller order than the corresponding contributions from $\epsilon_{\beta1}$. Therefore,
$$\nabla_{\siga^2} \epsilon_{\beta} = O_p(1), \qquad \nabla_{\siga^2}^2 \epsilon_{\beta} = O_p(1).$$ The same entry-wise argument, using the corresponding bounds for derivatives with respect to $\sige^2$, gives
$$\nabla_{\sige^2} \epsilon_{\beta} = O_p(1), \qquad \nabla_{\sige^2}^2 \epsilon_{\beta} = O_p(1).$$

Recall that $$U(\bsigma) = \log(1+\epsilon_{\beta}).$$ We have
\begin{equation}
\label{eq:order of nabla^2 U} 
\nabla_{\bsigma} U(\bsigma) = \frac{\nabla_{\bsigma}\epsilon_{\beta}}{1+\epsilon_{\beta}} = O_p(1),
\end{equation} 
and 
\begin{equation}
\label{eq:order of nabla U}   
\nabla_{\bsigma}^2 U(\bsigma) = \frac{\nabla_{\bsigma}^2\epsilon_{\beta}}{1+\epsilon_{\beta}} - \left(\frac{\nabla_{\bsigma} \epsilon_{\beta}}{1+\epsilon_{\beta}}\right)^2 = O_p(1).
\end{equation}

\section{Proof of Theorems}
\setcounter{equation}{0} 
The proofs of Theorems 1 and 2 are presented in Sections~\ref{subsec:proof of theorem 1} and \ref{subsec:proof of theorem 2}, respectively. We begin with Theorem 2, which establishes the asymptotic linear representation and limiting distribution of the exact ML estimator. The proof of Theorem 1 then builds on the ML results to derive the corresponding first-order properties of the REML estimator. The auxiliary lemmas used in these proofs are collected in Section S.4.3, and their detailed proofs are provided in Section S.5.
\subsection{Proof of Theorem 2}
\label{subsec:proof of theorem 2}
Let $\bxi$ be the leading stochastic term of $\bpsi(\btheta)$ at the true parameter with the form derived in \eqref{eq:expression of xi}. Write $\bpsi(\btheta) = \bpsi(\boldsymbol{\dot{\theta}}) + \bpsi(\btheta) - \bpsi(\dot{\btheta})$. 
Then
\begin{equation*}
    \bK^{-1/2}\bpsi(\btheta) = \bK^{-1/2}\xi - \bB\bK^{1/2}(\btheta -\dbtheta) + T_1 + T_2(\btheta) + T_3(\btheta),
\end{equation*}
where
\begin{align*}
    &\bB = \lim_{g, m_L \rightarrow \infty}\left\{-\bK^{-1/2}\mathbb{E}\nabla\bpsi(\dbtheta)\bK^{-1/2} \right\}, \\
    &T_1 = \bK^{-1/2}\{\bpsi(\dbtheta)-\bxi\}, \\
    &T_2(\btheta) = \bK^{-1/2}\mathbb{E}\{\bpsi(\btheta) - \bpsi(\dbtheta)\} + \bB\bK^{-1/2}(\btheta -\dbtheta),
\end{align*}
and 
$$T_3(\btheta) = \bK^{-1/2}\left[\bpsi(\btheta) - \bpsi(\dbtheta) - \mathbb{E}\left\{\bpsi(\btheta) - \bpsi(\dbtheta)\right\}\right].$$
For $0<M<\infty$, define the local shrinking neighborhood $\mathcal{N} = \{\btheta:\|\bK^{1/2}(\btheta - \dot{\btheta})\| \leq M\}$. We show below that 
$$
\|T_1\| = o_p(1), \quad \sup_{\btheta\in\mathcal{N}}\|T_2(\btheta)\| = o(1), \quad
\sup_{\btheta\in\mathcal{N}}\|T_3(\btheta)\| = o(1).
$$
It then follows that, uniformly on $\mathcal{N}$,
\begin{equation}
\label{eq:asymptotic linear representation}
\bK^{-1/2}\bpsi(\btheta) = \bK^{-1/2}\boldsymbol{\xi} - \mathbf{B}\bK^{1/2}(\btheta - \dot{\btheta}) + o_p(1).
\end{equation}
Since $\mathbf{B}$ is positive definite, the right-hand side is negative for $|\bK^{1/2}(\btheta - \dot{\btheta})|=M$ sufficiently large. Then, according
to Result 6.3.4 of \cite{ortega2000iterative}, a solution to the estimating equations exists in probability and satisfies $|\bK^{1/2}(\hat{\btheta} - \dot{\btheta})| = O_p(1)$, so $\hat{\btheta} \in \mathcal{N}$. This allows us to substitute $\hat{\btheta}$ for $\btheta$ in \eqref{eq:asymptotic linear representation} and rearrange the terms to obtain the asymptotic representation for $\hat{\btheta}$; the central limit theorem follows from the asymptotic representation and the central limit theorem for $\boldsymbol{\xi}$ that we establish in Lemma~\ref{lemma 1}.

It remains to establish the bounds of  $T_1$, $T_2(\btheta)$ and $T_3(\btheta)$. By Lemma~\ref{lemma 2}, $\|T_1\| = o_p(1)$. 

To handle $T_2(\btheta)$ and $T_3(\btheta)$, we adapt the one-step expansion argument originally used by \cite{bickel1975one} and later extended to maximum likelihood and REML estimation in LMMs by \cite{richardson1994asymptotic}. We further follow the technical procedure developed in \cite{lyu2022increasing}, where analogous plug-in terms are controlled in the score expansion for LMMs. For $T_2(\btheta)$, we have 
\begin{align*}
    \sup_{\btheta\in\mathcal{N}}\|T_2(\btheta)\|  &\leq  \sup_{\btheta\in\mathcal{N}} \left\| \bK^{-1/2}\left[\mathbb{E}\left\{\bpsi(\btheta) - \bpsi(\dot{\btheta})\right\} - \mathbb{E}\nabla \bpsi(\btheta)(\btheta - \dot{\btheta})\right] \right\| \\
    &+ \sup_{\btheta\in\mathcal{N}} \left\| \bK^{-1/2}\mathbb{E}\nabla \bpsi(\btheta)(\btheta - \dot{\btheta})+ \mathbf{B}\bK^{1/2}(\btheta - \dot{\btheta}) \right\| \\
    &\leq M \sup_{\btheta\in\mathcal{N}}\left\Vert \bK^{-1/2}\left\{\mathbb{E}\nabla\bpsi(\boldsymbol{\Omega}) - \mathbb{E}\nabla \bpsi(\dot{\btheta})\right\} \bK^{-1/2}\right\Vert \\ &
    + M \left\Vert -\bK^{-1/2}\mathbb{E}\nabla\bpsi(\dot{\btheta})\bK^{-1/2} - \mathbf{B} \right\Vert  \\
    &\leq M \sup_{\btheta\in\mathcal{N}}\left\Vert \bK^{-1/2}\left\{\mathbb{E}\nabla\bpsi(\boldsymbol{\Omega}) - \mathbb{E}\nabla \bpsi(\dot{\btheta})\right\} \bK^{-1/2}\right\Vert + M \left\Vert\mathbf{B}_n - \mathbf{B} \right\Vert, 
\end{align*}
where each of $\boldsymbol{\Omega}$ lies between $\btheta$ and $\dot{\btheta}$ and $\mathbf{B}_n = -\bK^{-1/2}\mathbb{E}\nabla\bpsi(\dot{\btheta})\bK^{-1/2}$. We show that $\Vert\mathbf{B}_n - \mathbf{B}\Vert = o(1)$ in Lemma~\ref{lemma 3} and in Lemma~\ref{lemma 4},
$$\sup_{\btheta \in \mathcal{N}}\left\Vert\bK^{-1/2}\mathbb{E}\left\{ \nabla \bpsi(\boldsymbol{\Omega}) - \nabla \bpsi(\dot{\btheta})\right\}\bK^{-1/2}\right\Vert = o(1).$$

To handle $T_3(\btheta)$, we partition $\mathcal{N} = \{\btheta:|\bK^{1/2}(\btheta - \dot{\btheta})| \leq M\}$ into the set of $N = O(g^{1/4})$ smaller cubes $\mathcal{C} = \{\mathcal{C}(\bt)\}$, where $\mathcal{C}(\bt) = \{\btheta:\|\bK^{1/2}(\btheta - \bt)\| \leq Mg^{-1/4}\}$. Then, we have
\begin{equation}
\begin{aligned}
\label{eq:H2}
\sup_{\btheta\in\mathcal N}
\|T_3(\btheta)\|
&\leq
\max_{1\leq k\leq N}\left\|\bK^{-1/2}\left[\bpsi(\bt_k)-\bpsi(\dot{\btheta})-\mathbb{E}\left\{\bpsi(\bt_k)-\bpsi(\dot{\btheta})\right\}\right]\right\| \\
&\quad+
\max_{1\leq k\leq N}\sup_{\btheta\in \mathcal{C}(\bt_k)}\left\|\bK^{-1/2}\left[\bpsi(\btheta)-\bpsi(\bt_k)-\mathbb{E}\left\{\bpsi(\btheta)-\bpsi(\bt_k)\right\}\right]\right\|,\end{aligned}
\end{equation}
where $\bt_k = (\bt_{k\bbeta^{(b)}}, \bt_{k\siga^2}, \bt_{k\bbeta^{(w)}}, \bt_{k\sige^2})^\top$ is the set of indices  for the cubes in $\mathcal{C}$.

We first show that $|T_3(\btheta)| = o(1)$ holds over the set of  $\bt_k$ and then that the difference between taking the supremum over a fine grid of points and over $\mathcal{N}$ is small.

For the first part on the right-hand side of \eqref{eq:H2}, applying Chebyshev's inequality, for any $\varepsilon > 0$, we have

\begin{align*}
&\Pr\left(
\max_{1\leq k\leq N}\left\|\bK^{-1/2}\left[\bpsi(\bt_k)-\bpsi(\dot{\btheta})-\mathbb{E}\left\{\bpsi(\bt_k)-\bpsi(\dot{\btheta})\right\}\right]\right\|>\varepsilon\right) \\
&\leq
\varepsilon^{-2}\sum_{k=1}^{N}\mathbb{E}\left\|\bK^{-1/2}\left[\bpsi(\bt_k)-\bpsi(\dot{\btheta})-\mathbb{E}\left\{\bpsi(\bt_k)-\bpsi(\dot{\btheta})\right\}\right]\right\|^2 \\
&= \varepsilon^{-2}g^{-1}
\sum_{k=1}^{N}\operatorname{tr}\left[\operatorname{Var}\left\{\bpsi_{\bbeta^{(b)}}(\bt_k)-\bpsi_{\bbeta^{(b)}}(\dot{\btheta})\right\}\right] \\
&\quad+
\varepsilon^{-2}n^{-1}\sum_{k=1}^{N}\operatorname{tr}\left[\operatorname{Var}\left\{\bpsi_{\bbeta^{(w)}}(\bt_k)-\bpsi_{\bbeta^{(w)}}(\dot{\btheta})\right\}\right] \\
&\quad+
\varepsilon^{-2}g^{-1}\sum_{k=1}^{N}\operatorname{Var}\left\{\psi_{\siga^2}(\bt_k)-\psi_{\siga^2}(\dot{\btheta})\right\} \\
&\quad+
\varepsilon^{-2}n^{-1}\sum_{k=1}^{N}\operatorname{Var}\left\{\psi_{\sige^2}(\bt_k)-\psi_{\sige^2}(\dot{\btheta})\right\}.
\end{align*}

We show in Lemma~\ref{lemma 5} that all variance terms are uniformly bounded by a finite constant L.
Therefore,
\begin{align*}
&\Pr\left(\max_{1\leq k\leq N}\left\|\bK^{-1/2}\left[\bpsi(\bt_k)-\bpsi(\dot{\btheta})-\mathbb{E}\left\{\bpsi(\bt_k)-\bpsi(\dot{\btheta})\right\}\right]\right\|>\varepsilon\right) \\
&\leq
\varepsilon^{-2}LN
\left\{g^{-1}(p_b+1)+n^{-1}(p_w+1)\right\} \\
&=o(1)
\end{align*} 
by the fact that $N=O(g^{1/4})$ 

For the second part on the right-hand side of \eqref{eq:H2}, Using Taylor expansion, we get
\begin{align*}
&\max_{1\leq k\leq N}
\sup_{\btheta\in C(\bt_k)}
\left\|\bK^{-1/2}\left[\bpsi(\btheta)-\bpsi(\bt_k)-\mathbb{E}\left\{\bpsi(\btheta)-\bpsi(\bt_k)\right\}\right]\right\| \\
&\leq
\max_{1\leq k\leq N}\sup_{\btheta\in C(\bt_k)}\left\|\bK^{-1/2}\left\{\nabla\bpsi(\boldsymbol{\Omega}_k)-\mathbb{E}\nabla\bpsi(\boldsymbol{\Omega}_k)\right\}\bK^{-1/2}\right\|\left\|\bK^{1/2}(\btheta-\bt_k)\right\| \\
&\leq
M g^{-1/4}\sup_{\tilde{\btheta}_k\in\mathcal N}\left\|\bK^{-1/2}\left\{\nabla\bpsi(\boldsymbol{\Omega}_k)-\mathbb{E}\nabla\bpsi(\boldsymbol{\Omega}_k)\right\}\bK^{-1/2}\right\|,
\end{align*}
where the rows of $\boldsymbol{\omega}_k$ are between $\bt_k$ and $\btheta$. By Lemma~\ref{lemma 6}, the right-hand side is $o_p(1)$. Combining the two bounds gives
$$
\sup_{\btheta\in\mathcal N}
\|T_3(\btheta)\| = o_p(1).
$$

\subsection{Proof of Theorem 1}
\label{subsec:proof of theorem 1}

Let $\bpsi_P(\bsigma)$ and $\bpsi_R(\bsigma)$ denote the score functions associated with $\ell_P(\bsigma)$ and $\ell_R(\sigma)$, respectively. By definition,
$$\bpsi_P(\hat{\bsigma}_P) = 0, \qquad \bpsi_R(\hat{\bsigma}_R) = 0.$$

The proof proceeds in three steps. We first establish the asymptotic equivalence of $\hat{\bsigma}_R$ and $\hat{\bsigma}_P$. We then transfer this equivalence to the corresponding fixed-effect estimators through the profiling map $\bbeta(\bsigma)$. Finally, we compare $\hat{\btheta}_P$ with the ML estimator $\hat{\btheta}$.

Applying the row-wise mean-value expansion to $\bpsi_R(\hat{\bsigma}_R)$ around $\hat{\bsigma}_P$, we obtain
$$0 = \bpsi_R(\hat{\bsigma}_R) = \bpsi_R(\hat{\bsigma}_P) + \nabla_{\bsigma}\bpsi(\tilde{\bsigma})(\hat{\bsigma}_R - \hat{\bsigma}_P),$$
where each row of $\tilde{\bsigma}$ lies between $\hat{\bsigma}_R$ and $\hat{\bsigma}_P$.
Since $\bpsi_P(\hat{\bsigma}_P)$ is 0, we have 
$$
0 = \bpsi_R(\hat{\bsigma}_P) -  \bpsi_P(\hat{\bsigma}_P) + \nabla_{\bsigma}\bpsi(\tilde{\bsigma})(\hat{\bsigma}_R - \hat{\bsigma}_P).$$
After inserting the normalization matrices $\bK_{\bsigma}$, the above expansion becomes 
\begin{equation}
\label{eq:taylor expansion for REML}
-\left\{\bK_{\bsigma}^{-1/2} \nabla_{\bsigma}\bpsi_R(\tilde{\bsigma}) \bK_{\bsigma}^{-1/2}\right\} \bK_{\bsigma}^{1/2}(\hat{\bsigma}_R - \hat{\bsigma}_P) = \bK_{\bsigma}^{-1/2}\{\bpsi_R(\hat{\bsigma}_P) - \bpsi_P(\hat{\bsigma}_P) \}.
\end{equation}
By Lemma~\ref{lemma 7}, the norm of the right-hand side of \eqref{eq:taylor expansion for REML} is $o_p(1)$. For the normalized derivative matrix on the left-hand side, write
\begin{equation}
\label{eq:decomposition of REML derivative matrix}
\begin{aligned}
-\bK_{\bsigma}^{-1/2} \nabla_{\bsigma}\bpsi_R(\tilde{\bsigma}) \bK_{\bsigma}^{-1/2} &= -\bK_{\bsigma}^{-1/2} \nabla_{\bsigma}\bpsi_P(\tilde{\bsigma}) \bK_{\bsigma}^{-1/2} \\
&\quad - \bK_{\bsigma}^{-1/2} \nabla_{\bsigma}\{\bpsi_R(\tilde{\bsigma}) - \bpsi_P(\tilde{\bsigma})\}\mathbf K_{\bsigma}^{-1/2}.
\end{aligned}
\end{equation}
By Lemma~\ref{lemma 8}, the first term on the right-hand side of \eqref{eq:decomposition of REML derivative matrix} converges to a finite positive definite matrix, whereas Lemma~\ref{lemma 9} shows that the norm the second term is $o_p(1)$. Hence, the normalized derivative matrix is asymptotically nonsingular. It follows from \eqref{eq:taylor expansion for REML} that $$\bK_{\bsigma}^{1/2}(\hat{\bsigma}_R - \hat{\bsigma}_P) = o_p(1),$$ which implies the $\hat{\bsigma}_R$ is asymptotically equivalent to $\hat{\bsigma}_P$. 

We next compare the corresponding fixed-effect estimators.By construction,
$$\hat{\bbeta}_R = \hat{\bbeta}(\hat{\bsigma}_R), \qquad \hat{\bbeta}_P = \hat{\bbeta}(\hat{\bsigma}_P),$$ We establish in Lemma~\ref{lemma 10} that $$\bK_{\bbeta}^{1/2}(\hat{\bbeta}_R - \hat{\bbeta}_P) = o_p(1),$$ where $\bK_{\bbeta} = \operatorname{diag}(g\boldsymbol{1}_{p_b}, n\boldsymbol{1}_{p_w})$. Combining this with asymptotic equivalence between $\hat{\bsigma}_R$ and $\hat{\bsigma}_P$, we have $$\bK^{1/2}(\hat{\btheta}_R - \hat{\btheta}_P) = o_p(1).$$

It remains to compare $\hat{\btheta}_P$ with $\hat{\btheta}$. By Lemma~\ref{lemma 11}, maximizing the $\ell_P(\bsigma)$ over $\bsigma$ and recovering $\bbeta$ through the profiling map $\bbeta(\bsigma)$ is equivalent to jointly maximizing $\ell(\bbeta, \sigma)$ over $(\bbeta, \bsigma)$. Hence $\hat{\btheta}_P$ is the joint maximizer of $\ell(\bbeta, \bsigma)$, which implies $\hat{\btheta}_P = \hat{\btheta}$
Consequently,
$$
\bK^{1/2}\left(\hat{\btheta}_R - \hat{\btheta}\right) = \bK^{1/2}\left(\hat{\btheta}_R - \hat{\btheta}_P\right) = o_p(1).
$$
Thus, the REML estimator is asymptotically equivalent to the ML estimator. It follows from Theorem 2 that
$$\bK^{1/2}(\hat{\btheta}_R - \dbtheta) = \bB^{-1}\bK^{-1/2}\bxi + o_p(1).$$
Restricting this relation to the variance-component block gives
$$\bK^{1/2}_{\bsigma}(\hat{\bsigma}_R - \dbsigma) = \bB^{-1}\bK^{-1/2}\bxi_{\sigma} + o_p(1).$$

\subsection{Lemmas for the estimating function.}
We collect below the lemmas used in the proofs of Theorems 1 and 2. Lemma~\ref{lemma 1}-\ref{lemma 6} show that $\bpsi(\dbtheta)$ approximates $\bpsi$, a central limit theorem holds for $\bpsi$, $T_1 = o_p(1)$, and the variances of $\bpsi(\btheta) - \bpsi(\dbtheta)$ are uniformly bounded. Lemma~\ref{lemma 7}-\ref{lemma 11} establish the asymptotic equivalence between the REML and profile estimating equations and transfer this equivalence to the corresponding estimators. Detailed proofs are provided in Section~\ref{sec:proofs of lemmas}.

\begin{lemma}
\label{lemma 1}
Suppose Condition A holds. Then, as $g,m_L\to\infty$, $\bK^{-1/2}\boldsymbol{\xi}
\overset{D}{\longrightarrow}
N(\boldsymbol{0},\mathbf{A})$, where
$$
\mathbf{A}
=
\begin{pmatrix}
\bC_2/\dot\siga^2
&
\mathbb{E}(\dot\alpha_i^3)\bc_1 / 2\dot\siga^6

& \boldsymbol{0}_{[p_b:p_w]} & \boldsymbol{0}_{[p_b:1]}
\\
\mathbb{E}(\dot\alpha_i^3)\bc_1^\top / 2\dot{\siga}^6
& \{\mathbb{E}(\dot\alpha_i^4)-\dot\siga^4\}/
4\dot\siga^8
& \boldsymbol{0}_{[1:p_w]} & 0
\\
\boldsymbol{0}_{[p_w:p_b]}
& \boldsymbol{0}_{[p_w:1]}
& \bC_3(\dbtheta) & \boldsymbol{0}_{[p_w:1]}
\\
\boldsymbol{0}_{[1:p_b]} & 0 & \boldsymbol{0}_{[1:p_w]} & c_4(\dbtheta)
\end{pmatrix}.
$$
\end{lemma}

\begin{lemma}
\label{lemma 2}
Suppose Condition A holds. Then $\|T_1\| = o_p(1).$  
\end{lemma}

\begin{lemma}
\label{lemma 3}
Suppose Condition A holds. Then, as $g,m_L\to\infty, \left\|\mathbf B_n- \mathbf B\right\|=o(1),$ where
$\mathbf{B}_n = -\bK^{-1/2} \mathbb{E}\{\nabla_{\btheta} \bpsi(\dot{\btheta})\}\bK^{-1/2} $
and 
\begin{equation}\label{matrix B}
\mathbf B
=
\begin{pmatrix}
\bC_2 / \dot\siga^2 & \boldsymbol 0_{[p_b:1]} & \boldsymbol 0_{[p_b:p_w]} & \boldsymbol 0_{[p_b:1]}
\\
\boldsymbol 0_{[1:p_b]} & 1/2\dot\siga^{4} & \boldsymbol 0_{[1:p_w]} & 0
\\
\boldsymbol 0_{[p_w:p_b]} & \boldsymbol 0_{[p_w:1]} & \bC_3(\dbtheta) & \boldsymbol 0_{[p_w:1]}
\\
\boldsymbol 0_{[1:p_b]} & 0 & \boldsymbol 0_{[1:p_w]} & c_4(\dbtheta)
\end{pmatrix}.
\end{equation}
\end{lemma}

\begin{lemma}
\label{lemma 4}
Suppose condition A holds. Then, as $g,m_L \rightarrow \infty$,
$$\sup_{\btheta \in \mathcal{N}}\left\Vert\bK^{-1/2}\mathbb{E}\left\{\nabla_{\btheta} \bpsi(\boldsymbol{\Omega}) - \nabla_{\btheta} \bpsi(\dot{\btheta})\right\}\bK^{-1/2}\right\Vert = o(1).$$
\end{lemma}

\begin{lemma}
\label{lemma 5}
Suppose Condition A holds. Then there exists a finite constant $L$ such that
\begin{align*}
    &\sup_{\btheta\in\mathcal N}\tr\var\left\{\bpsi_{\bbeta^{(b)}}(\btheta)-\bpsi_{\bbeta^{(b)}}(\dot{\btheta})\right\}\leq L,\\
    &\sup_{\btheta\in\mathcal N}\tr\var\left\{\bpsi_{\bbeta^{(w)}}(\btheta)-\bpsi_{\bbeta^{(w)}}(\dot{\btheta})\right\}\leq L, \\
    &\sup_{\btheta\in\mathcal N}\var\left\{\psi_{\siga^2}(\btheta)-\psi_{\siga^2}(\dot{\btheta})\right\}\leq L, \\
    &\sup_{\btheta\in\mathcal{N}}\var\left\{\psi_{\sige^2}(\btheta)-\psi_{\sige^2}(\dot{\btheta})\right\}\leq L. \\ 
\end{align*}
\end{lemma}

\begin{lemma}
\label{lemma 6}
Suppose Condition A holds. Then, as $g,m_L\to\infty$,
$$
\sup_{\tilde{\btheta}_k\in\mathcal N}
g^{-1/4}\left\|\bK^{-1/2}\left\{\nabla_{\btheta}\bpsi(\boldsymbol{\Omega}_k)-\mathbb{E}\nabla_{\btheta}\bpsi(\boldsymbol{\Omega}_k)\right\}\bK^{-1/2}\right\|=o_p(1).
$$
\end{lemma}

\begin{lemma}
\label{lemma 7}
Suppose Condition A holds. Then, as $g, m_L \rightarrow \infty$ 

$$\sup_{\bsigma \in \mathcal{N}_{\bsigma}}\left\|
\bK_{\bsigma}^{-1/2}
\left\{
\bpsi_R(\bsigma)
-
\bpsi_P(\bsigma)
\right\}\right\|
=
o_p(1).
$$
\end{lemma}

\begin{lemma}
\label{lemma 8}
Suppose Condition A holds. Define $\bB_{\bsigma n}(\bsigma)=-\bK_\sigma^{-1/2}\nabla_{\bsigma}\bpsi_P(\bsigma)\bK_\sigma^{-1/2}.$
Then, as $g,m_L\to\infty$,
$$
\sup_{\bsigma \in \mathcal{N}_{\bsigma}}\left\|\bB_{\bsigma n}(\bsigma)
-
\bB_{\bsigma}(\bsigma)\right\| = o_p(1),
$$
where $\bB_{\bsigma}(\bsigma)
= \diag\{1/(2\siga^4), c_4(\bbeta(\bsigma), \bsigma)\}$ is finite and positive definite.
\end{lemma}

\begin{lemma}
\label{lemma 9} Suppose Condition A holds. Then, as $g, m_L \rightarrow \infty$,
$$
\sup_{\bsigma \in \mathcal{N}_{\bsigma}} \left\|
\bK_{\bsigma}^{-1/2}
\nabla_{\bsigma}
\left\{
\bpsi_R(\bsigma) - \bpsi_P(\bsigma)
\right\}
\bK_{\bsigma}^{-1/2} \right\|
=
o_p(1).$$
\end{lemma}

\begin{lemma}
\label{lemma 10}
Suppose Condition A holds and the profile map
$\bbeta(\bsigma)$ is continuously differentiable on $\mathcal{N}_{\bsigma}$. 
Then,
$$
\bK_{\bbeta}^{1/2}
\left(\hat{\bbeta}_R-\hat{\bbeta}_P\right)
=o_p(1),
$$
provided that
$$
\bK_{\bsigma}^{1/2}
\left(\hat{\sigma}_R-\hat{\sigma}_P\right)
=o_p(1).
$$
\end{lemma}

\begin{lemma}
\label{lemma 11} For every fixed $\bsigma$, suppose that $\bbeta(\bsigma) = \arg\max_{\bbeta}\ell(\bbeta, \bsigma)$. Then  
$$\max_{\bsigma} \ell_P(\bsigma) = \max_{\bbeta, \bsigma}\ell(\bbeta, \bsigma).$$ If  $\hat{\bsigma} \in \arg\max_{\bsigma}\ell_P(\bsigma)$, then $\btheta_P = \left\{\hat{\bbeta}_P^\top, \hat{\bsigma}_P^\top\right\}^\top$ is a joint maximizer of $\ell(\bbeta, \bsigma)$, and hence is an ML estimator.
\end{lemma}

\section{Proofs of Lemmas}
\setcounter{equation}{0} 
\label{sec:proofs of lemmas}
\subsection{Detailed Proof of Lemma 1}
\label{subsec:proof of lemma 1}
\begin{proof}
Define $$\bA_n=\var\left(\bK^{-1/2}\bxi\right).$$ We first establish that $\bA_n \longrightarrow \bA$. The components of $\bxi$ are sums of contributions from independent clusters with zero means and finite variance. To verify the zero-mean property, consider $\bxi_{\bbetab}$, where the contribution of the $i$th clusters is
$$\bxi_{i, \bbetab} = \frac{1}{\dsiga^2}\bxbi\alpha_i.$$ Since the covariates are treated as fixed for each cluster and $\mathbb{E}(\alpha_i)=0$,
$$\mathbb{E}(\bxi_{i, \bbetab}) = \frac{1}{\dsiga^2}\bxbi\mathbb{E}(\alpha_i) = \boldsymbol{0}_{p_b}.$$
By linearity of expectation,
$$\mathbb{E}(\bxi_{\bbetab}) = \mathbb{E}\left\{\sumig\bxi_{i, \bbetab}\right\} = \boldsymbol{0}_{p_b}.$$ The zero-mean property of the remaining components of $\bxi$ can be proved by the same argument.

We next compute the covariance blocks of $\bA_n$ and show that $\bA_n$ has finite variances under Condition A. For illustration, consider the blocks involving $\bxi_{\bbetab}$ and $\xi_{\siga^2}$. Recall that
$$
\bxi_{\bbetab}
=
\frac{1}{\dot\siga^2}
\sumig
\bxbi\dot\alpha_i,
\qquad
\xi_{\siga^2}
=
\frac{1}{2\dot\siga^4}
\sumig
\left(
\dot\alpha_i^2-\dot\siga^2
\right).
$$
Using the independence of clusters and $\mathbb{E}(\alpha_i) = 0$, we obtain 
\begin{align*}
\var\left(g^{-1/2}\bxi_{\bbetab}\right)&=
\frac{1}{\dot\siga^4}
\left\{\frac{1}{g}
\sumig\bxbi\bxbit\right\}\mathbb{E}(\alpha_i^2) 
= \frac{\hat{\bC}_2}{\dsiga^2}, \\
\var\left(g^{-1/2}\xi_{\siga^2}\right)&=
\frac{\mathbb{E}(\dot\alpha_i^4)-\dot\siga^4}{4\dot\siga^8}, \\
\cov\left(g^{-1/2}\bxi_{\bbetab},g^{-1/2}\xi_{\siga^2}\right)&=
\frac{\mathbb{E}(\dot\alpha_i^3)}{2\dot\siga^6}
\left\{\frac{1}{g}\sumig\bxbi\right\}
=
\frac{\mathbb{E}(\dot\alpha_i^3)}
{2\dot\siga^6}
\hat{\bc}_1 .
\end{align*}
By condition A, as $g,m_L\to\infty$
\begin{align*}
    \lim_{g, m_L\to\infty} \var\left(g^{-1/2}\bxi_{\bbetab}\right) &= \frac{\bC_2}{\dsiga^2}, \\
    \lim_{g, m_L\to\infty} \var\left(g^{-1/2}\xi_{\siga^2}\right) &= \frac{\mathbb{E}(\dot\alpha_i^4)-\dot\siga^4}{4\dot\siga^8}, \\
    \lim_{g, m_L\to\infty} \cov\left(g^{-1/2}\bxi_{\bbetab},g^{-1/2}\xi_{\siga^2}\right) &= \frac{\mathbb{E}(\dot\alpha_i^3)}{2\dot\siga^6}\bc_1.
\end{align*}
The convergence of the remaining entries of $\bA_n$ can be proved by the same argument. Hence, as $g,m_L\to\infty$,
$$
\bA_n \longrightarrow \bA.
$$

It remains to establish the central limit theorem. Write $$\bK^{-1/2}\bxi = \sumig\bK^{-1/2}\bxi_i,$$ where $\bxi_i,\ldots, \bxi_g$ are independent and centered vectors. For any fixed vector $\ba$ with
$\ba^\top \ba=\boldsymbol{1}$,
$$\ba^\top\bK^{-1/2}\boldsymbol{\xi} = \sumig \ba^\top\bK^{-1/2}\bxi_i.$$ The summands form a triangular array of independent, centered scalar random variables. By the moment assumptions on the random effects and conditional responses, together with the bounded-design conditions in Condition A, there exists $\delta>0$ such that
$$\frac{\sumig\mathbb{E}\left|\ba^\top \bK^{-1/2}\bxi_i \right|^{2+\delta}}{\left\{\var\left(\ba^\top\bK^{-1/2}\bxi\right)\right\}^{1+\delta/2}} \longrightarrow \boldsymbol{0}.$$ Thus, Lyapunov’s central limit theorem gives
$$\ba^\top\bK^{-1/2}\bxi \overset{D}{\longrightarrow} N(\boldsymbol{0}, \ba^\top\bA\ba).$$ Since this holds for every fixed $\ba$, the Cram\'er-Wold device yields $$\bK^{-1/2}\bxi
\overset{D}{\longrightarrow}
N(\boldsymbol{0},\bA).$$
\end{proof}

\subsection{Detailed Proof of Lemma 2}
\begin{proof}
\label{subsec:proof of lemma 2}
We can establish the result component-wisely for $T_1 = \bK^{-1/2} (\bpsi(\dbtheta) - \bxi)$. For brevity, we give the details for the block corresponding to $\bbeta^{(b)}$; the remaining components follow from the same argument. By the expression in \eqref{eq:asymptotic score of betab},
$$\bpsi_{\bbetab}(\dbtheta) - \bxi_{\bbetab} = \dsiga^{-2}\sumig \left\{\frac{\dot{\boldsymbol{S}}_i^{(b)}}{\dot{A}_i} - \frac{\dsiga^{-2} \alpha_i \bxbi}{\dot{A}_i} -\frac{\dot{\btau}_{i,3}^{(b)}S_i^2}{2\dot{A}_i^3} - \frac{\dot{\btau}_{i,3}^{(b)}}{2\dot{A}_i^2} + O_p(m_i^{-3/2})\boldsymbol{1}_{[p_b:1]}\right\}.$$
Condition A.2 gives 
{$$\mathbb{E}\left\{\bpsi_{\bbetab}(\dbtheta)\right\} = \boldsymbol{0}_{[p_b:1]},$$}
and we have shown in Section~\ref{subsec:proof of lemma 1} that $\mathbb{E}(\bxi_{\bbeta}) = \boldsymbol{0}_{[p_b:1]}$. Consequently
{\begin{equation}
\label{eq:zero mean property of remainder}
\mathbb{E}\left\{\bpsi_{\bbetab}(\dbtheta) - \bxi_{\bbeta}\right\} = \mathbb{E}\left\{\bpsi_{\bbetab}(\dbtheta)\right\} - \mathbb{E}(\bxi_{\bbetab}) = \boldsymbol{0}_{[p_b:1]}.
\end{equation}}

We next control the stochastic fluctuation around this expectation. Because clusters are independent, the variance of the sum in \eqref{eq:asymptotic score of betab} is the sum of variances of each cluster. Applying the variance inequality gives
\begin{equation}
\label{eq:variance of remainder}
\begin{aligned}
\var\left\{\bpsi_{\bbetab}(\dbtheta) - \bxi_{\bbetab}\right\} &\leq 
5\sumig\left[\var\left\{\frac{\dot{\boldsymbol{S}}_i^{(b)}}{\dot{A}_i}\right\} + \var\left\{\frac{\dsiga^{-2} \alpha_i \bxbi}{\dot{A}_i}\right\} + \var\left\{\frac{\dot{\btau}_{i,3}^{(b)}\dot{S}_i^2}{2\dot{A}_i^3}\right\} \right.\\
&\qquad \qquad +\left.\operatorname{Var}\left(\frac{\dot{\btau}_{i,3}^{(b)}}{2\dot{A}_i^2}\right) + \var\left\{O_p(m_i^{-3/2})\boldsymbol{1}_{[p_b:1]}\right\} \right] \\
& \leq5\sumig\left\{O_p(m_i^{-1}) + O_p(m_i^{-2}) + O_p(m_i^{-2}) \right. \\
&\qquad \qquad+ \left. O_p(m_i^{-2}) + O_p(m_i^{-3})\right\}\boldsymbol{1}_{p_b}^{\otimes 2} \\
&\leq O(gm_L^{-1})\boldsymbol{1}_{p_b}^{\otimes 2}.
\end{aligned}
\end{equation}
It follows from \eqref{eq:zero mean property of remainder} and \eqref{eq:variance of remainder}, and Chebyshev's inequality that $$\bpsi_{\bbetab}(\dbtheta) - \bxi_{\bbetab} = O_p(g^{1/2}m_L^{-1/2})\boldsymbol{1}_{[p_b:1]}.$$
Following the same argument, one can verify that 
\begin{align*}
&\psi_{\siga^2}(\dbtheta) - \xi_{\siga^2} = O_p(g^{1/2}m_L^{-1/2}), \\
&\bpsi_{\bbetab}(\dbtheta) - \bxi_{\bbetab} = O_p(g^{1/2})\boldsymbol{1}_{[p_w:1]}, \\
&\psi_{\sige^2}(\dbtheta) - \xi_{\sige^2} = O_p(g^{1/2}).
\end{align*}
After applying the corresponding normalization, we have
\begin{align*}
g^{-1/2}\left\{\bpsi_{\bbetab}(\dbtheta) - \bxi_{\bbetab}\right\} &= O_p(m_L^{-1/2})\boldsymbol{1}_{[p_b:1]} = o_p(1)\boldsymbol{1}_{p_b}, \\
g^{-1/2}\left\{\psi_{\siga^2}(\dbtheta) - \xi_{\siga^2}\right\} &= O_p(m_L^{-1/2}) = o_p(1), \\
n^{-1/2}\left\{\bpsi_{\bbetaw}(\dbtheta) - \bxi_{\bbetaw}\right\} &= O_p(\sqrt{g/n})\boldsymbol{1}_{[p_w:1]} = o_p(1)\boldsymbol{1}_{p_w},\\
n^{-1/2}\left\{\psi_{\sige^2}(\dbtheta) - \xi_{\sige^2}\right\} &= O_p(\sqrt{g/n}) = o_p(1),
\end{align*}
Combining the four normalized blocks gives $$\left\|\bK^{-1/2} \left\{\bpsi(\dbtheta) - \bxi\right\}\right\| = o_p(1),$$
which completes the proof. 

\end{proof}

\subsection{Proof of Lemma 3}
\label{subsec:proof of lemma 3}
\begin{proof}
By definition, $\bB_n$ admits the following block representation 
$$\bB_n = - \begin{pmatrix}
    g^{-1}\mathbb{E}\nabla_{\bbetab}\bpsi_{\bbetab}^\top & g^{-1}\mathbb{E}\nabla_{\bbetab}\bpsi_{\siga^2} & (gn)^{-1/2}\mathbb{E}\nabla_{\bbetab}\bpsi_{\bbetaw}^\top & (gn)^{-1/2}\mathbb{E}\nabla_{\bbetab}\bpsi_{\sige^2} \\
    g^{-1}\mathbb{E}\nabla{\siga^2}\bpsi_{\bbetab}^\top & g^{-1}\mathbb{E}\nabla{\siga^2}\psi_{\siga^2} & (gn)^{-1/2}\mathbb{E}\nabla{\siga^2}\bpsi_{\bbetaw}^\top & (gn)^{-1/2}\mathbb{E}\nabla{\siga^2}\psi_{\sige^2} \\
    (gn)^{-1/2}\mathbb{E}\nabla_{\bbetaw}\bpsi_{\bbetab}^\top & (gn)^{-1/2}\mathbb{E}\nabla_{\bbetaw}\bpsi_{\siga^2} &
    n^{-1}\mathbb{E}\nabla_{\bbetaw}\bpsi_{\bbetaw}^\top &
    n^{-1}\mathbb{E}\nabla_{\bbetaw}\bpsi_{\sige^2} \\
    (gn)^{-1/2}\mathbb{E}\nabla_{\sige^2}\bpsi_{\bbetab}^\top & (gn)^{-1/2}\mathbb{E}\nabla_{\sige^2}\psi_{\siga^2} &
    n^{-1}\mathbb{E}\nabla_{\sige^2}\bpsi_{\bbetaw}^\top &
    n^{-1}\mathbb{E}\nabla_{\sige^2}\psi_{\sige^2}
\end{pmatrix},$$
where all score derivatives are estimated at $\dbtheta$. We establish the limit of this matrix based on the asymptotic expression of score derivatives we obtained in 
\eqref{eq:asymptotic expression of ML score derivative}. Since $\bB_n$ is symmetric, we show only the upper-triangular part.

For the $\bbetab\bbetabt$ block, we have
\begin{align*}
-g^{-1}\mathbb{E}\nabla_{\bbetab}\bpsi_{\bbetab}^\top &= g^{-1} \mathbb{E}\sum_{i=1}^g\left\{\frac{\dot{\sigma}^{-2}_\alpha \dot{\btau}_{i,2}^{(bb)}}{\dot A_i} + O_p(m_i^{-3/2}\boldsymbol{1}_{p_b}^{\otimes 2}) \right\}\\
&= \hat{\bC}_2/\dsiga^2 + o(1)\boldsymbol{1}_{p_b}^{\otimes 2} \longrightarrow \bC_2/\dsiga^2.
\end{align*}

For the 
$\bbetab\siga^2$ block, we have
\begin{align*}
-g^{-1}\mathbb{E}\nabla_{\bbetab}\bpsi_{\siga^2}
&=g^{-1}\mathbb{E}\left\{\dot\siga^{-4}\sum_{i=1}^{g}\boldsymbol{x}_{i}^{(b)}\frac{\dot\tau_{i,2}\alpha_i}{\dot A_i} + \sumig O(m_i^{-1/2})\boldsymbol{1}_{p_b}\right\}\\
&=g^{-1}\mathbb{E}\left\{\dsiga^{-4}\sumig \bxbi\dot\alpha_i \right\} - g^{-1}\mathbb{E}\left\{\dot\siga^{-4}\sumig \bxbi\frac{\dsiga^{-2}}{\dot A_i} \right\} + o(1)\boldsymbol{1}_{[p_b:1]} \\
&=g^{-1}\dsiga^{-4}\sumig \bxbi \mathbb{E}(\dot\alpha_i)  - O_p(g^{-1})\boldsymbol{1}_{[p_b:1]} + o(1)\boldsymbol{1}_{[p_b:1]} \\
&\longrightarrow \boldsymbol{0}_{[p_b:1]}. 
\end{align*}

For the 
$\bbetab
\bbetawt$ block, we have
\begin{align*}
-(gn)^{-1/2}
\mathbb{E}\nabla_{\bbetab}\bpsi_{\bbetaw}^\top &= (gn)^{-1/2}\mathbb{E}\left\{\dot\siga^{-2}\sum_{i=1}^{g}\frac{\dot{\btau}_{i,2}^{(bw)}}{\dot A_i} + \sumig O(m_i^{-1/2})\boldsymbol{1}_{[p_b:p_w]}\right\} \\
&= O(\sqrt{g/n})\boldsymbol{1}_{[p_b:p_w]} +o(1)\boldsymbol{1}_{[p_b:p_w]} \\
&\longrightarrow \boldsymbol{0}_{[p_b:p_w]}.
\end{align*}

For the 
$\bbetab\sige^2$ block, we have
\begin{align*}
-(gn)^{-1/2}\mathbb{E}\nabla_{\bbetab}\bpsi_{\sige^2} &= -(gn)^{-1/2}\mathbb{E}\left\{\dsiga^{-4}\dsige^{-2}\sumig\bxbi\frac{\dot\alpha_i}{\dot A_i} + \sumig O(m_i^{-1/2})\boldsymbol{1}_{[p_b:1]}\right\}  \\
&= -(gn)^{-1/2} \sumig O(m_i^{-1})\boldsymbol{1}_{[p_b:1]} + o(1)\boldsymbol{1}_{[p_b:1]} \\ 
&\longrightarrow \boldsymbol{0}_{[p_b:1]}.
\end{align*}

For the 
$\siga^2\siga^2$ block, we have 
\begin{align*}
-g^{-1}\mathbb{E}\nabla_{\siga^2}\psi_{\siga^2}
&=-g^{-1}\mathbb{E}\left\{\frac{g}{2}\dsiga^{-4}-\dsiga^{-6}\sumig \alpha_i^2 + \sumig O(m_i^{-1/2})\right\} \\
&= -\frac{1}{2}\dsiga^{-4}+\dsiga^{-6}\dsiga^2 + o(1) \\
&\longrightarrow \frac{1}{2\dsiga^4}.
\end{align*}

For the 
$\siga^2\bbetawt$ block, we have
\begin{align*}
-(gn)^{-1/2}
\mathbb{E}\nabla_{\siga^2}\bpsi_{\bbetaw}^\top&=(gn)^{-1/2}\mathbb{E}\left\{\dsiga^{-4}\sumig \frac{\alpha_i\dot{\btau}_{i,2}^{(w)\top}}{\dot A_i} + \sumig O(m_i^{-1/2})\boldsymbol{1}_{[1:p_w]} \right\} \\
&=O(\sqrt{g/n})\boldsymbol{1}_{[1:p_w]} + o(1)\boldsymbol{1}_{[1:p_w]} \\
&\longrightarrow \boldsymbol{0}_{[1:p_w]}.
\end{align*}

For the 
$\siga^2\sige^2$ block, the retained leading component is of order $\sumig O(m_i^{-1})$. Therefore, after the normalization, we obtain
$$
-(gn)^{-1/2}\mathbb{E}\nabla_{\siga^2}\psi_{\sige^2} = o(1) \longrightarrow 0.
$$

For the $\bbetaw\bbetawt$ block,
\begin{align*}
-n^{-1}\mathbb{E}\nabla_{\bbetaw}\bpsi_{\bbetaw}^\top
&=
n^{-1}\mathbb{E}\left[\sumig \left\{\dot{\btau}_{i,2}^{(ww)}-\frac{\dot{\btau}_{i,2}^{(w)}\dot{\btau}_{i,2}^{(w)\top}}{\dot A_i}\right\} + \sumig O(m_i^{1/2})\boldsymbol{1}_{p_w}^{\otimes 2}\right]\\
&= \hat{\bC}_3(\dbtheta) + o(1)\boldsymbol{1}_{p_w}^{\otimes 2} \\
&\longrightarrow \bC_3(\dbtheta).
\end{align*}

For the $\bbetaw\sige^2$ block,
\begin{align*}
-n^{-1}\mathbb{E}\nabla_{\bbetaw}\bpsi_{\sige^2}
&=
n^{-1}\mathbb{E}\left\{\dot\sigma_e^{-2}\sumig \dot{\boldsymbol{S}}_i^{(w)} + \sumig O(1)\boldsymbol{1}_{[p_w:1]}\right\} \\
&= \boldsymbol{0}_{[p_w:1]} + o(1)\boldsymbol{1}_{[p_w:1]} \\
& \longrightarrow \boldsymbol{0}_{[p_w:1]}.
\end{align*}

For the 
$\sige^2\sige^2$ block,
\begin{align*}
-n^{-1}\mathbb{E}\nabla_{\sige^2}\psi_{\sige^2}
&=
-n^{-1/2}\mathbb{E}\left[\dsige^{-4}\sumig\sumjmi\left\{\frac{2\{y_{ij}\eta_{ij}-b(\eta_{ij})\}}{\dsige^2}-\dsige^4\nabla_{\sige^2}^2c(y_{ij},\dsige^2)\right\} + O(g)\right]\\
&= \hat{c}_4(\dbtheta) + o(1) \\
&\longrightarrow c_4(\dbtheta).
\end{align*}
Collecting the above results, we have
$$\bB_n = \begin{pmatrix}
\hat{\bC}_2 / \dsiga^2 & \boldsymbol 0_{[p_b:1]} & \boldsymbol 0_{[p_b:p_w]} & \boldsymbol 0_{[p_b:1]}
\\
\boldsymbol 0_{[1:p_b]} & 1/2\dsiga^{4} & \boldsymbol 0_{[1:p_w]} & 0
\\
\boldsymbol 0_{[p_w:p_b]} & \boldsymbol 0_{[p_w:1]} & \hat{\bC}_3(\dbtheta) & \boldsymbol 0_{[p_w:1]}
\\
\boldsymbol 0_{[1:p_b]} & 0 & \boldsymbol 0_{[1:p_w]} & \hat{c}_4(\dbtheta)
\end{pmatrix} + o(1) \longrightarrow \bB.$$
Hence, $\|\bB_n -\bB\| = o(1)$, which completes the proof.
\end{proof}

\subsection{Proof of Lemma 4}
\label{subsec:proof of lemma 4}
\begin{proof}
We prove the asserted uniform convergence for the $\bbeta^{(b)}\bbeta^{(b)\top}$ block. The remaining blocks follow from the same argument, using their corresponding score-derivative expansions.
From \eqref{eq:asymptotic expression of ML score derivative}, we have
$$
\nabla_{\bbetab}\bpsi_{i,\bbetab}^\top(\btheta)
=-\frac{\siga^{-2}\btau_{i,2}^{(bb)}}{A_i}+O_p(m_i^{-3/2})\boldsymbol{1}_{p_b}^{\otimes 2},
$$
uniformly for $\boldsymbol{\theta}\in\mathcal{N}$. Hence,
\begin{align*}
&\sup_{\boldsymbol{\theta}\in\mathcal{N}}
\left\|g^{-1}\mathbb{E}\left\{\nabla_{\bbetab}\bpsi_{i,\bbetab}^\top(\btheta)-\nabla_{\bbetab}\bpsi_{i,\bbetab}^\top(\dbtheta)\right\}\right\| \\
&\quad \leq \sup_{\boldsymbol{\theta}\in\mathcal{N}} g^{-1} \sumig \mathbb{E}\left\|\frac{\sigma^{-2}_\alpha \btau_{i,2}^{(bb)}}{A_i} + \frac{\dot{\sigma}^{-2}_\alpha\dot{\btau}^{(bb)}_{i,2}}{\dot A_i}\right\| + o(1)\\
&\quad \leq\sup_{\boldsymbol{\theta}\in\mathcal{N}}g^{-1}\sumig \mathbb{E} \left\{\left\|\bxbi\boldsymbol{x}_i^{(b)\top}\right\|\left|\frac{\sigma^{-2}_\alpha\tau_{i,2}}{A_i}-\frac{\dot{\sigma}^{-2}_\alpha\dot{\tau}_{i,2}}{\dot{A}_i}\right|\right\}.
\end{align*}
It therefore suffices to control the scalar difference $$\frac{\siga^{-2}\tau_{i,2}}{A_i}-\frac{\dot\siga^{-2}\dot\tau_{i,2}}{\dot A_i}.$$
Expanding the denominator gives
\begin{align*}
\frac{\siga^{-2}\tau_{i,2}}{A_i}-\frac{\dsiga^{-2}\dot\tau_{i,2}}{\dot A_i}&=\left(\siga^{-2}-\dsiga^{-2}\right)-\left(\frac{\siga^{-4}}{A_i}-\frac{\dsiga^{-4}}{\dot A_i}\right)  \\
&=\left(\siga^{-2}-\dsiga^{-2}\right)-\frac{\siga^{-4}\dot A_i-\dsiga^{-4}A_i}{A_i\dot A_i}.
\end{align*}
Moreover,
\begin{align*}
\siga^{-4}\dot A_i-\dsiga^{-4}A_i
&=-\siga^{-4}(\tau_{i,2}-\dot\tau_{i,2})-(\siga^{-4}-\dsiga^{-4})\tau_{i,2} \\
&\quad-
\siga^{-2}\dsiga^{-2} \left(\siga^{-2} - \dsiga^{-2}\right).
\end{align*}
By the mean-value argument, uniformly for
$\btheta\in\mathcal N$,
\begin{align*}
\tau_{i,2}-\dot\tau_{i,2}&=O(m_i)(\btheta - \dbtheta)
\\
&=O_p(m_i)\left\{\left\|\bbetab -\dot{\bbeta}^{(b)}\right\|+\left\|\bbetaw-\dot{\bbeta}^{(w)}\right\|+\left|\sige^2-\dsige^2)\right|\right\}.
\end{align*}
The parameter $\siga^2$ does not appear in this bound because $\tau_{i,2}$ does not depend on $\siga^2$.
For $\btheta\in\mathcal N$, we have
\begin{align*}
&\left\|\bbetab-\dot{\bbeta}^{(b)}\right\|=O_p(g^{-1/2}),\qquad  \left|\siga^2-\dsiga^2)\right|=O(g^{-1/2}), \\
&\left\|\bbetaw-\dot{\bbeta}^{(w)}\right\|=O(n^{-1/2}), \qquad \left|\sige^2-\dsige^2)\right|=O(n^{-1/2}),
\end{align*}
which further leads to
\begin{align*}
&\tau_{i,2} - \dot\tau_{i,2} = O(g^{-1/2}m_i), \quad \siga^{-2} - \dsiga^{-2} = -\frac{\siga^{2} - \dsiga^{2}}{\siga^{2} \dsiga^{2}} = O_p(g^{-1/2}),\\
&\siga^{-4}-\dsiga^{-4}=-\frac{(\siga^{2} + \dsiga^{2})(\siga^{2} - \dsiga^{2})}{\siga^{4} \dsiga^{4}} = O_p(g^{-1/2}),
\end{align*}
and
$$
\frac{\siga^{-4}\dot A_i-\dot\siga^{-4}A_i}{A_i\dot A_i}=O_p(g^{-1/2}m_i^{-1}).
$$
Consequently,
$$
\sup_{\btheta \in \mathcal{N}}\left|\frac{\siga^{-2}\tau_{i,2}}{A_i}-\frac{\dsiga^{-2} \dot\tau_{i,2}}{\dot A_i}\right|=O_p(g^{-1/2})+O_p(g^{-1/2}m_i^{-1})=o(1).
$$
Under Condition A, the design quantities are uniformly bounded
$$
\sup_{1 \leq i \leq g}\left\|\bxbi\bxbit\right\|=O(1),
$$
we have
$$
\sup_{\boldsymbol{\theta}\in\mathcal{N}}
\left\|g^{-1}\mathbb{E}\left\{\nabla_{\bbetab}\bpsi_{i,\bbetab}^\top(\btheta)-\nabla_{\bbetab}\bpsi_{i,\bbetab}^\top(\dbtheta)\right\}\right\| \leq \frac{1}{g} \sumig O(1)\cdot o(1) = o(1).
$$
The same argument applies to the remaining score-derivative blocks, with the corresponding left and right normalizations determined by $\bK^{-1/2}$. Since the dimension of $\btheta$ is fixed, the block-wise bounds imply
$$\sup_{\btheta \in \mathcal{N}}\left\|\bK^{-1/2}\mathbb{E}\left\{\nabla_{\btheta}\bpsi(\btheta) - \nabla_{\btheta}\bpsi(\dbtheta)\right\}\bK^{-1/2} \right\| = o(1),$$
which completes the proof.
\end{proof}

\subsection{Proof of Lemma 5}
\label{subsec:proof of lemma 5}
\begin{proof}
We verify the result for the $\bbeta^{(b)}$ score block. The remaining blocks follow from the same argument using their corresponding score-derivative expansions. 

Write $$\bpsi_{\bbetab}(\btheta) - \bpsi_{\bbetab}(\dbtheta) = \sumig\left\{\bpsi_{i, \bbetab}(\btheta) - \bpsi_{i, \bbetab}(\dbtheta)\right\}.$$
For each cluster $i$, we have
$$\bpsi_{i, \bbetab}(\btheta) = \siga^{-2}\bxbi\alpha_i + \boldsymbol{r}_{i, \bbetab}(\btheta),$$ where $$\boldsymbol{r}_{i, \bbetab}(\btheta) = \bpsi_{i, \bbetab}(\btheta) - \siga^{-2}\bxbi\alpha_i.$$ collects all remaining terms in the cluster-wise score expansion. Therefore,
$$\bpsi_{i, \bbetab}(\btheta) - \bpsi_{i, \bbetab}(\dbtheta) = (\siga^{-2} - \dsiga^{-2})\bxbi \alpha_i + \boldsymbol{r}_{i, \bbetab}(\btheta) - \boldsymbol{r}_{i, \bbetab}(\dbtheta).$$

We now apply the multivariate mean-value argument only to the remainder difference. For each cluster $i$, there exists some $\tilde{\btheta}_i$ between $\btheta$ and $\dbtheta$ such that
\begin{equation}
\label{eq:taylor expansion of remainder substraction}
\begin{aligned}
\boldsymbol{r}_{i, \bbetab}(\btheta) - \boldsymbol{r}_{i, \bbetab}(\dbtheta) &= \nabla_{\bbetab}\boldsymbol{r}_{i, \bbetab}(\tilde{\btheta}_i)\left(\bbetab - \dot{\bbeta}^{(b)}\right) \\
&+ \nabla_{\siga^2}\boldsymbol{r}_{i, \bbetab}(\tilde{\btheta}_i)\left(\siga^2 - \dsiga^2\right) \\
&+ \nabla_{\bbetaw}\boldsymbol{r}_{i, \bbetab}(\tilde{\btheta}_i)\left(\bbetaw - \dot{\bbeta}^{(w)}\right) \\
&+\nabla_{\sige^2}\boldsymbol{r}_{i, \bbetab}(\tilde{\btheta}_i)\left(\sige^2 - \dsige^2\right).
\end{aligned}
\end{equation}
By the asymptotic expansions of ML score derivatives established in \eqref{eq:asymptotic expression of ML score derivative}, uniformly over $\boldsymbol{\theta}\in\mathcal{N}$, the remainder derivatives satisfy
\begin{align*}
    &\nabla_{\bbetab}\boldsymbol{r}_{i, \bbetab}(\btheta) = O_p(m_i^{-3/2})\boldsymbol{1}_{p_b}^{\otimes 2}, \\
    &\nabla_{\siga^2}\boldsymbol{r}_{i, \bbetab}(\btheta) = O_p(m_i^{-1/2})\boldsymbol{1}_{[p_b:1]}, \\
    &\nabla_{\bbetaw}\boldsymbol{r}_{i, \bbetab}(\btheta) = O_p(m_i^{-1/2})\boldsymbol{1}_{[p_b:p_w]}, \\
    &\nabla_{\sige^2}\boldsymbol{r}_{i, \bbetab}(\btheta) = O_p(m_i^{-1})\boldsymbol{1}_{[p_b:1]}.
\end{align*}
Substituting these bounds into \eqref{eq:taylor expansion of remainder substraction} and combining the leading part subtraction, we have
\begin{equation}
\label{eq:score function substraction}
\begin{aligned}
\bpsi_{i,\bbetab}(\btheta)-\bpsi_{i,\bbetab}(\dbtheta)
&=\left(\siga^{-2}-\dsiga^{-2}\right)\bxbi\alpha_i\\
&+O_p(m_i^{-3/2})\left\|\bbetab - \dot{\bbeta}^{(b)}\right\|\boldsymbol{1}_{[p_b:1]}\\
&+O_p(m_i^{-1/2})\left\|\bbetaw - \dot{\bbeta}^{(w)}\right\|\boldsymbol{1}_{[p_b:1]}\\
&+O_p(m_i^{-1/2})\left|\siga^2-\dsiga^2\right|\boldsymbol{1}_{[p_b:1]}\\
&+O_p(m_i^{-1})\left|\sige^2-\dsige^2\right| \boldsymbol{1}_{[p_b:1]}.
\end{aligned}
\end{equation}
For
$\boldsymbol{\theta}\in\mathcal{N}$,
\begin{align*}
&\left\|\bbetab - \dot{\bbeta}^{(b)}\right\| = O(g^{-1/2}), \qquad \left|\siga^2-\dsiga^2\right| = O(g^{-1/2}), \\
&\left\|\bbetaw - \dot{\bbeta}^{(w)}\right\| = O(n^{-1/2}), \qquad \left|\sige^2-\dsige^2\right| = O(n^{-1/2}), 
\end{align*}
and $$\siga^{-2} - \dsiga^{-2} = O(g^{-1/2}).$$
Substituting these bounds into \eqref{eq:score function substraction},
we obtain
\begin{equation}
\label{eq:bound of score function substraction}
\begin{aligned}
\bpsi_{i,\bbetab}(\btheta)
-
\bpsi_{i,\bbetab}(\dbtheta)
=&
O_p(g^{-1/2})\boldsymbol{1}_{[p_b:1]}
+
O_p(m_i^{-3/2}g^{-1/2})\boldsymbol{1}_{[p_b:1]}
\\
&+
O_p(m_i^{-1/2}n^{-1/2})\boldsymbol{1}_{[p_b:1]}
+
O_p(m_i^{-1/2}g^{-1/2})\boldsymbol{1}_{[p_b:1]}
\\
&+
O_p(m_i^{-1}n^{-1/2})\boldsymbol{1}_{[p_b:1]}.
\end{aligned}
\end{equation}

The independence across clusters yields
$$\var\left\{\bpsi_{\bbetab}(\btheta) - \bpsi_{\bbetab}(\dbtheta)\right\} = \sumig \var\left\{\bpsi_{i,\bbetab}(\btheta) - \bpsi_{i,\bbetab}(\dbtheta)\right\}.$$
Under the uniform moment bounds in Condition A, the stochastic bounds in \eqref{eq:bound of score function substraction} also yield the corresponding second-moment bounds. Applying the variance inequality, uniformly over $\btheta \in \mathcal{N}$,
\begin{align*}
\tr\var&
\left\{
\bpsi_{\bbetab}(\btheta)
-
\bpsi_{\bbetab}(\dbtheta)
\right\} \\
&\leq
\sumig
\left\{
g^{-1}
+
m_i^{-3}g^{-1}
+
m_i^{-1}n^{-1}
+
m_i^{-1}g^{-1}
+
m_i^{-2}n^{-1}
\right\}
\\
&=
O(1).
\end{align*}
It follows that,
$$
\sup_{\boldsymbol{\theta}\in\mathcal{N}}
\operatorname{Var}
\left\{
\bpsi_{\bbeta^{(b)}}(\boldsymbol{\theta})
-
\bpsi_{\bbeta^{(b)}}(\dot{\boldsymbol{\theta}})
\right\}
\leq L_{\bbetab},
$$ for some finite constant $L_{\bbetab}$. The same argument applies to the $\bbetaw$, $\siga^2$, and $\sige^2$ block. Taking $L$ to be the maximum of the resulting four finite constants, we complete the proof.
\end{proof}

\subsection{Proof of Lemma 6}
\label{subsec:proof of lemma 6}
\begin{proof}
It suffices to establish the result block by block. We give the details for the $\bbetab\bbetab$ block; the remaining blocks follow from the corresponding score-derivative expansions in \eqref{eq:asymptotic expression of ML score derivative}. Since the normalization associated with the $\bbetab\bbetab$ block is $g^{-1}$, it suffices to prove that
$$
\sup_{\tilde{\btheta}_k\in\mathcal{N}}
g^{-5/4}
\left\|
\nabla_{\bbetab}\bpsi_{\bbetab}^\top(\tilde{\btheta}_k)
-
\mathbb{E}\nabla_{\bbetab}\bpsi_{\bbetab}^\top(\tilde{\boldsymbol{\theta}}_k)
\right\|
=
o_p(1).
$$

By the score derivative expansion in \eqref{eq:asymptotic expression of ML score derivative}, uniformly for
$\tilde{\btheta}_k\in\mathcal{N}$,
$$
\nabla_{\bbetab}\bpsi_{i,\bbetab}^\top(\tilde{\btheta}_k)
=
-\siga^{-2}(\tilde{\btheta}_k)
\frac{\btau_{i,2}^{(bb)}(\tilde{\btheta}_k)}{A_i(\tilde{\boldsymbol{\theta}}_k)}+O_p(m_i^{-3/2})\boldsymbol{1}_{p_b}^{\otimes 2}.
$$
Therefore,
\begin{align*}
&g^{-5/4}
\left\|\nabla_{\bbetab}\bpsi_{\bbetab}^\top(\tilde{\btheta}_k)
-\mathbb{E}\nabla_{\bbetab}\bpsi_{\bbetab}^\top(\tilde{\btheta}_k)\right\|\\
&\quad \leq g^{-5/4}
\left\|\sumig\left[-\siga^{-2}(\tilde{\btheta}_k)
\frac{\btau_{i,2}^{(bb)}(\tilde{\btheta}_k)
}{A_i(\tilde{\btheta}_k)}
-\mathbb{E}
\left\{-\siga^{-2}(\tilde{\btheta}_k)
\frac{\btau_{i,2}^{(bb)}(\tilde{\btheta}_k)
}{A_i(\tilde{\btheta}_k)}
\right\}\right]\right\|
\\
&\qquad+g^{-5/4}\left\|\sumig\left[O_p(m_i^{-3/2})-\mathbb{E}\{O_p(m_i^{-3/2})\}
\right]\boldsymbol{1}_{p_b}^{\otimes 2}\right\|.
\end{align*}
For the leading term, by Condition A, uniformly over
$\tilde{\btheta}_k\in\mathcal{N}$,
$$
\left\|
\siga^{-2}(\tilde{\btheta}_k)
\frac{\btau_{i,2}^{(bb)}(\tilde{\btheta}_k)}{A_i(\tilde{\btheta}_k)}
\right\|=O_p(1).
$$
Thus, we obtain
\begin{align*}
\sup_{\tilde{\btheta}_k\in\mathcal{N}}&
g^{-5/4}\left\|\sumig\left[-\siga^{-2}(\tilde{\btheta}_k)\frac{\btau_{i,2}^{(bb)}(\tilde{\btheta}_k)}{A_i(\tilde{\btheta}_k)}
-
\mathbb{E}\left\{-\siga^{-2}(\tilde{\btheta}_k)\frac{\btau_{i,2}^{(bb)}(\tilde{\btheta}_k)}{A_i(\tilde{\btheta}_k)}\right\}\right]\right\| \\
&=O_p(g^{-1/4})=o_p(1).
\end{align*}
For the remainder terms, we have
$$
\sup_{\tilde{\btheta}_k\in\mathcal{N}}
g^{-5/4}\left\|\sumig\left[O_p(m_i^{-3/2})
-
\mathbb{E}\{O_p(m_i^{-3/2})\}
\right]\boldsymbol{1}_{p_b}^{\otimes 2}\right\|
\leq O_p(g^{-1/4}m_L^{-3/2})=o_p(1).
$$
Combining the two bounds yields
$$
\sup_{\tilde{\btheta}_k\in\mathcal{N}}
g^{-5/4}
\left\|
\nabla_{\bbetab}\bpsi_{\bbetab}^\top(\tilde{\btheta}_k)
-
\mathbb{E}\nabla_{\bbetab}\bpsi_{\bbetab}^\top(\tilde{\boldsymbol{\theta}}_k)
\right\|
=
o_p(1).
$$

The remaining blocks are treated analogously using their corresponding cluster-wise expansions in \eqref{eq:asymptotic expression of ML score derivative} and the left and right normalizations induced by $\bK^{-1/2}$. Hence,
$$
\sup_{\tilde{\btheta}_k\in\mathcal{N}}
g^{-1/4}\left\|\bK^{-1/2}
\left\{\nabla_{\btheta}\bpsi(\tilde{\btheta}_k)
-\mathbb{E}\nabla_{\btheta}\bpsi(\tilde{\btheta}_k)
\right\}\bK^{-1/2}\right\|
=
o_p(1).
$$
This completes the proof.
\end{proof}

\subsection{Proof of Lemma 7}
\label{subsec:proof of lemma 7}

\begin{proof}
From the exact REML criterion
$$
\ell_R(\bsigma)=\ell_P(\bsigma)-\frac12\log|B(\bsigma)|+U(\bsigma),
$$
we have
$$
\bpsi_{R}(\bsigma)-\bpsi_{P}(\bsigma)=-\frac12
\tr \left\{\bH(\bsigma)  \nabla_{\bsigma} \bD(\bsigma)\right\}+
\nabla_{\bsigma} U(\bsigma).
$$
For the variance component $\siga^2$ and dispersion parameter $\sige^2$, the above equation can be rewritten into 
\begin{align*}
\left|g^{-1/2}\{\psi_{R\siga^2}(\bsigma)
-
\psi_{P\siga^2}(\bsigma)\}\right| &\leq \frac{1}{2g^{1/2}}\tr\left|\left\{\bK_{\bbeta}^{1/2}\bH(\bsigma)\bK_{\bbeta}^{1/2} \right\}\left\{\bK_{\bbeta}^{-1/2} \nabla_{\siga^2}\bD(\bsigma)\bK_{\bbeta}^{-1/2}\right\}\right| \\
&+ \frac{1}{g^{1/2}}\left|\nabla_{\siga^2} U(\bsigma)\right|,
\end{align*}
and 
\begin{align*}
\left|n^{-1/2}\{\psi_{R\sige^2}(\bsigma)
-
\psi_{P\sige^2}(\bsigma)\}\right| &\leq \frac{1}{2n^{1/2}}\tr\left|\left\{\bK_{\bbeta}^{1/2}\bH(\bsigma)\bK_{\bbeta}^{1/2}\right\}\left\{ \bK_{\bbeta}^{-1/2} \nabla_{\sige^2} \bD(\bsigma) \bK_{\bbeta}^{-1/2}\right\}\right| \\
&+ \frac{1}{n^{1/2}}\left|\nabla_{\sige^2} U(\bsigma) \right|.
\end{align*}
Based on the orders we established in Section~\ref{subsec:first derivative of D} and \ref{subsec:second derivatives of D}. It is straightforward to show that \begin{align*}
&\sup_{\bsigma \in \mathcal{N}_{\bsigma}}\left\|\bK_{\bbeta}^{1/2}\bH(\bsigma)\bK_{\bbeta}^{1/2}\right\| = O_p(1), \\
&\sup_{\bsigma \in \mathcal{N}_{\bsigma}}\left\|\bK_{\bbeta}^{-1/2}\nabla_{\siga^2}\bD(\bsigma)\bK_{\bbeta}^{-1/2}\right\| = O_p(1), \\
&\sup_{\bsigma \in \mathcal{N}_{\bsigma}}\left\|\bK_{\bbeta}^{-1/2}\nabla_{\sige^2}\bD(\bsigma)\bK_{\bbeta}^{-1/2}\right\| = O_p(1).
\end{align*}
In addition, the results in \eqref{eq:order of nabla U} imply that
$$
\sup_{\bsigma \in \mathcal{N}_{\bsigma}}\left|\nabla_{\siga^2} U(\bsigma) \right| = O_p(1), \qquad \sup_{\bsigma \in \mathcal{N}_{\bsigma}}\left|\nabla_{\siga^2} U(\bsigma) \right| = O_p(1).
$$
Thus, we have
$$\left|g^{-1/2}\{\psi_{R\siga^2}(\bsigma)
-
\psi_{P\siga^2}(\bsigma)\}\right| \leq \frac{1}{2g^{1/2}}O_p(1) + \frac{1}{g^{1/2}}O_p(1) = o_p(1),$$
and 
$$\left|n^{-1/2}\{\psi_{R\sige^2}(\bsigma)
-
\psi_{P\sige^2}(\bsigma)\}\right| \leq \frac{1}{2n^{1/2}}O_p(1) + \frac{1}{n^{1/2}}O_p(1) = o_p(1),$$
which suggests that 
$$\sup_{\bsigma \in \mathcal{N}_{\bsigma}}\left\|\bK_{\sigma}^{-1/2}\{\bpsi_R(\bsigma) - \bpsi_P(\bsigma) \}\right\| = o_p(1).$$ This completes the proof.

\end{proof}

\subsection{Proof of Lemma 8}
\label{subsec:proof of lemma 8}
\begin{proof}
Along the profile path, denote the information matrix of the exact marginal log-likelihood by
\begin{align*}
    &\bI_{\bbeta\bbeta}(\bsigma) = -\sumig \begin{pmatrix}\nabla_{\bbetab}\bpsi_{i,\bbetab}^\top &  \nabla_{\bbetab}\bpsi_{i,\bbetaw}^\top\\
    \nabla_{\bbetaw}\bpsi_{i,\bbetab}^\top & \nabla_{\bbetaw}\bpsi_{i, \bbetaw}^\top \end{pmatrix}\bigg|_{\bbeta = \bbeta(\bsigma)} = \bD(\bsigma), \\
    &\bI_{\bbeta\bsigma}(\bsigma)  = \sumig \begin{pmatrix}\nabla_{\bbetab}\bpsi_{i,\siga^2} &  \nabla_{\bbetab}\bpsi_{i,\sige^2} \\
   \nabla_{\bbetaw}\bpsi_{i,\siga^2} & \nabla_{\bbetaw} \bpsi_{i,\sige^2}
    \end{pmatrix}\bigg|_{\bbeta = \bbeta(\bsigma)}, 
    \\
    &\bI_{\bsigma\bbeta}(\bsigma) = \sumig \begin{pmatrix}\nabla_{\siga^2}\bpsi_{i,\bbetab}^\top &  \nabla_{\siga^2}\bpsi_{i,\bbetaw}^\top\\
    \nabla_{\sige^2}\bpsi_{i,\bbetab}^\top & \nabla_{\sige^2}\bpsi_{i, \bbetaw}^\top 
    \end{pmatrix}\bigg|_{\bbeta = \bbeta(\bsigma)}, \\
    &\bI_{\bsigma\bsigma}(\bsigma) = \sumig \begin{pmatrix}\nabla_{\siga^2}\bpsi_{i,\siga^2} &  \nabla_{\siga^2}\bpsi_{i,\sige^2} \\
   \nabla_{\sige^2}\bpsi_{i,\siga^2} & \nabla_{\sige^2} \bpsi_{i,\sige^2}
    \end{pmatrix}\bigg|_{\bbeta = \bbeta(\bsigma)}.
\end{align*} 
By definition, the profile criterion is a likelihood function with the fixed
effects treated as nuisance parameters and replaced by their conditional maximizer.
Using the result of \cite{patefield1977maximized}, the curvature of the maximized
likelihood with respect to the remaining parameters is given by the corresponding
sub-formation matrix of the original log-likelihood after substituting the
conditional maximizer. Thus,
$$-\nabla_{\bsigma}\bpsi_P(\bsigma) = \bI_{\bsigma\bsigma}(\bsigma) - \bI_{\bsigma\bbeta}(\bsigma) \bI^{-1}_{\bbeta\bbeta}(\bsigma) \bI_{\bbeta\bsigma}(\bsigma).$$ 
It follows that
\begin{align*}
\bB_{\bsigma n}(\bsigma)
=&-\bK_{\bsigma}^{-1/2}\nabla_{\bsigma}\bpsi_P(\bsigma)\bK_{\bsigma}^{-1/2} \\
=&\bK_{\bsigma}^{-1/2}\bI_{\bsigma\bsigma}(\bsigma)\bK_{\bsigma}^{-1/2} \\
&-\bK_{\bsigma}^{-1/2}
\bI_{\bsigma\bbeta}(\bsigma)\bI_{\bbeta\bbeta}^{-1}(\bsigma)\bI_{\bbeta\bsigma}(\bsigma)
\bK_{\bsigma}^{-1/2}.
\end{align*}
Define the normalized Schur-complement by $$\bQ_n(\bsigma)
=
\bK_\sigma^{-1/2}
\bI_{\bsigma\bbeta}(\btheta)
\bI_{\bbeta\bbeta}^{-1}(\btheta)
\bI_{\bbeta\bsigma}(\btheta)
\bK_\sigma^{-1/2}.$$
Then $\bB_{\bsigma n}(\bsigma) = \bK_{\bsigma}^{-1/2}\bI_{\bsigma\bsigma}\bK_{\bsigma}^{-1/2} - \bQ_n(\btheta)$.
We rewrite the Schur-complement $\bQ_{n}(\bsigma)$ as
\begin{align*}
\bQ_n(\bsigma)&=\left(\bK_{\bsigma}^{-1/2}\bI_{\bsigma\bbeta}(\btheta)\bK_{\bbeta}^{-1/2}\right)
\left(\bK_{\bbeta}^{1/2}\bI_{\bbeta\bbeta}^{-1}(\btheta)\bK_{\bbeta}^{1/2}\right)
\left(\bK_{\bbeta}^{-1/2}\bI_{\bbeta\bsigma}(\btheta)\bK_{\bsigma}^{-1/2}
\right)\\
&=\left(\bK_{\bsigma}^{-1/2}\bI_{\bsigma\bbeta}(\btheta)\bK_{\bbeta}^{-1/2}\right)
\left(\bK_{\bbeta}^{1/2}\bH(\btheta)\bK_{\bbeta}^{1/2}\right)
\left(\bK_{\bbeta}^{-1/2}\bI_{\bbeta\bsigma}(\btheta)\bK_{\bsigma}^{-1/2}
\right).
\end{align*}

Based on the discussion in Section~\ref{subsec:property of D(sigma)}, as $g, m_L \to \infty$, we have $$\bK_{\bbeta}^{1/2}\bH(\btheta)\bK_{\bbeta}^{1/2} 
\longrightarrow O_p(1)\boldsymbol{1}_{p}^{\otimes 2}.$$
By \eqref{eq:asymptotic expression of ML score derivative}, uniformly over $\bsigma \in \mathcal{N}_{\bsigma}$, we have
\begin{align*} \lim_{g, m_L \rightarrow \infty}
g^{-1}
\sumig
\nabla_{\siga^2}\bpsi_{i,\bbetab}^\top\bigg|_{\bbeta = \bbeta(\bsigma)} &=-g^{-1}\siga^{-4}\sumig \alpha_i\bxbi + g^{-1}\sumig O_p(m_i^{-1/2}) \boldsymbol{1}_{p_b}^{\top} \\
&=-g^{-1} O_p(g^{1/2})\boldsymbol{1}_{[1:p_b]} +
o_p(1)\boldsymbol{1}_{[1:p_b]} = o_p(1)\boldsymbol{1}_{[1:p_b]}, \\
\end{align*} since $\mathbb{E}(\alpha_i) = 0$. Similarly, uniformly over $\bsigma \in \mathcal{N}_{\bsigma}$, we have
\begin{align*}
&\lim_{g, m_L \rightarrow \infty}(gn)^{-1/2}
\sumig
\nabla_{\siga^2}\bpsi_{i,\bbetaw}^\top\bigg|_{\bbeta = \bbeta(\bsigma)}
=
o_p(1)\boldsymbol{1}_{[1:p_w]}, \\
&\lim_{g, m_L \rightarrow \infty}(gn)^{-1/2}
\sumig
\nabla_{\sige^2}\bpsi_{i\bbetab}^\top \bigg|_{\bbeta = \bbeta(\bsigma)}
=
o_p(1)\boldsymbol{1}_{[1:p_b]},\\
&\lim_{g, m_L \rightarrow \infty} n^{-1}
\sumig
\nabla_{\sige^2} \bpsi_{i,\bbetaw}^\top \bigg|_{\bbeta = \bbeta(\bsigma)}
=
o_p(1)\boldsymbol{1}_{[1:p_w]}.
\end{align*}
Thus,
$$
\bK_{\bsigma}^{-1/2}
\bI_{\bsigma\bbeta}
\bK_{\bbeta}^{-1/2}
\longrightarrow o_p(1) \boldsymbol{1}_{[2:p]},
$$ and by symmetry,
$$
\bK_{\bbeta}^{-1/2}
\bI_{\bbeta\bsigma}
\bK_{\sigma}^{-1/2}
\longrightarrow o_p(1) \boldsymbol{1}_{[p:2]},
$$
Therefore,
$$
\lim_{g, m_L \rightarrow \infty} \bQ_{n}(\bsigma)
=
o_p(1)\boldsymbol{1}_{[2:p]}O_p(1)\boldsymbol{1}_{p}^{\otimes 2}o_p(1)\boldsymbol{1}_{[p:2]}
=
o_p(1)\boldsymbol{1}_{2}^{\otimes 2},
$$
which implies that $\bQ_n(\bsigma)$ is asymptotically negligible; the limiting behavior of $\bB_{\bsigma n}(\bsigma)$ is determined by the normalized
variance-component block
$\bK_{\bsigma}^{-1/2}
\bI_{\bsigma\bsigma}(\bsigma)
\bK_{\bsigma}^{-1/2}.$
Again, by \eqref{eq:asymptotic expression of ML score derivative}, uniformly over $\bsigma \in \mathcal{N}_{\bsigma}$,
\begin{align*}
&\lim_{g,m_L\rightarrow \infty} g^{-1}
\sumig
\nabla_{\siga^2}\psi_{i,\siga^2}\bigg|_{\bbeta = \bbeta(\bsigma)}
=
\frac{1}{2\siga^4}, 
\qquad
\lim_{g,m_L\rightarrow \infty} (gn)^{-1/2}
\sumig
\nabla_{\siga^2}\psi_{i,\sige^2}\bigg|_{\bbeta = \bbeta(\bsigma)}
=0, \\ 
&\lim_{g,m_L\rightarrow \infty} (gn)^{-1/2}
\sumig
\nabla_{\sige^2} \psi_{i,\siga^2}\bigg|_{\bbeta = \bbeta(\bsigma)}
=0, \qquad \lim_{g,m_L\rightarrow \infty} n^{-1}
\sumig
\nabla_{\sige^2} \psi_{i,\sige^2}\bigg|_{\bbeta = \bbeta(\bsigma)}
=
c_4(\bbeta(\bsigma), \bsigma).
\end{align*}
Therefore, uniformly for $\bsigma \in \mathcal{N}_{\bsigma}$,
$$\bB_{\bsigma n}(\bsigma) \overset{D}{\longrightarrow} \bB_{\bsigma}(\bsigma),$$
where $\bB_{\bsigma}(\bsigma) = \diag\{1/(2\siga^4), c_4(\bbeta(\bsigma), \bsigma)\}$.
By Condition~A, $c_4(\bbeta(\bsigma), \bsigma)$ is finite and positive. Hence, $\bB_{\bsigma}(\bsigma)$ is finite and positive definite. In particular, define the limiting matrix evaluated at the true parameter value by
$$\bB_{\bsigma}= \bB_{\bsigma}(\dbtheta) = \diag\{1/(2\dsiga^4), c_4(\dbtheta)\},$$
which is the matrix appearing in Theorem 1.
\end{proof}

\subsection{Proof of Lemma 9}

\begin{proof}
The argument follows the same strategy as that used in the proof of Lemma~\ref{lemma 7}. Following the Section~\ref{subsec:proof of lemma 7}, we have
$$
\bpsi_{R}(\bsigma)-\bpsi_{P=}(\bsigma)=-\frac12\tr\left\{\bH(\bsigma)\nabla_{\bsigma}\bD(\bsigma)\right\}+\nabla_{\bsigma} U(\bsigma).
$$
Differentiating this expression once more with respect to $\bsigma$ gives
$$
\begin{aligned}
\nabla_{\bsigma}
\left\{\bpsi_{R}(\bsigma)-\bpsi_{P}(\bsigma)\right\}
&=\frac12\tr\left\{\bH(\bsigma)\nabla_{\siga^2}\bD(\bsigma)\bH(\bsigma)\nabla_{\siga^2}\bD(\bsigma)\right\}
\\
&\quad
-
\frac12
\tr\left\{\bH(\bsigma)\nabla_{\bsigma}^2 \bD(\bsigma)\right\}
+
\nabla_{\bsigma}^2 U(\bsigma).
\end{aligned}
$$
Based on the order established in Section~\ref{subsec:property of D(sigma)}-\ref{subsec:second derivatives of D},
\begin{align*}
   &\sup_{\bsigma \in \mathcal{N}_{\bsigma}} \left\|\bK_{\bbeta}^{1/2}\bH(\bsigma)\bK_{\bbeta}^{1/2}\right\| = O_p(1), \\
   &\sup_{\bsigma \in \mathcal{N}_{\bsigma}} \left\|\bK_{\bbeta}^{-1/2}\nabla_{\bsigma}\bD(\bsigma)\bK_{\bbeta}^{-1/2}\right\| = O_p(1), \\
   &\sup_{\bsigma \in \mathcal{N}_{\bsigma}} \left\|\bK_{\bbeta}^{-1/2}\nabla_{\bsigma}^2\bD(\bsigma)\bK_{\bbeta}^{-1/2}\right\| = O_p(1), \\
   &\sup_{\bsigma \in \mathcal{N}_{\bsigma}} \left\|\bK_{\bsigma}^{-1/2}\nabla_{\bsigma}^2 U(\bsigma)\bK_{\bsigma}^{-1/2}\right\| = o_p(1).
\end{align*}

Thus, the two trace terms are $O_p(1)$ because
\begin{align*}
\tr\left\{\bH(\bsigma)\nabla_{\siga^2}\bD(\bsigma)\bH(\bsigma)\nabla_{\siga^2}\bD(\bsigma)\right\}&=
\tr\left[\left\{\bK_{\bbeta}^{1/2}\bH(\bsigma)\bK_{\bbeta}^{1/2}\right\}\right. \\
&\qquad \quad \left\{\bK_{\bbeta}^{-1/2} \nabla_{\siga^2}\bD(\bsigma) \bK_{\bbeta}^{-1/2}\right\} \\
&\qquad \quad \left\{\bK_{\bbeta}^{1/2}\bH(\bsigma)\bK_{\bbeta}^{1/2}\right\} \\
&\qquad \quad  \left.\left\{\bK_{\beta}^{-1/2} \nabla_{\siga^2}\bD(\bsigma) \bK_{\beta}^{-1/2}\right\}
\right] \\
&\leq p \left\|\bK_{\bbeta}^{1/2}\bH(\bsigma)\bK_{\bbeta}^{1/2}\right\|^2 \left\|\bK_{\beta}^{-1/2} \nabla_{\siga^2}\bD(\bsigma) \bK_{\beta}^{-1/2}\right\|^2 \\
&= O_p(1),
\end{align*}
and 
\begin{align*}
\tr
\left\{
\bH(\bsigma)
\nabla_{\siga^2}^2\bD(\bsigma)
\right\} &= \tr\left[\left\{\bK_{\bbeta}^{1/2}\bH(\bsigma)\bK_{\bbeta}^{1/2}\right\}\left\{\bK_{\beta}^{-1/2} \nabla_{\siga^2}\bD(\bsigma)^2 \bK_{\beta}^{-1/2}\right\}\right] \\
&\leq p\left\|\bK_{\bbeta}^{1/2}\bH(\bsigma)\bK_{\bbeta}^{1/2}\right\| \left\|\bK_{\beta}^{-1/2} \nabla_{\siga^2}\bD^2(\bsigma) \bK_{\beta}^{-1/2}\right\| = O_p(1).
\end{align*}
Combining the above results, uniformly in $\bsigma \in \mathcal{N}_{\bsigma}$, we have
$$\bK_{\bsigma}^{-1/2}\nabla_{\bsigma}
\left\{\bpsi_{R}(\bsigma)-\bpsi_{P}(\bsigma)\right\}\bK_{\bsigma}^{-1/2} = o_p(1)\boldsymbol{1}_{2}^{\otimes 2}.$$
Thus,
\begin{align*}
    \sup_{\bsigma \in \mathcal{N}_{\bsigma}} \left\|\bK_{\bsigma}^{-1/2}\nabla_{\bsigma}
\left\{\bpsi_{R}(\bsigma)-\bpsi_{P}(\bsigma)\right\}\bK_{\bsigma}^{-1/2}\right\| = o_p(1),
\end{align*}
which completes the proof.
\end{proof}

\subsection{Proof of Lemma 10}
\label{subsec:proof of lemma 10}
\begin{proof}
By definition,
$$
\hat{\bbeta}_R
-
\hat{\bbeta}_P
=
\bbeta(\hat{\bsigma}_R)
-
\bbeta(\hat{\bsigma}_P).
$$
By the mean-value argument, there exits a intermediate vector 
$\tilde{\bsigma} \in \mathcal{N}_{\bsigma}$ between 
$\hat{\bsigma}_R$ and $\hat{\bsigma}_P$ such that
$$
\hat{\bbeta}_R
-
\hat{\bbeta}_P
=
\nabla_{\bsigma} \bbeta(\tilde{\bsigma})
\left(
\hat{\bsigma}_R-\hat{\bsigma}_P
\right).
$$
After introducing the normalization matrices,
$$
\bK_{\bbeta}^{1/2}
\left(\hat{\bbeta}_R-\hat{\bbeta}_P\right)=\left[\bK_{\bbeta}^{1/2} \nabla_{\bsigma}\bbeta(\tilde{\bsigma})
\bK_{\bsigma}^{-1/2}\right]\left[\bK_{\bsigma}^{1/2}\left(\hat{\bsigma}_R-\hat{\bsigma}_P\right)\right].
$$
Based on \eqref{eq:order of nabla beta}, we have
$$\sup_{\bsigma \in \mathcal{N}_{\bsigma}} \left\|\bK_{\bbeta}^{1/2} \nabla_{\siga^2}\bbeta(\tilde{\bsigma})
\bK_{\bsigma}^{-1/2}\right\| = O_p(1),$$
and the same argument applies to differentiation with respect to $\sige^2$. Hence,
$$\sup_{\bsigma \in \mathcal{N}_{\bsigma}} \left\|\bK_{\bbeta}^{1/2} \nabla_{\bsigma} \hat{\bbeta}(\bsigma)
\bK_{\bsigma}^{-1/2}\right\| = O_p(1).$$ By assumption, $$\bK_{\bsigma}^{1/2}(\hat{\bsigma}_R - \hat{\bsigma}_P) = o_p(1).$$ Therefore, 
$$\bK_{\bbeta
}^{1/2}
\left(\hat{\bbeta}_R-\hat{\bbeta}_P\right) = o_p(1),$$ which completes the proof.
\end{proof}

\subsection{Proof of Lemma 11}
\label{subsec:proof of lemma 11}
\begin{proof}
By definition, for each fixed $\bsigma$, $$\ell_P(\bsigma) = \ell(\bbeta(\bsigma), \bsigma) = \max_{\bbeta}\ell(\bbeta, \bsigma).$$ Maximizing both sides over $\bsigma$ yields $$\max_{\bsigma} \ell_P(\bsigma) = \max_{\bsigma} \max_{\bbeta} \ell(\bbeta, \bsigma)= \max_{\bsigma,\bbeta} \ell(\bbeta, \bsigma).$$
Thus, the profile and joint maximization problems have the same maximum value. Let

Let $\hat{\bsigma}_P$ be the maximizer of $\ell_P(\bsigma)$. Since $\hat{\bbeta}_P = \bbeta(\hat{\bsigma}_P)$, we have 
$$\ell(\hat{\bbeta}_P, \hat{\bsigma}_P) = \ell_P(\hat{\bsigma}_P) = \max_{\bsigma}\ell_P(\bsigma) = \max_{\bbeta, \bsigma} \ell(\bbeta, \bsigma).$$ It follows that $(\hat{\bbeta}_P, \hat{\bsigma}_P)$ is the joint maximizer of $\ell(\bbeta,\bsigma)$, and therefore is an ML estimator. This completes the proof.
\end{proof}

\section{Bias Correction Property of REML}
\setcounter{equation}{0} 
\label{sec:bias_correction}
In this section, we make explicit the bias-correction mechanism of the REML criterion. Throughout, we retain only the leading contributions relevant to this correction; lower-order terms are controlled by the bounds established in Sections~S.3--S.4 and are omitted for brevity. Accordingly, the asymptotic expressions below display only the terms that contribute to the bias correction unless stated otherwise.

Although the marginal score evaluated at the true parameter is unbiased, replacing the fixed effects by their profile estimators introduces a nonzero leading bias into the variance-component score. We illustrate this mechanism for the random-effect variance $\siga^2$, derive the leading bias of its profile score, and show that the REML determinant adjustment contributes exactly the opposite term. This yields an explicit correction analogous to the familiar degrees-of-freedom adjustment in the LMM.

To characterize the effect of profiling the fixed effects, we first derive
the leading behavior of $\bbeta(\dbsigma)-\dbbeta$. At
$\bsigma=\dbsigma$, the profile estimator satisfies
$\psi_{\bbeta}\{\bbeta(\dbsigma),\dbsigma\}
=
\boldsymbol{0}.$
Expanding this estimating equation around $\dbbeta$ gives
$$\boldsymbol{0} = \bpsi_{\bbeta}(\dbtheta) - \bD(\dbsigma)\left\{\bbeta(\dbsigma) - \dbbeta\right\}.$$
Since $\bH(\bsigma) = \bD(\bsigma)^{-1}$, it follows that
\begin{equation}
\label{eq:difference of bbeta(bsigma) -bbeta}
\bbeta(\dbsigma)-\dbbeta
=
\bH(\dbsigma)\psi_{\bbeta}(\dbtheta).
\end{equation}
By \eqref{eq:asymptotic score of betab}, the leading part of the between-cluster score is
\begin{equation}
\label{eq:leading term of psi_bbetab}
\bpsi_{\bbetab}(\dbtheta)
=
\siga^{-2}\sumig\bxbi\alpha_i
=
O_p(g^{1/2})\boldsymbol{1}_{[p_b:1]}.
\end{equation}
Similarly, \eqref{eq:asymptotic score of betaw} gives $\bpsi_{\bbetaw}(\dbtheta)=O_p(n^{1/2})\boldsymbol{1}_{[p_w:1]}.$ The corresponding leading parts of the score-derivative blocks follow from \eqref{eq:asymptotic expression of ML score derivative}
\begin{align*}
\nabla_{\bbetab}\psi_{\bbetab}^{\top}(\dbtheta)
&=
-\siga^{-2}
\sumig\frac{\tau_{i,2}^{(bb)}}{A_i}
= O_p(g)\boldsymbol{1}_{p_b}^{\otimes2}, \\
\nabla_{\bbetaw}\psi_{\bbetab}^{\top}(\dbtheta)
&=
\siga^{-2}
\sumig\frac{\tau_{i,2}^{(bw)}}{A_i}
 = O_p(g)\boldsymbol{1}_{[p_b:p_w]},\\
\nabla_{\bbetaw}\psi_{\bbetaw}^{\top}(\dbtheta) 
&=
-\sumig
\left\{
\tau_{i,2}^{(ww)}
-
\frac{\tau_{i,2}^{(w)}
\tau_{i,2}^{(w)\top}}{A_i}
\right\}
 = O_p(n)\boldsymbol{1}_{p_w}^{\otimes2}.
\end{align*}
These orders agree with the block
structure of $\bD(\dbsigma)$ established in Section~\ref{subsec:property of D(sigma)}. Using the block inverse results in Section~\ref{subsec:property of D(sigma)},
\begin{align*}
\bH_{bb}(\dbsigma)&=O_p(g^{-1})\boldsymbol{1}_{p_b}^{\otimes 2},\qquad
\bH_{bw}(\dbsigma)=\bH_{wb}(\dbsigma)^{\top}=O_p(n^{-1})\boldsymbol{1}_{[p_b:p_w]},\\
\bH_{ww}(\dbsigma)&=O_p(n^{-1})\boldsymbol{1}_{p_w}^{\otimes 2}.
\end{align*}
Substituting these orders into \eqref{eq:difference of bbeta(bsigma) -bbeta}, the between-cluster component satisfies
\begin{align*}
\bbetab(\dbsigma)-\dbbetab
=&\bH_{bb}(\dbsigma)\bpsi_{\bbetab}(\dbtheta)+\bH_{bw}(\dbsigma)\bpsi_{\bbetaw}(\dbtheta)\\
=&O_p(g^{-1})O_p(g^{1/2})\boldsymbol{1}_{[p_b:1]}+O_p(n^{-1})O_p(n^{1/2})\boldsymbol{1}_{[p_b:1]}\\
=&O_p(g^{-1/2})\boldsymbol{1}_{[p_b:1]},
\end{align*}
and
\begin{align*}
\bbetaw(\dbsigma)-\dbbetaw
=&\bH_{wb}(\dbsigma)\bpsi_{\bbetab}(\dbtheta)+\bH_{ww}(\dbsigma)\bpsi_{\bbetaw}(\dbtheta)\\
=&O_p(n^{-1})O_p(g^{1/2})\boldsymbol{1}_{[p_w:1]}+O_p(n^{-1})O_p(n^{1/2})\boldsymbol{1}_{[p_w:1]}\\
=&O_p(n^{-1/2})\boldsymbol{1}_{[p_w:1]}.
\end{align*}
The preceding rates show that the dominant contribution to the between-cluster profile error arises from the $bb$ block. From the block inverse representation in Section~\ref{subsec:property of D(sigma)} the dominant part of $\bH_{bb}(\dbsigma)$ is determined by $\bD_{bb}(\dbsigma)^{-1}.$ The leading term of $\bD_{bb}(\dbsigma)$ is 
$$\bD_{bb}(\dbsigma) = \dbsigma^{-2}\sumig \bxbi\bxbit.$$ By the construction of $\bH(\bsigma)$ in Section~\ref{subsec:property of D(sigma)}, the leading part of $\bH_{bb}(\bsigma)$ is $\bD_{bb}(\bsigma)$, thus,
$$
\bH_{bb}(\dbsigma) = \dsiga^2 \left\{\sumig \bxbi\bxbit\right\}^{-1}.
$$

Combine this with \eqref{eq:leading term of psi_bbetab} gives 
\begin{equation}
\label{eq:leading term of bbeta(bsigma) - bbeta}
\bbetab(\dbsigma)-\dbbetab
=
\left\{
\sumig\bxbi\bxbit
\right\}^{-1}
\sumig\bxbi\alpha_i.
\end{equation}

We now consider the profile score for $\siga^2$. Since $\psi_{P,\siga^2}(\dbsigma)
=
\psi_{\siga^2}
\{\bbeta(\dbsigma),\dbsigma\},$
a second-order Taylor expansion around $\dbbeta$ gives
\begin{equation}
\label{eq:expansion of profile score}
\begin{aligned}
\psi_{P,\siga^2}(\dbsigma)
={}&
\psi_{\siga^2}(\dbtheta)
\\
&+
\{\nabla_{\bbetab}\psi_{\siga^2}(\dbtheta)\}^{\top}
\{\bbetab(\dbsigma)-\dbbetab\}
\\
&+
\{\nabla_{\bbetaw}\psi_{\siga^2}(\dbtheta)\}^{\top}
\{\bbetaw(\dbsigma)-\dbbetaw\}
\\
&+
\frac{1}{2}
\{\bbetab(\dbsigma)-\dbbetab\}^{\top}
\nabla_{\bbetab}^{2}\psi_{\siga^2}(\dbtheta)
\{\bbetab(\dbsigma)-\dbbetab\}
\\
&+
\{\bbetab(\dbsigma)-\dbbetab\}^{\top}
\nabla_{\bbetab}\nabla_{\bbetaw}
\psi_{\siga^2}(\dbtheta)
\{\bbetaw(\dbsigma)-\dbbetaw\}
\\
&+
\frac{1}{2}
\{\bbetaw(\dbsigma)-\dbbetaw\}^{\top}
\nabla_{\bbetaw}^{2}\psi_{\siga^2}(\dbtheta)
\{\bbetaw(\dbsigma)-\dbbetaw\}.
\\
\end{aligned}
\end{equation}
Since the marginal score is unbiased at the true parameter,
$\mathbb{E}\{\psi_{\siga^2}(\dbtheta)\}=0$. Hence, the bias of the profile score is determined by the terms induced by replacing $\bbeta(\dbsigma)$. From \eqref{eq:asymptotic expression of ML score derivative},
$$\nabla_{\bbetab}\psi_{i,\siga^2}
=
-\frac{\siga^{-4}\alpha_i\btau_{i,2}^{(b)}}{A_i}
+
O_p(m_i^{-1/2})\boldsymbol{1}_{p_b}.$$
Since $\btau_{i,2}^{(b)}=\bxbi\tau_{i,2}$ and
$\tau_{i,2}/A_i=1+O_p(m_i^{-1})$, the dominant part is
$$
\nabla_{\bbetab}\psi_{\siga^2}(\dbtheta)
=
-\siga^{-4}\sumig\bxbi\alpha_i.
$$
Combining this expression with the leading representation in \eqref{eq:leading term of bbeta(bsigma) - bbeta}, we have
\begin{equation}
\label{eq:leading term of linear bias}
\begin{aligned}
&
\mathbb{E}\left[
\{\nabla_{\bbetab}\psi_{\siga^2}(\dbtheta)\}^{\top}
\{\bbetab(\dbsigma)-\dbbetab\}
\right]
\\
&\quad=-\siga^{-4}\mathbb{E}\left[
\left(\sumig\bxbi\alpha_i\right)^{\top}
\left\{\sumig\bxbi\bxbit\right\}^{-1}
\left(\sumig\bxbi\alpha_i\right)\right]
\\
&\quad= -\frac{p_b}{\siga^2},
\end{aligned}
\end{equation}
where the last equality follows from independence across clusters and $\mathbb{E}(\alpha_i^2) = \siga^2$. For the quadratic between-cluster term, differentiating the exact expression
for $\nabla_{\bbetab}\psi_{i,\siga^2}$ with respect to $\bbetab$, its dominant part is
$$\nabla_{\bbetab}^{2}\psi_{\siga^2}(\dbtheta)
=
\siga^{-4}
\sumig\bxbi\bxbit.$$
Therefore
\begin{equation}
\label{eq:leading term of quadratic bias}
\begin{aligned}
\frac{1}{2}\mathbb{E}&\left[
\{\bbetab(\dbsigma)-\dbbetab\}^{\top}
\nabla_{\bbetab}^{2}\psi_{\siga^2}(\dbtheta)
\{\bbetab(\dbsigma)-\dbbetab\}\right] \\
&= \frac{\siga^{-4}}{2}
\mathbb{E}\left[\left(\sumig\bxbi\alpha_i\right)^{\top}
\left\{\sumig\bxbi\bxbit\right\}^{-1}
\left(\sumig\bxbi\alpha_i\right)\right] \\
&=\frac{p_b}{2\siga^2}.
\end{aligned}
\end{equation}

The remaining terms in \eqref{eq:expansion of profile score} involve the within-cluster profile error $\bbetaw(\dbsigma)-\dbbetaw$, which is of order $O_p(n^{-1/2})\boldsymbol{1}_{[p_w:1]}$. Together with the asymptotic expression in \eqref{eq:asymptotic expression of ML score derivative} and the
block inverse orders above, these terms are of smaller order than the two
between-cluster contributions in \eqref{eq:leading term of linear bias} and \eqref{eq:leading term of quadratic bias}. Consequently, the leading bias of the profile score is $$-\frac{p_b}{\siga^2} + \frac{p_b}{2\siga^2} = -\frac{p_b}{2\siga^2}.$$

For the REML adjustment. As shown in Section~\ref{subsec:proof of lemma 7}, the adjustment term contributes $-\frac{1}{2}
\tr\left\{
\bH(\dbsigma)
\nabla_{\siga^2}\bD(\dbsigma)
\right\}$
to the $\siga^2$ score, apart from terms of smaller order. Partitioning the
trace conformably with the fixed-effect blocks gives
\begin{equation}
\label{eq:expansion of REML adjustment}
\begin{aligned}
\tr\{\bH(\dbsigma)\nabla_{\siga^2}\bD(\dbsigma)\}
=&
\tr\{\bH_{bb}(\dbsigma)\nabla_{\siga^2}\bD_{bb}(\dbsigma)\}
+
\tr\{\bH_{bw}(\dbsigma)\nabla_{\siga^2}\bD_{wb}(\dbsigma)\}
\\
&+
\tr\{\bH_{wb}(\dbsigma)\nabla_{\siga^2}\bD_{bw}(\dbsigma)\}
+
\tr\{\bH_{ww}(\dbsigma)\nabla_{\siga^2}\bD_{ww}(\dbsigma)\}.
\end{aligned}
\end{equation}
By the block orders established in Section~~\ref{subsec:property of D(sigma)} and ~\ref{subsec:first derivative of D}, the last three terms are of smaller order, so the dominant contribution is determined by the $bb$ block. The leading term of $\nabla_{\siga^2}\bD_{bb}(\dbsigma)$ is
$$
\nabla_{\siga^2}\bD_{bb}(\dbsigma)
=
-\siga^{-4}
\sumig\bxbi\bxbit.
$$
Thus, the leading part of the REML adjustment to the profile score $\psi_{P, \siga^2}(\bsigma)$ is
\begin{align*}
-\frac{1}{2}
\tr\left\{
\bH(\dbsigma)
\nabla_{\siga^2}\bD(\dbsigma)
\right\}
&=
\frac{1}{2\siga^2}
\tr\left[
\left\{
\sumig\bxbi\bxbit
\right\}^{-1}
\left\{
\sumig\bxbi\bxbit
\right\}
\right]
\\
&=
\frac{p_b}{2\siga^2}.
\end{align*}
Hence, the REML determinant adjustment contributes
$$\frac{p_b}{2\siga^2},$$
which exactly cancels the leading profile-score bias.

For $\sige^2$, an analogous calculation gives the leading profile-score bias $$-\frac{p_w}{2\sige^2},$$ while the corresponding REML adjustment contributes $$\frac{p_w}{2\sige^2}.$$ Thus, the leading contributions also cancel for $\sige^2$. The derivation follows the same argument as for $\siga^2$, using the within-cluster score and curvature expansions, and is omitted for brevity.

Taken together, these results show that the REML adjustment corrects the leading bias induced by profiling the fixed effects at the corresponding information scale:
$$\siga^2: -\frac{p_b}{2\siga^2} + \frac{p_b}{2\siga^2} = 0,$$
and 
$$\sige^2: -\frac{p_w}{2\sige^2} + \frac{p_w}{2\sige^2} = 0,$$

This provides a direct analogue of the classical bias-correction mechanism of REML in the LMM. In the present hierarchical GLMM, however, the correction separates according to the two information scales: the between-cluster fixed effects contribute to the correction for the random-effect variance $\siga^2$, whereas the within-cluster fixed effects contribute to the correction for the dispersion parameter $\sige^2$. Thus, the familiar degrees-of-freedom adjustment in LMMs is reflected here through parameter-specific corrections determined by the corresponding fixed-effect dimensions $p_b$ and $p_w$.

\section{Additional Numerical Study}
\renewcommand{\thetable}{S.\arabic{table}}
This section reports the simulation results under the two random-effect specifications considered in Section 4. Section~\ref{subsec:Simulation for Gaussian Random Effects} presents the results for Gaussian random effects, whereas Section 8.2 considers the asymmetric two-component Gaussian mixture. The numerical findings are consistent with the discussion in the main text, and we therefore do not repeat the detailed interpretation here.

\subsection{Simulation for Gaussian Random Effects}
\setcounter{table}{0}
\label{subsec:Simulation for Gaussian Random Effects}
\begin{table}[!h]
\centering
\caption{Simulated coverage probabilities and average confidence interval lengths for the Poisson GLMM When $\alpha_i$ has a Gaussian distribution with $\siga^2 = 4$.}
\small
\setlength{\tabcolsep}{4pt}
\renewcommand{\arraystretch}{1.05}

\begin{tabular}{llcccccccc}
\toprule
\multirow{2}{*}{$g$}
& \multirow{2}{*}{Estimator}
& \multicolumn{2}{c}{$m_L=10$}
& \multicolumn{2}{c}{$m_L=20$}
& \multicolumn{2}{c}{$m_L=50$}
& \multicolumn{2}{c}{$m_L=100$} \\
\cmidrule(lr){3-4}
\cmidrule(lr){5-6}
\cmidrule(lr){7-8}
\cmidrule(lr){9-10}
&
& Cvge & Len
& Cvge & Len
& Cvge & Len
& Cvge & Len \\
\midrule

\multirow{6}{*}{$10$}
& $SE(\hat\beta^{(b)}_0)$ & 0.87 & 7.23 & 0.87 & 7.13 & 0.89 & 7.27 & 0.88 & 7.30 \\
& $SE(\hat\beta^{(b)}_1)$ & 0.88 & 1.72 & 0.87 & 1.70 & 0.89 & 1.73 & 0.88 & 1.75 \\
& $SE(\hat\beta^{(w)}_0)$ & 0.95 & 0.07 & 0.95 & 0.05 & 0.95 & 0.03 & 0.94 & 0.02 \\
& $SE(\hat\sigma_\alpha^2)$ & 0.60 & 3.87 & 0.64 & 4.16 & 0.70 & 4.74 & 0.71 & 4.79 \\
& $SE_{C}(\hat\sigma_\alpha^2)$ & 0.68 & 4.75 & 0.70 & 4.81 & 0.74 & 5.11 & 0.73 & 5.01 \\
& $SE_{N}(\hat\sigma_\alpha^2)$ & 0.80 & 6.63 & 0.81 & 6.40 & 0.84 & 6.51 & 0.84 & 6.39 \\
\midrule

\multirow{6}{*}{$20$}
& $SE(\hat\beta^{(b)}_0)$ & 0.91 & 5.05 & 0.92 & 5.08 & 0.91 & 5.16 & 0.92 & 5.18 \\
& $SE(\hat\beta^{(b)}_1)$ & 0.91 & 1.19 & 0.92 & 1.20 & 0.91 & 1.22 & 0.91 & 1.24 \\
& $SE(\hat\beta^{(w)}_0)$ & 0.93 & 0.04 & 0.95 & 0.03 & 0.95 & 0.02 & 0.95 & 0.01 \\
& $SE(\hat\sigma_\alpha^2)$ & 0.72 & 3.22 & 0.76 & 3.53 & 0.81 & 3.86 & 0.82 & 4.00 \\
& $SE_{C}(\hat\sigma_\alpha^2)$ & 0.78 & 3.82 & 0.81 & 3.93 & 0.83 & 4.10 & 0.83 & 4.16 \\
& $SE_{N}(\hat\sigma_\alpha^2)$ & 0.87 & 4.79 & 0.87 & 4.76 & 0.90 & 4.78 & 0.89 & 4.74 \\
\midrule

\multirow{6}{*}{$50$}
& $SE(\hat\beta^{(b)}_0)$ & 0.93 & 3.19 & 0.93 & 3.25 & 0.95 & 3.27 & 0.93 & 3.28 \\
& $SE(\hat\beta^{(b)}_1)$ & 0.93 & 0.75 & 0.93 & 0.76 & 0.95 & 0.77 & 0.93 & 0.78 \\
& $SE(\hat\beta^{(w)}_0)$ & 0.94 & 0.02 & 0.94 & 0.02 & 0.94 & 0.01 & 0.96 & 0.01 \\
& $SE(\hat\sigma_\alpha^2)$ & 0.78 & 2.28 & 0.82 & 2.48 & 0.85 & 2.66 & 0.87 & 2.75 \\
& $SE_{C}(\hat\sigma_\alpha^2)$ & 0.83 & 2.62 & 0.86 & 2.72 & 0.87 & 2.80 & 0.87 & 2.84 \\
& $SE_{N}(\hat\sigma_\alpha^2)$ & 0.89 & 3.09 & 0.91 & 3.10 & 0.92 & 3.09 & 0.92 & 3.08 \\
\midrule

\multirow{6}{*}{$100$}
& $SE(\hat\beta^{(b)}_0)$ & 0.94 & 2.27 & 0.96 & 2.30 & 0.94 & 2.34 & 0.95 & 2.35 \\
& $SE(\hat\beta^{(b)}_1)$ & 0.93 & 0.53 & 0.95 & 0.54 & 0.94 & 0.55 & 0.95 & 0.55 \\
& $SE(\hat\beta^{(w)}_0)$ & 0.94 & 0.02 & 0.96 & 0.01 & 0.96 & 0.01 & 0.95 & 0.00 \\
& $SE(\hat\sigma_\alpha^2)$ & 0.80 & 1.69 & 0.88 & 1.81 & 0.90 & 1.97 & 0.90 & 2.02 \\
& $SE_{C}(\hat\sigma_\alpha^2)$ & 0.85 & 1.92 & 0.90 & 1.97 & 0.91 & 2.06 & 0.91 & 2.08 \\
& $SE_{N}(\hat\sigma_\alpha^2)$ & 0.91 & 2.22 & 0.93 & 2.20 & 0.93 & 2.23 & 0.93 & 2.21 \\
\bottomrule
\end{tabular}
\end{table}

\begin{table}[!h]
\centering
\caption{Simulated coverage probabilities and average confidence interval lengths for the Binomial GLMM When $\alpha_i$ has a Gaussian distribution with number of trials $k =5$ and $\siga^2 = 1$.}
\small
\setlength{\tabcolsep}{4pt}
\renewcommand{\arraystretch}{1.05}

\begin{tabular}{llcccccccc}
\toprule
\multirow{2}{*}{$g$}
& \multirow{2}{*}{Estimator}
& \multicolumn{2}{c}{$m_L=10$}
& \multicolumn{2}{c}{$m_L=20$}
& \multicolumn{2}{c}{$m_L=50$}
& \multicolumn{2}{c}{$m_L=100$} \\
\cmidrule(lr){3-4}
\cmidrule(lr){5-6}
\cmidrule(lr){7-8}
\cmidrule(lr){9-10}
&
& Cvge & Len
& Cvge & Len
& Cvge & Len
& Cvge & Len \\
\midrule

\multirow{6}{*}{$10$}
& $SE(\hat\beta^{(b)}_0)$ & 0.86 & 3.62 & 0.88 & 3.62 & 0.88 & 3.63 & 0.90 & 3.70 \\
& $SE(\hat\beta^{(b)}_1)$ & 0.87 & 0.85 & 0.88 & 0.87 & 0.89 & 0.86 & 0.89 & 0.88 \\
& $SE(\hat\beta^{(w)}_0)$ & 0.95 & 0.17 & 0.95 & 0.12 & 0.95 & 0.08 & 0.95 & 0.05 \\
& $SE(\hat\sigma_\alpha^2)$ & 0.63 & 1.07 & 0.69 & 1.21 & 0.72 & 1.25 & 0.73 & 1.30 \\
& $SE_{C}(\hat\sigma_\alpha^2)$ & 0.73 & 1.30 & 0.74 & 1.33 & 0.73 & 1.31 & 0.75 & 1.32 \\
& $SE_{N}(\hat\sigma_\alpha^2)$ & 0.82 & 1.61 & 0.82 & 1.62 & 0.83 & 1.59 & 0.85 & 1.63 \\
\midrule

\multirow{6}{*}{$20$}
& $SE(\hat\beta^{(b)}_0)$ & 0.91 & 2.52 & 0.90 & 2.57 & 0.92 & 2.62 & 0.90 & 2.57 \\
& $SE(\hat\beta^{(b)}_1)$ & 0.91 & 0.59 & 0.91 & 0.61 & 0.91 & 0.62 & 0.90 & 0.61 \\
& $SE(\hat\beta^{(w)}_0)$ & 0.94 & 0.12 & 0.95 & 0.08 & 0.96 & 0.05 & 0.95 & 0.04 \\
& $SE(\hat\sigma_\alpha^2)$ & 0.77 & 0.94 & 0.81 & 0.99 & 0.83 & 1.05 & 0.84 & 1.04 \\
& $SE_{C}(\hat\sigma_\alpha^2)$ & 0.83 & 1.09 & 0.85 & 1.07 & 0.85 & 1.09 & 0.84 & 1.06 \\
& $SE_{N}(\hat\sigma_\alpha^2)$ & 0.88 & 1.21 & 0.89 & 1.19 & 0.89 & 1.21 & 0.89 & 1.17 \\
\midrule

\multirow{6}{*}{$50$}
& $SE(\hat\beta^{(b)}_0)$ & 0.93 & 1.61 & 0.93 & 1.62 & 0.94 & 1.64 & 0.94 & 1.64 \\
& $SE(\hat\beta^{(b)}_1)$ & 0.93 & 0.38 & 0.93 & 0.38 & 0.93 & 0.39 & 0.93 & 0.39 \\
& $SE(\hat\beta^{(w)}_0)$ & 0.95 & 0.07 & 0.96 & 0.05 & 0.95 & 0.03 & 0.96 & 0.02 \\
& $SE(\hat\sigma_\alpha^2)$ & 0.83 & 0.64 & 0.88 & 0.68 & 0.89 & 0.72 & 0.88 & 0.72 \\
& $SE_{C}(\hat\sigma_\alpha^2)$ & 0.88 & 0.73 & 0.90 & 0.73 & 0.90 & 0.74 & 0.89 & 0.73 \\
& $SE_{N}(\hat\sigma_\alpha^2)$ & 0.91 & 0.78 & 0.93 & 0.77 & 0.92 & 0.77 & 0.93 & 0.77 \\
\midrule

\multirow{6}{*}{$100$}
& $SE(\hat\beta^{(b)}_0)$ & 0.95 & 1.14 & 0.95 & 1.15 & 0.94 & 1.16 & 0.94 & 1.18 \\
& $SE(\hat\beta^{(b)}_1)$ & 0.94 & 0.27 & 0.94 & 0.27 & 0.95 & 0.27 & 0.95 & 0.28 \\
& $SE(\hat\beta^{(w)}_0)$ & 0.95 & 0.05 & 0.94 & 0.04 & 0.94 & 0.02 & 0.95 & 0.02 \\
& $SE(\hat\sigma_\alpha^2)$ & 0.86 & 0.47 & 0.90 & 0.50 & 0.93 & 0.52 & 0.93 & 0.53 \\
& $SE_{C}(\hat\sigma_\alpha^2)$ & 0.90 & 0.53 & 0.92 & 0.53 & 0.93 & 0.53 & 0.93 & 0.54 \\
& $SE_{N}(\hat\sigma_\alpha^2)$ & 0.93 & 0.55 & 0.93 & 0.55 & 0.95 & 0.55 & 0.94 & 0.55 \\
\bottomrule
\end{tabular}
\end{table}

\begin{table}[!h]
\centering
\caption{Simulated coverage probabilities and average confidence interval lengths for the Binomial GLMM When $\alpha_i$ has a Gaussian distribution with number of trials $k =5$ and $\siga^2 = 4$.}
\small
\setlength{\tabcolsep}{4pt}
\renewcommand{\arraystretch}{1.05}

\begin{tabular}{llcccccccc}
\toprule
\multirow{2}{*}{$g$}
& \multirow{2}{*}{Estimator}
& \multicolumn{2}{c}{$m_L=10$}
& \multicolumn{2}{c}{$m_L=20$}
& \multicolumn{2}{c}{$m_L=50$}
& \multicolumn{2}{c}{$m_L=100$} \\
\cmidrule(lr){3-4}
\cmidrule(lr){5-6}
\cmidrule(lr){7-8}
\cmidrule(lr){9-10}
&
& Cvge & Len
& Cvge & Len
& Cvge & Len
& Cvge & Len \\
\midrule

\multirow{6}{*}{$10$}
& $SE(\hat\beta^{(b)}_0)$ & 0.87 & 7.39 & 0.88 & 7.34 & 0.89 & 7.31 & 0.90 & 7.44 \\
& $SE(\hat\beta^{(b)}_1)$ & 0.87 & 1.74 & 0.89 & 1.76 & 0.89 & 1.74 & 0.89 & 1.77 \\
& $SE(\hat\beta^{(w)}_0)$ & 0.97 & 0.16 & 0.96 & 0.11 & 0.95 & 0.07 & 0.95 & 0.05 \\
& $SE(\hat\sigma_\alpha^2)$ & 0.61 & 4.01 & 0.68 & 4.55 & 0.71 & 4.85 & 0.73 & 5.13 \\
& $SE_{C}(\hat\sigma_\alpha^2)$ & 0.68 & 4.91 & 0.72 & 5.12 & 0.73 & 5.12 & 0.74 & 5.29 \\
& $SE_{N}(\hat\sigma_\alpha^2)$ & 0.81 & 6.74 & 0.83 & 6.64 & 0.83 & 6.45 & 0.85 & 6.64 \\
\midrule

\multirow{6}{*}{$20$}
& $SE(\hat\beta^{(b)}_0)$ & 0.92 & 5.07 & 0.91 & 5.18 & 0.91 & 5.25 & 0.91 & 5.15 \\
& $SE(\hat\beta^{(b)}_1)$ & 0.92 & 1.19 & 0.91 & 1.22 & 0.92 & 1.24 & 0.90 & 1.23 \\
& $SE(\hat\beta^{(w)}_0)$ & 0.94 & 0.11 & 0.96 & 0.08 & 0.96 & 0.05 & 0.95 & 0.04 \\
& $SE(\hat\sigma_\alpha^2)$ & 0.76 & 3.38 & 0.79 & 3.74 & 0.82 & 4.06 & 0.83 & 4.09 \\
& $SE_{C}(\hat\sigma_\alpha^2)$ & 0.81 & 3.96 & 0.82 & 4.10 & 0.83 & 4.24 & 0.84 & 4.19 \\
& $SE_{N}(\hat\sigma_\alpha^2)$ & 0.89 & 4.88 & 0.89 & 4.82 & 0.89 & 4.88 & 0.89 & 4.73 \\
\midrule

\multirow{6}{*}{$50$}
& $SE(\hat\beta^{(b)}_0)$ & 0.93 & 3.23 & 0.93 & 3.25 & 0.94 & 3.28 & 0.94 & 3.28 \\
& $SE(\hat\beta^{(b)}_1)$ & 0.94 & 0.76 & 0.93 & 0.77 & 0.94 & 0.78 & 0.93 & 0.78 \\
& $SE(\hat\beta^{(w)}_0)$ & 0.95 & 0.07 & 0.95 & 0.05 & 0.96 & 0.03 & 0.95 & 0.02 \\
& $SE(\hat\sigma_\alpha^2)$ & 0.81 & 2.31 & 0.85 & 2.51 & 0.88 & 2.76 & 0.88 & 2.82 \\
& $SE_{C}(\hat\sigma_\alpha^2)$ & 0.87 & 2.66 & 0.88 & 2.73 & 0.89 & 2.87 & 0.88 & 2.88 \\
& $SE_{N}(\hat\sigma_\alpha^2)$ & 0.92 & 3.14 & 0.93 & 3.10 & 0.92 & 3.11 & 0.93 & 3.09 \\
\midrule

\multirow{6}{*}{$100$}
& $SE(\hat\beta^{(b)}_0)$ & 0.95 & 2.28 & 0.95 & 2.31 & 0.94 & 2.33 & 0.95 & 2.35 \\
& $SE(\hat\beta^{(b)}_1)$ & 0.95 & 0.53 & 0.94 & 0.54 & 0.95 & 0.55 & 0.94 & 0.56 \\
& $SE(\hat\beta^{(w)}_0)$ & 0.94 & 0.05 & 0.96 & 0.03 & 0.94 & 0.02 & 0.96 & 0.02 \\
& $SE(\hat\sigma_\alpha^2)$ & 0.82 & 1.68 & 0.87 & 1.84 & 0.92 & 1.98 & 0.93 & 2.07 \\
& $SE_{C}(\hat\sigma_\alpha^2)$ & 0.86 & 1.92 & 0.90 & 1.99 & 0.93 & 2.05 & 0.93 & 2.11 \\
& $SE_{N}(\hat\sigma_\alpha^2)$ & 0.91 & 2.23 & 0.92 & 2.21 & 0.95 & 2.20 & 0.94 & 2.21 \\
\bottomrule
\end{tabular}
\end{table}

\begin{table}[!h]
\centering
\caption{Simulated coverage probabilities and average confidence interval lengths for the Gamma GLMM When $\alpha_i$ has a Gaussian distribution with $\siga^2 = 1$ and $\sige^2 = 0.5$.}
\small
\setlength{\tabcolsep}{4pt}
\renewcommand{\arraystretch}{1.05}

\begin{tabular}{llcccccccc}
\toprule
\multirow{2}{*}{$g$}
& \multirow{2}{*}{Estimator}
& \multicolumn{2}{c}{$m_L=10$}
& \multicolumn{2}{c}{$m_L=20$}
& \multicolumn{2}{c}{$m_L=50$}
& \multicolumn{2}{c}{$m_L=100$} \\
\cmidrule(lr){3-4}
\cmidrule(lr){5-6}
\cmidrule(lr){7-8}
\cmidrule(lr){9-10}
&
& Cvge & Len
& Cvge & Len
& Cvge & Len
& Cvge & Len \\
\midrule

\multirow{7}{*}{$10$}
& $SE(\hat\beta^{(b)}_0)$ & 0.87 & 3.54 & 0.87 & 3.58 & 0.89 & 3.67 & 0.88 & 3.73 \\
& $SE(\hat\beta^{(b)}_1)$ & 0.88 & 0.84 & 0.87 & 0.85 & 0.90 & 0.87 & 0.89 & 0.89 \\
& $SE(\hat\beta^{(w)}_0)$ & 0.94 & 0.09 & 0.95 & 0.07 & 0.95 & 0.04 & 0.95 & 0.03 \\
& $SE(\hat\sigma_e^2)$ & 0.96 & 0.18 & 0.94 & 0.12 & 0.95 & 0.08 & 0.96 & 0.06 \\
& $SE(\hat\sigma_\alpha^2)$ & 0.71 & 1.20 & 0.72 & 1.25 & 0.74 & 1.29 & 0.73 & 1.25 \\
& $SE_{C}(\hat\sigma_\alpha^2)$ & 0.75 & 1.29 & 0.74 & 1.30 & 0.74 & 1.31 & 0.74 & 1.26 \\
& $SE_{N}(\hat\sigma_\alpha^2)$ & 0.84 & 1.58 & 0.83 & 1.59 & 0.86 & 1.64 & 0.87 & 1.65 \\

\midrule

\multirow{7}{*}{$20$}
& $SE(\hat\beta^{(b)}_0)$ & 0.92 & 2.50 & 0.93 & 2.54 & 0.92 & 2.59 & 0.91 & 2.65 \\
& $SE(\hat\beta^{(b)}_1)$ & 0.92 & 0.59 & 0.92 & 0.60 & 0.91 & 0.61 & 0.92 & 0.63 \\
& $SE(\hat\beta^{(w)}_0)$ & 0.95 & 0.07 & 0.95 & 0.05 & 0.95 & 0.03 & 0.95 & 0.02 \\
& $SE(\hat\sigma_e^2)$ & 0.95 & 0.12 & 0.97 & 0.09 & 0.95 & 0.06 & 0.94 & 0.04 \\
& $SE(\hat\sigma_\alpha^2)$ & 0.82 & 1.01 & 0.83 & 1.03 & 0.85 & 1.05 & 0.86 & 1.06 \\
& $SE_{C}(\hat\sigma_\alpha^2)$ & 0.84 & 1.06 & 0.84 & 1.06 & 0.85 & 1.06 & 0.87 & 1.07 \\
& $SE_{N}(\hat\sigma_\alpha^2)$ & 0.89 & 1.17 & 0.90 & 1.18 & 0.91 & 1.20 & 0.92 & 1.23 \\
\midrule

\multirow{7}{*}{$50$}
& $SE(\hat\beta^{(b)}_0)$ & 0.93 & 1.59 & 0.92 & 1.63 & 0.94 & 1.64 & 0.91 & 1.66 \\
& $SE(\hat\beta^{(b)}_1)$ & 0.93 & 0.37 & 0.93 & 0.38 & 0.93 & 0.39 & 0.92 & 0.39 \\
& $SE(\hat\beta^{(w)}_0)$ & 0.95 & 0.04 & 0.96 & 0.03 & 0.94 & 0.02 & 0.96 & 0.01 \\
& $SE(\hat\sigma_e^2)$ & 0.95 & 0.08 & 0.94 & 0.06 & 0.95 & 0.04 & 0.94 & 0.03 \\
& $SE(\hat\sigma_\alpha^2)$ & 0.88 & 0.71 & 0.90 & 0.73 & 0.92 & 0.74 & 0.89 & 0.73 \\
& $SE_{C}(\hat\sigma_\alpha^2)$ & 0.89 & 0.74 & 0.91 & 0.75 & 0.92 & 0.75 & 0.89 & 0.74 \\
& $SE_{N}(\hat\sigma_\alpha^2)$ & 0.90 & 0.76 & 0.93 & 0.78 & 0.94 & 0.78 & 0.94 & 0.79 \\
\midrule

\multirow{7}{*}{$100$}
& $SE(\hat\beta^{(b)}_0)$ & 0.94 & 1.13 & 0.95 & 1.15 & 0.93 & 1.17 & 0.90 & 1.19 \\
& $SE(\hat\beta^{(b)}_1)$ & 0.94 & 0.27 & 0.94 & 0.27 & 0.93 & 0.28 & 0.92 & 0.28 \\
& $SE(\hat\beta^{(w)}_0)$ & 0.95 & 0.03 & 0.95 & 0.02 & 0.96 & 0.01 & 0.95 & 0.01 \\
& $SE(\hat\sigma_e^2)$ & 0.97 & 0.06 & 0.95 & 0.04 & 0.95 & 0.03 & 0.95 & 0.02 \\
& $SE(\hat\sigma_\alpha^2)$ & 0.91 & 0.53 & 0.94 & 0.53 & 0.93 & 0.54 & 0.90 & 0.54 \\
& $SE_{C}(\hat\sigma_\alpha^2)$ & 0.91 & 0.55 & 0.95 & 0.54 & 0.93 & 0.54 & 0.90 & 0.54 \\
& $SE_{N}(\hat\sigma_\alpha^2)$ & 0.93 & 0.55 & 0.95 & 0.55 & 0.93 & 0.56 & 0.92 & 0.57 \\
\bottomrule
\end{tabular}
\end{table}

\begin{table}[!h]
\centering
\caption{Simulated coverage probabilities and average confidence interval lengths for the Gamma GLMM When $\alpha_i$ has a Gaussian distribution with $\siga^2 = 0.5$ and $\sige^2 = 1$.}
\small
\setlength{\tabcolsep}{4pt}
\renewcommand{\arraystretch}{1.05}

\begin{tabular}{llcccccccc}
\toprule
\multirow{2}{*}{$g$}
& \multirow{2}{*}{Estimator}
& \multicolumn{2}{c}{$m_L=10$}
& \multicolumn{2}{c}{$m_L=20$}
& \multicolumn{2}{c}{$m_L=50$}
& \multicolumn{2}{c}{$m_L=100$} \\
\cmidrule(lr){3-4}
\cmidrule(lr){5-6}
\cmidrule(lr){7-8}
\cmidrule(lr){9-10}
&
& Cvge & Len
& Cvge & Len
& Cvge & Len
& Cvge & Len \\
\midrule

\multirow{7}{*}{$10$}
& $SE(\hat\beta^{(b)}_0)$ & 0.86 & 2.47 & 0.86 & 2.52 & 0.89 & 2.57 & 0.89 & 2.63 \\
& $SE(\hat\beta^{(b)}_1)$ & 0.87 & 0.59 & 0.86 & 0.60 & 0.90 & 0.61 & 0.89 & 0.63 \\
& $SE(\hat\beta^{(w)}_0)$ & 0.94 & 0.13 & 0.94 & 0.09 & 0.96 & 0.06 & 0.94 & 0.04 \\
& $SE(\hat\sigma_e^2)$ & 0.64 & 0.53 & 0.67 & 0.58 & 0.71 & 0.62 & 0.72 & 0.62 \\
& $SE(\hat\sigma_\alpha^2)$ & 0.72 & 0.65 & 0.74 & 0.66 & 0.74 & 0.66 & 0.74 & 0.63 \\
& $SE_{C}(\hat\sigma_\alpha^2)$ & 0.80 & 0.77 & 0.82 & 0.79 & 0.86 & 0.81 & 0.87 & 0.81 \\
& $SE_{N}(\hat\sigma_\alpha^2)$ & 0.96 & 0.33 & 0.96 & 0.24 & 0.95 & 0.15 & 0.96 & 0.11 \\
\midrule

\multirow{7}{*}{$20$}
& $SE(\hat\beta^{(b)}_0)$ & 0.91 & 1.75 & 0.92 & 1.79 & 0.93 & 1.82 & 0.91 & 1.85 \\
& $SE(\hat\beta^{(b)}_1)$ & 0.90 & 0.41 & 0.91 & 0.42 & 0.92 & 0.43 & 0.92 & 0.44 \\
& $SE(\hat\beta^{(w)}_0)$ & 0.95 & 0.09 & 0.93 & 0.07 & 0.95 & 0.04 & 0.94 & 0.03 \\
& $SE(\hat\sigma_e^2)$ & 0.95 & 0.23 & 0.96 & 0.17 & 0.95 & 0.11 & 0.94 & 0.08 \\
& $SE(\hat\sigma_\alpha^2)$ & 0.74 & 0.44 & 0.79 & 0.48 & 0.83 & 0.51 & 0.85 & 0.52 \\
& $SE_{C}(\hat\sigma_\alpha^2)$ & 0.83 & 0.53 & 0.84 & 0.53 & 0.86 & 0.53 & 0.87 & 0.54 \\
& $SE_{N}(\hat\sigma_\alpha^2)$ & 0.87 & 0.58 & 0.88 & 0.58 & 0.91 & 0.59 & 0.92 & 0.60 \\
\midrule

\multirow{7}{*}{$50$}
& $SE(\hat\beta^{(b)}_0)$ & 0.93 & 1.13 & 0.92 & 1.15 & 0.94 & 1.16 & 0.93 & 1.17 \\
& $SE(\hat\beta^{(b)}_1)$ & 0.93 & 0.26 & 0.93 & 0.27 & 0.94 & 0.27 & 0.94 & 0.28 \\
& $SE(\hat\beta^{(w)}_0)$ & 0.95 & 0.06 & 0.94 & 0.04 & 0.95 & 0.03 & 0.96 & 0.02 \\
& $SE(\hat\sigma_e^2)$ & 0.95 & 0.15 & 0.96 & 0.11 & 0.96 & 0.07 & 0.94 & 0.05 \\
& $SE(\hat\sigma_\alpha^2)$ & 0.81 & 0.32 & 0.87 & 0.35 & 0.90 & 0.36 & 0.90 & 0.36 \\
& $SE_{C}(\hat\sigma_\alpha^2)$ & 0.87 & 0.37 & 0.90 & 0.37 & 0.91 & 0.38 & 0.91 & 0.37 \\
& $SE_{N}(\hat\sigma_\alpha^2)$ & 0.89 & 0.38 & 0.92 & 0.39 & 0.93 & 0.39 & 0.94 & 0.39 \\
\midrule

\multirow{7}{*}{$100$}
& $SE(\hat\beta^{(b)}_0)$ & 0.94 & 0.80 & 0.94 & 0.82 & 0.94 & 0.83 & 0.95 & 0.83 \\
& $SE(\hat\beta^{(b)}_1)$ & 0.92 & 0.19 & 0.94 & 0.19 & 0.95 & 0.19 & 0.95 & 0.20 \\
& $SE(\hat\beta^{(w)}_0)$ & 0.93 & 0.04 & 0.96 & 0.03 & 0.96 & 0.02 & 0.95 & 0.01 \\
& $SE(\hat\sigma_e^2)$ & 0.95 & 0.10 & 0.94 & 0.07 & 0.96 & 0.05 & 0.95 & 0.03 \\
& $SE(\hat\sigma_\alpha^2)$ & 0.85 & 0.24 & 0.91 & 0.25 & 0.92 & 0.26 & 0.91 & 0.26 \\
& $SE_{C}(\hat\sigma_\alpha^2)$ & 0.91 & 0.27 & 0.93 & 0.27 & 0.92 & 0.27 & 0.92 & 0.27 \\
& $SE_{N}(\hat\sigma_\alpha^2)$ & 0.91 & 0.28 & 0.93 & 0.28 & 0.93 & 0.28 & 0.93 & 0.28 \\
\bottomrule
\end{tabular}
\end{table}

\begin{table}[!h]
\centering
\caption{Simulated coverage probabilities and average confidence interval lengths for the Gamma GLMM When $\alpha_i$ has a Gaussian distribution with $\siga^2 = 1$ and $\sige^2 = 1$.}
\small
\setlength{\tabcolsep}{4pt}
\renewcommand{\arraystretch}{1.05}

\begin{tabular}{llcccccccc}
\toprule
\multirow{2}{*}{$g$}
& \multirow{2}{*}{Estimator}
& \multicolumn{2}{c}{$m_L=10$}
& \multicolumn{2}{c}{$m_L=20$}
& \multicolumn{2}{c}{$m_L=50$}
& \multicolumn{2}{c}{$m_L=100$} \\
\cmidrule(lr){3-4}
\cmidrule(lr){5-6}
\cmidrule(lr){7-8}
\cmidrule(lr){9-10}
&
& Cvge & Len
& Cvge & Len
& Cvge & Len
& Cvge & Len \\
\midrule

\multirow{7}{*}{$10$}
& $SE(\hat\beta^{(b)}_0)$ & 0.87 & 3.51 & 0.86 & 3.57 & 0.89 & 3.64 & 0.88 & 3.72 \\
& $SE(\hat\beta^{(b)}_1)$ & 0.88 & 0.83 & 0.87 & 0.85 & 0.90 & 0.87 & 0.89 & 0.89 \\
& $SE(\hat\beta^{(w)}_0)$ & 0.94 & 0.13 & 0.94 & 0.09 & 0.96 & 0.06 & 0.94 & 0.04 \\
& $SE(\hat\sigma_e^2)$ & 0.96 & 0.33 & 0.96 & 0.24 & 0.95 & 0.15 & 0.96 & 0.11 \\
& $SE(\hat\sigma_\alpha^2)$ & 0.67 & 1.15 & 0.70 & 1.22 & 0.72 & 1.28 & 0.72 & 1.24 \\
& $SE_{C}(\hat\sigma_\alpha^2)$ & 0.73 & 1.30 & 0.74 & 1.31 & 0.74 & 1.31 & 0.73 & 1.26 \\
& $SE_{N}(\hat\sigma_\alpha^2)$ & 0.82 & 1.56 & 0.82 & 1.59 & 0.86 & 1.62 & 0.87 & 1.64 \\
\midrule

\multirow{7}{*}{$20$}
& $SE(\hat\beta^{(b)}_0)$ & 0.91 & 2.49 & 0.93 & 2.53 & 0.93 & 2.59 & 0.92 & 2.63 \\
& $SE(\hat\beta^{(b)}_1)$ & 0.91 & 0.59 & 0.92 & 0.60 & 0.93 & 0.61 & 0.92 & 0.63 \\
& $SE(\hat\beta^{(w)}_0)$ & 0.95 & 0.09 & 0.93 & 0.07 & 0.95 & 0.04 & 0.94 & 0.03 \\
& $SE(\hat\sigma_e^2)$ & 0.95 & 0.23 & 0.96 & 0.17 & 0.95 & 0.11 & 0.94 & 0.08 \\
& $SE(\hat\sigma_\alpha^2)$ & 0.79 & 0.96 & 0.82 & 1.00 & 0.85 & 1.04 & 0.86 & 1.05 \\
& $SE_{C}(\hat\sigma_\alpha^2)$ & 0.84 & 1.07 & 0.85 & 1.06 & 0.86 & 1.06 & 0.86 & 1.06 \\
& $SE_{N}(\hat\sigma_\alpha^2)$ & 0.89 & 1.16 & 0.89 & 1.17 & 0.90 & 1.19 & 0.91 & 1.21 \\
\midrule

\multirow{7}{*}{$50$}
& $SE(\hat\beta^{(b)}_0)$ & 0.93 & 1.59 & 0.92 & 1.63 & 0.95 & 1.64 & 0.92 & 1.65 \\
& $SE(\hat\beta^{(b)}_1)$ & 0.94 & 0.37 & 0.93 & 0.38 & 0.94 & 0.39 & 0.92 & 0.39 \\
& $SE(\hat\beta^{(w)}_0)$ & 0.95 & 0.06 & 0.94 & 0.04 & 0.95 & 0.03 & 0.96 & 0.02 \\
& $SE(\hat\sigma_e^2)$ & 0.95 & 0.15 & 0.96 & 0.11 & 0.96 & 0.07 & 0.94 & 0.05 \\
& $SE(\hat\sigma_\alpha^2)$ & 0.86 & 0.68 & 0.90 & 0.72 & 0.92 & 0.74 & 0.90 & 0.73 \\
& $SE_{C}(\hat\sigma_\alpha^2)$ & 0.89 & 0.74 & 0.91 & 0.75 & 0.92 & 0.75 & 0.90 & 0.73 \\
& $SE_{N}(\hat\sigma_\alpha^2)$ & 0.90 & 0.77 & 0.93 & 0.78 & 0.94 & 0.78 & 0.94 & 0.78 \\
\midrule

\multirow{7}{*}{$100$}
& $SE(\hat\beta^{(b)}_0)$ & 0.94 & 1.13 & 0.95 & 1.15 & 0.94 & 1.17 & 0.93 & 1.17 \\
& $SE(\hat\beta^{(b)}_1)$ & 0.93 & 0.27 & 0.94 & 0.27 & 0.95 & 0.28 & 0.93 & 0.28 \\
& $SE(\hat\beta^{(w)}_0)$ & 0.93 & 0.04 & 0.96 & 0.03 & 0.96 & 0.02 & 0.95 & 0.01 \\
& $SE(\hat\sigma_e^2)$ & 0.95 & 0.11 & 0.94 & 0.07 & 0.96 & 0.05 & 0.96 & 0.03 \\
& $SE(\hat\sigma_\alpha^2)$ & 0.90 & 0.51 & 0.93 & 0.52 & 0.92 & 0.54 & 0.91 & 0.53 \\
& $SE_{C}(\hat\sigma_\alpha^2)$ & 0.92 & 0.55 & 0.94 & 0.54 & 0.93 & 0.54 & 0.91 & 0.54 \\
& $SE_{N}(\hat\sigma_\alpha^2)$ & 0.92 & 0.55 & 0.94 & 0.55 & 0.94 & 0.56 & 0.93 & 0.55 \\
\bottomrule
\end{tabular}
\end{table}

\FloatBarrier

\subsection{Simulation for Mixture Random Effects}
\label{subsec:Simulation for Mixture Random Effects}

\begin{table}[!h]
\centering
\caption{Simulated coverage probabilities and average confidence interval lengths for the Poisson GLMM When $\alpha_i$ has a mixture distribution with $\siga^2 = 1$.}
\small
\setlength{\tabcolsep}{4pt}
\renewcommand{\arraystretch}{1.05}

\begin{tabular}{llcccccccc}
\toprule
\multirow{2}{*}{$g$}
& \multirow{2}{*}{Estimator}
& \multicolumn{2}{c}{$m_L=10$}
& \multicolumn{2}{c}{$m_L=20$}
& \multicolumn{2}{c}{$m_L=50$}
& \multicolumn{2}{c}{$m_L=100$} \\
\cmidrule(lr){3-4}
\cmidrule(lr){5-6}
\cmidrule(lr){7-8}
\cmidrule(lr){9-10}
&
& Cvge & Len
& Cvge & Len
& Cvge & Len
& Cvge & Len \\
\midrule

\multirow{6}{*}{$10$}
& SE$(\hat\beta^{(b)}_0)$ & 0.87 & 3.55 & 0.87 & 3.50 & 0.88 & 3.60 & 0.88 & 3.64 \\
& SE$(\hat\beta^{(b)}_1)$ & 0.87 & 0.84 & 0.86 & 0.83 & 0.88 & 0.86 & 0.87 & 0.87 \\
& SE$(\hat\beta^{(w)}_0)$ & 0.94 & 0.10 & 0.94 & 0.07 & 0.95 & 0.05 & 0.94 & 0.03 \\
& SE$(\hat\sigma_\alpha^2)$ & 0.61 & 1.00 & 0.62 & 1.06 & 0.71 & 1.19 & 0.73 & 1.23 \\
& SE$_{C}(\hat\sigma_\alpha^2)$ & 0.71 & 1.27 & 0.70 & 1.24 & 0.75 & 1.28 & 0.74 & 1.27 \\
& SE$_{N}(\hat\sigma_\alpha^2)$ & 0.81 & 1.59 & 0.80 & 1.55 & 0.82 & 1.60 & 0.83 & 1.58 \\
\midrule

\multirow{6}{*}{$20$}
& SE$(\hat\beta^{(b)}_0)$ & 0.89 & 2.49 & 0.91 & 2.54 & 0.92 & 2.56 & 0.91 & 2.58 \\
& SE$(\hat\beta^{(b)}_1)$ & 0.90 & 0.59 & 0.91 & 0.60 & 0.92 & 0.61 & 0.92 & 0.62 \\
& SE$(\hat\beta^{(w)}_0)$ & 0.96 & 0.07 & 0.96 & 0.05 & 0.94 & 0.03 & 0.95 & 0.02 \\
& SE$(\hat\sigma_\alpha^2)$ & 0.70 & 0.80 & 0.76 & 0.93 & 0.82 & 1.00 & 0.83 & 1.01 \\
& SE$_{C}(\hat\sigma_\alpha^2)$ & 0.80 & 0.99 & 0.81 & 1.04 & 0.84 & 1.05 & 0.84 & 1.04 \\
& SE$_{N}(\hat\sigma_\alpha^2)$ & 0.85 & 1.16 & 0.86 & 1.19 & 0.89 & 1.18 & 0.89 & 1.18 \\
\midrule

\multirow{6}{*}{$50$}
& SE$(\hat\beta^{(b)}_0)$ & 0.92 & 1.57 & 0.94 & 1.61 & 0.95 & 1.63 & 0.93 & 1.64 \\
& SE$(\hat\beta^{(b)}_1)$ & 0.93 & 0.37 & 0.93 & 0.38 & 0.94 & 0.39 & 0.93 & 0.39 \\
& SE$(\hat\beta^{(w)}_0)$ & 0.94 & 0.04 & 0.94 & 0.03 & 0.93 & 0.02 & 0.95 & 0.01 \\
& SE$(\hat\sigma_\alpha^2)$ & 0.74 & 0.56 & 0.83 & 0.63 & 0.87 & 0.69 & 0.89 & 0.71 \\
& SE$_{C}(\hat\sigma_\alpha^2)$ & 0.83 & 0.67 & 0.87 & 0.70 & 0.88 & 0.73 & 0.90 & 0.73 \\
& SE$_{N}(\hat\sigma_\alpha^2)$ & 0.86 & 0.75 & 0.90 & 0.76 & 0.91 & 0.77 & 0.92 & 0.77 \\
\midrule

\multirow{6}{*}{$100$}
& SE$(\hat\beta^{(b)}_0)$ & 0.92 & 1.11 & 0.94 & 1.14 & 0.94 & 1.16 & 0.95 & 1.17 \\
& SE$(\hat\beta^{(b)}_1)$ & 0.92 & 0.26 & 0.95 & 0.27 & 0.95 & 0.27 & 0.95 & 0.28 \\
& SE$(\hat\beta^{(w)}_0)$ & 0.94 & 0.03 & 0.94 & 0.02 & 0.95 & 0.01 & 0.95 & 0.01 \\
& SE$(\hat\sigma_\alpha^2)$ & 0.80 & 0.41 & 0.87 & 0.46 & 0.90 & 0.50 & 0.91 & 0.52 \\
& SE$_{C}(\hat\sigma_\alpha^2)$ & 0.87 & 0.49 & 0.90 & 0.51 & 0.91 & 0.52 & 0.92 & 0.53 \\
& SE$_{N}(\hat\sigma_\alpha^2)$ & 0.91 & 0.53 & 0.92 & 0.54 & 0.94 & 0.55 & 0.94 & 0.55 \\
\bottomrule
\end{tabular}
\end{table}

\begin{table}[!h]
\centering
\caption{Simulated coverage probabilities and average confidence interval lengths for the Poisson GLMM When $\alpha_i$ has a mixture distribution with $\siga^2 = 4$.}
\small
\setlength{\tabcolsep}{4pt}
\renewcommand{\arraystretch}{1.05}

\begin{tabular}{llcccccccc}
\toprule
\multirow{2}{*}{$g$}
& \multirow{2}{*}{Estimator}
& \multicolumn{2}{c}{$m_L=10$}
& \multicolumn{2}{c}{$m_L=20$}
& \multicolumn{2}{c}{$m_L=50$}
& \multicolumn{2}{c}{$m_L=100$} \\
\cmidrule(lr){3-4}
\cmidrule(lr){5-6}
\cmidrule(lr){7-8}
\cmidrule(lr){9-10}
&
& Cvge & Len
& Cvge & Len
& Cvge & Len
& Cvge & Len \\
\midrule

\multirow{6}{*}{$10$}
& SE$(\hat\beta^{(b)}_0)$ & 0.88 & 6.89 & 0.88 & 6.90 & 0.88 & 7.09 & 0.88 & 7.22 \\
& SE$(\hat\beta^{(b)}_1)$ & 0.87 & 1.64 & 0.87 & 1.65 & 0.88 & 1.69 & 0.87 & 1.73 \\
& SE$(\hat\beta^{(w)}_0)$ & 0.95 & 0.07 & 0.95 & 0.05 & 0.95 & 0.03 & 0.94 & 0.02 \\
& SE$(\hat\sigma_\alpha^2)$ & 0.53 & 3.82 & 0.58 & 4.18 & 0.66 & 4.68 & 0.69 & 5.11 \\
& SE$_{C}(\hat\sigma_\alpha^2)$ & 0.62 & 4.73 & 0.65 & 4.93 & 0.69 & 5.18 & 0.71 & 5.43 \\
& SE$_{N}(\hat\sigma_\alpha^2)$ & 0.73 & 6.16 & 0.75 & 6.16 & 0.76 & 6.35 & 0.78 & 6.38 \\
\midrule

\multirow{6}{*}{$20$}
& SE$(\hat\beta^{(b)}_0)$ & 0.90 & 4.80 & 0.92 & 4.97 & 0.92 & 5.04 & 0.92 & 5.11 \\
& SE$(\hat\beta^{(b)}_1)$ & 0.91 & 1.13 & 0.91 & 1.17 & 0.92 & 1.19 & 0.92 & 1.22 \\
& SE$(\hat\beta^{(w)}_0)$ & 0.94 & 0.04 & 0.95 & 0.03 & 0.95 & 0.02 & 0.93 & 0.01 \\
& SE$(\hat\sigma_\alpha^2)$ & 0.62 & 3.26 & 0.71 & 3.67 & 0.75 & 4.11 & 0.80 & 4.30 \\
& SE$_{C}(\hat\sigma_\alpha^2)$ & 0.70 & 3.85 & 0.77 & 4.15 & 0.78 & 4.42 & 0.81 & 4.51 \\
& SE$_{N}(\hat\sigma_\alpha^2)$ & 0.78 & 4.38 & 0.82 & 4.60 & 0.82 & 4.62 & 0.83 & 4.66 \\
\midrule

\multirow{6}{*}{$50$}
& SE$(\hat\beta^{(b)}_0)$ & 0.93 & 3.05 & 0.94 & 3.15 & 0.95 & 3.23 & 0.94 & 3.24 \\
& SE$(\hat\beta^{(b)}_1)$ & 0.94 & 0.72 & 0.93 & 0.74 & 0.94 & 0.76 & 0.94 & 0.77 \\
& SE$(\hat\beta^{(w)}_0)$ & 0.94 & 0.02 & 0.94 & 0.02 & 0.96 & 0.01 & 0.95 & 0.01 \\
& SE$(\hat\sigma_\alpha^2)$ & 0.70 & 2.35 & 0.77 & 2.62 & 0.82 & 2.91 & 0.86 & 3.09 \\
& SE$_{C}(\hat\sigma_\alpha^2)$ & 0.76 & 2.68 & 0.81 & 2.88 & 0.85 & 3.08 & 0.87 & 3.21 \\
& SE$_{N}(\hat\sigma_\alpha^2)$ & 0.79 & 2.82 & 0.83 & 2.92 & 0.84 & 3.01 & 0.86 & 3.02 \\
\midrule

\multirow{6}{*}{$100$}
& SE$(\hat\beta^{(b)}_0)$ & 0.94 & 2.16 & 0.95 & 2.24 & 0.94 & 2.29 & 0.94 & 2.32 \\
& SE$(\hat\beta^{(b)}_1)$ & 0.93 & 0.51 & 0.95 & 0.53 & 0.94 & 0.54 & 0.94 & 0.55 \\
& SE$(\hat\beta^{(w)}_0)$ & 0.95 & 0.02 & 0.95 & 0.01 & 0.94 & 0.01 & 0.94 & 0.00 \\
& SE$(\hat\sigma_\alpha^2)$ & 0.71 & 1.77 & 0.81 & 1.94 & 0.86 & 2.13 & 0.88 & 2.25 \\
& SE$_{C}(\hat\sigma_\alpha^2)$ & 0.77 & 1.99 & 0.84 & 2.11 & 0.88 & 2.24 & 0.90 & 2.33 \\
& SE$_{N}(\hat\sigma_\alpha^2)$ & 0.79 & 2.01 & 0.83 & 2.08 & 0.87 & 2.13 & 0.87 & 2.17 \\
\bottomrule
\end{tabular}
\end{table}

\begin{table}[!h]
\centering
\caption{Simulated coverage probabilities and average confidence interval lengths for the Binomial GLMM when $\alpha_i$ has a mixture distribution with number of trials $k=5$ and $\siga^2 = 1$.}
\small
\setlength{\tabcolsep}{4pt}
\renewcommand{\arraystretch}{1.05}

\begin{tabular}{llcccccccc}
\toprule
\multirow{2}{*}{$g$}
& \multirow{2}{*}{Estimator}
& \multicolumn{2}{c}{$m_L=10$}
& \multicolumn{2}{c}{$m_L=20$}
& \multicolumn{2}{c}{$m_L=50$}
& \multicolumn{2}{c}{$m_L=100$} \\
\cmidrule(lr){3-4}
\cmidrule(lr){5-6}
\cmidrule(lr){7-8}
\cmidrule(lr){9-10}
&
& Cvge & Len
& Cvge & Len
& Cvge & Len
& Cvge & Len \\
\midrule

\multirow{6}{*}{$10$}
& SE$(\hat\beta^{(b)}_0)$ & 0.89 & 3.59 & 0.87 & 3.61 & 0.88 & 3.69 & 0.91 & 3.63 \\
& SE$(\hat\beta^{(b)}_1)$ & 0.87 & 0.85 & 0.86 & 0.86 & 0.89 & 0.88 & 0.89 & 0.87 \\
& SE$(\hat\beta^{(w)}_0)$ & 0.95 & 0.13 & 0.96 & 0.09 & 0.95 & 0.06 & 0.93 & 0.04 \\
& SE$(\hat\sigma_\alpha^2)$ & 0.62 & 1.06 & 0.70 & 1.17 & 0.74 & 1.28 & 0.73 & 1.26 \\
& SE$_{C}(\hat\sigma_\alpha^2)$ & 0.73 & 1.30 & 0.75 & 1.30 & 0.75 & 1.33 & 0.74 & 1.29 \\
& SE$_{N}(\hat\sigma_\alpha^2)$ & 0.82 & 1.60 & 0.83 & 1.59 & 0.85 & 1.64 & 0.83 & 1.59 \\
\midrule

\multirow{6}{*}{$20$}
& SE$(\hat\beta^{(b)}_0)$ & 0.90 & 2.55 & 0.91 & 2.58 & 0.90 & 2.60 & 0.91 & 2.59 \\
& SE$(\hat\beta^{(b)}_1)$ & 0.90 & 0.60 & 0.91 & 0.61 & 0.91 & 0.62 & 0.92 & 0.62 \\
& SE$(\hat\beta^{(w)}_0)$ & 0.94 & 0.09 & 0.96 & 0.07 & 0.95 & 0.04 & 0.95 & 0.03 \\
& SE$(\hat\sigma_\alpha^2)$ & 0.77 & 0.94 & 0.79 & 0.99 & 0.81 & 1.03 & 0.85 & 1.06 \\
& SE$_{C}(\hat\sigma_\alpha^2)$ & 0.84 & 1.09 & 0.83 & 1.07 & 0.83 & 1.07 & 0.85 & 1.07 \\
& SE$_{N}(\hat\sigma_\alpha^2)$ & 0.89 & 1.23 & 0.88 & 1.20 & 0.90 & 1.20 & 0.91 & 1.19 \\
\midrule

\multirow{6}{*}{$50$}
& SE$(\hat\beta^{(b)}_0)$ & 0.94 & 1.60 & 0.94 & 1.62 & 0.95 & 1.64 & 0.93 & 1.64 \\
& SE$(\hat\beta^{(b)}_1)$ & 0.93 & 0.38 & 0.94 & 0.38 & 0.94 & 0.39 & 0.93 & 0.39 \\
& SE$(\hat\beta^{(w)}_0)$ & 0.95 & 0.06 & 0.95 & 0.04 & 0.95 & 0.03 & 0.94 & 0.02 \\
& SE$(\hat\sigma_\alpha^2)$ & 0.80 & 0.63 & 0.87 & 0.68 & 0.88 & 0.72 & 0.91 & 0.73 \\
& SE$_{C}(\hat\sigma_\alpha^2)$ & 0.86 & 0.72 & 0.89 & 0.73 & 0.90 & 0.74 & 0.91 & 0.74 \\
& SE$_{N}(\hat\sigma_\alpha^2)$ & 0.90 & 0.78 & 0.91 & 0.77 & 0.92 & 0.77 & 0.92 & 0.77 \\
\midrule

\multirow{6}{*}{$100$}
& SE$(\hat\beta^{(b)}_0)$ & 0.93 & 1.13 & 0.95 & 1.15 & 0.94 & 1.16 & 0.94 & 1.17 \\
& SE$(\hat\beta^{(b)}_1)$ & 0.93 & 0.27 & 0.95 & 0.27 & 0.95 & 0.27 & 0.94 & 0.28 \\
& SE$(\hat\beta^{(w)}_0)$ & 0.96 & 0.04 & 0.96 & 0.03 & 0.96 & 0.02 & 0.94 & 0.01 \\
& SE$(\hat\sigma_\alpha^2)$ & 0.86 & 0.46 & 0.90 & 0.50 & 0.92 & 0.52 & 0.93 & 0.53 \\
& SE$_{C}(\hat\sigma_\alpha^2)$ & 0.90 & 0.52 & 0.92 & 0.53 & 0.92 & 0.53 & 0.93 & 0.53 \\
& SE$_{N}(\hat\sigma_\alpha^2)$ & 0.92 & 0.55 & 0.93 & 0.55 & 0.93 & 0.55 & 0.94 & 0.55 \\
\bottomrule
\end{tabular}
\end{table}

\begin{table}[!h]
\centering
\caption{Simulated coverage probabilities and average confidence interval lengths for the Binomial GLMM when $\alpha_i$ has a mixture distribution with number of trials $k=5$ and $\siga^2 = 4$.}
\small
\setlength{\tabcolsep}{4pt}
\renewcommand{\arraystretch}{1.05}

\begin{tabular}{llcccccccc}
\toprule
\multirow{2}{*}{$g$}
& \multirow{2}{*}{Estimator}
& \multicolumn{2}{c}{$m_L=10$}
& \multicolumn{2}{c}{$m_L=20$}
& \multicolumn{2}{c}{$m_L=50$}
& \multicolumn{2}{c}{$m_L=100$} \\
\cmidrule(lr){3-4}
\cmidrule(lr){5-6}
\cmidrule(lr){7-8}
\cmidrule(lr){9-10}
&
& Cvge & Len
& Cvge & Len
& Cvge & Len
& Cvge & Len \\
\midrule

\multirow{6}{*}{$10$}
& SE$(\hat\beta^{(b)}_0)$ & 0.89 & 7.08 & 0.88 & 7.18 & 0.89 & 7.32 & 0.90 & 7.24 \\
& SE$(\hat\beta^{(b)}_1)$ & 0.88 & 1.67 & 0.87 & 1.72 & 0.89 & 1.74 & 0.89 & 1.73 \\
& SE$(\hat\beta^{(w)}_0)$ & 0.95 & 0.15 & 0.95 & 0.11 & 0.95 & 0.07 & 0.95 & 0.05 \\
& SE$(\hat\sigma_\alpha^2)$ & 0.59 & 4.00 & 0.65 & 4.66 & 0.71 & 5.21 & 0.69 & 5.46 \\
& SE$_{C}(\hat\sigma_\alpha^2)$ & 0.66 & 4.98 & 0.69 & 5.32 & 0.73 & 5.60 & 0.70 & 5.68 \\
& SE$_{N}(\hat\sigma_\alpha^2)$ & 0.77 & 6.38 & 0.77 & 6.43 & 0.79 & 6.59 & 0.78 & 6.50 \\
\midrule

\multirow{6}{*}{$20$}
& SE$(\hat\beta^{(b)}_0)$ & 0.90 & 4.97 & 0.91 & 5.10 & 0.90 & 5.13 & 0.92 & 5.14 \\
& SE$(\hat\beta^{(b)}_1)$ & 0.91 & 1.17 & 0.90 & 1.21 & 0.91 & 1.22 & 0.92 & 1.22 \\
& SE$(\hat\beta^{(w)}_0)$ & 0.95 & 0.11 & 0.96 & 0.08 & 0.94 & 0.05 & 0.95 & 0.03 \\
& SE$(\hat\sigma_\alpha^2)$ & 0.71 & 3.52 & 0.75 & 3.97 & 0.81 & 4.34 & 0.82 & 4.57 \\
& SE$_{C}(\hat\sigma_\alpha^2)$ & 0.79 & 4.16 & 0.79 & 4.40 & 0.82 & 4.57 & 0.83 & 4.71 \\
& SE$_{N}(\hat\sigma_\alpha^2)$ & 0.83 & 4.74 & 0.82 & 4.75 & 0.84 & 4.71 & 0.85 & 4.73 \\
\midrule

\multirow{6}{*}{$50$}
& SE$(\hat\beta^{(b)}_0)$ & 0.94 & 3.12 & 0.93 & 3.19 & 0.94 & 3.26 & 0.93 & 3.26 \\
& SE$(\hat\beta^{(b)}_1)$ & 0.93 & 0.73 & 0.93 & 0.75 & 0.94 & 0.77 & 0.94 & 0.77 \\
& SE$(\hat\beta^{(w)}_0)$ & 0.96 & 0.07 & 0.95 & 0.05 & 0.95 & 0.03 & 0.94 & 0.02 \\
& SE$(\hat\sigma_\alpha^2)$ & 0.74 & 2.40 & 0.83 & 2.72 & 0.87 & 3.06 & 0.87 & 3.19 \\
& SE$_{C}(\hat\sigma_\alpha^2)$ & 0.81 & 2.76 & 0.86 & 2.95 & 0.89 & 3.19 & 0.87 & 3.28 \\
& SE$_{N}(\hat\sigma_\alpha^2)$ & 0.83 & 2.96 & 0.85 & 2.99 & 0.87 & 3.07 & 0.86 & 3.06 \\
\midrule

\multirow{6}{*}{$100$}
& SE$(\hat\beta^{(b)}_0)$ & 0.94 & 2.19 & 0.94 & 2.27 & 0.95 & 2.30 & 0.92 & 2.32 \\
& SE$(\hat\beta^{(b)}_1)$ & 0.94 & 0.52 & 0.94 & 0.53 & 0.96 & 0.54 & 0.93 & 0.55 \\
& SE$(\hat\beta^{(w)}_0)$ & 0.96 & 0.05 & 0.96 & 0.03 & 0.97 & 0.02 & 0.95 & 0.02 \\
& SE$(\hat\sigma_\alpha^2)$ & 0.77 & 1.75 & 0.83 & 1.98 & 0.87 & 2.20 & 0.89 & 2.31 \\
& SE$_{C}(\hat\sigma_\alpha^2)$ & 0.82 & 1.99 & 0.86 & 2.15 & 0.88 & 2.29 & 0.90 & 2.37 \\
& SE$_{N}(\hat\sigma_\alpha^2)$ & 0.83 & 2.07 & 0.85 & 2.13 & 0.86 & 2.15 & 0.88 & 2.16 \\
\bottomrule
\end{tabular}
\end{table}

\begin{table}[!h]
\centering
\caption{Simulated coverage probabilities and average confidence interval lengths for the Gamma GLMM when $\alpha_i$ has a mixture distribution with $\siga^2 = 0.5$ and $\sige^2 = 0.5$.}
\label{tab:coverage_gamma}
\small
\setlength{\tabcolsep}{4pt}
\renewcommand{\arraystretch}{1.05}

\begin{tabular}{llcccccccc}
\toprule
\multirow{2}{*}{$g$}
& \multirow{2}{*}{Estimator}
& \multicolumn{2}{c}{$m_L=10$}
& \multicolumn{2}{c}{$m_L=20$}
& \multicolumn{2}{c}{$m_L=50$}
& \multicolumn{2}{c}{$m_L=100$} \\
\cmidrule(lr){3-4}
\cmidrule(lr){5-6}
\cmidrule(lr){7-8}
\cmidrule(lr){9-10}
&
& Cvge & Len
& Cvge & Len
& Cvge & Len
& Cvge & Len \\
\midrule

\multirow{7}{*}{$10$}
& SE$(\hat\beta^{(b)}_0)$ & 0.89 & 2.50 & 0.88 & 2.50 & 0.88 & 2.57 & 0.89 & 2.63 \\
& SE$(\hat\beta^{(b)}_1)$ & 0.87 & 0.59 & 0.88 & 0.59 & 0.88 & 0.61 & 0.88 & 0.63 \\
& SE$(\hat\beta^{(w)}_0)$ & 0.95 & 0.09 & 0.95 & 0.07 & 0.96 & 0.04 & 0.95 & 0.03 \\
& SE$(\hat\sigma_e^2)$ & 0.95 & 0.18 & 0.94 & 0.12 & 0.95 & 0.08 & 0.96 & 0.06 \\
& SE$(\hat\sigma_\alpha^2)$ & 0.68 & 0.59 & 0.68 & 0.60 & 0.72 & 0.64 & 0.72 & 0.62 \\
& SE$_{C}(\hat\sigma_\alpha^2)$ & 0.74 & 0.67 & 0.72 & 0.65 & 0.74 & 0.65 & 0.73 & 0.63 \\
& SE$_{N}(\hat\sigma_\alpha^2)$ & 0.83 & 0.79 & 0.81 & 0.78 & 0.85 & 0.81 & 0.86 & 0.82 \\
\midrule

\multirow{7}{*}{$20$}
& SE$(\hat\beta^{(b)}_0)$ & 0.90 & 1.77 & 0.92 & 1.80 & 0.92 & 1.82 & 0.91 & 1.85 \\
& SE$(\hat\beta^{(b)}_1)$ & 0.90 & 0.42 & 0.91 & 0.43 & 0.92 & 0.43 & 0.91 & 0.44 \\
& SE$(\hat\beta^{(w)}_0)$ & 0.95 & 0.07 & 0.95 & 0.05 & 0.95 & 0.03 & 0.94 & 0.02 \\
& SE$(\hat\sigma_e^2)$ & 0.96 & 0.12 & 0.97 & 0.09 & 0.95 & 0.06 & 0.95 & 0.04 \\
& SE$(\hat\sigma_\alpha^2)$ & 0.78 & 0.50 & 0.81 & 0.53 & 0.84 & 0.54 & 0.86 & 0.53 \\
& SE$_{C}(\hat\sigma_\alpha^2)$ & 0.83 & 0.55 & 0.84 & 0.56 & 0.85 & 0.55 & 0.86 & 0.54 \\
& SE$_{N}(\hat\sigma_\alpha^2)$ & 0.87 & 0.59 & 0.89 & 0.59 & 0.90 & 0.59 & 0.91 & 0.60 \\
\midrule

\multirow{7}{*}{$50$}
& SE$(\hat\beta^{(b)}_0)$ & 0.93 & 1.12 & 0.94 & 1.15 & 0.95 & 1.16 & 0.93 & 1.17 \\
& SE$(\hat\beta^{(b)}_1)$ & 0.93 & 0.26 & 0.95 & 0.27 & 0.94 & 0.27 & 0.93 & 0.28 \\
& SE$(\hat\beta^{(w)}_0)$ & 0.95 & 0.04 & 0.95 & 0.03 & 0.95 & 0.02 & 0.96 & 0.01 \\
& SE$(\hat\sigma_e^2)$ & 0.95 & 0.08 & 0.95 & 0.06 & 0.95 & 0.04 & 0.94 & 0.03 \\
& SE$(\hat\sigma_\alpha^2)$ & 0.85 & 0.35 & 0.87 & 0.37 & 0.91 & 0.38 & 0.91 & 0.38 \\
& SE$_{C}(\hat\sigma_\alpha^2)$ & 0.88 & 0.38 & 0.89 & 0.38 & 0.92 & 0.38 & 0.92 & 0.38 \\
& SE$_{N}(\hat\sigma_\alpha^2)$ & 0.90 & 0.38 & 0.90 & 0.38 & 0.93 & 0.39 & 0.93 & 0.39 \\
\midrule

\multirow{7}{*}{$100$}
& SE$(\hat\beta^{(b)}_0)$ & 0.93 & 0.80 & 0.94 & 0.82 & 0.94 & 0.82 & 0.94 & 0.83 \\
& SE$(\hat\beta^{(b)}_1)$ & 0.93 & 0.19 & 0.95 & 0.19 & 0.95 & 0.19 & 0.94 & 0.20 \\
& SE$(\hat\beta^{(w)}_0)$ & 0.96 & 0.03 & 0.94 & 0.02 & 0.95 & 0.01 & 0.96 & 0.01 \\
& SE$(\hat\sigma_e^2)$ & 0.96 & 0.06 & 0.96 & 0.04 & 0.95 & 0.03 & 0.95 & 0.02 \\
& SE$(\hat\sigma_\alpha^2)$ & 0.89 & 0.26 & 0.92 & 0.27 & 0.93 & 0.27 & 0.93 & 0.28 \\
& SE$_{C}(\hat\sigma_\alpha^2)$ & 0.91 & 0.28 & 0.93 & 0.28 & 0.93 & 0.28 & 0.93 & 0.28 \\
& SE$_{N}(\hat\sigma_\alpha^2)$ & 0.92 & 0.27 & 0.93 & 0.28 & 0.94 & 0.28 & 0.94 & 0.28 \\
\bottomrule
\end{tabular}
\end{table}

\begin{table}[!h]
\centering
\caption{Simulated coverage probabilities and average confidence interval lengths for the Gamma GLMM when $\alpha_i$ has a mixture distribution with $\siga^2 = 1$ and $\sige^2 = 0.5$.}
\label{tab:coverage_gamma}
\small
\setlength{\tabcolsep}{4pt}
\renewcommand{\arraystretch}{1.05}

\begin{tabular}{llcccccccc}
\toprule
\multirow{2}{*}{$g$}
& \multirow{2}{*}{Estimator}
& \multicolumn{2}{c}{$m_L=10$}
& \multicolumn{2}{c}{$m_L=20$}
& \multicolumn{2}{c}{$m_L=50$}
& \multicolumn{2}{c}{$m_L=100$} \\
\cmidrule(lr){3-4}
\cmidrule(lr){5-6}
\cmidrule(lr){7-8}
\cmidrule(lr){9-10}
&
& Cvge & Len
& Cvge & Len
& Cvge & Len
& Cvge & Len \\
\midrule

\multirow{7}{*}{$10$}
& SE$(\hat\beta^{(b)}_0)$ & 0.90 & 3.56 & 0.88 & 3.55 & 0.88 & 3.65 & 0.90 & 3.74 \\
& SE$(\hat\beta^{(b)}_1)$ & 0.88 & 0.85 & 0.87 & 0.84 & 0.88 & 0.87 & 0.88 & 0.90 \\
& SE$(\hat\beta^{(w)}_0)$ & 0.95 & 0.09 & 0.94 & 0.07 & 0.96 & 0.04 & 0.95 & 0.03 \\
& SE$(\hat\sigma_e^2)$ & 0.96 & 0.18 & 0.95 & 0.12 & 0.95 & 0.08 & 0.96 & 0.06 \\
& SE$(\hat\sigma_\alpha^2)$ & 0.70 & 1.23 & 0.69 & 1.24 & 0.74 & 1.26 & 0.74 & 1.26 \\
& SE$_{C}(\hat\sigma_\alpha^2)$ & 0.73 & 1.32 & 0.72 & 1.29 & 0.75 & 1.28 & 0.75 & 1.27 \\
& SE$_{N}(\hat\sigma_\alpha^2)$ & 0.83 & 1.59 & 0.82 & 1.57 & 0.86 & 1.63 & 0.87 & 1.66 \\
\midrule

\multirow{7}{*}{$20$}
& SE$(\hat\beta^{(b)}_0)$ & 0.90 & 2.50 & 0.92 & 2.56 & 0.92 & 2.58 & 0.92 & 2.64 \\
& SE$(\hat\beta^{(b)}_1)$ & 0.90 & 0.59 & 0.91 & 0.60 & 0.93 & 0.61 & 0.92 & 0.63 \\
& SE$(\hat\beta^{(w)}_0)$ & 0.95 & 0.07 & 0.95 & 0.05 & 0.95 & 0.03 & 0.94 & 0.02 \\
& SE$(\hat\sigma_e^2)$ & 0.96 & 0.12 & 0.97 & 0.09 & 0.95 & 0.06 & 0.95 & 0.04 \\
& SE$(\hat\sigma_\alpha^2)$ & 0.81 & 1.02 & 0.82 & 1.06 & 0.83 & 1.05 & 0.87 & 1.06 \\
& SE$_{C}(\hat\sigma_\alpha^2)$ & 0.83 & 1.08 & 0.83 & 1.08 & 0.84 & 1.06 & 0.87 & 1.07 \\
& SE$_{N}(\hat\sigma_\alpha^2)$ & 0.88 & 1.17 & 0.90 & 1.19 & 0.90 & 1.19 & 0.92 & 1.22 \\
\midrule

\multirow{7}{*}{$50$}
& SE$(\hat\beta^{(b)}_0)$ & 0.93 & 1.59 & 0.94 & 1.62 & 0.94 & 1.65 & 0.93 & 1.66 \\
& SE$(\hat\beta^{(b)}_1)$ & 0.93 & 0.37 & 0.94 & 0.38 & 0.94 & 0.39 & 0.93 & 0.39 \\
& SE$(\hat\beta^{(w)}_0)$ & 0.95 & 0.04 & 0.95 & 0.03 & 0.95 & 0.02 & 0.96 & 0.01 \\
& SE$(\hat\sigma_e^2)$ & 0.95 & 0.08 & 0.95 & 0.06 & 0.95 & 0.04 & 0.94 & 0.03 \\
& SE$(\hat\sigma_\alpha^2)$ & 0.88 & 0.71 & 0.89 & 0.73 & 0.91 & 0.75 & 0.92 & 0.75 \\
& SE$_{C}(\hat\sigma_\alpha^2)$ & 0.90 & 0.74 & 0.90 & 0.74 & 0.91 & 0.75 & 0.92 & 0.76 \\
& SE$_{N}(\hat\sigma_\alpha^2)$ & 0.92 & 0.76 & 0.92 & 0.77 & 0.94 & 0.78 & 0.95 & 0.78 \\
\midrule

\multirow{7}{*}{$100$}
& SE$(\hat\beta^{(b)}_0)$ & 0.94 & 1.13 & 0.94 & 1.16 & 0.94 & 1.16 & 0.92 & 1.19 \\
& SE$(\hat\beta^{(b)}_1)$ & 0.93 & 0.26 & 0.95 & 0.27 & 0.94 & 0.27 & 0.93 & 0.28 \\
& SE$(\hat\beta^{(w)}_0)$ & 0.96 & 0.03 & 0.94 & 0.02 & 0.95 & 0.01 & 0.96 & 0.01 \\
& SE$(\hat\sigma_e^2)$ & 0.97 & 0.06 & 0.96 & 0.04 & 0.95 & 0.03 & 0.95 & 0.02 \\
& SE$(\hat\sigma_\alpha^2)$ & 0.92 & 0.52 & 0.93 & 0.54 & 0.93 & 0.54 & 0.94 & 0.55 \\
& SE$_{C}(\hat\sigma_\alpha^2)$ & 0.94 & 0.54 & 0.93 & 0.55 & 0.93 & 0.54 & 0.94 & 0.56 \\
& SE$_{N}(\hat\sigma_\alpha^2)$ & 0.95 & 0.55 & 0.93 & 0.55 & 0.94 & 0.55 & 0.95 & 0.56 \\
\bottomrule
\end{tabular}
\end{table}

\begin{table}[!h]
\centering
\caption{Simulated coverage probabilities and average confidence interval lengths for the Gamma GLMM when $\alpha_i$ has a mixture distribution with $\siga^2 = 0.5$ and $\sige^2 = 1$.}
\label{tab:coverage_gamma}
\small
\setlength{\tabcolsep}{4pt}
\renewcommand{\arraystretch}{1.05}

\begin{tabular}{llcccccccc}
\toprule
\multirow{2}{*}{$g$}
& \multirow{2}{*}{Estimator}
& \multicolumn{2}{c}{$m_L=10$}
& \multicolumn{2}{c}{$m_L=20$}
& \multicolumn{2}{c}{$m_L=50$}
& \multicolumn{2}{c}{$m_L=100$} \\
\cmidrule(lr){3-4}
\cmidrule(lr){5-6}
\cmidrule(lr){7-8}
\cmidrule(lr){9-10}
&
& Cvge & Len
& Cvge & Len
& Cvge & Len
& Cvge & Len \\
\midrule

\multirow{7}{*}{$10$}
& SE$(\hat\beta^{(b)}_0)$ & 0.89 & 2.48 & 0.87 & 2.50 & 0.88 & 2.56 & 0.90 & 2.63 \\
& SE$(\hat\beta^{(b)}_1)$ & 0.87 & 0.59 & 0.87 & 0.59 & 0.87 & 0.61 & 0.89 & 0.63 \\
& SE$(\hat\beta^{(w)}_0)$ & 0.93 & 0.13 & 0.93 & 0.09 & 0.93 & 0.06 & 0.96 & 0.04 \\
& SE$(\hat\sigma_e^2)$ & 0.96 & 0.33 & 0.95 & 0.24 & 0.95 & 0.15 & 0.96 & 0.11 \\
& SE$(\hat\sigma_\alpha^2)$ & 0.65 & 0.55 & 0.67 & 0.57 & 0.72 & 0.62 & 0.73 & 0.62 \\
& SE$_{C}(\hat\sigma_\alpha^2)$ & 0.73 & 0.67 & 0.74 & 0.65 & 0.75 & 0.66 & 0.74 & 0.63 \\
& SE$_{N}(\hat\sigma_\alpha^2)$ & 0.80 & 0.78 & 0.82 & 0.78 & 0.84 & 0.81 & 0.86 & 0.82 \\
\midrule

\multirow{7}{*}{$20$}
& SE$(\hat\beta^{(b)}_0)$ & 0.89 & 1.76 & 0.92 & 1.79 & 0.92 & 1.83 & 0.91 & 1.84 \\
& SE$(\hat\beta^{(b)}_1)$ & 0.89 & 0.42 & 0.90 & 0.42 & 0.92 & 0.43 & 0.91 & 0.44 \\
& SE$(\hat\beta^{(w)}_0)$ & 0.94 & 0.09 & 0.95 & 0.07 & 0.95 & 0.04 & 0.96 & 0.03 \\
& SE$(\hat\sigma_e^2)$ & 0.95 & 0.23 & 0.96 & 0.17 & 0.95 & 0.11 & 0.94 & 0.08 \\
& SE$(\hat\sigma_\alpha^2)$ & 0.74 & 0.45 & 0.78 & 0.50 & 0.83 & 0.53 & 0.85 & 0.53 \\
& SE$_{C}(\hat\sigma_\alpha^2)$ & 0.83 & 0.55 & 0.83 & 0.55 & 0.85 & 0.55 & 0.86 & 0.54 \\
& SE$_{N}(\hat\sigma_\alpha^2)$ & 0.86 & 0.59 & 0.88 & 0.59 & 0.91 & 0.60 & 0.91 & 0.59 \\
\midrule

\multirow{7}{*}{$50$}
& SE$(\hat\beta^{(b)}_0)$ & 0.92 & 1.12 & 0.94 & 1.15 & 0.95 & 1.16 & 0.93 & 1.17 \\
& SE$(\hat\beta^{(b)}_1)$ & 0.92 & 0.26 & 0.93 & 0.27 & 0.95 & 0.27 & 0.93 & 0.28 \\
& SE$(\hat\beta^{(w)}_0)$ & 0.95 & 0.06 & 0.95 & 0.04 & 0.95 & 0.03 & 0.97 & 0.02 \\
& SE$(\hat\sigma_e^2)$ & 0.95 & 0.15 & 0.95 & 0.11 & 0.96 & 0.07 & 0.94 & 0.05 \\
& SE$(\hat\sigma_\alpha^2)$ & 0.81 & 0.32 & 0.85 & 0.35 & 0.90 & 0.37 & 0.91 & 0.38 \\
& SE$_{C}(\hat\sigma_\alpha^2)$ & 0.87 & 0.38 & 0.88 & 0.38 & 0.90 & 0.38 & 0.91 & 0.38 \\
& SE$_{N}(\hat\sigma_\alpha^2)$ & 0.89 & 0.38 & 0.90 & 0.38 & 0.93 & 0.39 & 0.93 & 0.39 \\
\midrule

\multirow{7}{*}{$100$}
& SE$(\hat\beta^{(b)}_0)$ & 0.92 & 0.80 & 0.94 & 0.82 & 0.94 & 0.82 & 0.94 & 0.83 \\
& SE$(\hat\beta^{(b)}_1)$ & 0.92 & 0.19 & 0.94 & 0.19 & 0.95 & 0.19 & 0.95 & 0.20 \\
& SE$(\hat\beta^{(w)}_0)$ & 0.95 & 0.04 & 0.95 & 0.03 & 0.94 & 0.02 & 0.94 & 0.01 \\
& SE$(\hat\sigma_e^2)$ & 0.95 & 0.10 & 0.94 & 0.07 & 0.96 & 0.05 & 0.95 & 0.03 \\
& SE$(\hat\sigma_\alpha^2)$ & 0.85 & 0.24 & 0.90 & 0.26 & 0.93 & 0.27 & 0.92 & 0.27 \\
& SE$_{C}(\hat\sigma_\alpha^2)$ & 0.90 & 0.28 & 0.93 & 0.28 & 0.93 & 0.28 & 0.92 & 0.28 \\
& SE$_{N}(\hat\sigma_\alpha^2)$ & 0.90 & 0.27 & 0.93 & 0.28 & 0.94 & 0.28 & 0.94 & 0.28 \\
\bottomrule
\end{tabular}
\end{table}

\begin{table}[!h]
\centering
\caption{Simulated coverage probabilities and average confidence interval lengths for the Gamma GLMM when $\alpha_i$ has a mixture distribution with $\siga^2 = 1$ and $\sige^2 = 1$.}
\label{tab:coverage_gamma}
\small
\setlength{\tabcolsep}{4pt}
\renewcommand{\arraystretch}{1.05}

\begin{tabular}{llcccccccc}
\toprule
\multirow{2}{*}{$g$}
& \multirow{2}{*}{Estimator}
& \multicolumn{2}{c}{$m_L=10$}
& \multicolumn{2}{c}{$m_L=20$}
& \multicolumn{2}{c}{$m_L=50$}
& \multicolumn{2}{c}{$m_L=100$} \\
\cmidrule(lr){3-4}
\cmidrule(lr){5-6}
\cmidrule(lr){7-8}
\cmidrule(lr){9-10}
&
& Cvge & Len
& Cvge & Len
& Cvge & Len
& Cvge & Len \\
\midrule

\multirow{7}{*}{$10$}
& SE$(\hat\beta^{(b)}_0)$ & 0.89 & 3.54 & 0.87 & 3.55 & 0.88 & 3.64 & 0.90 & 3.73 \\
& SE$(\hat\beta^{(b)}_1)$ & 0.88 & 0.84 & 0.87 & 0.85 & 0.88 & 0.87 & 0.88 & 0.89 \\
& SE$(\hat\beta^{(w)}_0)$ & 0.93 & 0.13 & 0.93 & 0.09 & 0.93 & 0.06 & 0.96 & 0.04 \\
& SE$(\hat\sigma_e^2)$ & 0.96 & 0.33 & 0.95 & 0.24 & 0.95 & 0.15 & 0.96 & 0.11 \\
& SE$(\hat\sigma_\alpha^2)$ & 0.67 & 1.17 & 0.69 & 1.20 & 0.73 & 1.24 & 0.74 & 1.24 \\
& SE$_{C}(\hat\sigma_\alpha^2)$ & 0.74 & 1.33 & 0.73 & 1.29 & 0.74 & 1.28 & 0.75 & 1.26 \\
& SE$_{N}(\hat\sigma_\alpha^2)$ & 0.83 & 1.58 & 0.82 & 1.57 & 0.85 & 1.62 & 0.87 & 1.64 \\
\midrule

\multirow{7}{*}{$20$}
& SE$(\hat\beta^{(b)}_0)$ & 0.89 & 2.50 & 0.91 & 2.55 & 0.92 & 2.59 & 0.92 & 2.62 \\
& SE$(\hat\beta^{(b)}_1)$ & 0.89 & 0.59 & 0.90 & 0.60 & 0.92 & 0.61 & 0.91 & 0.62 \\
& SE$(\hat\beta^{(w)}_0)$ & 0.95 & 0.09 & 0.95 & 0.07 & 0.95 & 0.04 & 0.96 & 0.03 \\
& SE$(\hat\sigma_e^2)$ & 0.95 & 0.23 & 0.96 & 0.17 & 0.95 & 0.11 & 0.94 & 0.08 \\
& SE$(\hat\sigma_\alpha^2)$ & 0.79 & 0.98 & 0.80 & 1.03 & 0.83 & 1.05 & 0.86 & 1.04 \\
& SE$_{C}(\hat\sigma_\alpha^2)$ & 0.83 & 1.08 & 0.83 & 1.09 & 0.84 & 1.07 & 0.86 & 1.05 \\
& SE$_{N}(\hat\sigma_\alpha^2)$ & 0.88 & 1.17 & 0.88 & 1.19 & 0.91 & 1.19 & 0.92 & 1.20 \\
\midrule

\multirow{7}{*}{$50$}
& SE$(\hat\beta^{(b)}_0)$ & 0.93 & 1.59 & 0.95 & 1.62 & 0.95 & 1.64 & 0.93 & 1.65 \\
& SE$(\hat\beta^{(b)}_1)$ & 0.93 & 0.37 & 0.94 & 0.38 & 0.95 & 0.39 & 0.94 & 0.39 \\
& SE$(\hat\beta^{(w)}_0)$ & 0.95 & 0.06 & 0.95 & 0.04 & 0.95 & 0.03 & 0.97 & 0.02 \\
& SE$(\hat\sigma_e^2)$ & 0.95 & 0.15 & 0.95 & 0.11 & 0.96 & 0.07 & 0.93 & 0.05 \\
& SE$(\hat\sigma_\alpha^2)$ & 0.85 & 0.68 & 0.88 & 0.71 & 0.91 & 0.74 & 0.92 & 0.74 \\
& SE$_{C}(\hat\sigma_\alpha^2)$ & 0.89 & 0.74 & 0.90 & 0.74 & 0.91 & 0.75 & 0.93 & 0.75 \\
& SE$_{N}(\hat\sigma_\alpha^2)$ & 0.90 & 0.76 & 0.91 & 0.77 & 0.93 & 0.78 & 0.95 & 0.77 \\
\midrule

\multirow{7}{*}{$100$}
& SE$(\hat\beta^{(b)}_0)$ & 0.93 & 1.13 & 0.94 & 1.16 & 0.95 & 1.16 & 0.94 & 1.17 \\
& SE$(\hat\beta^{(b)}_1)$ & 0.93 & 0.26 & 0.94 & 0.27 & 0.94 & 0.27 & 0.94 & 0.28 \\
& SE$(\hat\beta^{(w)}_0)$ & 0.94 & 0.04 & 0.94 & 0.03 & 0.95 & 0.02 & 0.94 & 0.01 \\
& SE$(\hat\sigma_e^2)$ & 0.95 & 0.11 & 0.94 & 0.07 & 0.96 & 0.05 & 0.95 & 0.03 \\
& SE$(\hat\sigma_\alpha^2)$ & 0.90 & 0.51 & 0.92 & 0.53 & 0.93 & 0.53 & 0.93 & 0.54 \\
& SE$_{C}(\hat\sigma_\alpha^2)$ & 0.93 & 0.54 & 0.93 & 0.55 & 0.94 & 0.54 & 0.94 & 0.54 \\
& SE$_{N}(\hat\sigma_\alpha^2)$ & 0.93 & 0.55 & 0.93 & 0.55 & 0.95 & 0.55 & 0.95 & 0.55 \\
\bottomrule
\end{tabular}
\end{table}

\FloatBarrier

\bibliographystyle{chicago}
\bibliography{reference}